\documentclass[preprint,11pt]{aastex}

\shorttitle{Envelopes Kinematics}
\shortauthors{Tobin et al.}

\newcommand{\mum}{\mbox{$\mu$m}}
\newcommand{\nthp}{\mbox{N$_2$H$^+$}}

\newcommand{\nht}{\mbox{NH$_3$}}
\newcommand{\hcop}{\mbox{HCO$^+$}}
\newcommand{\htcop}{\mbox{H$^{13}$CO$^+$}}
\newcommand{\kmspc}{\mbox{km s$^{-1}$ pc$^{-1}$ }}
\newcommand{\kms}{\mbox{km s$^{-1}$}}

\begin{document}

\title{Complex Structure in Class 0 Protostellar Envelopes II: Kinematic Structure from Single-Dish and Interferometric Molecular Line Mapping\footnotemark}
\author{John J. Tobin\altaffilmark{2}, Lee Hartmann\altaffilmark{2}, Hsin-Fang Chiang\altaffilmark{3}, Leslie W. Looney\altaffilmark{3},
 Edwin A. Bergin\altaffilmark{2}, Claire J. Chandler\altaffilmark{4}, Josep M. Masqu\'e\altaffilmark{5}, S\'ebastien Maret\altaffilmark{6},  
Fabian Heitsch\altaffilmark{7}}

\begin{abstract}
We present a study of dense molecular gas kinematics in seventeen nearby protostellar systems  
using single-dish and interferometric molecular line observations. The 
non-axisymmetric envelopes around a sample of Class 0/I protostars were mapped in the
\nthp\ ($J=1\rightarrow0$) tracer with the IRAM 30m, CARMA and PdBI as well as
\nht\ (1,1) with the VLA. The molecular line emission is used
to construct line-center velocity and linewidth maps for all sources to 
examine the kinematic structure in the envelopes on spatial scales from 0.1 pc to $\sim$1000 AU. 
The direction of the large-scale velocity gradients from single-dish mapping
is within 45$^{\circ}$ of normal to the outflow axis in more than half the sample. Furthermore, the
velocity gradients are often quite substantial, the average being $\sim$2.3 \kmspc. 
The interferometric data often reveal small-scale velocity structure, departing
from the more gradual large-scale velocity gradients. In some cases, this likely indicates 
accelerating infall and/or rotational spin-up in the inner envelope; the median velocity gradient from the
interferometric data is $\sim$10.7 \kmspc. 
 In two systems, we detect high-velocity HCO$^+$ ($J=1\rightarrow0$) emission inside the highest-velocity
\nthp\ emission. This enables us to study the infall and
rotation close to the disk and estimate the central object masses. 
The velocity fields observed on large and small-scales
are more complex than would be expected from rotation alone, suggesting that complex envelope
structure enables other dynamical processes (i.e. infall) to affect the velocity field.

\end{abstract}
\footnotetext[1]{Based on observations carried out with the IRAM 30m Telescope and IRAM Plateau de Bure Interferometer.
 IRAM is supported by INSU/CNRS (France), MPG (Germany) and IGN (Spain).}
\altaffiltext{2}{Department of Astronomy, University of Michigan, Ann
Arbor, MI 48109; jjtobin@umich.edu}
\altaffiltext{3}{Department of Astronomy, University of Illinois, Urbana, IL 61801 }
\altaffiltext{4}{National Radio Astronomy Observatory, P.O. Box O, Socorro, NM 87801}
\altaffiltext{5}{Departament d'Astronomia i Meteorologia, Universitat de
Barcelona, Mart\'i i Franqu\`es 1, 08028 Barcelona, Catalunya, Spain}
\altaffiltext{6}{UJF-Grenoble 1 / CNRS-INSU, Institut de Planétologie et d'Astrophysique de Grenoble (IPAG) UMR 5274, Grenoble, F-38041, France}
\altaffiltext{7}{Department of Physics and Astronomy, University of North Carolina, Chapel Hill, NC 27599}

\section{Introduction}

Infall and rotation in dense cores and protostellar envelopes both play important roles in the formation of protostars
and their surrounding disks. Infall must be taking place in the
envelopes because we observe newborn protostars embedded within their natal 
clouds. The two classic analytic theories of collapse
describe infall as either being outside-in \citep{larson1969} or inside-out \citep{shu1977}.
The initial angular momentum of the protostellar cloud governs the formation and
sizes of the proto-planetary disks \citep[e.g.][]{cassen1981,tsc1984} 
and will affect the ability of the cloud core to fragment into multiple stellar 
systems \citep[e.g.][]{burkert1993,bonnell1994a}. The ubiquitous formation of proto-planetary
disks \citep{haisch2001,hernandez2007} and the prevalence of binary systems \citep{raghavan2010}
lends strong indirect evidence for the presence of rotation.  Therefore, observing these two
processes in protostellar envelopes has enormous potential for constraining star formation theory through
comparisons to analytic models and numerical simulations.
Dense protostellar cores are well known as sites of isolated, low-mass star
formation \citep[e.g.][]{shu1987,bm1989,mckeeostriker2007} and are the ideal place to study
the kinematic structure from large, core/envelope scales ($\sim$0.1 pc) down to scales near 
the disk radius. 

Several previous studies have attempted to characterize the rotation in dark clouds and dense cores.
\citet{arquilla1986} examined the kinematic structure of dark clouds using CO on $\sim$10\arcmin\ scales,
 finding that some exhibit possible rotation signatures derived from velocity gradients.
Later, \citet{goodman1993} and \citet{caselli2002} used the dense molecular tracers \nht\ and \nthp\ to
examine rotation in the dense cores within dark clouds ($\sim$2-3\arcmin\ scales).
\citet{goodman1993} and \citet{caselli2002} found typical velocity gradients of
$\sim$1 to 2 \kmspc, which were interpreted to be slow, solid-body core rotation.
However, the observations had limited resolution (60\arcsec-90\arcsec), nearly the size of the 
envelopes in some cases. Rotation is not necessarily solid-body; however, this assumption
simplifies analysis in terms of equipartition and the early data did not warrant more sophisticated
models.

\citet{caselli2002} noted that their finer resolution as compared to \citet{bm1989} revealed clear
deviations from linearity in the velocity gradients. This indicates that the velocity structure
of cores may be more complex than simply axisymmetric, solid-body rotation. \citet{chen2007}
carried out a higher resolution study using interferometric observations of \nthp\ from OVRO
on a sample of protostars in the Class 0 \citep{andre1993} and Class I phases \citep{lada1987}.
They found substantially larger velocity gradients in the inner envelopes,
with complex velocity fields, only showing probable rotation signatures in a few cases.

While the possible rotation has been probed using optically thin tracers, the study of infall
in protostellar envelopes has been limited to observations of optically thick tracers toward the envelope
center. \citet{zhou1993} and \citet{myers1995} used observations of CS and H$_2$CO to infer the detection of infall in the envelope
of the Class 0 protostar B335. The optically thick molecular lines are self
absorbed at the line-center and the key signature of infall is the blue-shifted side of the line being
brighter than the red-shifted side. Later, \citet{difrancesco2001} found inverse-P Cygni line profiles in
H$_2$CO which are interpreted as infall. Furthermore, infall at large-scales 
may have been detected in the pre-stellar core L1544 \citep{tafalla1998}.

These previous studies have interpreted the kinematic structure in terms of axisymmetric
envelopes. However, our recent studies \cite[i.e.][]{tobin2010a,looney2007} have revealed that
the dense envelopes surrounding the youngest, generally Class 0,
protostars often have complex, non-axisymmetric morphological structure 
\citep[also see][]{bm1989,myers1991,stutz2009}. Out of the 22 protostellar systems exhibiting
extinction at 8\mum, only three appeared to be roughly axisymmetric. We suggested that 
the asymmetric envelope structure may play a role in the formation of binary systems
by making fragmentation easier and that the initial disk structure may be perturbed from
uneven mass loading. However, in order to understand the effects that complex envelope 
structure has on disk formation and fragmentation, the kinematics of the dense gas must be
characterized.

Given the apparent prevalence of complex morphological structure in protostellar envelopes, 
the kinematics of the envelopes must also be studied on scales such that the morphological
structure is spatially resolved. To examine the kinematic structure of morphologically complex envelopes, we have undertaken
a molecular line survey focusing on nearby (d $<$ 500 pc), embedded protostars drawn from 
our sample of envelopes in \citet{tobin2010a}. We have approached this study
from two directions. First, we obtained new single-dish \nthp\ mapping of sixteen systems with
the IRAM 30m, a factor of two improvement in resolution over \citet{caselli2002}. Secondly,
we obtained new interferometric \nthp\ and \nht\ observations of fourteen systems with resolutions
between 3.5\arcsec\ and 6\arcsec. The analysis of both the single-dish and interferometric observations
offers a comprehensive view of the kinematic and morphological structure from 0.1 pc scales down to 
$\sim$1000 AU. Complementary \textit{Spitzer} imaging gives us a clear view of the outflow angular extent
and direction in all of these objects.

In this paper, we are primarily presenting the dataset as a whole and give a basic analysis of each object;
an upcoming paper will give a more detailed interpretation of the velocity structures.
The data for each object are quite rich and we find many envelopes with velocity gradients normal to the outflow
on large-scales in the single-dish maps, persisting down to small-scales in the interferometer maps.
We have organized the paper as follows: Section 2 discusses the
sample, observations, data reduction, and analysis, Section 3
presents our general results and discusses each object in detail, and Section 4 discusses the basic
overall properties of the sample as a whole.

\section{Observations and Data Reduction}

We mapped the protostellar envelopes in the dense gas tracers \nht\ and \nthp, which
are known to be present over a wide range of spatial scales and preferentially trace high
density regions (critical densities of $\sim$2$\times$10$^3$ cm$^{-3}$
\citep{danby1988,schoier2005} and $\sim$1.4$\times$10$^5$ cm$^{-3}$ \citep{schoier2005} respectively),
 where CO has depleted \citep[e.g.][]{bergin2002, tafalla2004}.
The single-dish data were all obtained using the IRAM 30m telescope, mapping \nthp\ ($J=1\rightarrow 0$).
The single-dish observations were followed by interferometric observations of \nthp\ ($J=1\rightarrow 0$) 
using CARMA and the Plateau de Bure Interferometer, in addition to \nht\ (1,1) 
observations using the VLA\footnote[1]{The National Radio
Astronomy Observatory is a facility of the National Science Foundation operated under
cooperative agreement by Associated Universities, Inc.}, including VLA archival data for some objects.
We will briefly describe the observations and the data analysis procedure for each data set.

\subsection{The Sample}

Our sample of protostellar envelopes selected for kinematic study is directly drawn
from the objects presented in \citet[][hereafter Paper I]{tobin2010a}. 
All protostellar systems observed have a surrounding envelope visible in 8\mum\ extinction.
Much of the envelope sample was found within
archival \textit{Spitzer} data from the cores2disks (c2d) legacy program \citep{evans2009}. Given that c2d
and most other archival observations were short integrations, we were only able to detect very dense
structures which happened to have a bright backgrounds at 8\mum. Thus, there may be a bias towards denser envelopes, but
the range of bolometric luminosity is fairly broad with objects $<$1 $L_{\sun}$ and as much as $\sim$14 $L_{\sun}$.
These protostars are listed in Table 1 and are mostly Class 0 systems with a few Class Is.
The requirement of having an envelope
visible in extinction enables us to consider the envelope morphology in our interpretation of
the kinematic data. We regard the 8\mum\ extinction to more robustly reflect the structure
of high density material around the protostars on scales $>$ 1000 AU, as molecular tracers may have abundance variations
which affect their spatial distribution. The requirement of 8\mum\ extinction also biases the sample to more isolated systems, only
Serpens MMS3 and L673 have multiple neighbors $<$ 0.1 pc away.

\subsection{IRAM 30m Observations}
We observed our sample of protostellar envelopes in two observing runs at the IRAM 30m radio telescope on
Pico Veleta in the Spanish Sierra Nevada. The first observing run took place between 2008 December 26 - 30;
during this run we only observed four protostars due to poor weather.
We used the AB receiver system and observed the protostars in the \nthp\ ($J=1\rightarrow0$) transition
($\nu_{\rm rest}$= 93.1737637 GHz; $JF_1F=123\rightarrow012$ component \citep{keto2010});
the half power beam width (HPBW) of the 30m at this
frequency is $\sim$27\arcsec. The second observing run took place between 2009 October 22-26 using the new
Eight MIxer Receiver (EMIR). We observed thirteen protostars and revisited L1527 from the first run, using 
the VESPA auto-correlation spectrometer as the backend for all
observations. Single-sideband system temperatures of 130 K at $\nu$= 93 GHz were typical for both observing runs.
 During the 2008 run we used the 20 MHz bandwidth mode with 20kHz channels and in 2009
we used 40 MHz with 20kHz channels; see Table 2 the list of sources observed and more detail.

We conducted our observations using frequency-switched on-the-fly (OTF) mapping mode. The maps varied
in size depending on the extent of the source being observed, most being 3$^{\prime}$ $\times$ 3$^{\prime}$.
Most maps were integrated down to at least $\sigma_T$ $\sim$ 150mK for the \nthp\ ($J=1\rightarrow0$)
transition, noise levels for each map are listed in Table 2. We mapped the sources by
scanning in the north-south direction and again in the east-west direction to minimize 
striping in the final map. The scan legs were stepped by 5\arcsec\ and we repeated the
maps to gain a higher signal-to-noise ratio. Calibration scans were taken about every 10
minutes between scan legs and the final maps took approximately 2 hours to complete. Pointing was checked 
about every two hours, azimuth and elevation offsets were typically $\pm$5\arcsec; the pointing offset
remained stable, typically within $\sim$2\arcsec\ during an observation. These
values agree well with the rms pointing accuracy of $\sim$2\arcsec.

The initial calibration of the OTF data to the antenna temperature scale and CLASS
data format was performed automatically on-site by the Multichannel Imaging and calibration software
for Receiver Arrays (MIRA)\footnote[2]{http://www.iram.fr/IRAMFR/GILDAS} package. Further data reduction was done using CLASS
(part of GILDAS\footnotemark[2]). For all molecular lines observed, the frequency switched spectra were folded
and baseline subtracted using a second order polynomial.
We then reconstructed the spectral map on a grid such that the FWHM of the beam
was spanned by 3 pixels and each pixel is the average of all measurements within the FWHM of the beam.

\subsection{CARMA Observations}

The CARMA observations were taken in four observing semesters; most sources were observed solely
in the D-array configuration (except L1527 and L1157 see below) which yields $\sim$6\arcsec\ resolution.
Three objects were observed in 2009 July and August. The
correlator at this time only had three bands, operating dual side-band mode, giving
6 spectral windows and was configured for an IF frequency of 91.181 GHz. 
\nthp\ ($J=1\rightarrow0$) and \hcop\ ($J=1\rightarrow0$) ($\nu_{\rm rest}$=89.188518 GHz;
\citep{lovas}) were observed in opposite side-bands of
one spectral window with 2MHz bandwidth and 63 channels
giving $\sim$0.1 km/s velocity resolution. The \nthp\ ($J=1\rightarrow0$) emission spectrum
comprises 7 hyperfine lines over $\sim$ 17 \kms, consisting of two groups of three lines and
a single isolated line. The 2MHz window is too narrow
to observe all the lines; therefore, we observed the strongest 
set of three lines, rather than observing the isolated line to maximize our signal-to-noise. The multiple hyperfine
transitions enables the optical depth and excitation temperature to be determined.
The second band was also configured for 2 MHz bandwidth and was centered on the main
hyperfine component of the HCN ($J=1\rightarrow0$) transition ($\nu_{\rm rest}$=88.6318473 GHz; \citep{lovas}),
and the third band was configured for continuum observations with 500 MHz bandwidth (1 GHz dual-side band).

Another set of sources was observed in 2010 April and May. During this time,
a new correlator was being completed with higher velocity
resolution over a wider bandwidth and more spectral windows; most data were taken with six bands (twelve spectral windows).
For these observations, the IF frequency was set to 90.9027 GHz such that
\nthp\ ($J=1\rightarrow0$) and HCN ($J=1\rightarrow0$)
could be observed in opposite side-bands of the same correlator band. We
also observed \hcop\ ($J=1\rightarrow0$), 
H$^{13}$CO$^+$ ($J=1\rightarrow0$) ($\nu_{\rm rest}$= 86.754294 GHz), ortho-NH$_2$D
($1_{1,1}\rightarrow1_{0,1}$), ($\nu_{\rm rest}$= 85.926263 GHz), and continuum; 
rest frequencies for these transitions are taken from \citet{lovas}. All spectral line observations 
used 8 MHz bandwidth with 385 channels yielding $\sim$ 0.06 \kms\ velocity resolution,
during reduction this was rebinned to 0.1 \kms\ resolution to reduce noise. The continuum
observations again had 500 MHz bandwidth (1 GHz dual-side band). See Table 3 for exact 
dates of observation for particular sources.

In addition, L1527 and L1157 were observed in E-array configuration in 2008 October and L1157 was again
observed in D-array in 2009 March. These observations were taken
in a three-point mosaic pattern to better recover the large-scale emission from the envelopes. The correlator
was configured with one band for continuum, the other two bands were set to observe \nthp\ ($J=1\rightarrow0$) 
and \hcop\ ($J=1\rightarrow0$). One band had 2MHz bandwidth and was centered on the isolated \nthp\ component.
The other band was configured with 8 MHz bandwidth and 63 channels with 0.4 \kms\ resolution to
cover all 7 \nthp\ ($J=1\rightarrow0$) hyperfine lines. However, we only use the 0.1 \kms\ velocity resolution data for
kinematics. The observations of L1157 are further detailed in \citet{chiang2010}.

All datasets were observed in a standard loop (calibrator--source--calibrator), a bright quasar within 15$^{\circ}$
was used for phase and amplitude calibration.
The calibrator was integrated for 3 minutes while the source was integrated for 15 minutes in each cycle.
Absolute flux calibration was obtained by observing standard flux calibration sources, see Table 3.
Bandpass calibration for the continuum bands was accomplished by observing a 
bright quasar, generally 3C454.3; the spectral line bands were bandpass corrected using the noise source.

Each dataset was processed using the MIRIAD software package \citep{sault1995}. The raw visibilities
were corrected for refined antenna baseline solutions and transmission line-length
variation. The data were then edited to remove uncalibratable data (i.e. poor phase coherence, phase jumps, 
anomalous system temperatures/amplitudes). The bandpass corrections
were computed using the \textit{mfcal} routine. The absolute flux calibration was derived
using the \textit{bootflux} routine which determines the flux density of the gain calibrator
relative to the flux calibration source (absolute calibration uncertainty is typically $\sim$10\%).
The phases and amplitudes were
calibrated using the \textit{mselfcal} routine. The phase and amplitude solution
calculated for the continuum bands was then transferred to the spectral line bands.
Continuum images and spectral line cubes were generated by 
inverting the corrected visibilities with natural weighting,
creating the dirty map. Then the dirty map is CLEANed using the \textit{mossdi} routine using
a clean box of 60$^{\prime\prime}$ x 60$^{\prime\prime}$ which fits within
the primary beam of the 10.4m dishes ($\sim$72\arcsec\ at $\lambda$=3.2mm). In this paper,
we will mainly interpret the \nthp\ data but will comment on the other molecules when relevant and
the 3mm continuum data are presented in the Appendix.

\subsection{VLA Observations}

The \nht\ (1,1) observations were taken with EVLA transition system 
during the final semester of VLA correlator operation in D configuration. Eight-hour tracks were taken
for 5 sources during 2009 October, November, and 2010 January (see Table 4). We used the 4 IF mode providing two
tunings and dual-polarization to observe the \nht\ (1,1) and (2,2) inversion
transitions ($\nu$=23.6944955, 23.7226336 GHz respectively \citep{ho1983}) 
with 1.5 MHz bandwidth and 127 channels yielding
$\sim$0.15 km/s velocity resolution. This configuration was able to observe
the main component and two sets of satellite lines for 
the \nht\ (1,1) transition. We alternately observed the source and a calibrator
within 15$^{\circ}$. Two minutes were spent integrating on the
calibrator while ten minutes were spent on the target during each cycle. Pointing
was updated every hour, 3C84 was observed as the bandpass calibrator and
3C48 or 3C286 was used for absolute flux calibration. 

The raw visibility data from the VLA were reduced and calibrated using the CASA
(Common Astronomy Software Applications)\footnote[3]{http://casa.nrao.edu} package. The task
\textit{importvla} was used to convert the VLA data to a CASA measurement set.
The visibility data were then inspected and edited, specifically flagging shadowed data
and any VLA antennas. Only EVLA antennas were used
in our final dataset because we used doppler tracking during the observations, which
caused phase jumps between VLA and EVLA antennas.
 Antenna positions were corrected using the \textit{gencal} task when necessary
and the absolute flux scale was set using the \textit{setjy} task 
(absolute calibration uncertainty is typically 10\%). The phases and
amplitudes were calibrated using the \textit{gaincal} task; the bandpass correction was
determined using the \textit{bandpass} task. All the calibrations were then 
applied using the \textit{applycal} task. We then inspected the corrected data to ensure
 proper phase correction; any severely outlying amplitude points in the
source data were also flagged.

The final spectral datacubes were generated using the \textit{clean} task. The clean 
task encompasses several individual processes including inverting the
visibilities, CLEANing the image, and restoring the image. Since the VLA provides 
a 2$^{\prime}$ diameter primary beam (field of view), we took several additional
steps to increase our final image fidelity. We first performed a first-pass 
CLEANing of a spectral data cube over the 2$^{\prime}$
primary beam. Then we calculated the integrated intensity of the main component 
of the \nht\ (1,1) or (2,2) transition by summing the spectral line channels.
Next, we created a mask image in which pixels below a 2$\sigma$ intensity 
threshold were rejected and those above were kept. This isolated the area of
\nht\ emission around a particular protostar enabling CLEANing down to near 
a 1$\sigma$ threshold. Then for the \nht\ (1,1) images, we employed the multi-scale
CLEAN algorithm \citep{rich2008}; this algorithm models they sky as the sum
of Gaussian components of various widths, which for extended sources
works better than modeling the sky as the sum of many point source
CLEAN components. In this paper, we will only present the \nht\ (1,1) data.

In addition to our own observations, we include VLA archival data\footnote[4]{http://archive.nrao.edu}
for the sources L1527, L1521F, and L483. The data for L1521F and L483 were taken in different correlator
configurations than our observations, see Table 4. The same reduction procedure was applied
for these archival data as for our observations.

\subsection{Plateau de Bure Interferometer Observations}

L1157 was observed with the Plateau de Bure Interferometer on 2009 June 17
and 2009 July 8 in the D-array configuration with 5 antennas
operating. It was observed again on 2009 November 13 in C-array
configuration with 6 antennas. During the first track the weather
conditions were average, with 5 to 10~mm of precipitable water vapor
(pwv), respectively and a RMS phase noise lower than 64\degr. During
the last two tracks the weather conditions were better, with 3-6~mm
and 5-7~mm of pwv and a RMS noise phase lower than 41\degr\ and 24\degr,
respectively

The 3mm receivers were tuned to the \nthp\ ($J=1\rightarrow0$)
transition in the lower-sideband. The correlator was configured with
two windows for continuum observation, each having 320 MHz of
bandwidth and one window with 20 MHz bandwidth for the
 \nthp\ ($J=1\rightarrow0$), yielding a velocity
resolution of 0.125 km/s. Both the spectral line and continuum
observations were taken in dual-polarization mode.

The raw data were calibrated using the CLIC program from the GILDAS
software package. A standard calibration procedure was
used. Visibilities amplitudes and phases for each baseline were
inspected and bad data (e.g. affected by phase jumps or antenna
shadowing) were flagged. Phases were then calibrated using
observations of the 1927+739 calibrator. Absolute flux was derived from observations of
MWC349, assuming a flux of 1.15~Jy at 3~mm for that source; uncertainty in the absolute
flux is $\sim$10\%.

The calibrated data were then reduced using the MAPPING program from
GILDAS. A UV table was created for the 3mm continuum as well as for
the \nthp\ ($J=1\rightarrow0$) line. The 3mm continuum visibilities
were then subtracted from the line visibilities. Finally, a deconvolved
map was produced using the the CLEAN algorithm. The synthesized beam
in the final map is roughly circular with a FWHM of $\sim 3.4 \times
3.3\arcsec$ at PA 114$^{\circ}$; further details of the observations are listed
in Table 5. 

\subsection{Data Analysis}
\subsubsection{Hyperfine Fitting}
The \nthp\ and \nht\ molecular lines both have a hyperfine emission line
spectrum. This allows us to robustly determine the line-center velocity
and full-width half-maximum (FWHM) linewidths in each pixel by fitting
all the hyperfine components simultaneously. We can fit the line-center
velocities substantially better than native resolution of the observations.
\citet{goodman1993} approximates the observed velocity accuracy
as
\begin{equation}
\sigma_v= 1.15\left(\frac{\sigma_T}{T_{peak}}\right)(1.21\delta_v\Delta v)^{1/2}.
\end{equation}
This assumes that the lines have a Gaussian shape, $\sigma_T$ is the rms noise, $T_{peak}$ is the
peak line intensity, $\Delta v$ is the FWHM linewidth, and $\delta_v$ is the velocity width of
the channels. The additional factor 1.21 is the intrinsic error in the autocorrelation spectrometer
with unity weighting \citep{thompsonmoranswenson}. Using this relationship for velocity accuracy, 
if we had a signal-to-noise of 5, assuming $\Delta v$ = 0.3 \kms and $\delta_v$=0.1 \kms, our
velocity accuracy would be 0.043 \kms. Note that this is
the velocity accuracy for one line, using the multiple hyperfine components of \nthp\ and \nht\ we can
obtain even greater accuracy.

To fit the lines, we applied the
CLASS hyperfine fitting routines to each line.
The hyperfine frequencies and line ratios for \nthp\ ($J=1\rightarrow0$) were taken from \citet{keto2010} and we use 
the \nht\ frequencies and line ratios that are built into CLASS \citep{rydbeck1977,ho1983}.
A semi-automated routine was used for fitting the hyperfine structure
across the entire spectral map. We first generated an integrated intensity map of the central three hyperfine lines,
from this map we selected each pixel with 3$\sigma$ detection of \nthp\ or \nht\ and then
generated a CLASS script to fit the hyperfine structure at each point and write out a table containing these data. 
The fitting does not require all hyperfine components to have 3$\sigma$ detections, only that there are clear
detections of the strongest hyperfine lines. Points where the fitting failed due to inadequate signal-to-noise
were removed from the final table.
The table is then used for plotting maps of the velocity field, linewidth, optical depth, and excitation temperature, all of 
which are all determined from fitting the hyperfine structure. The zeroth-moment maps (integrated intensity)
are computed simplistically by measuring the intensity of the hyperfine lines within a given 
velocity range, summing the emission, and multiplying by the channel velocity width. The basic emission properties
of \nthp and \nht\ are separated into three tables: Table \ref{sdnthp} for the single-dish \nthp, Table \ref{intnthp} for
the interferometric \nthp, and Table \ref{intnht} for the VLA \nht\ data. These tables list the line center velocities,
linewidths, maximum integrated intensities, total optical depth of the transition (sum of optical depths for each component), excitation temperature, column density
 (following \citet{goldsmith1999} for \nthp\ and \citep{bourke1995b} for \nht), envelope mass as a function of assumed 
abundance, and approximate envelope radii.

\subsubsection{Velocity Gradient Fitting}

Using the line center velocities computed from the hyperfine fitting, the velocity gradients are computed for each object in three ways.
First we simply computed the velocity difference between two points offset from the protostar
by 10000 AU, normal to the outflow direction, and divided by the distance. Second, we computed a linear fits to cuts 
through the velocity data, taken normal to the outflow using points
within $\pm$30\arcsec\ of the protostar and/or within the region
of the cloud directly associated with the protostar.
Lastly, we fit a plane to the entire velocity field using the method described in \citet{goodman1993},
but using our own IDL implementation coupled with the MPFIT routines \citep{markwardt2009}. We fit the 
line-center velocity field with the function
\begin{equation}
v_{LSR} = v_{0} + a\Delta\alpha + b\Delta\delta
\end{equation}
where $v_{0}$ is the systemic velocity, $\Delta\alpha$ and $\Delta\delta$ are offsets in right ascension
and declination (in arcseconds) and $a$ and $b$ are the velocity gradients per arcseconds in the $\alpha$
and $\delta$ directions respectively. The total velocity gradient is then given by
\begin{equation}
g = (a^2 + b^2)^{1/2} \times \frac{206264.98}{D}
\end{equation}
with a position angle (PA) east of north (toward increasing velocity) given by
\begin{equation}
\theta_g = tan^{-1}\frac{b}{a}
\end{equation}
where D is the distance in parsecs and the constant is the number of arcseconds per radian. We use all three 
methods on the single-dish data and only the latter two methods on the interferometer data since the extent of
\nthp\ and \nht\ emission varies widely from object to object.

\subsection{Extended Structure Sensitivity}

As a general rule, interferometers filter-out large-scale emission; however, 
if there is a velocity gradient and the lines are well-resolved,
the largest scale of emission in a given channel may only be a fraction of the full structure.
Thus, different portions of the envelopes become visible in different velocity channels and the 
integrated intensity maps (zeroth-moment) build a picture of the envelope emission from the multiple
velocity components. If all of this emission had resided in one velocity channel, substantially more emission would
have been filtered-out. Furthermore, the filamentary nature of many envelopes enabled
them to be viewed over a larger area than an envelope that fills the primary beam more evenly. 

We do not combine the single-dish and interferometer data because
each probe different size scales and it is often advantageous to have the large-scale emission
resolved-out of the interferometer maps. Resolving-out the large-scale emission
isolates the regions of compact emission in the inner envelope where there may
be significant detail in the velocity field. Furthermore, our analysis of only the kinematic structure does
not necessitate recovery of all flux; a detailed summary of the steps needed to combine single-dish
data with CARMA observations is given in \citet{koda2011}.

\section{Results}

We mapped the regions around each protostar where we detected the presence of 8\mum\ 
extinction in Paper I. Each envelope was observed to be a bright source of \nthp\ ($J=1\rightarrow0$)
(hereafter \nthp) emission in the single-dish data; \nthp\ emission was present over much 
of the area where we detect extinction at 8$\mu$m in each map, shown in Figures 1 through 23. 
\nthp\ and \nht\ (1,1) (hereafter \nht) is also detected toward all sources in the interferometer
data. The interferometer observations select out the regions of brightest, compact emission which are
usually associated with the densest regions of the protostellar envelope. 
Furthermore, there are many cases where the \nthp\ or \nht\
emission peak is not centered on the protostar in the single-dish and/or interferometer
observations (see Section 4.5). The protostar positions in Figures 1 through 23 are derived from
their 3mm continuum source, which is often coincident with the 8\mum\ point source (see Appendix), or the 24\mum\
source where continuum data were not available.

We know the outflow direction and the angular width of the cavity for all objects in the sample from
the \textit{Spitzer} IRAC data. This gives an observational
constraint on the region in which the outflow may impact the envelope. These data enable the
characterization of kinematic properties of the envelopes and
determination of the origin of the kinematic structure. We can determine whether 
the kinematics reflect the intrinsic velocity structure of the core/envelope
or if the outflow is likely affecting the observed kinematics. Such distinction is critical
to ensure that we are not misled in further interpretation. While kinematic information 
is missing from the \textit{Spitzer} images (i.e. blue and red-shifted sides traced by CO emission), this information
is readily available in the literature for most objects (Table 1). Note that
we can also often infer the blue and red-shifted outflow directions from the scattered
light morphology and intensity \citep{whitney2003b}.

\subsection{Similarity of \nthp\ and \nht\ Emission Properties}
Since our interferometric observations mapped either \nht\ or \nthp\ for most
protostars and our single-dish data solely mapped \nthp, it is important to demonstrate that
 the kinematic structure observed in the two tracers is consistent.
We have observed the envelope around L1157 in \nthp\ with the IRAM 30m, CARMA
\citep{chiang2010}, and the PdBI, while observing it in \nht\ with the VLA.
The \nht\ and \nthp\ emission both closely follow the regions of 8\mum\ extinction
as shown in Figures \ref{L1157} and \ref{L1157-2} and the velocity maps
have a very similar structure. The similar spatial emission and kinematic properties
indicate that emission from these molecular species arises
from approximately the same region of the envelope (from 0.1 pc to 1000 AU). Further comparisons
of our sample can be made to data in the literature: \citet{chen2007} for CB230 and IRAS 03282+3035,
L1527 with \citet{goodman1993}, HH211 with \citet{tanner2011}, and L483 with \citet{fuller2000}.
In all these cases, the \nthp\ and \nht\ emission is detected in the same regions of the envelope
with similar kinematic properties. This confirms that
\nthp\ and \nht\ trace similar physical conditions at the level of precision we are probing,
in agreement with the results from \citet{johnstone2010}. Section 4.5 further discusses
the impact of chemistry on these species.

\subsection{Velocity Gradients}
\label{results:vgrad}

We have computed velocity gradients for all objects using 
both the single-dish and interferometer data with the methods
described in Section 2.6.2. The velocity gradients calculated from the single-dish data are given in Table 9
and the interferometric (\nthp\ and \nht) velocity gradients are listed in Table 10. 
The gradient directions from the 2D fitting are plotted in Figures 1-23 and also listed in Tables 9 and 10.
The one-dimensional (1D) cuts through the velocity fields, normal to the outflow and
across equatorial plane of each envelope, are shown in Figure \ref{profiles} presenting an alternative view of
the envelope velocity structure. Linear fits to the single-dish and interferometer data are overlaid on the plots.
Notice that in some cases the velocity of the interferometer
data diverges from the single-dish data. This results from the interferometer
filtering-out larger-scale emission that dominated the single-dish data and the increased 
resolution picking out smaller-scale velocity structure. 

The gradients calculated for the single-dish data with each method are comparable. The median single-dish velocity
gradients from the different fitting methods are: 2.1 \kmspc\ (1D fitting), 1.7 \kmspc\ (1D two points), and 2.2 \kmspc\ (2D fitting); 
the mean gradients are 2.3, 2.2, and 2.04 \kmspc\ respectively. The distribution of single-dish velocity gradients
from the three methods is shown in Figure \ref{sdgradienthistos}.
 The lower values of the two point method reflect that the region
inside the 10000 AU radius of some sources has a higher velocity gradient; there are velocity decreases toward the 
edges of some envelopes that are reflected in the plots in Figure \ref{profiles}, yielding a preference toward lower
gradients in Figure \ref{sdgradienthistos}.

The mean velocity gradient of the single-dish sample ($\sim$2.2 \kmspc) is about twice the
average gradient in \citet{goodman1993} and slightly higher than \citet{caselli2002}, but our sample of 16
objects is smaller than their larger samples. We can expect to observe larger velocity gradients
with our higher resolution data because the lower resolution data in \citet{goodman1993} and \citet{caselli2002}
tend to smear velocity components together.
L483 was common between our work and the two previous studies,
with very similar gradient magnitudes and PAs. L1527 and L1152 were also
common between our work and \citet{goodman1993}. The gradient directions fit for these sources were similar, but
the magnitude of the gradients are different. We regard our values as being more reliable because our
maps are comprised of substantially more independent points.

The interferometric sample has a median velocity gradient of 8.1 \kmspc\ from 2D fitting and 10.7 \kmspc\ from 1D fitting,
both having a mean gradient of 8.6 \kmspc. The distribution of interferometric velocity gradients
from the two methods is shown in Figure \ref{intgradienthistos}. The interferometric gradients are often larger than the single-dish
gradients by factors of several. The 2D fitting method was less reliable for the interferometric data given
the often complex velocity fields present within the data. Reliable 2D fits could not be obtained 
for L1157 (PdBI data) or Serpens MMS3 due to lack of convergence on their complex velocity fields.
Our range of observed velocity gradients is comparable to what \citet{chen2007} found; however,
they found three protostellar systems with gradients $>$20 \kmspc, we do not find such large gradients in our data.

Using the outflow PAs derived from the \textit{Spitzer} imaging and CO data from the literature,
we have compared the gradient PAs (the angle toward increasing velocities)
with the outflow PAs in Figure \ref{paoffset}. The gradient position angles are also marked
in Figures 1-23 with solid arrow; the majority of the velocity
gradients are within 45\degr\ of normal to the outflow. 
The distribution of interferometric velocity directions strongly shows a trend
for being oriented normal to the outflow axis.
The gradient directions also generally reflect what is seen 
at large-scales in the single-dish data. Table 10 and Figure \ref{sdintoffset} show that only three systems 
have velocity gradient directions which differ by more than 45\degr\ between the single-dish and interferometric
measurements.

The majority of envelopes in the interferometric sample have an ordered velocity structure, despite their often 
complex morphological structure. In contrast to \citet{volgenau2006} and \citet{chen2007}, 
many systems have velocity gradients roughly normal to the 
outflow direction, this likely results from our larger sample of observations as compared to
the \citet{chen2007} and especially \citet{volgenau2006}. In addition, visual inspection
of the data shows that the large and small-scale
velocity gradient directions are generally consistent with one another, a feature which 
\citet{volgenau2006} also sees in their sample.
The interferometer observations often reveal small-scale kinematic detail
near the protostar that is smeared-out in the lower-resolution single-dish data.

\subsection{Description of Individual Sources}

We will describe the dataset for each source individually in the following subsections
and the discuss of our results as a whole is in Section 4.

\subsubsection{L1157}
The flattened, filamentary envelope of L1157 has been extensively studied in recent years 
\citep[e.g.][Paper I]{looney2007,chiang2010}. Its velocity field
was first studied in \citet{chiang2010}, which showed a weak velocity gradient along the filament,
normal to the outflow. We subsequently observed L1157 with the IRAM 30m, the PdBI,
and the VLA. Figure \ref{L1157} shows the data from the IRAM 30m and PdBI and Figure \ref{L1157-2}
shows the data from the VLA and CARMA. The 30m data detects a
large-scale velocity gradient normal to the outflow and this gradient follows
the long axis of the envelope. On the east side of the envelope,
where the emission curves downward, the \nthp\ continues to trace dense material as it becomes
more blue-shifted. 

The PdBI data are shown in the bottom panels of Figure \ref{L1157}. The \nthp\
emission appears double-peaked on small-scales. The 3mm
 dust continuum shows that the protostar resides
between the peaks (see Appendix); the peaks are at radii of $\sim$1000 AU ($\sim$3.5\arcsec).
The velocity field on scales $>$15\arcsec\
reflects what has been seen with the CARMA, 30m, and VLA data. On the other hand,
the velocity of the gas becomes highly red-shifted ($\sim$1 \kms) at 
the \nthp\ peaks as compared to the surrounding gas at $\sim$1000 AU from the protostar. The
high-velocity gas was observed in \citet{chiang2010}; however, its spatial location
was not well-resolved in the CARMA data due to having a factor of $\sim$2 
lower resolution than the PdBI data. Note that the 2D velocity gradient fits for the VLA and CARMA
data have position angles that differ by 80\degr. This results from the north-south gradient
being more prominent in the CARMA \nthp\ data while the east-west gradient appears more prominent
in the \nht\ data. A 2D gradient could not be fit to the PdBI data due to the highly complex velocity field
present on small-scales.

The single-dish velocity and linewidth maps show a gradual large-scale velocity gradient in L1157
with broad linewidth in the inner envelope, consistent with the PdBI linewidth map in Figure \ref{L1157}.
The broad linewidths were also shown in lower resolution data from \citet{chiang2010}. Both the
east and west peaks in the PdBI maps have velocity wings toward the red and are
not significantly extended toward blue-shifted velocities.
\citet{chiang2010} attributed the broad inner envelope line wings to infall. However,
close examination of our higher-resolution PdBI data indicate that the blue and red-shifted line 
wings may result from outflow interaction effects. Figure \ref{L1157-outflow} shows that the most 
red-shifted emission is slightly shifted to the southeast, along the outflow and 
traces one edge of the northern outflow cavity. Furthermore,
the blue-shifted emission also seems to outline the southern outflow cavity quite well.
 Thus, it appears that the outflow is may be entraining inner envelope material,
while at larger scales the velocity structure appears unaffected by the outflow, tracing
the intrinsic kinematic structure of the envelope. The spatial overlap of red and blue-shifted
\nthp\ southeast of the protostar on the blue-shifted side of the outflow can understood
if the outflow is entraining material within a symmetric cavity in the inner envelope, producing both a blue
and red-shifted component. There also appear to be outflow effects in the \nthp\ emission
on the northwest side of the protostar but to a lesser extent. This is the first example of possible outflow
entrainment in such a dense gas tracer. However, we note that rather than entrainment, the broad \nthp\ emission
could also result from an outflow shock at that location.

\subsubsection{L1165}

The protostar L1165IRS is located within a $\sim$1.5 pc (17\arcmin) long filamentary dark cloud (Paper I).
The narrow filament, from which the protostar has formed, is normal to the protostellar outflow. We mapped a 4$^{\prime}$ section of the 
dark cloud surrounding the protostar, as shown in the left panels of Figure \ref{L1165}. The \nthp\ emission is highly peaked
very near the protostar, it is slightly offset 4.87\arcsec ($\sim$1462 AU) to the southeast, and there is low-level extended emission associated with
regions of 8$\mu$m extinction. The velocity field from the single-dish \nthp\ data shows a fairly linear
gradient nearly normal to the outflow; in areas away from the protostar the filament generally seems to have a fairly
constant velocity with little variation and small linewidth.

The \nthp\ data from CARMA reveal a small-scale structure that is extended in the direction of the filament axis and
the emission is strongly correlated with the small-scale 8$\mu$m extinction shown in Figure \ref{L1165}.
At the edges of detected emission, the velocity field of the interferometer map is consistent with the single-dish data. However,
near the protostar there is a 0.35 km/s velocity shift between the blue and red-shifted velocity peaks that are on opposite sides of the 
protostar. We note that this velocity gradient is not perfectly normal to the outflow, but rather offset by about $\sim$30$^{\circ}$.
However, the most red-shifted emission is slightly extended in the direction of the outflow, but the linewidth
peak is extended normal to the outflow and the red-shifted emission at this location does not appear to be outflow affected.
Thus, the gradient from the envelope itself appears to be normal to the outflow and tracing gravitationally dominated motion.

We show the \hcop\ emission in Figure \ref{L1165-pv-hco}, in which the red and blue-shifted emission
is confined to two clumps, that are oriented normal to the outflow and offset from the protostar by 3$^{\prime\prime}$.
The position-velocity plot of these data show the blue and red-shifted emission extending $\sim$2
and $\sim$1.5 \kms\ away from the systemic velocity respectively.
Given the orientation of the blue and red-shifted emission normal to the outflow and
a lack of emission along the outflow, the \hcop\ emission appears to be originating from the inner
envelope. Assuming that the \hcop\ emission indicates rotationally
supported motion, we can calculate the enclosed mass using $M = R \Delta v^2/G$ which gives M$_{enc}$ $\sim$ 2.0$M_{\sun}$.
This is not unreasonable for this source which has $L_{bol}$ $\sim$14$L_{\sun}$ assuming
a distance of 300 pc. However, if the velocities result from equal contributions of rotation and infall,
then the enclosed mass would be a more modest 0.5 $M_{\sun}$. We have overlaid
lines representing Keplerian rotation (or infall) on Figure \ref{L1165-pv-hco}, the 0.5$M_{\sun}$ curve
matches the data much better than the 2.0$M_{\sun}$ curve.

\subsubsection{CB230}
CB230 is an isolated protostar that formed at one end of its natal globule (Paper I).
This protostar was discovered to be a wide binary system by \citet{yun1996}
with a separation of $\sim$10$^{\prime\prime}$, the companion is evident in the 8$\mu$m images shown
in Figure \ref{CB230}. The envelope around CB230 was classified as a ``one-sided'' envelope in Paper I
due to the 60$^{\prime\prime}$ (30000 AU) extension of the extinction envelope to the west, while the 8$\mu$m
extinction terminates just $\sim$25$^{\prime\prime}$ (6000 AU) on the eastern side. The single-dish
\nthp\ emission is consistent with the 8\mum\ extinction observations, the emission
falling off steeply to the east and more extended to the west. The large-scale extension of material beyond the
region of detected \nthp\ emission suggests that it has lower density and has not formed \nthp\ at detectable
levels, see Section 4.5 for further discussion.

The \nht\ emission from the VLA is strongest in a ``bar,'' about 10$^{\circ}$ from normal to the
outflow. Notably, at the location of the protostar there is a "hole" in the ammonia emission; a similar
depression of emission is seen in \nthp\ ($J=1\rightarrow0$) by \citet{chen2007} and \citet{launhardt2001}.
The region of decreased emission is $\sim$2200AU in diameter, similar in size to the double-peaked \nthp\ 
emission in L1157. The lower level emission on the eastern side
of the protostar extends northward along the outflow cavity. Incidentally, the northward extension
is also where the scattered light emission in the near-IR and \textit{Spitzer} 3.6$\mu$m is brightest
\citep[Paper I;][]{launhardt2010}.

The velocity field of the single-dish \nthp\ shows a fairly linear gradient,
normal to the outflow, with a ``plateau'' of the most highly red/blue-shifted emission $\pm$30$^{\prime\prime}$
from the protostar. South of the protostar, the \nthp\ linewidth peaks, similar to L1157.
The \nht\ velocity gradient from the VLA data is in the same direction as the single-dish gradient and similar to the
interferometric \nthp\ map from \citet{chen2007}. While the single-dish gradient is fairly gradual, the gradient
from the VLA observations has an abrupt shift from blue to red-shifted emission coincident with the 
protostar. The VLA \nht\ data also show a velocity ``plateau'' in the blue and red-shifted emission,
with the highest relative-velocity emission being $\pm$15$^{\prime\prime}$ from the protostar.
The linewidth remains fairly constant throughout the regions near the protostar,
peaking at 0.5 km/s. The region of highest linewidth also corresponds to the region of strongest ammonia emission. The 
large linewidth seen near the protostar in the single-dish data is not reflected in these \nht\ data nor the \nthp\ 
data of \citet{chen2007}. The line-center velocity changes quite rapidly at the location of the large single-dish
linewidth; therefore, the linewidth peak is likely due to the unresolved velocity gradient. This means 
that the outflow is not likely to be affecting the kinematics of the inner envelope in this protostar.

\subsubsection{HH108IRS}

The protostar HH108IRS, the driving source of HH108, is located within a large-scale filament, $\sim$0.5 pc in length,
1.75$^{\circ}$ south of the Serpens star forming region \citep{harvey2006}. 
There are at least two protostars forming in the filament: the higher luminosity object HH108IRS and the
deeply embedded source HH108MMS \citep{chini2001}. The single-dish \nthp\ map in Figure \ref{HH108IRS} 
shows an emission peak coincident with HH108IRS,but slightly offset from the protostar $\sim$5\arcsec\ (1500 AU).
The \nthp\ map from CARMA reveals that the \nthp\ peak emission is truly offset from the protostar and emission is
extended normal to the outflow direction, forming a flattened structure across the protostar. 

The single-dish \nthp\ velocity map indicates that there may be a slight gradient normal to the
outflow. The CARMA \nthp\ velocity map reveals that there is indeed
a velocity gradient normal to the outflow, though its structure is complex. Southeast
of the protostar the velocities are red-shifted and moving toward the protostar the velocities
are becoming more blue-shifted. This trend continues after crossing the protostar and moving
northwest, but then the trend reverses itself rapidly and becomes more red-shifted. We also note that the linewidth
is $\sim$0.6 km/s within $\pm$10$^{\prime\prime}$ (3000 AU) around the protostar,
indicative of a dynamic environment near the protostar. The single-dish
linewidth is consistent with this value as well.

\subsubsection{HH108MMS}

HH108MMS is the nearby neighbor to HH108IRS, separated by $\sim$60\arcsec\ ($\sim$0.09 pc). 
This protostar is deeply embedded and invisible at 24\mum, 
only becoming visible at 70\mum\ (Paper I). There are no IRAC data available for this object, therefore
we are showing an ISPI Ks-band (2.15$\mu$m) image from Paper I. Despite the lack of 8\mum\ extinction
observations, we can clearly see the dense material of the envelope blocking out the rich background 
star field. The single-dish \nthp\ in Figure \ref{HH108MMS} shows a slight extension
toward the location of the protostar (derived from 70\mum\ and 3mm continuum), while the CARMA \nthp\ observations
clearly show the \nthp\ emission centrally peaked on the protostar.

The single-dish velocity field does not show much structure, the region within $\pm$30\arcsec\ (0.045 pc) of
the protostar has a roughly constant velocity. The filament that HH108MMS is forming within
has a velocity gradient of 1.6 \kmspc\ running from southwest to northeast. The CARMA \nthp\ velocity map on the other
hand shows significant structure on $\sim$10\arcsec\ scales within $\pm$15\arcsec\ ($\sim$0.02 pc) of 
the protostar. Southeast of the protostar the emission is red-shifted and northwest
there is blue-shifted emission, along the presumptive outflow axis and HCN emission mapped with CARMA
(Tobin et al. 2011 in preparation). The outflow axis 
is determined from faint diffuse emission seen in Ks-band. The CARMA map also shows increased linewidth
along the outflow; there is no indication of such an increase in the single-dish map, likely due to beam
dilution. HH108MMS appears to be a prime example of the outflow impacting
the envelope of a deeply embedded protostar.

\subsubsection{Serpens MMS3}

Serpens MMS3 is located within a complex network of filamentary structure in the Serpens B
cluster \citep{dab2006,harvey2006}, shown in Figure \ref{serpmms3}. One of the most
prominent filaments runs $\sim$0.1 pc in length into Serpens MMS3, see left panels of
Figure \ref{serpmms3}. Directly west of the protostar
the filament turns southward toward a small clustering of bright young stars. We also noticed that
Serpens MMS3 has a faint companion separated by $\sim$7$^{\prime\prime}$.

The single-dish \nthp\ map shows that the emission 
is highly pervasive throughout the region. The emission is peaked near the clustering
of young stars in the southwest corner of the image in Figure \ref{serpmms3}. However, the emission
is extended toward Serpens MMS3 and there is enhancement emission coincident with the large scale filament seen
in 8$\mu$m extinction. The VLA \nht\ map reveals the structure of the region in substantially
more detail. The interferometer resolved out whatever diffuse \nht\ emission was in the region and
the remaining emission directly correlates to the highest extinction regions seen in the
8$\mu$m image, with a peak coincident with the Serpens MMS3 protostar. The \nht\ emission
is still extended toward the clustering of young stars in the southwest, but it is at the edge
of the primary beam.

The overall velocity structure of this region is confusing in the single-dish \nthp\ velocity
map because of the multitude of high-density structures in the region; however, there
is a large scale gradient along the filament that Serpens MMS3 resides in and there
is also a pocket of red-shifted emission next to the protostar.
The VLA \nht\ (1,1) map also shows this large scale gradient along the filament of
 Serpens MMS3 and red-shifted emission next to the protostar. The red-shifted emission
appears over a region 15\arcsec\ from the protostar and is extended in the direction of the
outflow. We also  detected an increased linewidth ($\sim$1 \kms) next to the protostar.
Directly southwest of the protostar the velocity gradient appears
to resume the large-scale velocity trend exhibited northeast of the protostar. It is presently
unclear if the kinematic structure near the protostar is related to the outflow or infall, but
its proximity and extension along the outflow makes us suspicious. However, the broad linewidth
is quite localized to the east of the protostar.

\subsubsection{HH211}

HH211MMS is a deeply embedded protostar on the outskirts of the IC348 cluster in the Perseus molecular
cloud; emission from the central protostar itself only becomes evident at 70$\mu$m \citep{rebull2007}
and has been found to be a proto-binary in the submillimeter \citep{lee2009}.
We see a large absorbing structure in the 8$\mu$m extinction map shown in 
Figure \ref{HH211}, as well as its powerful outflow \citep{mccaughrean1994,gueth1999}.
The single-dish \nthp\ emission
associated with HH211MMS is very strong, peaked to the southwest of the protostar itself.
The \nthp\ emission also appears extended in the direction of the higher extinction areas.
The \nthp\ emission mapped with CARMA detects emission on small scales around the protostar, with the emission peak
offset $\sim$2$^{\prime\prime}$ southwest of the protostar (not coincident with
the single-dish \nthp\ peaks). The emission is more extended along the
northwestern side of the outflow, consistent with the extinction seen in the 8$\mu$m image.

The single-dish \nthp\ velocity map shows a linear velocity gradient
normal to the outflow and south of the protostar there is another velocity component in the
dense gas. The transition between these two velocity components
appears as an area of artificially large linewidth (an artifact from fitting); however, there are two
sets of narrow emission lines present, not broad lines. The CARMA \nthp\ velocity map also finds a
linear gradient normal to the outflow as well as the second velocity component to the south.
We also note that near the protostar the gradient is not perfectly linear at all scales. The deviance
from a linear gradient is slight; however, it is present where we also have excellent signal-to-noise and
this agrees with the velocity map by \citet{tanner2011}.

The linewidths in the single-dish data were quite low across the source, only 0.3 - 0.4 \kms\
with similar levels seen in the CARMA \nthp\ map. We note that there is an area of increased linewidth
just southeast of the protostar, apparently at the base of the outflow. We suggest that the increased linewidth
in this region is due to outflow interaction, in agreement with \citet{tanner2011}. In addition,
the filament northeast of the protostar has a very narrow linewidth, $<$0.2 \kms, appearing both the 
single-dish and interferometer maps.

\subsubsection{IRAS 16253-2429}
IRAS 16253-2429 is a low-luminosity ($L_{Bol}=0.25L_{\sun}$)
Class 0 protostar in the $\rho$ Ophiuchus star forming region; it is
also identified as Oph MMS 126 \citep{stanke2006}. We noted in Paper I that
this was one of the more ``symmetric'' envelopes seen in our 8$\mu$m extinction study. Its 
symmetric bipolar outflow has been traced in CO by \citet{stanke2006} as well as in shocked 
H$_2$ emission from \textit{Spitzer} IRS spectral mapping \citep{barsony2010}.

The single-dish \nthp\ shown in Figure \ref{IRAS16253}
correlates quite well with the 8$\mu$m extinction. The emission peak
is slightly offset from the location of the protostar and the \nthp\ emission 
appears to be depressed at the location of the outflow cavities. 
The CARMA \nthp\ emission shows similar features in that it strongly
correlates with the regions of 8$\mu$m extinction and there is less emission in regions occupied by the
outflow cavities. The lack of emission within the outflow cavities is likely due to evacuation of envelope
material and/or destruction of \nthp\ by CO in the outflow (section 4.2); there may also be some interferometric
filtering-out of emission in this region. Furthermore, there appears to be a deficit of \nthp\
emission near the protostar.

The velocity field of the single-dish \nthp\ map indicates that there is a very small velocity gradient 
across the envelope, approximately normal to the outflow; note the small velocity range occupied by the
envelope and velocity gradient. The velocity map from the CARMA \nthp\ data is more
complex with several gradient reversals throughout the emitting region. However, the global gradient still
seems to be present in interferometer data. Furthermore, a VLA \nht\ map
shows a velocity structure very similar to our CARMA \nthp\ map (J. Wiseman, private communication).

The single-dish \nthp\ linewidth is quite small and constant across the envelope, whereas many other objects
in our sample have linewidths which peak near the protostar. We also note that the linewidth is increasing in this source
toward the edge of \nthp\ emission along the outflow; this is likely an outflow interaction
effect. The CARMA \nthp\ map shows a similar small linewidth
across the most of the envelope; however, there is a region of increased linewidth east and south of the protostar
associated with an area of strong \nthp\ emission; at this location there is a slight enhancement of linewidth 
in the single-dish map. 

\subsubsection{L1152}

The L1152 dark cloud is located in Cepheus, about 1.7 pc (20$^{\prime}$) away from L1157 on the sky.
L1152 hosts three young stars; however, only one (IRAS 20353+6742) is classified as 
a Class 0 object and it is the only one embedded in the main core of L1152 \citep{chapman2009}. 
Paper I found that the main core of L1152 appears to have a ``dumbbell'' morphology in which
the northeastern core (see Figure \ref{L1152}) appears to be starless and the southeastern core
harbors IRAS 20353+6742 (hereafter L1152). These two concentrations are connected by what appears to be
a thinner filament of high density material.

The single-dish \nthp\ ($J=1\rightarrow0$) emission shown in Figure \ref{L1152} exactly matches the
morphology of the extinction in the 8$\mu$m images. However, the peak \nthp\ emission in the southwestern
core is offset from the protostar by $\sim$20$^{\prime\prime}$ (6000 AU). The \nthp\
map from CARMA shown in the right panel of Figure \ref{L1152} only observed the southeastern core. The
map confirms that the \nthp\ is substantially offset from the protostar and there is no sub-peak
at its location. However, the \nthp\ emission appears to extend toward the protostar.

The single-dish \nthp\ velocity field exhibits a velocity gradient normal to the outflow
of L1152, noting that the protostar appears at the edge of the region exhibiting the gradient.
The rest of the cloud, including the star-less core, has a fairly constant velocity; only varying by $\sim$ 
0.1 km/s. However, we do notice increased linewidths northeast and southwest of the protostar.
Southwest of the protostar we can clearly see the jet from the protostar, possibly
interacting with envelope material, then in the northeast
there is nothing obvious happening at this linewidth peak in the 8$\mu$m image. However, the northeast linewidth
peak is near the outflow axis and this could be the cause of the
increased linewidth at this location.

The velocity map from the CARMA \nthp\ data tells a remarkably similar story to the single-dish data; the
velocity gradient is only slightly better resolved. However, the most remarkable feature is in the linewidth
map, where we clearly see an increase in linewidth on the axis of the jet that is visible in the 8$\mu$m maps.
This appears to be a another very clear example of the outflow interacting with the envelope material,
though the velocity field does not seem to show outflow effects. 

\subsubsection{L1527}

L1527 (IRAS 04368+2557) is an extensively studied protostar in Taurus. \citet{bm1989}
observed its compact \nht\ core, from which \citet{goodman1993} derived its velocity gradient. Subsequent observations
indicated the possibility of infall in the envelope from H$_2$CO observations by \cite{myers1995}. Furthermore,
detailed modeling of its scattered light cavities observed in \textit{Spitzer} IRAC imaging have
been done by \citet{tobin2008} and \citet{gramajo2010}. High-resolution mid-infrared imaging by \citet{tobin2010b}
found the signature of a large (R$\sim$200 AU) disk in scattered light. 

IRAC 8$\mu$m imaging of this source revealed an 
asymmetric distribution of extinction, the northern side of the envelope is substantially more extended than
the southern side. This asymmetry is also exhibited in our single-dish \nthp\ shown in 
Figure \ref{L1527}; in addition, the peak emission is also offset to the north of the protostar
by $\sim$25$^{\prime\prime}$ (3500 AU). The VLA \nht\ and CARMA \nthp\ maps
both show emission associated with the protostar, but the maps 
are somewhat difficult to interpret due to the likelihood of spatial filtering. In addition,
the CARMA observation was done as a mosaic in order to cover the entire region of emission as the primary beam
is only 70$^{\prime\prime}$; both maps seem to detect
emission in the same general areas.

The velocity field from the single-dish \nthp\ map has a complicated morphology. There appear to be
two velocity gradients in the map, one along the outflow (pointed out by \citet{myers1995}) and another
normal to the outflow isolated by \citet{goodman1993}. However, the gradient along the outflow
is not linear, the velocities go from red to blue and back to red. The linewidth remains fairly constant throughout
the map, with a minimum at the northeast and southwest edges of the map.

The velocity fields from both the VLA and CARMA reveal further kinematic complexity in this system. 
We can see the consistency with the single-dish velocity map on large-scales; however, the \nthp\ 
and \nht\ maps show that there is a small-scale velocity gradient 
near the protostar. Notice that this small-scale
velocity gradient is in the opposite direction as compared to the large-scale gradient. The linewidths
of the \nthp\ and \nht\ exhibit a corresponding increase in the inner envelope, near these small-scale
velocity gradients. This is the only protostellar envelope that where a velocity gradient reversal is seen going
from large to small-scales.

\subsubsection{RNO43}

RNO43 is protostar forming within the $\lambda$ Ori ring. On large scales the envelope is quite asymmetric, with
several filamentary structures appearing to converge at the location of the protostar as shown
in the left panels of Figure \ref{RNO43}. RNO43 also drives a powerful,
parsec-scale outflow; CO emission has been mapped on small scales by \citet{arce2005} tracing 
an outflow cavity and on large-scales, tracing a $\sim$5 pc long outflow \citep{bence1996}.
The \nthp\ emission is mostly unresolved in the single-dish map as shown in Figure \ref{RNO43}.
The peak emission is located near the location of the protostar and there are slight extensions 
in the direction of the outflow. 

The CARMA \nthp\ ($J=1\rightarrow0$) map traces the small-scale structure seen in 8$\mu$m extinction very well. 
We also note a depression of \nthp\ emission 
at the location of the protostar, consistent with observations of other protostars in our sample; see
Section 4.5 for further discussion of this feature.
East of the protostar there is a ridge of \nthp\ emission which is composed of the three bright
knots almost running north-south in the image extending $\sim$35$^{\prime\prime}$. The southern-most knot
is associated with the highest column density region east of the protostar and the two northern knots
correlate well with an extinction filament running from the north into the envelope of RNO43. In addition,
this filament of 8$\mu$m extinction and \nthp\ emission are coincident with the brightest part of
the outflow cavity in the \textit{Spitzer} 3.6$\mu$m image (Paper I).  Directly
west of the protostar, there is another peak of \nthp\ emission and weaker
\nthp\ emission extended further west, in agreement with the 8$\mu$m extinction. 
\citet{chen2007} mapped this region in \nthp\ using OVRO, the data agree quite
well with our observations. However, our map appears to have recovered more large-scale emission, 
likely due to better uv-coverage at short spacings.

The velocity field from the single-dish \nthp\ map shows a large scale velocity gradient that is nearly
normal to the outflow axis and there is an area of enhanced linewidth southeast of the protostar. 
The CARMA data reveal significant kinematic detail in the velocity field of the \nthp\ gas. 
The CARMA velocity maps in Figure \ref{RNO43} clearly show red and blue-shifted sides of the envelope;
however, separating those sides
of the envelope is a sharp velocity jump from blue to red by $\sim$0.7 \kms. Due to the overlapping lines
at the location of the protostar, the \nthp\ linewidth forms a line marking the jump in velocity. There also
appears to be a north-south gradient in the interferometer data as well (the single-dish map hints at this).
\citet{chen2007} ignored the western, red-shifted portion of the envelope thinking that it
was a line-of-sight alignment with another clump; however, the envelope
has density increasing in 8\mum\ extinction toward the protostar on both sides (shown in Paper I),
suggesting that the western side is indeed part of the same structure. 

We note that the most highly blue-shifted gas is not located directly adjacent to the 
protostar; this is likely due to the absence of \nthp\ near the protostar, as
mentioned earlier. Furthermore, small-scale emission of HCO$^+$ was also detected with
similar morphology to L1165 (Figure \ref{L1165-pv-hco}). Figure \ref{RNO43-pv-hco} shows that the
centroid of blue and red-shifted emission are located normal to the outflow 
and are offset from each other by $\sim$3$^{\prime\prime}$ (1400 AU). In RNO43, the \hcop\
line wings extend $\pm\sim$2 km/s from the systemic velocity. If we assume that the HCO$^+$ emission 
reflects only rotation, its velocities would imply an enclosed mass of 2.7$M_{\sun}$.
If only half of this velocity is due to rotation then the enclosed mass would be 0.67$M_{\sun}$. The bolometric
luminosity of RNO43 is $\sim$8.0$L_{\sun}$; comparable to L1165 in both luminosity and mass. We have overlaid
lines representing Keplerian rotation (or infall) on Figure \ref{RNO43-pv-hco}, the 0.67$M_{\sun}$ curve
matches the data much better that the 2.67$M_{\sun}$ curve.

Note that we have redefined the outflow position axis to be 20$^{\circ}$ east of north
in contrast to the 54$^{\circ}$ found by \citet{arce2005}; our value is more accurate taking into account
the outflow cavity observed by \textit{Spitzer} (Figure \ref{RNO43}) and CO maps from both \citet{arce2005} and 
\citet{bence1996}.
Furthermore, \citet{chen2007} assumed the 54$^{\circ}$ outflow position axis, leading them to
interpret the velocity gradient along the eastern ridge as symmetric rotation. The \nthp\ gradient across
the protostar has a very similar direction to the \htcop\ and C$^{18}$O velocity gradients found by
\citet{arce2005}. However, our revised outflow axis and the observed \nthp\ velocity structure,
in conjunction with the \htcop\ and C$^{18}$O  data, lead us to suggest that we are likely not seeing
 envelope material being ``pushed out'' in this system, as suggested by \citet{arce2005}. Thus, the \nthp\
velocity structure appears to reflect kinematic structure intrinsic to the envelope.

\subsubsection{IRAS 04325+2402}

IRAS 04325+2402, sometimes referred to as L1535, harbors a multiple Class I
protostellar system in the Taurus star forming region. The primary is possibly a sub-arcsecond
binary with a wider companion separated by 8.2$^{\prime\prime}$ \citep{hartmann1999}.
The 8$\mu$m extinction around IRAS 04325 was found in our envelope study but not published
in Paper I due to its low signal-to-noise; however, \citet{scholz2010} noticed
the 8$\mu$m extinction in their study of the system. These authors pointed out that
there is a bright diffuse region of emission at 3.6 and 4.5$\mu$m, at the peak of 
the 8$\mu$m extinction. Furthermore, they noticed a dark band between the protostar and the 
 4.5$\mu$m diffuse emission, suggesting a dense cloud; however, there is a lack of
8$\mu$m extinction at the location of the dark band.

Our single-dish \nthp\ map finds emission throughout the core surrounding the protostar
with the peak emission coincident with the 8$\mu$m extinction peak and the 
4.5$\mu$m diffuse scattered light peak. In fact, the
\nthp\ is peaked $\sim$60\arcsec\ northeast of the protostar, but the map does show a
slight enhancement of \nthp\ emission west of the protostar. Given this emission morphology,
we suggest that the dark band see at 4.5$\mu$m is really just a lack of material and that 
the diffuse emission is light from the protostar shining onto the neighboring
star-less core. We have no interferometry data for this object, however we would not expect
to observe substantial \nthp\ emission peaked around
the protostar, based on the single-dish map.

The velocity structure of the \nthp\ shows that there is a relatively smooth velocity
gradient across the entire object with an increased gradient just southeast
of the protostar. The linewidth of \nthp\ however shows large increase along
the outflow axis of the protostar. Thus, the outflow from the protostar may be interacting with 
the dense material in the surrounding core producing the increased linewidth. On the other
hand, the velocity field does not seem to show effects from the outflow, similar to L1152.

\subsubsection{L483}

L483 is an isolated globule harboring a Class 0 protostar \citep{tafalla2000}. The envelope surrounding the
protostar is quite large, $\sim$0.15 pc in diameter with at least 10-20$M_{\sun}$ of material
measured from 8$\mu$m extinction in Paper I. The densest 
regions seen in 8$\mu$m extinction form a ``tri-lobed'' pattern that
is also traced by 850$\mu$m emission \citep{jorgensen2004}.
The single-dish \nthp\ also follows this same pattern with the peak emission 
coincident with the protostar, see Figure \ref{L483}. The VLA \nht\ emission is not peaked on the protostar,
but also follows the ''tri-lobed'' morphology. The \nthp\ emission mapped with OVRO by
\citet{jorgensen2004} is extended along the outflow; however, this observation
appears to have resolved out a significant amount of extended emission.

The velocity gradient from the single-dish \nthp\ map is not normal to the outflow but
is at an angle of $\sim$45$^{\circ}$. The VLA \nht\ map shows a velocity gradient in
the same direction as the single-dish data; however, the protostar is located in a pocket of blue-shifted 
emission. This is consistent with what \citet{jorgensen2004} observed. Furthermore,
directly north of the protostar, there is an area of highly red-shifted emission seen
in the \nht\ map. The single-dish \nthp\ also shows red-shifted emission in this region, but
it is not as prominent due to the larger beamsize. This emission appears to come from another
distinct velocity component in the cloud, as evidenced by the large linewidths in the \nht\
map at the transition to the red-shifted emission. 

We also noticed that the \nthp\ linewidth map shows a region of enhanced linewidth running 
across the envelope, nearly normal to the outflow. This region connects to where there is the second
velocity component in the VLA \nht\ map and this is also where the velocity field is most
rapidly changing in the single-dish \nthp\ map. Since the increased linewidth appears to be a global feature
we do not attribute it to outflow effects and could be related to the initial formation of the dense core.

\subsubsection{L673}
The L673 dark cloud in the constellation Aquila has been the subject of a
SCUBA survey by \citet{visser2002} and two \textit{Spitzer} studies by \citet{tsitali2010} and \citet{dunham2010}.
In Paper I, we highlighted a small region of the cloud exhibiting highly filamentary 8$\mu$m
extinction associated with L673-SMM2 as identified in \citet{visser2002}. There are
more regions with 8$\mu$m extinction within the cloud that we did not focus on in Paper I,
but are apparent in the images shown by \citet{tsitali2010}.
The \textit{Spitzer} IRAC data around L673-SMM2, show four point sources closely associated with the sub-millimeter
emission peak and another 70$\mu$m source which may be a Class 0 protostar \citep{tsitali2010}.

The filamentary region around L673-SMM2 is shown in Figure \ref{L673}. The \nthp\ emission
maps closely to the 8$\mu$m extinction and the \nthp\ peak is centered on the small clustering of protostars.
The peak \nht\ emission from the VLA is located very near the \nthp\ peak and the dense filament 
is further traced by the low-level \nht\ emission; a substantial amount of extended emission is likely
resolved out by the interferometer.

The velocity field traced by the single-dish \nthp\ appears to show a gradient along the filament
going from north to south and there is an area of blue-shifted emission coincident with the
southern protostar marked with an X in Figure \ref{L673}. This southern-most protostellar
source is comprised of three sources in higher-resolution Ks-band imaging (Tobin et al. in preparation).
In the northeast part of the image, there
is another velocity component of \nthp\ present. The linewidths are fairly low across the filament with about
a factor of two increase at the location of the protostars; there is an area of artificially large
linewidth due to the second velocity component. 

The VLA \nht\ map shows similar velocity structures that were present in 
the single-dish map; however, it is now clear that the protostar near $\Delta \delta$=0\arcsec\
is located in an area of red-shifted emission while the southern protostar
is still in a localized area of blue-shifted emission. The line-center velocity shift
between these components is 0.4-0.5 km/s. The linewidth peak falls between the
two main protostars, coincident with the region of peak \nht\ emission.  Also,
the northernmost, deeply embedded protostar appears to be associated with a fairly ordered
velocity gradient, north of the two more obvious protostars.

\subsubsection{L1521F}

L1521F is a dense core found in the Taurus star forming region. \citet{bourke2006} found a deeply embedded
protostar within what was previously considered a star-less core \citep{crapsi2004,crapsi2005}.
An approximately symmetric extinction envelope was found around L1521F, elongated
normal to the outflow in Paper I. The \nthp\ integrated intensity correlates very
well with the 8$\mu$m extinction. The \nht\ observations from the VLA
are also centrally peaked and show a flattened structure normal to the outflow.
However, there is an extension to the east, along the outflow.

The velocity structure of the core is complex and appears similar to that of L1527. The \nthp\
velocity field shows that there is emission blue-shifted relative to the protostar
normal to the outflow. Along the outflow there is red-shifted emission toward the 
edge of the envelope. \citet{crapsi2004} examined the velocity structure of L1521F finding that
the average gradient across the core was 0.37 \kmspc\ with a position angle of 180\degr. With our
two dimensional fitting we derive a gradient of 0.76 \kmspc\ and a position axis of 239\degr.
The differences between out results likely come from mapping a larger area around the core,
which detects more red-shifted emission in the western side of the map, influencing the gradient
fit. Otherwise, the emission and velocity structure are quite similar.

The \nht\ velocity map shows similar structure to the \nthp\ map, and
near the protostar there appears to be a gradient emerging normal to the outflow
on small-scales. However, the \nht\ emission is optically thick toward the
center of L1521F: the satellite and main lines have approximately equivalent intensities. Therefore, we cannot
obtain a better measure of the small-scale kinematic structure. The \nthp\ linewidth map shows a roughly constant
0.2 - 0.3 km/s linewidth across the map. The \nht\ linewidth is similarly low, except at the southern end of 
the envelope where the blue-shifted emission is present.

\subsubsection{Perseus 5}

Perseus 5 is a relatively isolated core in the Perseus molecular cloud, just northeast of NGC1333
and observed by \citet{caselli2002}. The protostar is deeply embedded and is obscured shortward
of 8\mum\ with only its outflow as a prominent signpost. It was discovered to have an asymmetric
extinction envelope around it in Paper I. We did not have the opportunity
to take single-dish \nthp\ observations of this object, but we did take data with 
CARMA as shown in Figure \ref{Per5}. The \nthp\ intensity image shows that
the entire extinction region is not well traced by the interferometric \nthp. The data
indicate that substantial emission around this source is resolved-out indicated
by the strong negative bowls in the image. However, NH$_2$D (another molecule we observed)
does seem to fully trace the envelope seen in 8$\mu$m extinction, since emission from this molecule 
is more spatially compact and not filtered-out by the interferometer.

The velocity structure is complex in both \nthp\ and NH$_2$D, showing a blue-shifted
feature east of the protostar along the outflow (Figure \ref{Per5}). 
Furthermore, both tracers show a similar gradient along
the outflow direction; NH$_2$D shows increased linewidth through the envelope, close to the
outflow direction and there are several regions of enhanced linewidth in \nthp\ along
the outflow. Furthermore, there may be a gradient normal to the outflow as seen in both the \nthp\
and NH$_2$D velocity maps. However, the outflow seems to be significantly influencing the kinematics
of the dense gas.

\subsubsection{IRAS 03282+3035}

IRAS 03282+3035 is an isolated, deeply embedded Class 0 protostar located in 
the B1-ridge of the Perseus star forming region \citep{jorgensen2006}.
Mid-infrared emission from the protostar itself is quite faint, but appears as a point-source
at 8$\mu$m and \citet{chen2007} identified it as a binary in millimeter continuum emission. 
The IRAC 8$\mu$m extinction toward this object in Paper I
highlights a rather complex morphology on large scales; however, near the protostar the extinction
appears to be concentrated into a filamentary structure. 

The single-dish \nthp\ observations in Figure \ref{IRAS03282} 
trace the large-scale extinction morphology very well and the emission
is observed to be quite extended, with the emission peak slightly offset from the protostar
along the outflow. The \nthp\ emission also ends at the northeast edge of the core where
the extinction rapidly falls off. The VLA \nht\ map traces 
a filamentary structure on large-scales north and south of the protostar. Furthermore, the 
emission double-peaked, with the individual peaks located north and south of the protostar, in agreement with
the \nthp\ emission shown by \citet{chen2007}.

The velocity field derived from the single-dish \nthp\ map shows a strong velocity gradient in the direction of the
outflow; however, there also appears to be another gradient that is normal to the outflow on the southeast
side of the envelope. We also note that there is a strong linewidth gradient in the direction of the outflow (same direction
as the line-center velocity gradient) with the largest linewidths appearing on the west side of the envelope.
The \nht\ velocity map from the VLA again finds the velocity gradient in the direction of the outflow along
with blue-shifted emission north and south of the protostar. The linewidth of the \nht\ emission is
peaked just north and south of the protostar indicative of dynamic motion in the line of sight.

\citet{chen2007} suggested that the outflow may be interacting with the envelope, causing the gradient along
the outflow. However, the IRAC data show that the outflow is well-collimated and appears only able to affect
the material extended to the northwest of the protostar.
To further examine the kinematic structure of this system, we have overlaid the \nht\ channel map contours on
the CO map from \citet{arce2006} in Figure \ref{IRAS03282-chanmaps}. 
The top and bottom rows show the portions of the \nht\ lines that are not blended while
in the middle panel the two lines are blended. At red-shifted velocities in the top panel, notice how
the different parts of the emission come into view going toward lower velocities. The top right
panel shows the red-shifted emission being emitted from two very thin structures running
north-south. In the bottom panels of Figure \ref{IRAS03282-chanmaps}, we can see that the more blue-shifted emission
is still confined to a very thin structure running north-south; however, its location has shifted further to the
east compared to the upper panels. Thus, we suggest that rather than an outflow interaction giving rise
to the global velocity structure, a more probable explanation is that the envelope is
filamentary and that the velocity shifts are flows along the filament toward the protostar.
However, certain regions of the envelope that are spatially coincident with the outflow cavities 
may be affected by the outflow.

\subsubsection{HH270 VLA1}

HH270 VLA 1 is located in the L1617 cloud near Orion and is the driving
source of the HH270 outflow \citep{reipurth1996,rodriguez1998}. This outflow in particular appears to be 
deflected and colliding with a neighboring core (the driving source of HH110 \citep{rodriguez1998}).
In Paper I, we found that this protostar
exhibits 8$\mu$m extinction in a filamentary envelope that is extended along
the outflow and only on one side of the protostar. Furthermore, it was noted that
the 4.5$\mu$m scattered light is brighter at the edge of the extincting envelope, indicating
that extinction structure is indeed extended along the outflow and
not due to a complex line of sight projection effect. The single-dish \nthp\ emission
from this source is peaked on the 8$\mu$m extinction
with no peak coincident with the protostar. The CARMA \nthp\ map
is still tracing the 8$\mu$m extinction and shows a sharp decline of emission
at the location of the protostar; the emission still appears one-sided even at high
resolution.

The single-dish \nthp\ velocity map shows a gradient along the outflow 
with a linewidth enhancement just southwest of the protostar. 
This region of high linewidth is likely an unresolved velocity gradient, which is then resolved
in the CARMA velocity map.
The pattern in the CARMA velocity map closely matches the shape of the
scattered light outflow cavity visible at 8\mum\ in Figure \ref{HH270}. We suggest that these
features are due to an outflow interaction with the surrounding material. We note that the
data do not show a linewidth increase; therefore the outflow seems to be inducing bulk motion, but not 
an increase in linewidth.

\section{Discussion}

\subsection{Large-Scale Velocity Field}
The large-scale velocity gradients in dense molecular cores have long
been interpreted as rotation (possibly solid-body) \citep[e.g.][]{goodman1993,
caselli2002, belloche2002, belloche2004, chen2007}. If the large-scale
motions are indeed rotation, then the circumstellar disk forming from the collapse of the surrounding
envelope should be rotating in the same direction as the surrounding cloud, assuming axisymmetric
collapse \citep[e.g.][]{shu1987,bodenheimer1995}.
However, if the core has fragmented, then the angular momenta of the individual
disks could be misaligned, but the sum of the angular momenta would still be aligned with the cloud rotation.
The direction of the jet and outflow of the system is commonly thought to
reflect the current angular momentum vector of the protostar and disk system. This is because
outflows and jets are thought to be related to the magnetic field of the protostar and disk which
should be along the rotation axis \citep{pudritz1983,shu1994} and resolved observations of disks have shown that
jets are oriented normal to the disk midplane. 
Thus, we have used the outflow direction and protostar locations as guideposts as to where
we should measure the velocity gradients relative to and in which direction because the equatorial plane
of the envelope is where we are most likely to observe rotation.

The 2D fitting method (see Section 2.6.2) shows that eleven out of sixteen systems have 
a gradient direction that is within 45\degr\ of normal to the outflow axis
in the single-dish data; however, there is considerable spread in this distribution. We caution that the PA of
the gradients calculated by this method may have systematic error due to the velocity fields not being uniform and the
velocities are sampled over a region that is not symmetric, leading to a bias of data points in a particular direction.
Visual inspection of the velocity fields does yield a similar result to the 2D fitting,
with eleven normal to the outflow and five not normal; IRAS 03282+3035, HH270 VLA1, L1527, L1521F, and L483 were not
 normal to the outflow. In contrast to the 2D fitting, we visually identify L483 and 
HH270 VLA1  as not normal to the outflow and L1157 and Serpens MMS3 to be normal. We further caution that
most velocity fields have structure not easily described by a single position angle or velocity gradient.
Nevertheless, as a simple significance test of the gradient-outflow direction relation, we can apply
the binomial distribution if we consider the $<$45\degr\ and $\ge$45\degr\ as two bins and that the gradient direction 
relative to the outflow may be oriented randomly between 0 and 90\degr\ with a mean of 45\degr\ \citep{bevington1969}.
Thus, the probability of the relative directions falling within $<$45\degr\ or $>$45\degr\ should be equal.
The chance of eleven or more objects out of sixteen with randomly oriented velocity gradients falling 
within the $\ge$45\degr\ bin is only $\sim$10\%. 

We also attempted to see if there is any trend with envelope mass and velocity gradient, 
we compared the observed velocity gradients and linewidths with the escape
velocity in Figure \ref{bound}. The envelope mass is calculated from the dust
mass measured from 8\mum\ extinction (Paper I, Table 1) plus 1 $M_{\sun}$ to account for the
central object; note that the central object masses may be overestimated
while the envelope masses are likely underestimated (Paper I).
There is no apparent trend of velocity gradients or linewidth with increasing envelope
mass (in terms of escape velocity). This figure further demonstrates that these protostellar systems are
consistent with being gravitationally bound and are not supported by rotation or turbulence,
consistent with many previous studies \citep[e.g.][]{goodman1993,caselli2002,chen2007}.

Furthermore, the smallest velocity gradients in our sample appear to be coming from the 
most symmetric envelopes L1521F, IRAS 16253-2429, and L1157, and the largest gradients are found
in the morphologically complex HH211 and IRAS 03282+3035 systems.
This could be taken to mean that the velocity structure is no strongly influenced by complex 
projection effects in the symmetric systems and are observing slow envelope rotation, while
the more complex systems have projection effects altering the observed velocity structure.
However, despite this trend, we note that the envelope of the 
low-luminosity source IRAM 04191 is approximately symmetric (Paper I) and
 has a velocity gradient of 17 \kmspc\ \citep{belloche2002, belloche2004}, a clear counter-example
to this morphological trend. 

The simple interpretation of our velocity gradient data as a whole
is that we are observing core rotation in these systems and that they
have a variety of angular momenta ; some angular momenta being quite large (Figure \ref{sdgradienthistos}).
However, if we were observing pure rotation on large-scales, then one would expect the
velocity gradient directions to be much more clustered toward being orthogonal to the outflow
rather than the broad distribution shown in Figure \ref{paoffset}.
Therefore, either the complex morphology makes the rotation ambiguous,
 causing the velocity gradients to not be orthogonal to the outflow or
we are not observing pure rotation. In either case, the angular momenta derived from the velocity gradients will be suspect at best.
Given these complicating factors, we regard the velocity gradients taken as 1D cuts normal to the outflow
as most likely to be probing the velocity gradients due to rotation, but these values should be regarded 
as upper limits.

If we are not observing pure rotation, the only other dynamical process which should give rise to
ordered velocity structures are infall and/or outflow entrainment. The outflows from the protostars in 
our sample are highly collimated, limiting their ability to affect the large-scale kinematic 
structure, see Section 4.3 for further discussion, which leaves infall as the only other mechanism 
to contribute to the velocity field. This would require that infall is happening on large-scales and that 
collapse would need to be outside-in and not inside-out. Large-scale infall is shown to be possible in
numerical simulations of complex, filamentary cores forming within a molecular cloud by \citet{smith2011}. 
Their simulations show that there is infall onto filaments (scales of 0.1 pc to 0.01 pc) from the surrounding molecular cloud, with
subsequent infall from the filament to the protostar (sink particle) ($<$ 0.01 pc). 
Thus, infall convolved with rotation in real envelopes
could lead to some of the complex velocity structures that we observe. 

In summary, it is difficult to interpret the large-scale
velocity gradients wholly being due to rotation since the envelopes are asymmetric and the gradients
are often not normal to the outflow. Furthermore, flows or infall along a filament
could give a similar signature to rotation simply from geometric projection, and if an envelope is filamentary,
infall along the envelope would also produce a velocity gradient normal to the outflow. We therefore suggest 
that a component of infall velocity could be projected along our line of sight, entangled with the rotation
velocity, resulting in the large velocity gradients. This issue of rotation versus
projected infall will be further explored in an upcoming paper.

\subsection{Small-scale Velocity Structure}

The high-resolution interferometer data are essential for probing the kinematics of the 
envelope at scales smaller than 5000 AU. These data enable us to 
localize the gas in the envelope and to assess the dynamical processes 
at work. Small-scale velocity sub-structure, beyond an ordered/linear velocity gradient,
is found in the interferometric observations of most envelopes, including L1157, L1165,
HH108IRS, Serpens MMS3, L1527, CB230, IRAS 03282+3035, and RNO43. These features 
are sometimes apparent in the velocity maps or profiles shown in Figure \ref{profiles},
but they also appear as increased linewidth in the inner envelope.
The small-scale velocity structure is generally found on $\sim$2000 AU scales in the 
envelopes. This radius could possibly be the centrifugal radius where material 
can be rotationally supported against gravity and the increased rotation velocity makes this
region stand out against the rest of the envelope in velocity. However, assuming the large-scale 
velocity gradients reflect rotation, there would not be enough angular momentum for material to be
rotationally supported at this radius.

We found that the relationship between outflow axis and velocity gradient direction
in the interferometer data trends even more strongly toward being normal to the outflow
that in the single-dish data. Twelve of fourteen systems have gradients within 45\degr\ of the outflow
in the right panel Figure \ref{paoffset}, as found by the 2D fitting technique.
Using the same statistical analysis as the single-dish gradients,
this result being due to chance is $\sim$0.6\%. Note that this plot is missing L1157 and Serpens MMS3 since their
complex velocity fields could not be reliably fit. Visually, we find that twelve envelopes clearly have 
gradients normal to the outflow and four do not (HH108MMS, L483, IRAS03282, and HH270VLA1); L483 and HH270 VLA1
were counted as within 45\degr\ of normal to the outflow. Note that RNO43, Perseus 5, and L1521F appear to have 
gradients normal to the outflow, but they also have complex velocity fields. L1157 is left out of this 
analysis because the gradient directions from the CARMA and VLA data differ by $\sim$80\degr.

The velocity gradient direction is generally 
consistent between both large and small scales in the single-dish and interferometer data as shown in
Figure \ref{sdintoffset}. This could mean that the dynamical processes observed at large-scales are 
also responsible for the kinematics observed at small-scales. The protostellar systems
which show substantial deviation from large to small-scales are L1527 and L1521F.
The difference in L1527 marks a velocity gradient reversal from large to small-scale (as also shown
in Figure \ref{profiles}) and the difference in L1521F reflects that the kinematic
structure was not well-resolved in the single-dish data. The \nthp\ and \nht\ lines in L1521F are also
extremely optically thick making the velocity field derived from fitting uncertain since we cannot probe
all the gas along the line of sight.

The preference for the vast majority of velocity gradients to be within 45\degr\ of normal to the 
outflow contrasts with \citet{chen2007}, where only two of nine targets had this feature. Moreover,
in \citet{volgenau2006}, only one protostellar core of three observed had a well-ordered velocity structure. This result
may be due to environment since the ordered structure was found in L1448 IRS3, a more isolated system,
and the other sources were located in the more complex environment of NGC1333. Our sources in
contrast are generally isolated, like those in \citet{chen2007}. As noted in the preceding paragraphs,
we do see several objects with similarly complex velocity fields in our data; however, our greater
number of observed systems likely enabled us to find more with ordered velocity fields.

The velocity gradients fit for the interferometer data are systematically larger than those found for the single-dish data, with
the average being 8.6 \kmspc\ from 1D fitting, the distribution is shown in Figure \ref{intgradienthistos}. This factor is nearly equal to our increase 
in resolution in the interferometer data as compared
to the IRAM 30m, but the object-to-object increase is more varied.
Some of the small-scale velocity structure is due to outflow interactions in the inner envelope
as we will discuss in more detail in the following section; however, we have attempted to mask out 
these regions when fitting the velocity gradients.  Like the single-dish 
data, we regard the 1D fitting method to more accurately reflect the velocity gradient intrinsic to the envelope
itself since the 2D method will be more susceptible to outflow effects on the envelope kinematics.
The only objects not showing large increase in the velocity gradient on small-scales are HH211 and IRAS 16253-2429
whose small-scale velocity gradient may be slightly overestimated. 

We strongly cautioned
in the previous section about interpreting the velocity gradients as rotation on large-scales
and we are again hesitant to interpret the small-scale gradients as rotation and/or spin-up.
This is because most envelopes are highly filamentary and on small-scales projections effects
on the velocity structure will be even more apparent since
both infall and rotation velocities increase at small radii. The velocity fields themselves
on small-scales are not well-ordered as one might expect from rotationally dominated motion. Highlighting a
few examples from Section 3.2: Serpens MMS3 shows
deep red-shifted emission and increased linewidth only on one side of the protostar, the velocity gradient in HH108IRS
reverses itself just past the protostar, the gradient in L1527 on small-scales is opposite of large scales, and RNO43
has an abrupt velocity jump across the envelope. The complex nature in the velocity fields of many sources
do not necessitate an interpretation as rotation. Furthermore, infall velocities
will always be necessarily be larger than rotation since the envelopes are not rotationally supported \citep{chen2007}.
Thus, the only way to robustly separate infall velocities from rotation is at $<$ 1000 AU scales, 
near the centrifugal radius where material can be rotationally supported. 

Two systems in our sample (L1165 and RNO43) show high-velocity line wings in the inner envelope that could 
indeed reflect significant rotation. In both sources, the emission appears to be coming from a radius 
of $\sim$600 AU. Our data indicate that the observed velocities
could reflect rotationally supported motion around $\sim$0.5$M_{\sun}$ central objects.
Higher resolution and higher signal-to-noise data are needed to accurately centroid the high velocity
emission in order to more precisely constrain the enclosed masses. In the future, ALMA could be used
to spatially resolve the high-velocity emission in order to fully separate rotation from infall.
Further analysis and modeling of the velocity structures observed with the interferometric data with 
the goal of determining the kinematic processes at work in the envelopes will be 
presented in an upcoming paper (Tobin et al. 2011 in preparation).

\subsection{Outflow Induced Kinematic Structure}

Our knowledge of protostellar outflows has been greatly enhanced in recent years
\citep[e.g.][and references therein]{bachiller1996,arce2007} and their possible effects on the surrounding 
envelope have been characterized \citep{arce2006}. Furthermore, IRAC imaging from 
the \textit{Spitzer Space Telescope}, in addition to near-IR
imaging from the ground, can give a strong constraint on the outflow axis and 
cavity width \citep[e.g.][]{seale2008}. These data complement observations of outflow tracers such as
CO and give a more complete picture of how the outflow may be impacting the protostellar envelope.

Several protostars in the sample have velocity structures that are strongly suggestive of outflow
effects on the \nthp\ and \nht\ gas kinematics.  L1157, HH108MMS, and Perseus 5 appear to be
significantly affected by the outflow, while HH108IRS, HH211, IRAS 03282+3035, L1152, and IRAS 04325+2402 only appear
to be mildly affected. In the mildly affected cases, the large-scale bulk motion of the
envelopes (line-center velocity) does not appear to be affected by the outflow, rather we see
the effects in the linewidth at both large-scales and small-scales. This probably means that only
a small portion of the total cloud mass is being affected by the outflow. The bulk motion effects
are revealed in the strongly affected cases, generally on small-scales and only visible in the interferometer data.

\citet{arce2006} presented an empirical model for 
how the outflow will affect the envelope during protostellar evolution, concluding that the outflow
is ultimately responsible for disrupting the protostellar envelopes. The number of objects we find showing
outflow effects in the kinematic data strongly support 
the outflow-envelope interaction framework put forward by \citet{arce2006} and enable us
to offer some further input to this empirical framework. 

Perseus 5 and HH108MMS appear to be some of the youngest objects in our sample, as evidenced by
their deeply embedded nature and lack of visible outflow cavities in 3.6\mum\ or Ks-band imaging.
In these systems, the outflow seems to be having the greatest impact on the kinematics in both
line-center velocity and linewidth. Therefore, the effects of the outflow on the kinematic
structure may be most prominent during its initial breakout of the
envelope, early in the Class 0 phase. Smaller-scale effects on the envelope can 
clearly be seen in the case of L1157 where the envelope material may be entrained at 1000 AU scales.
Thus, the outflows could be carrying significant momentum at wide angles near the protostar in order to be actively
forcing inner envelope material out. If the outflows are impacting the envelope at small-scales and wide angles,
an important question remains as to how much the outflow can quench infall in filamentary envelopes. In L1157,
the amount of entrained gas appears tenuous, only $\sim$0.04 $M_{\sun}$ out of 0.7 $M_{\sun}$ in the inner 
envelope, following a simple analysis \citep{goldsmith1999} assuming an \nthp\ abundance of 10$^{-9}$ per H$_2$.
On larger-scales ($>$ 1000AU), we expect that material extended normal to the outflow in filamentary
envelopes will not be strongly influenced by the outflow, as the outflows appear to stay well
collimated throughout the Class 0 phase.

\subsection{Linewidths}

The linewidths in the envelopes are generally quite small away from the protostar. In most cases, the 
linewidths are 0.2 - 0.5 \kms\ in the single-dish data; linewidths averaged
over the entire source are given Table 3. Note that these linewidths are substantially broader
than the 10K thermal linewidth of \nthp\ which is $\sim$0.13 km/s, and is generally attributed to
turbulent motions and/or unresolved velocity gradients along the line of sight. However, these
linewidths are not large enough to make the envelope unbounded, as shown Figure \ref{bound}.

The interferometer data 
find even smaller linewidths toward the outer edges of the envelopes; there 
were several cases where the lines are less than 2 channels wide. The narrow line-widths in
the interferometer data likely stem from the larger scale structure being resolved out; the 
larger scale emission may have some turbulent or large-scale infall velocity component.
The regions of broad linewidth in the interferometer observations appear to be
directly related to the outflow, increased line-of-sight motion, and/or heating from the protostar;
not a transition to a turbulent core \citep[cf.][]{chen2007}.

\nht\ linewidths were used by \citet{pineda2010} to probe the kinematics
of the large-scale molecular core, detecting a transition from the
quiescent core to the turbulent cloud. We looked for such an effect in our sensitive \nthp\ data but did not
find similar structure. We also did not detect \nthp\ emission
on the scales for which \citet{pineda2010} were able to detect
\nht, despite our high sensitivity. We believe that this is likely due to the differing critical densities between \nthp\ and \nht\
($\sim$2$\times$10$^3$ cm$^{-3}$ for \nht\ (1,1) and $\sim$1.4$\times$10$^5$ cm$^{-3}$ for \nthp\ ($J=1\rightarrow0$)).

We did, however, find that some maps (i.e. L673, HH211, HH108, RNO43) had regions with two distinct velocity
components. The regions of overlap between the components appear as artificially large linewidths in the
maps generated by the hyperfine fitting routine (section 2.6.1). We
show three \nthp\ spectra from HH211 taken at three positions showing the different velocity components
in Figure \ref{lineprofiles}; this is indicative of what takes place in the other regions showing this
feature. Notably, the additional velocity components tend to appear toward
the edge of the maps, except in RNO43 where
the transition takes place near the protostar. The distinct velocity components are located
in regions that appear form a contiguous structure when viewed in 8\mum\ extinction. The second velocity component
generally appears about $\sim$0.05 pc from the nearest protostar. Thus, the reason for multiple components could
have to do with the initial conditions of the clouds themselves. Recent simulations have suggested that
colliding clouds could be an important component to setting up the initial conditions for star formation
\citep[e.g.][]{heitsch2006}.

\subsection{Chemical Effects on Molecular Tracers}

The kinematic data presented are based on the molecular tracers \nthp, \nht, and \hcop.
We have used \nthp\ and \nht\ relatively interchangeably, since they appear to trace the same
kinematics and physical conditions (see Section 3.1 and \citet{johnstone2010}). This makes sense
because we know that in the pre-stellar phase, the formation of \nthp\ and \nht\ appear to be linked
given their similar abundance distributions and only deplete onto dust grains at very high densities
\citep{bergin2007}. In many of our observations, \nthp\ and \nht\ only appear to trace the gas on scales 
$>$1000 AU from the protostar; the emission generally peaks near the 
protostar, but not directly on it. This indicates a drop in abundance
either due to depletion or destruction of the molecules.
Several of the protostars for which there are both interferometric \nht\ and \nthp\ observations (in this
paper or in the literature) show decreased emission at the location of the protostar in both \nthp\ and \nht.
Observations of CB230 and IRAS 03282+3035 in \nht\ (Figures \ref{CB230} and \ref{IRAS03282}) 
and \nthp\ \citep{chen2007} are clear examples of \nht\ and \nthp\ not being peaked coincident with the 
protostar. L1157 also exhibits this effect in \nthp\ from the PdBI observations and also does in \nht\ if the longer baselines 
are given more weight in the VLA map. Finally, the low-luminosity source IRAM 04191 shows a
similar depletion pattern in both of these tracers
\citep[][J. Mangum, Private Communication]{belloche2004}.

We can understand the decrease in \nthp\ emission in terms of molecular destruction by reactions with
other molecules. In the inner
envelope, where the temperatures rise above 20K, the CO that has depleted onto the dust grains
\citep{bergin2002} is released back into the gas phase. 
CO and \nthp\ rapidly react to form \hcop; this is the 
dominant destruction mechanism for \nthp\ \citep{aikawa2001,lee2004}. This is the same reason that \nthp\
is not seen in outflows and why the \nthp\ is often not centrally peaked on the protostars 
in our interferometric observations. \nht\, however, is not directly destroyed by CO, but
rather \hcop\ \citep{lee2004}. \hcop\ will readily form within the region of CO evaporation making
it available to react with \nht. Alternatively, \nht\ could also become depleted onto dust grains
 in an ice mantle, and if it is well-mixed with the water ice, 
then it would only be evaporated at temperatures $\geq$100K. \nht\ ice is 
frequently observed in the envelopes surrounding protostars \citep{bottinelli2010} via mid-infrared spectroscopy.
The absorbing \nht\ ice should be in the inner envelope since \nht\ only depletes onto grains at high densities. If
any of the \nht\ ice were released into the gas phase, then \hcop\ would be present to destroy it.

Since both \nht\ and \nthp\ are not present on scales $<$1000 AU, we must look for other
tracers to probe the kinematic structure on these scales. In two protostars, we have been able to use \hcop\
to trace the small-scale kinematic structure. The \hcop\ emission from L1165 and RNO43 show
high-velocity wings on small-scales inside the innermost \nthp\ emission; the centroids of the
red and blue-shifted emission are offset from the protostar, normal to the outflow direction, at radii
of $\sim$600AU.

Chemical models of an infalling protostellar envelope have been calculated by
\citet{lee2004}, showing that \hcop\ becomes enhanced at small-scales after formation
of the protostar. This reflects the evaporation of CO ice, releasing CO back into the gas phase where \hcop\ is readily
formed. In addition, the primary destruction pathway of \nthp\ from CO results in the formation of
 HCO$^+$ and N$_2$. Thus, we can see why the \nthp\ is tracing the gas at larger radii and lower velocities, while
in the inner envelope HCO$^+$ is readily being formed and can then trace the small-scale high-velocity gas.

Our dataset explicitly shows how multiple tracers can be used to gain a more complete
picture of the kinematics in protostellar envelopes. \nthp\
and \nht\ are excellent tracers of the cold, dense gas on scales from $\sim$1000-2000 AU
out to $\sim$10000-20000 AU. Inside of 1000AU, an abundant tracer of the warm, inner envelope tracer must be observed;
HCO$^+$ works quite well in two out of nine sources observed with CARMA and other tracers and/or higher-J transitions
of \hcop\ may work as well (see \citet{lee2009} and \citet{brinch2007}).
However, these other tracers are often found in outflows and they must be
observed with sufficiently high resolution to confirm their origin in the envelope 
and not the outflow. In the future, ALMA may be able to observe inner envelope
tracers to resolve the motion of the dense gas in the inner envelope, tracing infall onto 
the disk.

\section{Summary}
We have conducted a single-dish and interferometric survey  
mapping emission of the dense gas tracers \nthp\ ($J=1\rightarrow0$) and/or \nht\ (1,1)
in envelopes around low-mass protostars. Many of these envelopes are known to be morphologically complex
from 8\mum\ extinction mapping. We used these data to
map the line-center velocity and linewidth across these envelopes. We quantitatively measured the 
velocity gradients and their directions in order 
to characterize the dense gas kinematics in the complex protostellar envelopes, our specific results are as follows.

1. Ordered velocity fields are present on large-scales in most protostellar envelopes from the single-dish 
sample. In eleven out of sixteen cases, the velocity gradients appear to be within 45\degr\ of normal to the outflow axis
with an average gradient of $\sim$ 2.3 \kmspc, depending on fitting method. The
velocity gradients could be due to core rotation; however, the velocity gradient position angles do have a broad
distribution with respect to the outflow direction. Furthermore, the strongly asymmetric nature of the envelopes and the 
fact that most envelopes in the sample are substantially bound leads us to suggest that
we may be seeing a component of infall projected along our line of sight
entangled with rotation. We find evidence of multiple components of \nthp\ emission in several clouds, possibly relating
our observations to a colliding-cloud formation scenario.

2. The small-scale kinematic structure observed by the interferometers appears to be
gravitationally dominated by the central protostar and it likely originates from a combination
of infall and rotation in some cases. The average velocity gradient in the interferometer
data is 8.6 \kmspc, with gradient directions within 45$^{\degr}$ of normal to the outflow in
twelve out of fourteen cases. The complex velocity fields in many systems suggest that interpreting 
the velocity gradients as pure rotation is incorrect. Only on the smallest scales ($<$1000 AU)
will rotation differentiate itself from infall.
Furthermore, multiple tracers must be used to gain a complete picture of the kinematic structure of the envelope
down to sub-1000 AU scales due to depletion of the cold gas tracers within $\sim$1000 AU. In the cases of RNO43 and
L1165, we were able to use \hcop\ $(J=1\rightarrow0)$ to trace inner envelope kinematics on scales of $\sim$600 AU.

3. Outflows do impact the envelope kinematics derived from \nht\ and \nthp\ in some systems and
their effects are most prominent at $\sim$1000-2000 AU scales.
Clear outflow effects on the large-scale kinematics in the single-dish data are only seen for
five systems, but the only evident kinematic effect is increased linewidth along the outflow.
The effects are most pronounced in the interferometer data for the deeply embedded sources HH108MMS and Perseus 5.
Furthermore, we see a remarkable case of the outflow possibly entraining the inner envelope
of L1157 on $\sim$1000 AU scales.

The authors thank the anonymous referee for a thorough report which improved the paper as a whole and 
H. Arce for providing the OVRO CO data of IRAS 03282+3035. We would also like to thank A. Goodman, 
S. Offner, Y. Shirley, A. Stutz, W. Kwon, J. Lee, and P. Myers
for useful discussions. We wish to thank the CARMA observers for carrying out the observations. 
We thank V. Pi\'etu for assistance with the PdBI data reduction and
C. Buchbender for assistance in conducting observations at the 30m in addition to the IRAM staff as a whole.
Support for CARMA construction was derived from the states of Illinois, California,
and Maryland, the James S. McDonnell Foundation, the Gordon and Betty Moore Foundation,
the Kenneth T. and Eileen L. Norris Foundation, the University of Chicago, the Associates of the
California Institute of Technology, and the National Science Foundation. Ongoing CARMA development
and operations are supported by the National Science Foundation under a cooperative agreement,
and by the CARMA partner universities. The National Radio Astronomy Observatory is a facility of
the National Science Foundation operated under cooperative agreement by Associated Universities, Inc. 
IRAM is supported by INSU/CNRS (France), MPG (Germany) and IGN (Spain). J. J. Tobin acknowledges support
from HST-GO-11548.04-A, the University of Michigan Rackham Dissertation Fellowship,
and \textit{Spitzer} archival research program 50668.

{\it Facilities:} \facility{IRAM:30m}, \facility{CARMA}, \facility{VLA}, \facility{IRAM:Interferometer}, \facility{Spitzer (IRAC)}, \facility{Blanco (ISPI)}

\appendix
\section{Continuum Data}

In conjunction with our CARMA and PdBI \nthp\ observations, the 3mm continuum emission was observed for those 11 sources. The 
continuum data are overlaid on the IRAC 8\mum\ images for all but HH108IRS/MMS (we shown 70\mum\ data due to lack of
IRAC imaging) in Figure \ref{cont8um}. The 3mm continuum sources
are point-like in most cases, HH108IRS and HH211 appear to show extended structure at the 3$\sigma$ level, consistent
with envelope dust emission. L1157, L1165, and HH270 are also slightly extended, but in the direction of their outflows, indicating
that there may be a component of free-free jet emission or heated dust along the outflow 
in the 3mm continuum data. In all cases, the continuum sources are coincident
with the 8\mum/70\mum\ point sources. There are a few cases (L1157, HH211, and RNO43) where
there is not a clear point source at 8\mum\ and the 3mm emission is located between the outflow cavities where there is significant
extinction. In these cases, the 24\mum\ point source is then found to peak at the same location of the 3mm source. We note that 
the protostar IRAS 16253-2429 shows a slight offset between the 8\mum\ point source and the continuum detection; however, the detection
is barely 3$\sigma$, thus we do not believe that this offset is real. The continuum fluxes at 3mm are listed in Table 11.

\begin{small}
\bibliographystyle{apj}
\bibliography{ms}

\begin{thebibliography}{115}
\expandafter\ifx\csname natexlab\endcsname\relax\def\natexlab#1{#1}\fi

\bibitem[{{Aikawa} {et~al.}(2001){Aikawa}, {Ohashi}, {Inutsuka}, {Herbst}, \&
  {Takakuwa}}]{aikawa2001}
{Aikawa}, Y., {Ohashi}, N., {Inutsuka}, S., {Herbst}, E., \& {Takakuwa}, S.
  2001, \apj, 552, 639

\bibitem[{{Andr{\'e}} {et~al.}(1993){Andr{\'e}}, {Ward-Thompson}, \&
  {Barsony}}]{andre1993}
{Andr{\'e}}, P., {Ward-Thompson}, D., \& {Barsony}, M. 1993, \apj, 406, 122

\bibitem[{{Arce} \& {Sargent}(2005)}]{arce2005}
{Arce}, H.~G., \& {Sargent}, A.~I. 2005, \apj, 624, 232

\bibitem[{{Arce} \& {Sargent}(2006)}]{arce2006}
---. 2006, \apj, 646, 1070

\bibitem[{{Arce} {et~al.}(2007){Arce}, {Shepherd}, {Gueth}, {Lee}, {Bachiller},
  {Rosen}, \& {Beuther}}]{arce2007}
{Arce}, H.~G., {Shepherd}, D., {Gueth}, F., {Lee}, C., {Bachiller}, R.,
  {Rosen}, A., \& {Beuther}, H. 2007, Protostars and Planets V, 245

\bibitem[{{Arquilla} \& {Goldsmith}(1986)}]{arquilla1986}
{Arquilla}, R., \& {Goldsmith}, P.~F. 1986, \apj, 303, 356

\bibitem[{{Bachiller}(1996)}]{bachiller1996}
{Bachiller}, R. 1996, \araa, 34, 111

\bibitem[{{Barsony} {et~al.}(2010){Barsony}, {Wolf-Chase}, {Ciardi}, \&
  {O'Linger}}]{barsony2010}
{Barsony}, M., {Wolf-Chase}, G.~A., {Ciardi}, D.~R., \& {O'Linger}, J. 2010,
  \apj, 720, 64

\bibitem[{{Belloche} \& {Andr{\'e}}(2004)}]{belloche2004}
{Belloche}, A., \& {Andr{\'e}}, P. 2004, \aap, 419, L35

\bibitem[{{Belloche} {et~al.}(2002){Belloche}, {Andr{\'e}}, {Despois}, \&
  {Blinder}}]{belloche2002}
{Belloche}, A., {Andr{\'e}}, P., {Despois}, D., \& {Blinder}, S. 2002, \aap,
  393, 927

\bibitem[{{Bence} {et~al.}(1996){Bence}, {Richer}, \& {Padman}}]{bence1996}
{Bence}, S.~J., {Richer}, J.~S., \& {Padman}, R. 1996, \mnras, 279, 866

\bibitem[{{Benson} \& {Myers}(1989)}]{bm1989}
{Benson}, P.~J., \& {Myers}, P.~C. 1989, \apjs, 71, 89

\bibitem[{{Bergin} {et~al.}(2002){Bergin}, {Alves}, {Huard}, \&
  {Lada}}]{bergin2002}
{Bergin}, E.~A., {Alves}, J., {Huard}, T., \& {Lada}, C.~J. 2002, \apjl, 570,
  L101

\bibitem[{{Bergin} \& {Tafalla}(2007)}]{bergin2007}
{Bergin}, E.~A., \& {Tafalla}, M. 2007, ArXiv e-prints, 705

\bibitem[{{Bevington}(1969)}]{bevington1969}
{Bevington}, P.~R. 1969, {Data reduction and error analysis for the physical
  sciences}, ed. {Bevington, P.~R.}

\bibitem[{{Bodenheimer}(1995)}]{bodenheimer1995}
{Bodenheimer}, P. 1995, \araa, 33, 199

\bibitem[{{Bonnell} \& {Bate}(1994)}]{bonnell1994a}
{Bonnell}, I.~A., \& {Bate}, M.~R. 1994, \mnras, 269, L45

\bibitem[{{Bottinelli} {et~al.}(2010){Bottinelli}, {Adwin Boogert}, {Bouwman},
  {Beckwith}, {van Dishoeck}, {{\"O}berg}, {Pontoppidan}, {Linnartz}, {Blake},
  {Evans}, \& {Lahuis}}]{bottinelli2010}
{Bottinelli}, S., {Adwin Boogert}, A.~C., {Bouwman}, J., {Beckwith}, M., {van
  Dishoeck}, E.~F., {{\"O}berg}, K.~I., {Pontoppidan}, K.~M., {Linnartz}, H.,
  {Blake}, G.~A., {Evans}, N.~J., \& {Lahuis}, F. 2010, \apj, 718, 1100

\bibitem[{{Bourke} {et~al.}(1995){Bourke}, {Hyland}, {Robinson}, {James}, \&
  {Wright}}]{bourke1995b}
{Bourke}, T.~L., {Hyland}, A.~R., {Robinson}, G., {James}, S.~D., \& {Wright},
  C.~M. 1995, \mnras, 276, 1067

\bibitem[{{Bourke} {et~al.}(2006){Bourke}, {Myers}, {Evans}, {Dunham},
  {Kauffmann}, {Shirley}, {Crapsi}, {Young}, {Huard}, {Brooke}, {Chapman},
  {Cieza}, {Lee}, {Teuben}, \& {Wahhaj}}]{bourke2006}
{Bourke}, T.~L., {Myers}, P.~C., {Evans}, II, N.~J., {Dunham}, M.~M.,
  {Kauffmann}, J., {Shirley}, Y.~L., {Crapsi}, A., {Young}, C.~H., {Huard},
  T.~L., {Brooke}, T.~Y., {Chapman}, N., {Cieza}, L., {Lee}, C.~W., {Teuben},
  P., \& {Wahhaj}, Z. 2006, \apjl, 649, L37

\bibitem[{{Brinch} {et~al.}(2007){Brinch}, {Crapsi}, {J{\o}rgensen},
  {Hogerheijde}, \& {Hill}}]{brinch2007}
{Brinch}, C., {Crapsi}, A., {J{\o}rgensen}, J.~K., {Hogerheijde}, M.~R., \&
  {Hill}, T. 2007, \aap, 475, 915

\bibitem[{{Burkert} \& {Bodenheimer}(1993)}]{burkert1993}
{Burkert}, A., \& {Bodenheimer}, P. 1993, \mnras, 264, 798

\bibitem[{{Caselli} {et~al.}(2002){Caselli}, {Benson}, {Myers}, \&
  {Tafalla}}]{caselli2002}
{Caselli}, P., {Benson}, P.~J., {Myers}, P.~C., \& {Tafalla}, M. 2002, \apj,
  572, 238

\bibitem[{{Cassen} \& {Moosman}(1981)}]{cassen1981}
{Cassen}, P., \& {Moosman}, A. 1981, \icarus, 48, 353

\bibitem[{{Chapman} \& {Mundy}(2009)}]{chapman2009}
{Chapman}, N.~L., \& {Mundy}, L.~G. 2009, \apj, 699, 1866

\bibitem[{{Chen} {et~al.}(2007){Chen}, {Launhardt}, \& {Henning}}]{chen2007}
{Chen}, X., {Launhardt}, R., \& {Henning}, T. 2007, \apj, 669, 1058

\bibitem[{{Chiang} {et~al.}(2010){Chiang}, {Looney}, {Tobin}, \&
  {Hartmann}}]{chiang2010}
{Chiang}, H., {Looney}, L.~W., {Tobin}, J.~J., \& {Hartmann}, L. 2010, \apj,
  709, 470

\bibitem[{{Chini} {et~al.}(2001){Chini}, {Ward-Thompson}, {Kirk}, {Nielbock},
  {Reipurth}, \& {Sievers}}]{chini2001}
{Chini}, R., {Ward-Thompson}, D., {Kirk}, J.~M., {Nielbock}, M., {Reipurth},
  B., \& {Sievers}, A. 2001, \aap, 369, 155

\bibitem[{{Crapsi} {et~al.}(2005){Crapsi}, {Caselli}, {Walmsley}, {Myers},
  {Tafalla}, {Lee}, \& {Bourke}}]{crapsi2005}
{Crapsi}, A., {Caselli}, P., {Walmsley}, C.~M., {Myers}, P.~C., {Tafalla}, M.,
  {Lee}, C.~W., \& {Bourke}, T.~L. 2005, \apj, 619, 379

\bibitem[{{Crapsi} {et~al.}(2004){Crapsi}, {Caselli}, {Walmsley}, {Tafalla},
  {Lee}, {Bourke}, \& {Myers}}]{crapsi2004}
{Crapsi}, A., {Caselli}, P., {Walmsley}, C.~M., {Tafalla}, M., {Lee}, C.~W.,
  {Bourke}, T.~L., \& {Myers}, P.~C. 2004, \aap, 420, 957

\bibitem[{{Danby} {et~al.}(1988){Danby}, {Flower}, {Valiron}, {Schilke}, \&
  {Walmsley}}]{danby1988}
{Danby}, G., {Flower}, D.~R., {Valiron}, P., {Schilke}, P., \& {Walmsley},
  C.~M. 1988, \mnras, 235, 229

\bibitem[{{Di Francesco} {et~al.}(2001){Di Francesco}, {Myers}, {Wilner},
  {Ohashi}, \& {Mardones}}]{difrancesco2001}
{Di Francesco}, J., {Myers}, P.~C., {Wilner}, D.~J., {Ohashi}, N., \&
  {Mardones}, D. 2001, \apj, 562, 770

\bibitem[{{Djupvik} {et~al.}(2006){Djupvik}, {Andr{\'e}}, {Bontemps}, {Motte},
  {Olofsson}, {G{\aa}lfalk}, \& {Flor{\'e}n}}]{dab2006}
{Djupvik}, A.~A., {Andr{\'e}}, P., {Bontemps}, S., {Motte}, F., {Olofsson}, G.,
  {G{\aa}lfalk}, M., \& {Flor{\'e}n}, H. 2006, \aap, 458, 789

\bibitem[{{Dunham} {et~al.}(2010){Dunham}, {Evans}, {Terebey}, {Dullemond}, \&
  {Young}}]{dunham2010}
{Dunham}, M.~M., {Evans}, N.~J., {Terebey}, S., {Dullemond}, C.~P., \& {Young},
  C.~H. 2010, \apj, 710, 470

\bibitem[{{Dzib} {et~al.}(2010){Dzib}, {Loinard}, {Mioduszewski}, {Boden},
  {Rodr{\'{\i}}guez}, \& {Torres}}]{dzib2010}
{Dzib}, S., {Loinard}, L., {Mioduszewski}, A.~J., {Boden}, A.~F.,
  {Rodr{\'{\i}}guez}, L.~F., \& {Torres}, R.~M. 2010, \apj, 718, 610

\bibitem[{{Enoch} {et~al.}(2009){Enoch}, {Evans}, {Sargent}, \&
  {Glenn}}]{enoch2009}
{Enoch}, M.~L., {Evans}, N.~J., {Sargent}, A.~I., \& {Glenn}, J. 2009, \apj,
  692, 973

\bibitem[{{Enoch} {et~al.}(2007){Enoch}, {Glenn}, {Evans}, {Sargent}, {Young},
  \& {Huard}}]{enoch2007}
{Enoch}, M.~L., {Glenn}, J., {Evans}, II, N.~J., {Sargent}, A.~I., {Young},
  K.~E., \& {Huard}, T.~L. 2007, \apj, 666, 982

\bibitem[{{Evans} {et~al.}(2009){Evans}, {Dunham}, {J{\o}rgensen}, {Enoch},
  {Mer{\'{\i}}n}, {van Dishoeck}, {Alcal{\'a}}, {Myers}, {Stapelfeldt},
  {Huard}, {Allen}, {Harvey}, {van Kempen}, {Blake}, {Koerner}, {Mundy},
  {Padgett}, \& {Sargent}}]{evans2009}
{Evans}, N.~J., {Dunham}, M.~M., {J{\o}rgensen}, J.~K., {Enoch}, M.~L.,
  {Mer{\'{\i}}n}, B., {van Dishoeck}, E.~F., {Alcal{\'a}}, J.~M., {Myers},
  P.~C., {Stapelfeldt}, K.~R., {Huard}, T.~L., {Allen}, L.~E., {Harvey}, P.~M.,
  {van Kempen}, T., {Blake}, G.~A., {Koerner}, D.~W., {Mundy}, L.~G.,
  {Padgett}, D.~L., \& {Sargent}, A.~I. 2009, \apjs, 181, 321

\bibitem[{{Froebrich}(2005)}]{froebrich2005}
{Froebrich}, D. 2005, \apjs, 156, 169

\bibitem[{{Fuller} {et~al.}(1995){Fuller}, {Lada}, {Masson}, \&
  {Myers}}]{fuller1995}
{Fuller}, G.~A., {Lada}, E.~A., {Masson}, C.~R., \& {Myers}, P.~C. 1995, \apj,
  453, 754

\bibitem[{{Fuller} \& {Wootten}(2000)}]{fuller2000}
{Fuller}, G.~A., \& {Wootten}, A. 2000, \apj, 534, 854

\bibitem[{{Goldsmith} \& {Langer}(1999)}]{goldsmith1999}
{Goldsmith}, P.~F., \& {Langer}, W.~D. 1999, \apj, 517, 209

\bibitem[{{Goodman} {et~al.}(1993){Goodman}, {Benson}, {Fuller}, \&
  {Myers}}]{goodman1993}
{Goodman}, A.~A., {Benson}, P.~J., {Fuller}, G.~A., \& {Myers}, P.~C. 1993,
  \apj, 406, 528

\bibitem[{{Gramajo} {et~al.}(2010){Gramajo}, {Whitney}, {G{\'o}mez}, \&
  {Robitaille}}]{gramajo2010}
{Gramajo}, L.~V., {Whitney}, B.~A., {G{\'o}mez}, M., \& {Robitaille}, T.~P.
  2010, \aj, 139, 2504

\bibitem[{{Gueth} \& {Guilloteau}(1999)}]{gueth1999}
{Gueth}, F., \& {Guilloteau}, S. 1999, \aap, 343, 571

\bibitem[{{Gueth} {et~al.}(1996){Gueth}, {Guilloteau}, \&
  {Bachiller}}]{gueth1996}
{Gueth}, F., {Guilloteau}, S., \& {Bachiller}, R. 1996, \aap, 307, 891

\bibitem[{{Haisch} {et~al.}(2001){Haisch}, {Lada}, \& {Lada}}]{haisch2001}
{Haisch}, Jr., K.~E., {Lada}, E.~A., \& {Lada}, C.~J. 2001, \apjl, 553, L153

\bibitem[{{Hartmann} {et~al.}(1999){Hartmann}, {Calvet}, {Allen}, {Chen}, \&
  {Jayawardhana}}]{hartmann1999}
{Hartmann}, L., {Calvet}, N., {Allen}, L., {Chen}, H., \& {Jayawardhana}, R.
  1999, \aj, 118, 1784

\bibitem[{{Harvey} {et~al.}(2006){Harvey}, {Chapman}, {Lai}, {Evans}, {Allen},
  {J{\o}rgensen}, {Mundy}, {Huard}, {Porras}, {Cieza}, {Myers}, {Mer{\'{\i}}n},
  {van Dishoeck}, {Young}, {Spiesman}, {Blake}, {Koerner}, {Padgett},
  {Sargent}, \& {Stapelfeldt}}]{harvey2006}
{Harvey}, P.~M., {Chapman}, N., {Lai}, S., {Evans}, II, N.~J., {Allen}, L.~E.,
  {J{\o}rgensen}, J.~K., {Mundy}, L.~G., {Huard}, T.~L., {Porras}, A., {Cieza},
  L., {Myers}, P.~C., {Mer{\'{\i}}n}, B., {van Dishoeck}, E.~F., {Young},
  K.~E., {Spiesman}, W., {Blake}, G.~A., {Koerner}, D.~W., {Padgett}, D.~L.,
  {Sargent}, A.~I., \& {Stapelfeldt}, K.~R. 2006, \apj, 644, 307

\bibitem[{{Heitsch} {et~al.}(2006){Heitsch}, {Slyz}, {Devriendt}, {Hartmann},
  \& {Burkert}}]{heitsch2006}
{Heitsch}, F., {Slyz}, A.~D., {Devriendt}, J.~E.~G., {Hartmann}, L.~W., \&
  {Burkert}, A. 2006, \apj, 648, 1052

\bibitem[{{Herbig} \& {Jones}(1983)}]{herbig1983}
{Herbig}, G.~H., \& {Jones}, B.~F. 1983, \aj, 88, 1040

\bibitem[{{Hern{\'a}ndez} {et~al.}(2007){Hern{\'a}ndez}, {Hartmann}, {Megeath},
  {Gutermuth}, {Muzerolle}, {Calvet}, {Vivas}, {Brice{\~n}o}, {Allen},
  {Stauffer}, {Young}, \& {Fazio}}]{hernandez2007}
{Hern{\'a}ndez}, J., {Hartmann}, L., {Megeath}, T., {Gutermuth}, R.,
  {Muzerolle}, J., {Calvet}, N., {Vivas}, A.~K., {Brice{\~n}o}, C., {Allen},
  L., {Stauffer}, J., {Young}, E., \& {Fazio}, G. 2007, \apj, 662, 1067

\bibitem[{{Hirota} {et~al.}(2011){Hirota}, {Honma}, {Imai}, {Sunada}, {Ueno},
  {Kobayashi}, \& {Kawaguchi}}]{hirota2011}
{Hirota}, T., {Honma}, M., {Imai}, H., {Sunada}, K., {Ueno}, Y., {Kobayashi},
  H., \& {Kawaguchi}, N. 2011, \pasj, 63, 1

\bibitem[{{Ho} \& {Townes}(1983)}]{ho1983}
{Ho}, P.~T.~P., \& {Townes}, C.~H. 1983, \araa, 21, 239

\bibitem[{{Hogerheijde} {et~al.}(1998){Hogerheijde}, {van Dishoeck}, {Blake},
  \& {van Langevelde}}]{hoger1998}
{Hogerheijde}, M.~R., {van Dishoeck}, E.~F., {Blake}, G.~A., \& {van
  Langevelde}, H.~J. 1998, \apj, 502, 315

\bibitem[{{Johnstone} {et~al.}(2010){Johnstone}, {Rosolowsky}, {Tafalla}, \&
  {Kirk}}]{johnstone2010}
{Johnstone}, D., {Rosolowsky}, E., {Tafalla}, M., \& {Kirk}, H. 2010, \apj,
  711, 655

\bibitem[{{J{\o}rgensen}(2004)}]{jorgensen2004}
{J{\o}rgensen}, J.~K. 2004, \aap, 424, 589

\bibitem[{{J{\o}rgensen} {et~al.}(2007){J{\o}rgensen}, {Bourke}, {Myers}, {Di
  Francesco}, {van Dishoeck}, {Lee}, {Ohashi}, {Sch{\"o}ier}, {Takakuwa},
  {Wilner}, \& {Zhang}}]{jorgensen2007}
{J{\o}rgensen}, J.~K., {Bourke}, T.~L., {Myers}, P.~C., {Di Francesco}, J.,
  {van Dishoeck}, E.~F., {Lee}, C., {Ohashi}, N., {Sch{\"o}ier}, F.~L.,
  {Takakuwa}, S., {Wilner}, D.~J., \& {Zhang}, Q. 2007, \apj, 659, 479

\bibitem[{{J{\o}rgensen} {et~al.}(2006){J{\o}rgensen}, {Harvey}, {Evans},
  {Huard}, {Allen}, {Porras}, {Blake}, {Bourke}, {Chapman}, {Cieza}, {Koerner},
  {Lai}, {Mundy}, {Myers}, {Padgett}, {Rebull}, {Sargent}, {Spiesman},
  {Stapelfeldt}, {van Dishoeck}, {Wahhaj}, \& {Young}}]{jorgensen2006}
{J{\o}rgensen}, J.~K., {Harvey}, P.~M., {Evans}, II, N.~J., {Huard}, T.~L.,
  {Allen}, L.~E., {Porras}, A., {Blake}, G.~A., {Bourke}, T.~L., {Chapman}, N.,
  {Cieza}, L., {Koerner}, D.~W., {Lai}, S., {Mundy}, L.~G., {Myers}, P.~C.,
  {Padgett}, D.~L., {Rebull}, L., {Sargent}, A.~I., {Spiesman}, W.,
  {Stapelfeldt}, K.~R., {van Dishoeck}, E.~F., {Wahhaj}, Z., \& {Young}, K.~E.
  2006, \apj, 645, 1246

\bibitem[{{Kauffmann} {et~al.}(2008){Kauffmann}, {Bertoldi}, {Bourke}, {Evans},
  \& {Lee}}]{kauffmann2008}
{Kauffmann}, J., {Bertoldi}, F., {Bourke}, T.~L., {Evans}, II, N.~J., \& {Lee},
  C.~W. 2008, \aap, 487, 993

\bibitem[{{Keto} \& {Rybicki}(2010)}]{keto2010}
{Keto}, E., \& {Rybicki}, G. 2010, \apj, 716, 1315

\bibitem[{{Kirk} {et~al.}(2009){Kirk}, {Ward-Thompson}, {Di Francesco},
  {Bourke}, {Evans}, {Mer{\'{\i}}n}, {Allen}, {Cieza}, {Dunham}, {Harvey},
  {Huard}, {J{\o}rgensen}, {Miller}, {Noriega-Crespo}, {Peterson}, {Ray}, \&
  {Rebull}}]{kirk2009}
{Kirk}, J.~M., {Ward-Thompson}, D., {Di Francesco}, J., {Bourke}, T.~L.,
  {Evans}, N.~J., {Mer{\'{\i}}n}, B., {Allen}, L.~E., {Cieza}, L.~A., {Dunham},
  M.~M., {Harvey}, P., {Huard}, T., {J{\o}rgensen}, J.~K., {Miller}, J.~F.,
  {Noriega-Crespo}, A., {Peterson}, D., {Ray}, T.~P., \& {Rebull}, L.~M. 2009,
  \apjs, 185, 198

\bibitem[{{Koda} {et~al.}(2011){Koda}, {Sawada}, {Wright}, {Teuben}, {Corder},
  {Patience}, {Scoville}, {Donovan Meyer}, \& {Egusa}}]{koda2011}
{Koda}, J., {Sawada}, T., {Wright}, M.~C.~H., {Teuben}, P., {Corder}, S.~A.,
  {Patience}, J., {Scoville}, N., {Donovan Meyer}, J., \& {Egusa}, F. 2011,
  \apjs, 193, 19

\bibitem[{{Lada}(1987)}]{lada1987}
{Lada}, C.~J. 1987, in IAU Symp. 115: Star Forming Regions, ed. M.~{Peimbert}
  \& J.~{Jugaku}, 1--17

\bibitem[{{Larson}(1969)}]{larson1969}
{Larson}, R.~B. 1969, \mnras, 145, 271

\bibitem[{{Launhardt} {et~al.}(2010){Launhardt}, {Nutter}, {Ward-Thompson},
  {Bourke}, {Henning}, {Khanzadyan}, {Schmalzl}, {Wolf}, \&
  {Zylka}}]{launhardt2010}
{Launhardt}, R., {Nutter}, D., {Ward-Thompson}, D., {Bourke}, T.~L., {Henning},
  T., {Khanzadyan}, T., {Schmalzl}, M., {Wolf}, S., \& {Zylka}, R. 2010, \apjs,
  188, 139

\bibitem[{{Launhardt} {et~al.}(2001){Launhardt}, {Sargent}, \&
  {Zinnecker}}]{launhardt2001}
{Launhardt}, R., {Sargent}, A., \& {Zinnecker}, H. 2001, in Astronomical
  Society of the Pacific Conference Series, Vol. 235, Science with the Atacama
  Large Millimeter Array, ed. {A.~Wootten}, 134--+

\bibitem[{{Lee} {et~al.}(2009){Lee}, {Hirano}, {Palau}, {Ho}, {Bourke},
  {Zhang}, \& {Shang}}]{lee2009}
{Lee}, C., {Hirano}, N., {Palau}, A., {Ho}, P.~T.~P., {Bourke}, T.~L., {Zhang},
  Q., \& {Shang}, H. 2009, \apj, 699, 1584

\bibitem[{{Lee} {et~al.}(2004){Lee}, {Bergin}, \& {Evans}}]{lee2004}
{Lee}, J., {Bergin}, E.~A., \& {Evans}, II, N.~J. 2004, \apj, 617, 360

\bibitem[{{Loinard} {et~al.}(2008){Loinard}, {Torres}, {Mioduszewski}, \&
  {Rodr{\'{\i}}guez}}]{loinard2008}
{Loinard}, L., {Torres}, R.~M., {Mioduszewski}, A.~J., \& {Rodr{\'{\i}}guez},
  L.~F. 2008, \apjl, 675, L29

\bibitem[{{Loinard} {et~al.}(2007){Loinard}, {Torres}, {Mioduszewski},
  {Rodr{\'{\i}}guez}, {Gonz{\'a}lez-L{\'o}pezlira}, {Lachaume}, {V{\'a}zquez},
  \& {Gonz{\'a}lez}}]{loinard2007}
{Loinard}, L., {Torres}, R.~M., {Mioduszewski}, A.~J., {Rodr{\'{\i}}guez},
  L.~F., {Gonz{\'a}lez-L{\'o}pezlira}, R.~A., {Lachaume}, R., {V{\'a}zquez},
  V., \& {Gonz{\'a}lez}, E. 2007, \apj, 671, 546

\bibitem[{{Looney} {et~al.}(2007){Looney}, {Tobin}, \& {Kwon}}]{looney2007}
{Looney}, L.~W., {Tobin}, J.~J., \& {Kwon}, W. 2007, \apjl, 670, L131

\bibitem[{{Lovas}(1992)}]{lovas}
{Lovas}, F.~J. 1992, Journal of Physical and Chemical Reference Data, 21, 181

\bibitem[{{Markwardt}(2009)}]{markwardt2009}
{Markwardt}, C.~B. 2009, in Astronomical Society of the Pacific Conference
  Series, Vol. 411, Astronomical Data Analysis Software and Systems XVIII, ed.
  {D.~A.~Bohlender, D.~Durand, \& P.~Dowler}, 251--+

\bibitem[{{McCaughrean} {et~al.}(1994){McCaughrean}, {Rayner}, \&
  {Zinnecker}}]{mccaughrean1994}
{McCaughrean}, M.~J., {Rayner}, J.~T., \& {Zinnecker}, H. 1994, \apjl, 436,
  L189

\bibitem[{{McKee} \& {Ostriker}(2007)}]{mckeeostriker2007}
{McKee}, C.~F., \& {Ostriker}, E.~C. 2007, \araa, 45, 565

\bibitem[{{Menten} {et~al.}(2007){Menten}, {Reid}, {Forbrich}, \&
  {Brunthaler}}]{menten2007}
{Menten}, K.~M., {Reid}, M.~J., {Forbrich}, J., \& {Brunthaler}, A. 2007, \aap,
  474, 515

\bibitem[{{Myers} {et~al.}(1995){Myers}, {Bachiller}, {Caselli}, {Fuller},
  {Mardones}, {Tafalla}, \& {Wilner}}]{myers1995}
{Myers}, P.~C., {Bachiller}, R., {Caselli}, P., {Fuller}, G.~A., {Mardones},
  D., {Tafalla}, M., \& {Wilner}, D.~J. 1995, \apjl, 449, L65+

\bibitem[{{Myers} {et~al.}(1991){Myers}, {Fuller}, {Goodman}, \&
  {Benson}}]{myers1991}
{Myers}, P.~C., {Fuller}, G.~A., {Goodman}, A.~A., \& {Benson}, P.~J. 1991,
  \apj, 376, 561

\bibitem[{{Pineda} {et~al.}(2010){Pineda}, {Goodman}, {Arce}, {Caselli},
  {Foster}, {Myers}, \& {Rosolowsky}}]{pineda2010}
{Pineda}, J.~E., {Goodman}, A.~A., {Arce}, H.~G., {Caselli}, P., {Foster},
  J.~B., {Myers}, P.~C., \& {Rosolowsky}, E.~W. 2010, \apjl, 712, L116

\bibitem[{{Pudritz} \& {Norman}(1983)}]{pudritz1983}
{Pudritz}, R.~E., \& {Norman}, C.~A. 1983, \apj, 274, 677

\bibitem[{{Raghavan} {et~al.}(2010){Raghavan}, {McAlister}, {Henry}, {Latham},
  {Marcy}, {Mason}, {Gies}, {White}, \& {ten Brummelaar}}]{raghavan2010}
{Raghavan}, D., {McAlister}, H.~A., {Henry}, T.~J., {Latham}, D.~W., {Marcy},
  G.~W., {Mason}, B.~D., {Gies}, D.~R., {White}, R.~J., \& {ten Brummelaar},
  T.~A. 2010, \apjs, 190, 1

\bibitem[{{Rebull} {et~al.}(2007){Rebull}, {Stapelfeldt}, {Evans},
  {J{\o}rgensen}, {Harvey}, {Brooke}, {Bourke}, {Padgett}, {Chapman}, {Lai},
  {Spiesman}, {Noriega-Crespo}, {Mer{\'{\i}}n}, {Huard}, {Allen}, {Blake},
  {Jarrett}, {Koerner}, {Mundy}, {Myers}, {Sargent}, {van Dishoeck}, {Wahhaj},
  \& {Young}}]{rebull2007}
{Rebull}, L.~M., {Stapelfeldt}, K.~R., {Evans}, II, N.~J., {J{\o}rgensen},
  J.~K., {Harvey}, P.~M., {Brooke}, T.~Y., {Bourke}, T.~L., {Padgett}, D.~L.,
  {Chapman}, N.~L., {Lai}, S., {Spiesman}, W.~J., {Noriega-Crespo}, A.,
  {Mer{\'{\i}}n}, B., {Huard}, T., {Allen}, L.~E., {Blake}, G.~A., {Jarrett},
  T., {Koerner}, D.~W., {Mundy}, L.~G., {Myers}, P.~C., {Sargent}, A.~I., {van
  Dishoeck}, E.~F., {Wahhaj}, Z., \& {Young}, K.~E. 2007, \apjs, 171, 447

\bibitem[{{Reipurth} {et~al.}(1996){Reipurth}, {Raga}, \&
  {Heathcote}}]{reipurth1996}
{Reipurth}, B., {Raga}, A.~C., \& {Heathcote}, S. 1996, \aap, 311, 989

\bibitem[{{Rich} {et~al.}(2008){Rich}, {de Blok}, {Cornwell}, {Brinks},
  {Walter}, {Bagetakos}, \& {Kennicutt}}]{rich2008}
{Rich}, J.~W., {de Blok}, W.~J.~G., {Cornwell}, T.~J., {Brinks}, E., {Walter},
  F., {Bagetakos}, I., \& {Kennicutt}, R.~C. 2008, \aj, 136, 2897

\bibitem[{{Rodr{\'{\i}}guez} {et~al.}(1998){Rodr{\'{\i}}guez}, {Reipurth},
  {Raga}, \& {Cant{\'o}}}]{rodriguez1998}
{Rodr{\'{\i}}guez}, L.~F., {Reipurth}, B., {Raga}, A.~C., \& {Cant{\'o}}, J.
  1998, \rmxaa, 34, 69

\bibitem[{{Rydbeck} {et~al.}(1977){Rydbeck}, {Sume}, {Hjalmarson}, {Ellder},
  {Ronnang}, \& {Kollberg}}]{rydbeck1977}
{Rydbeck}, O.~E.~H., {Sume}, A., {Hjalmarson}, A., {Ellder}, J., {Ronnang},
  B.~O., \& {Kollberg}, E. 1977, \apjl, 215, L35

\bibitem[{{Sault} {et~al.}(1995){Sault}, {Teuben}, \& {Wright}}]{sault1995}
{Sault}, R.~J., {Teuben}, P.~J., \& {Wright}, M.~C.~H. 1995, in Astronomical
  Society of the Pacific Conference Series, Vol.~77, Astronomical Data Analysis
  Software and Systems IV, ed. {R.~A.~Shaw, H.~E.~Payne, \& J.~J.~E.~Hayes},
  433--+

\bibitem[{{Sch{\"o}ier} {et~al.}(2005){Sch{\"o}ier}, {van der Tak}, {van
  Dishoeck}, \& {Black}}]{schoier2005}
{Sch{\"o}ier}, F.~L., {van der Tak}, F.~F.~S., {van Dishoeck}, E.~F., \&
  {Black}, J.~H. 2005, \aap, 432, 369

\bibitem[{{Scholz} {et~al.}(2010){Scholz}, {Wood}, {Wilner}, {Jayawardhana},
  {Delorme}, {Caratti O Garatti}, {Ivanov}, {Saviane}, \&
  {Whitney}}]{scholz2010}
{Scholz}, A., {Wood}, K., {Wilner}, D., {Jayawardhana}, R., {Delorme}, P.,
  {Caratti O Garatti}, A., {Ivanov}, V.~D., {Saviane}, I., \& {Whitney}, B.
  2010, \mnras, 409, 1557

\bibitem[{{Seale} \& {Looney}(2008)}]{seale2008}
{Seale}, J.~P., \& {Looney}, L.~W. 2008, \apj, 675, 427

\bibitem[{{Shirley} {et~al.}(2000){Shirley}, {Evans}, {Rawlings}, \&
  {Gregersen}}]{shirley2000}
{Shirley}, Y.~L., {Evans}, II, N.~J., {Rawlings}, J.~M.~C., \& {Gregersen},
  E.~M. 2000, \apjs, 131, 249

\bibitem[{{Shu} {et~al.}(1994){Shu}, {Najita}, {Ostriker}, {Wilkin}, {Ruden},
  \& {Lizano}}]{shu1994}
{Shu}, F., {Najita}, J., {Ostriker}, E., {Wilkin}, F., {Ruden}, S., \&
  {Lizano}, S. 1994, \apj, 429, 781

\bibitem[{{Shu}(1977)}]{shu1977}
{Shu}, F.~H. 1977, \apj, 214, 488

\bibitem[{{Shu} {et~al.}(1987){Shu}, {Adams}, \& {Lizano}}]{shu1987}
{Shu}, F.~H., {Adams}, F.~C., \& {Lizano}, S. 1987, \araa, 25, 23

\bibitem[{{Smith} {et~al.}(2011){Smith}, {Glover}, {Bonnell}, {Clark}, \&
  {Klessen}}]{smith2011}
{Smith}, R.~J., {Glover}, S.~C.~O., {Bonnell}, I.~A., {Clark}, P.~C., \&
  {Klessen}, R.~S. 2011, \mnras, 411, 1354

\bibitem[{{Stanke} {et~al.}(2006){Stanke}, {Smith}, {Gredel}, \&
  {Khanzadyan}}]{stanke2006}
{Stanke}, T., {Smith}, M.~D., {Gredel}, R., \& {Khanzadyan}, T. 2006, \aap,
  447, 609

\bibitem[{{Stutz} {et~al.}(2009){Stutz}, {Rieke}, {Bieging}, {Balog},
  {Heitsch}, {Kang}, {Peters}, {Shirley}, \& {Werner}}]{stutz2009}
{Stutz}, A.~M., {Rieke}, G.~H., {Bieging}, J.~H., {Balog}, Z., {Heitsch}, F.,
  {Kang}, M., {Peters}, W.~L., {Shirley}, Y.~L., \& {Werner}, M.~W. 2009, \apj,
  707, 137

\bibitem[{{Tafalla} {et~al.}(1998){Tafalla}, {Mardones}, {Myers}, {Caselli},
  {Bachiller}, \& {Benson}}]{tafalla1998}
{Tafalla}, M., {Mardones}, D., {Myers}, P.~C., {Caselli}, P., {Bachiller}, R.,
  \& {Benson}, P.~J. 1998, \apj, 504, 900

\bibitem[{{Tafalla} {et~al.}(2004){Tafalla}, {Myers}, {Caselli}, \&
  {Walmsley}}]{tafalla2004}
{Tafalla}, M., {Myers}, P.~C., {Caselli}, P., \& {Walmsley}, C.~M. 2004, \aap,
  416, 191

\bibitem[{{Tafalla} {et~al.}(2000){Tafalla}, {Myers}, {Mardones}, \&
  {Bachiller}}]{tafalla2000}
{Tafalla}, M., {Myers}, P.~C., {Mardones}, D., \& {Bachiller}, R. 2000, \aap,
  359, 967

\bibitem[{{Tanner} \& {Arce}(2011)}]{tanner2011}
{Tanner}, J.~D., \& {Arce}, H.~G. 2011, \apj, 726, 40

\bibitem[{{Terebey} {et~al.}(2009){Terebey}, {Fich}, {Noriega-Crespo},
  {Padgett}, {Fukagawa}, {Audard}, {Brooke}, {Carey}, {Evans}, {Guedel},
  {Hines}, {Huard}, {Knapp}, {McCabe}, {Menard}, {Monin}, \&
  {Rebull}}]{terebey2009}
{Terebey}, S., {Fich}, M., {Noriega-Crespo}, A., {Padgett}, D.~L., {Fukagawa},
  M., {Audard}, M., {Brooke}, T., {Carey}, S., {Evans}, N.~J., {Guedel}, M.,
  {Hines}, D., {Huard}, T., {Knapp}, G.~R., {McCabe}, C., {Menard}, F.,
  {Monin}, J., \& {Rebull}, L. 2009, \apj, 696, 1918

\bibitem[{{Terebey} {et~al.}(1984){Terebey}, {Shu}, \& {Cassen}}]{tsc1984}
{Terebey}, S., {Shu}, F.~H., \& {Cassen}, P. 1984, \apj, 286, 529

\bibitem[{{Thompson} {et~al.}(2001){Thompson}, {Moran}, \&
  {Swenson}}]{thompsonmoranswenson}
{Thompson}, A.~R., {Moran}, J.~M., \& {Swenson}, Jr., G.~W. 2001,
  {Interferometry and Synthesis in Radio Astronomy, 2nd Edition}, ed.
  {Thompson, A.~R., Moran, J.~M., \& Swenson, G.~W., Jr.}

\bibitem[{{Tobin} {et~al.}(2008){Tobin}, {Hartmann}, {Calvet}, \&
  {D'Alessio}}]{tobin2008}
{Tobin}, J.~J., {Hartmann}, L., {Calvet}, N., \& {D'Alessio}, P. 2008, \apj,
  679, 1364

\bibitem[{{Tobin} {et~al.}(2010{\natexlab{a}}){Tobin}, {Hartmann}, \&
  {Loinard}}]{tobin2010b}
{Tobin}, J.~J., {Hartmann}, L., \& {Loinard}, L. 2010{\natexlab{a}}, \apjl,
  722, L12

\bibitem[{{Tobin} {et~al.}(2010{\natexlab{b}}){Tobin}, {Hartmann}, {Looney}, \&
  {Chiang}}]{tobin2010a}
{Tobin}, J.~J., {Hartmann}, L., {Looney}, L.~W., \& {Chiang}, H.
  2010{\natexlab{b}}, \apj, 712, 1010

\bibitem[{{Tsitali} {et~al.}(2010){Tsitali}, {Bourke}, {Peterson}, {Myers},
  {Dunham}, {Evans}, \& {Huard}}]{tsitali2010}
{Tsitali}, A.~E., {Bourke}, T.~L., {Peterson}, D.~E., {Myers}, P.~C., {Dunham},
  M.~M., {Evans}, N.~J., \& {Huard}, T.~L. 2010, \apj, 725, 2461

\bibitem[{{Visser} {et~al.}(2002){Visser}, {Richer}, \&
  {Chandler}}]{visser2002}
{Visser}, A.~E., {Richer}, J.~S., \& {Chandler}, C.~J. 2002, \aj, 124, 2756

\bibitem[{{Volgenau} {et~al.}(2006){Volgenau}, {Mundy}, {Looney}, \&
  {Welch}}]{volgenau2006}
{Volgenau}, N.~H., {Mundy}, L.~G., {Looney}, L.~W., \& {Welch}, W.~J. 2006,
  \apj, 651, 301

\bibitem[{{Whitney} {et~al.}(2003){Whitney}, {Wood}, {Bjorkman}, \&
  {Cohen}}]{whitney2003b}
{Whitney}, B.~A., {Wood}, K., {Bjorkman}, J.~E., \& {Cohen}, M. 2003, \apj,
  598, 1079

\bibitem[{{Young} {et~al.}(2006){Young}, {Enoch}, {Evans}, {Glenn}, {Sargent},
  {Huard}, {Aguirre}, {Golwala}, {Haig}, {Harvey}, {Laurent}, {Mauskopf}, \&
  {Sayers}}]{young2006}
{Young}, K.~E., {Enoch}, M.~L., {Evans}, II, N.~J., {Glenn}, J., {Sargent}, A.,
  {Huard}, T.~L., {Aguirre}, J., {Golwala}, S., {Haig}, D., {Harvey}, P.,
  {Laurent}, G., {Mauskopf}, P., \& {Sayers}, J. 2006, \apj, 644, 326

\bibitem[{{Yun}(1996)}]{yun1996}
{Yun}, J.~L. 1996, \aj, 111, 930

\bibitem[{{Zhou} {et~al.}(1993){Zhou}, {Evans}, {Koempe}, \&
  {Walmsley}}]{zhou1993}
{Zhou}, S., {Evans}, II, N.~J., {Koempe}, C., \& {Walmsley}, C.~M. 1993, \apj,
  404, 232

\end{thebibliography}

\clearpage

\begin{figure}
\begin{center}
\includegraphics[scale=0.75]{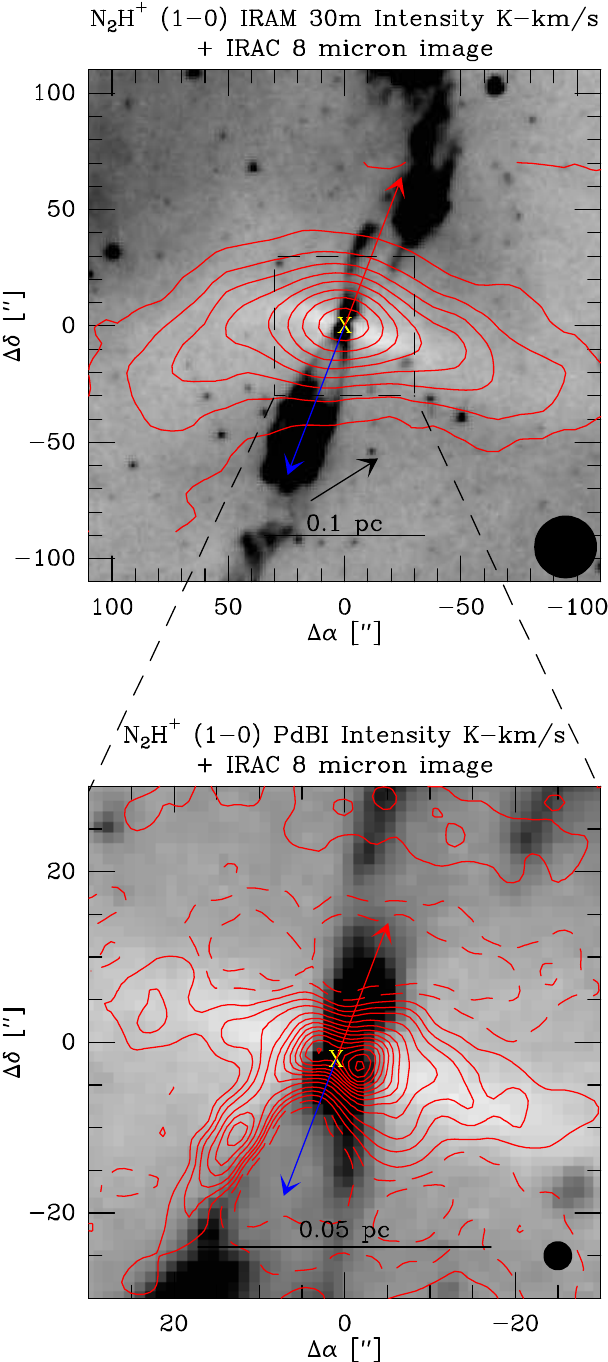}
\includegraphics[scale=0.75]{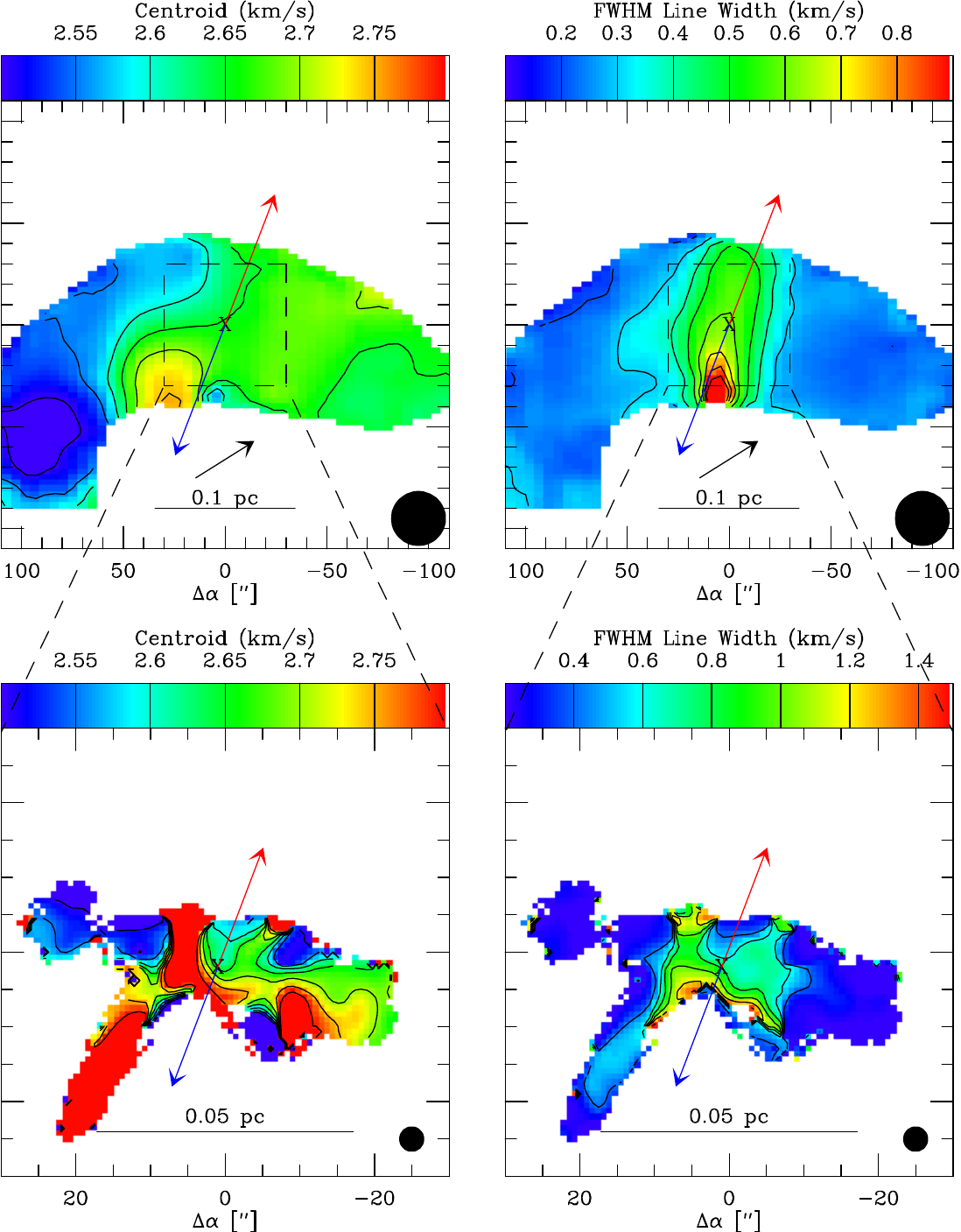}
\end{center}
\caption{
L1157-- The \textit{top row} shows the data from the IRAM 30m and the \textit{bottom row} shows
the PdBI data. In the \textit{left column}, the 8\mum\ IRAC images with \nthp\ ($J=1\rightarrow0$) integrated intensity
contours overlaid. The IRAM 30m contours start at the 10$\sigma$ level and increase in 20$\sigma$ increments while
the PdBI data start at the $\pm$5$\sigma$ level and increase in 20$\sigma$ increments,
see Tables 6 - 8 for values of $\sigma_I$. The \textit{middle column} shows the line-center velocity fit
of the \nthp\ emission across the envelope and the \textit{right column} shows the FWHM linewidth.
The \textit{red and blue lines} mark the central axis of the outflow and their respective orientation in the plane
of the sky; the X marks the position of the protostar from dust continuum and/or 24\mum\
emission and the \textit{black arrows} indicate the direction of the velocity gradient derived
from the velocity field. The velocity gradient direction for the PdBI data could not be fit
for L1157. The single-dish data show an ordered velocity gradient along the filamentary envelope, following the high density
region as it curves south. The interferometer map also reflects this velocity gradient, but there is a reversal with red-shifted
emission just east of the protostar. There is also a red-shifted feature that appears in the interferometer and single-dish
map just southeast of the protostar, this appears to be due to outflow interaction given its location along the cavity wall.
Both the single-dish and interferometer maps have large linewidths near the protostar, which appears to be due to 
outflow interaction. Reference positions for the observations are listed in Tables 2 and 5.}
\label{L1157}
\end{figure}
\clearpage

\begin{figure}
\begin{center}
\includegraphics[scale=0.75]{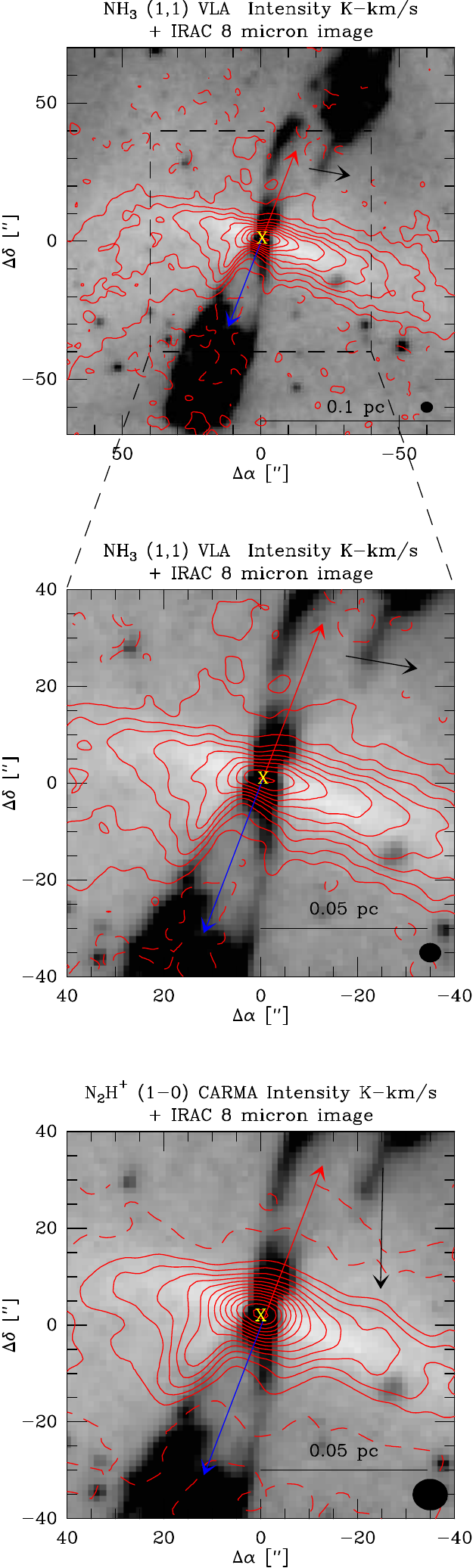}
\includegraphics[scale=0.75]{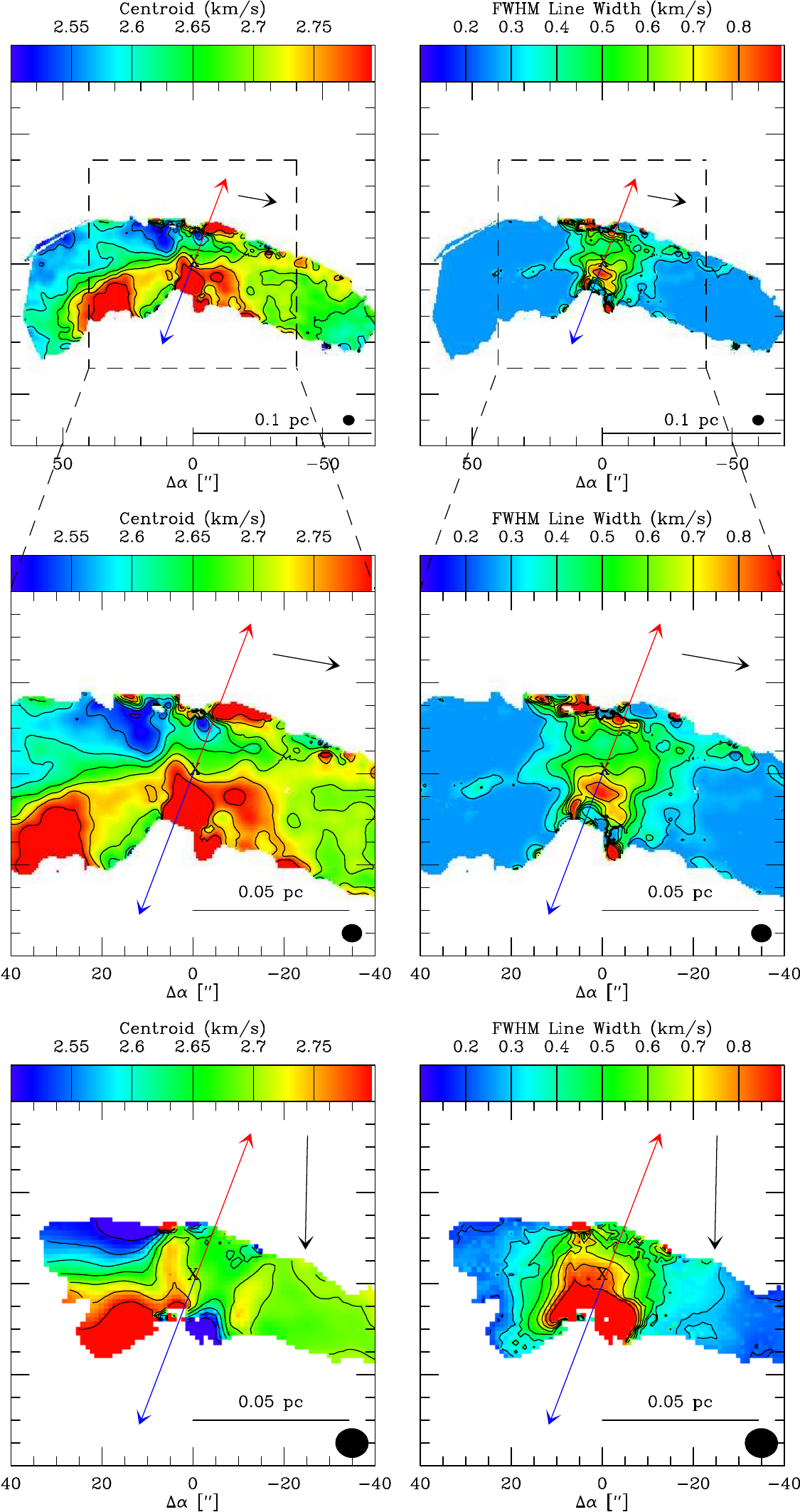}
\end{center}
\caption{L1157-- The \textit{top and middle rows} show the VLA \nht\ (1,1) data while the \textit{bottom row}
shows the CARMA \nthp\ ($J=1\rightarrow0$) data. In the \textit{left column}, the 8\mum\ IRAC images with integrated
intensity contours overlaid. The VLA contours start at the $\pm$3$\sigma$ level and increase in 6$\sigma$ increments while
the CARMA data start at the $\pm$3$\sigma$ level and increase in 3$\sigma$ increments,
see Tables 6 - 8 for values of $\sigma_I$. The \textit{middle column} shows
the line-center velocity of the gas across the envelope and the \textit{right column} shows the FWHM linewidth.
The \textit{red and blue lines} mark the central axis of the outflow and their respective orientation in the plane
of the sky; the X marks the position of the protostar from dust continuum and 24$\mum$
emission and the \textit{black arrows} indicate the direction of the velocity gradient derived
from the velocity field. Notice the close
correspondence of the intensity distribution and line kinematics between the \nht\ and \nthp\ maps indicating that they
are likely tracing the same material. Reference positions for the observations are listed in Tables 2 - 5.}
\label{L1157-2}
\end{figure}
\clearpage

\begin{figure}
\begin{center}
\includegraphics[scale=0.75,angle=-90]{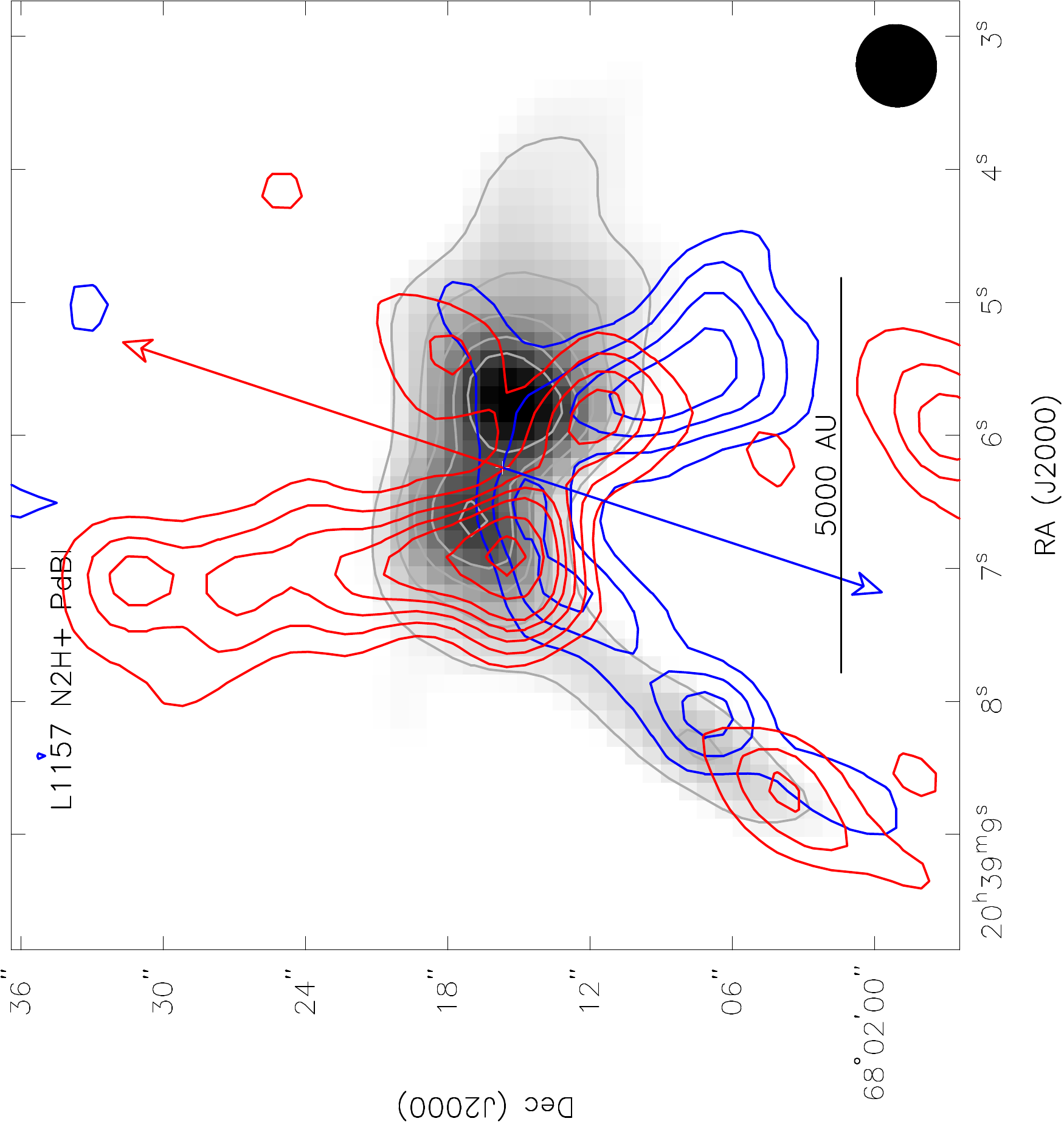}
\end{center}
\caption{L1157-- Maps of \nthp\ emission from the PdBI in L1157 in three velocity ranges; the
blue and red lines denote the blue and red-shifted sides of the outflow \citep{jorgensen2007}. 
The line-center emission from 2.33-3.1 \kms\ is plotted as grayscale
with light gray contours in units of 15, 30, 25, 60, 75$\sigma$ where $\sigma$=0.027 K \kms.
The red contours are emission between 3.2 and 4 \kms\ and the blue contours are emission
between 1.57 and 2.2 \kms and plotted in units of 3, 6, 9, 12, 15, 21, 27$\sigma$. The red- and blue-shifted emission 
near the protostar is clearly shifted along the blue-shifted side of the outflow and both the blue and red-shifted emission 
seem to outline the outflow cavity wall. The spatial shifts along the outflow in the blue and red-shifted emission strongly
suggest that the outflow is entraining material from the inner envelope.} 
\label{L1157-outflow}
\end{figure}

\clearpage

\begin{figure}
\begin{center}
\includegraphics[scale=0.75]{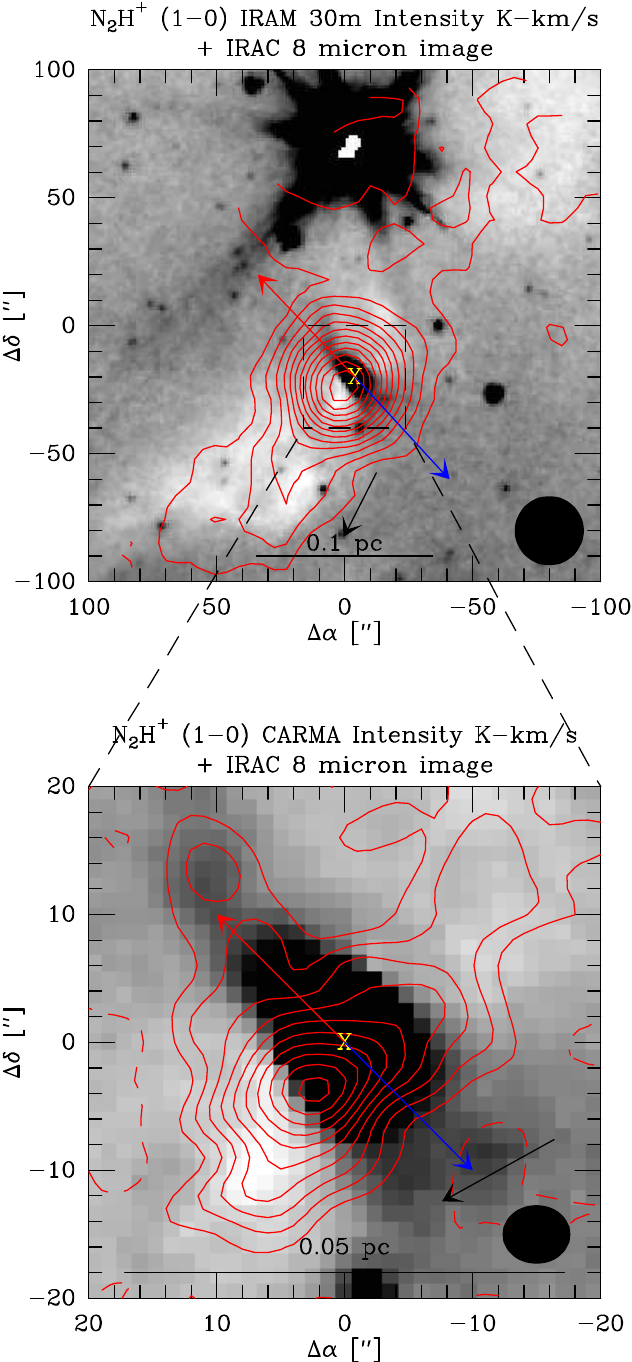}
\includegraphics[scale=0.75]{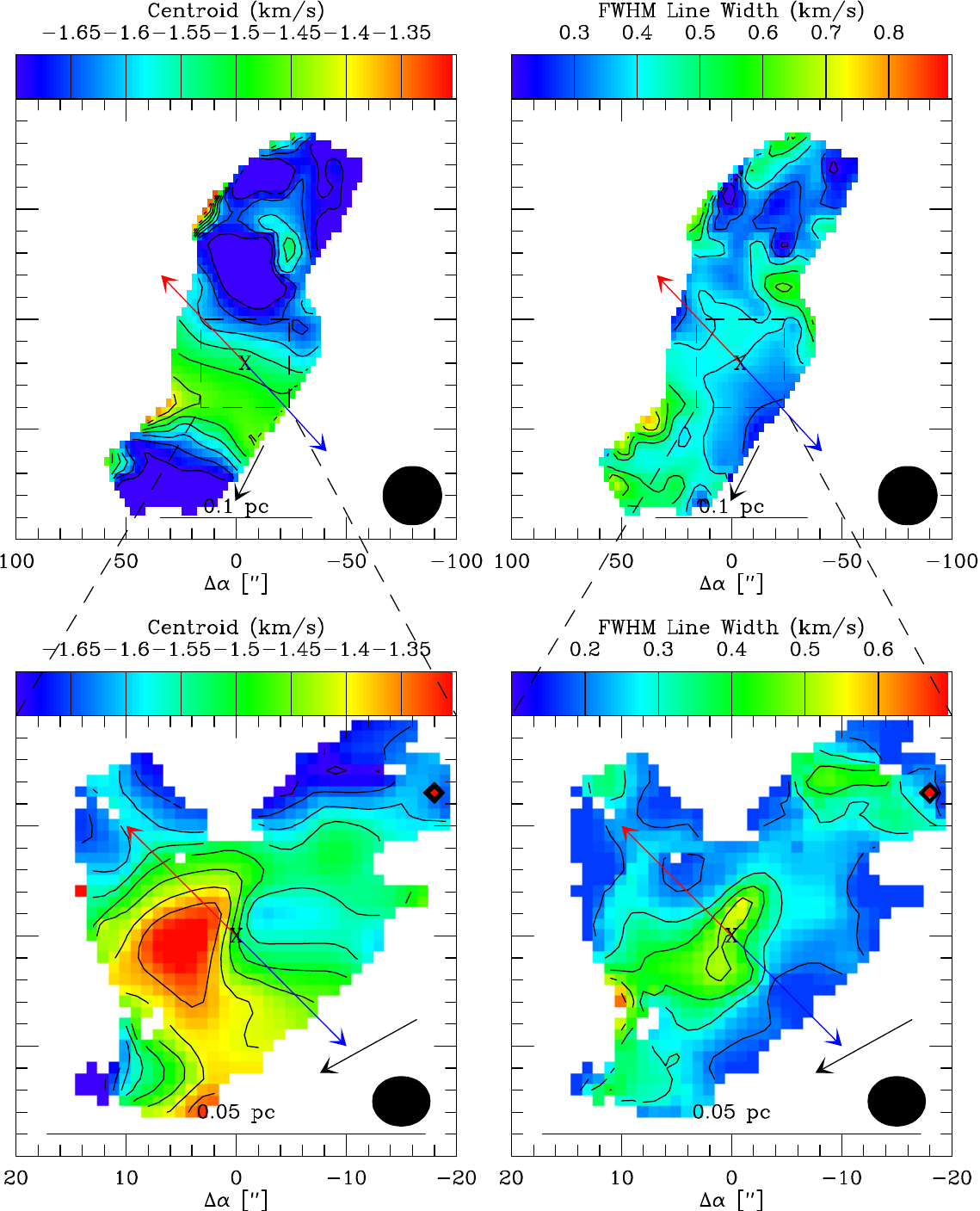}
\end{center}
\caption{L1165-- Same as Figure \ref{L1157} but with CARMA data in the \textit{bottom row}. The IRAM 30m contours
start at 5$\sigma$ with 5$\sigma$ intervals; the CARMA data start at $\pm$3$\sigma$ with 3$\sigma$ intervals.
The single-dish data show that the cloud has a fairly constant velocity away from the protostar, but near
the protostar the kinematic structure is distinct; however, the single-dish linewidth map shows
no indication of enhancement near the protostar. The velocity field from the interferometer map shows considerable
detail, with a velocity gradient across the protostar nearly normal to the outflow. There is also enhanced linewidth near the
protostar in the central envelope. Reference positions for the observations are listed in Tables 2 - 5.}
\label{L1165}
\end{figure}
\clearpage

\begin{figure}
\begin{center}
\includegraphics[scale=0.75,angle=-90]{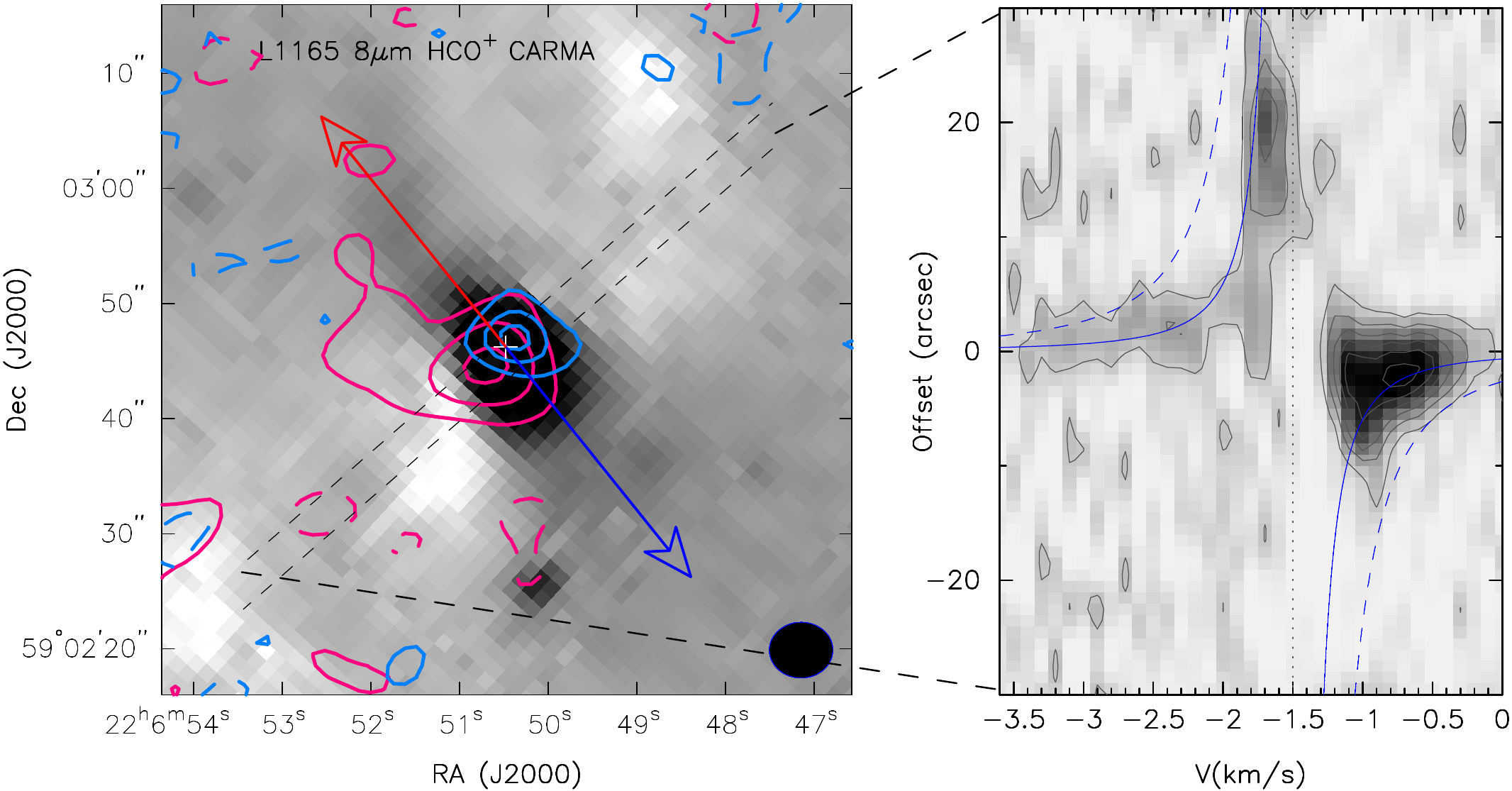}
\end{center}
\caption{L1165-- The \textit{left panel} shows the IRAC 8\mum\ image with CARMA HCO$^+$
blue and red-shifted emission, summed over -3.5 to -2 \kms\ and -1.0 to 0 \kms, 
plotted as blue and red contours respectively. The contours levels are $\pm$3, 6, and 8.25$\sigma$ ($\sigma$=0.175 K)
for the blue-shifted emission and $\pm$3, 9, 18, and 27$\sigma$ ($\sigma$=0.212 K) for the red-shifted emission.
 The blue and red-shifted emission from HCO$^+$ is located symmetrically about
the protostar, normal to the outflow. The dashed lines mark the regions where
the position-velocity cut was taken and point to respective ends of the PV plot in
the \textit{right panel}. The position of the protostar/continuum source is marked
with a white cross. The position-velocity cut shows that the blue and red-shifted
emission traces higher velocity material and there is a slight gradient of material
going to higher velocity closer to the continuum source. The PV plot contours start at 3$\sigma$
and increase in 3$\sigma$ intervals ($\sigma$=0.2). The \textit{solid-blue} curve represents Keplerian
rotation (or infall) for a 0.5 $M_{\sun}$ central object and the \textit{dashed-blue} curve is for
a 2.0 $M_{\sun}$ central object.  }
\label{L1165-pv-hco}
\end{figure}
\clearpage

\begin{figure}
\begin{center}
\includegraphics[scale=0.75]{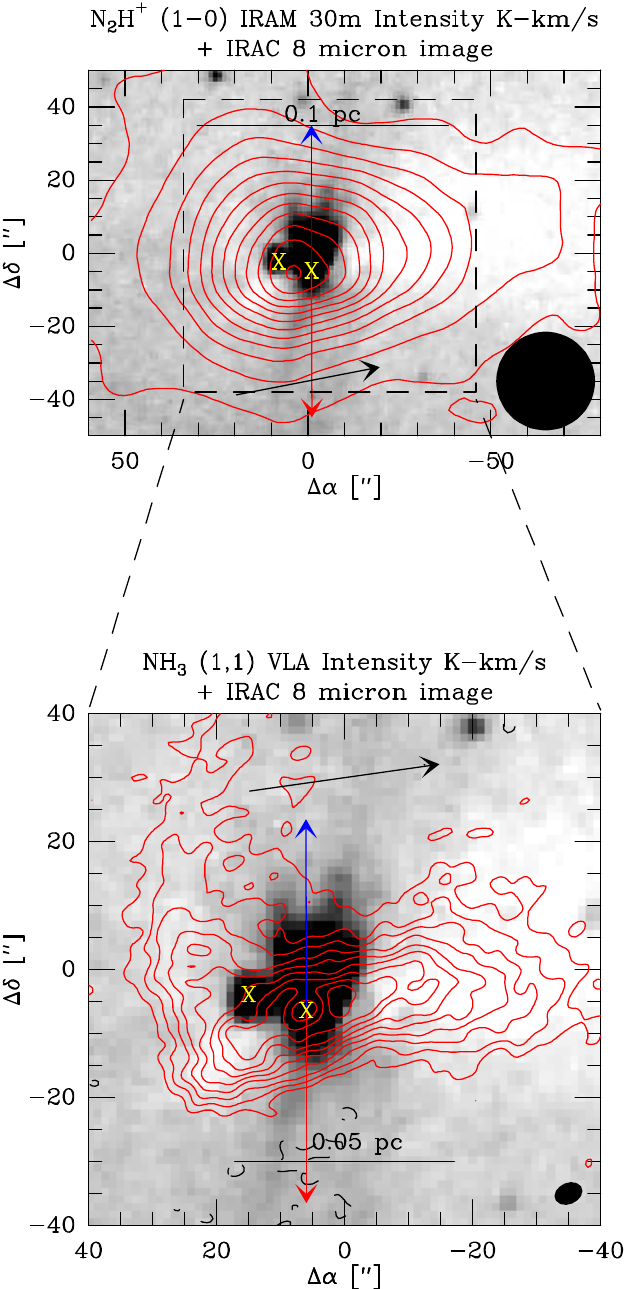}
\includegraphics[scale=0.75]{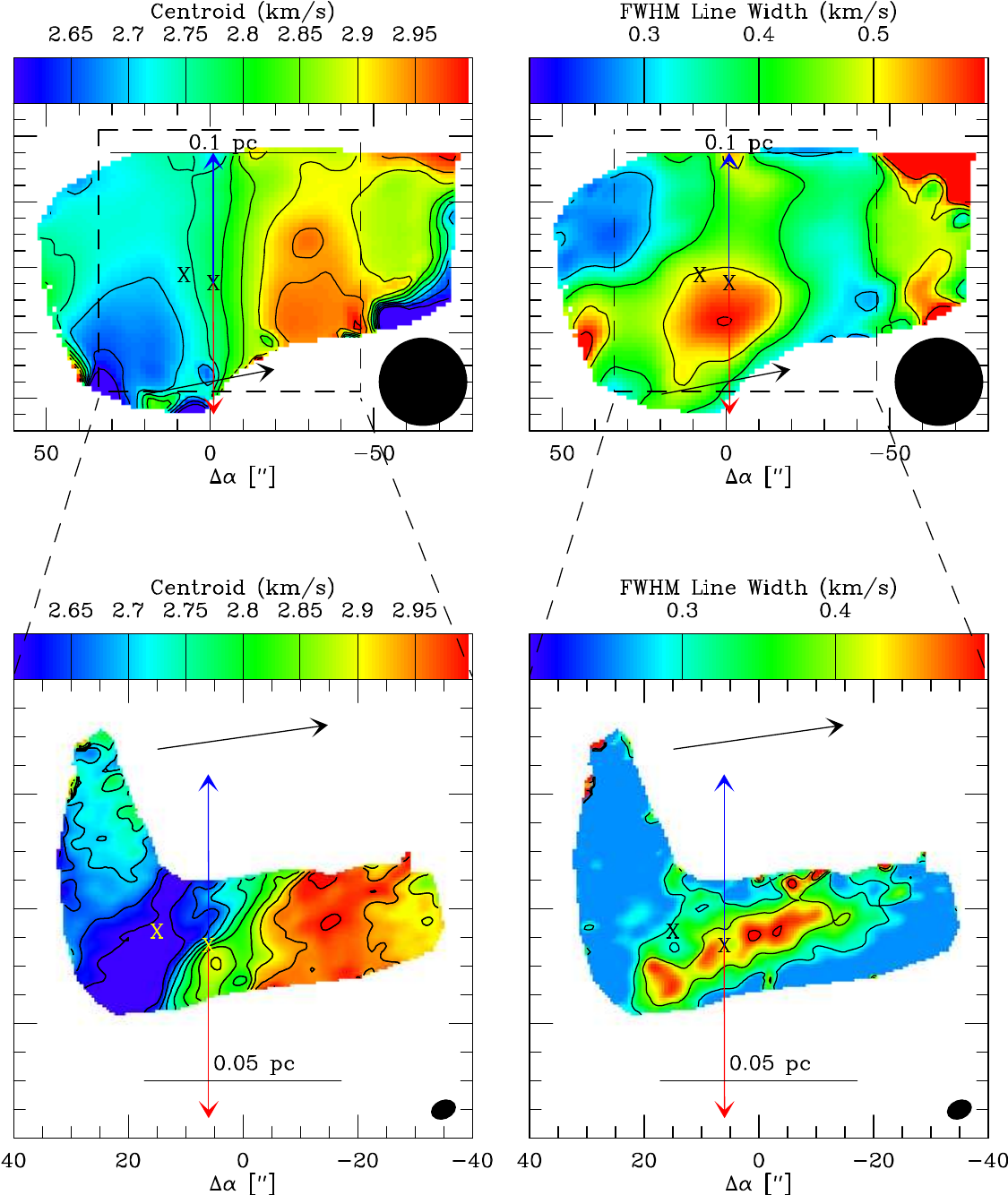}
\end{center}
\caption{CB230-- Same as Figure \ref{L1157}, but the interferometer data are VLA \nht\ (1,1) observations.
The IRAM 30m contours start at 3$\sigma$ with 10$\sigma$ intervals; the VLA data start at $\pm$3$\sigma$ with 3$\sigma$ intervals.
The line-center velocity from the single-dish data traces a fairly smooth velocity gradient across the envelope with enhanced
linewidth near the protostar. The \nht\ intensity is less extended directly east of the protostar associated
with the cutoff of 8\mum\ extinction, the rest of the envelope appears fairly flat. The line-center velocity
from the \nht\ emission traces a velocity gradient similar to the single-dish data; however, the shift
from red to blue-shifted emission is quite abrupt and the transition region itself is curved. The 
\nht\ line-width does not show much detail other than having its peak coincident with the highest intensity
\nht\ emission.}
\label{CB230}
\end{figure}
\clearpage

\begin{figure}
\begin{center}
\includegraphics[scale=0.75]{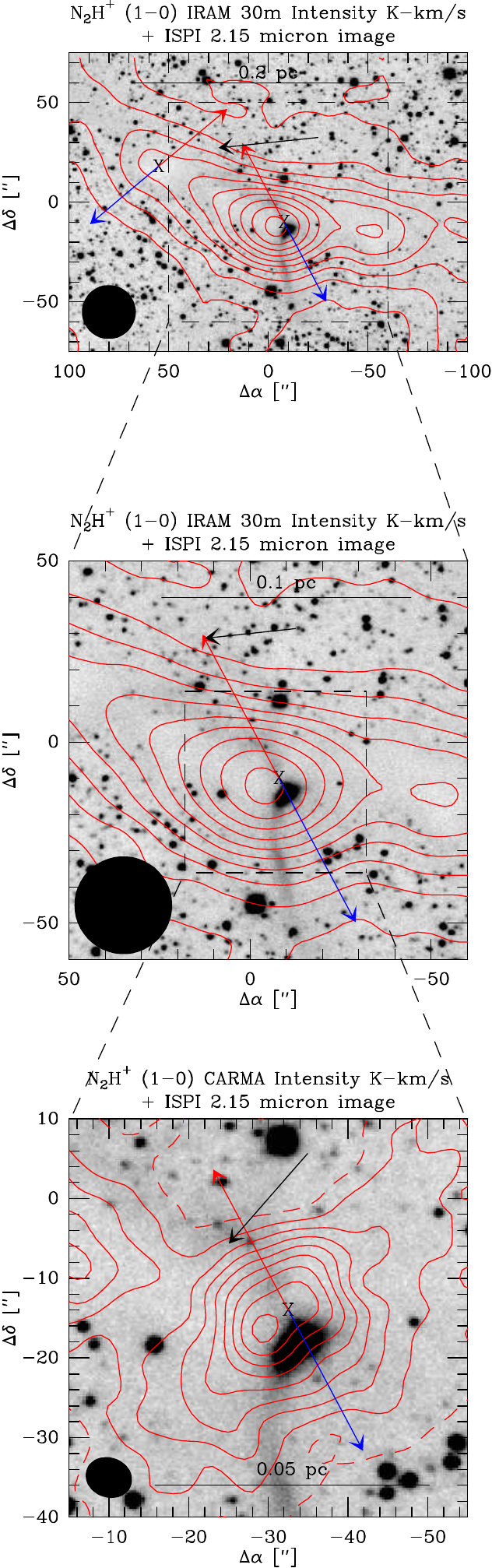}
\includegraphics[scale=0.75]{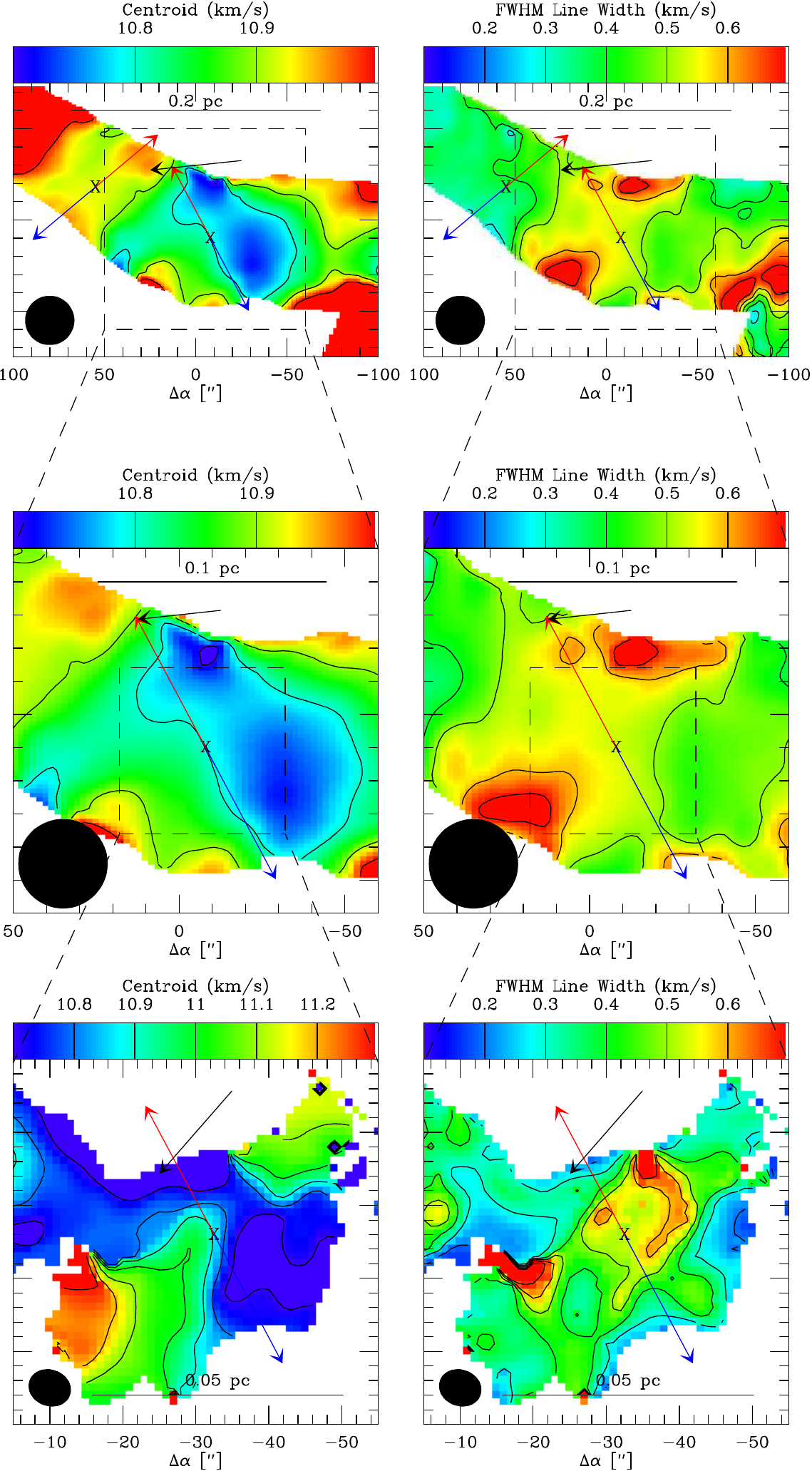}
\end{center}
\caption{HH108IRS-- Same as Figure \ref{L1165}, but zooming in on the single-dish data in the \textit{middle row} and then the
interferometer data in the \textit{bottom row}. The IRAM 30m contours
start at 3$\sigma$ with 10$\sigma$ intervals; the CARMA data start at $\pm$3$\sigma$ with 6$\sigma$ intervals.
 The \nthp\ peak in this object appears offset from the protostar 
in the single-dish and interferometer map. The interferometer map further appears to be elongated normal
to the outflow. The single-dish line-center velocity map appears to show a slight gradient normal to the outflow,
but the interferometric velocity field starts red-shifted, becomes
blue-shifted and then goes back to red-shifted. The single-dish linewidth maps show a slight enhancement near the protostar;
the interferometer data on the other hand show a highly increased linewidth near the protostar as compared to the surrounding
region. }
\label{HH108IRS}
\end{figure}
\clearpage

\begin{figure}
\begin{center}
\includegraphics[scale=0.75]{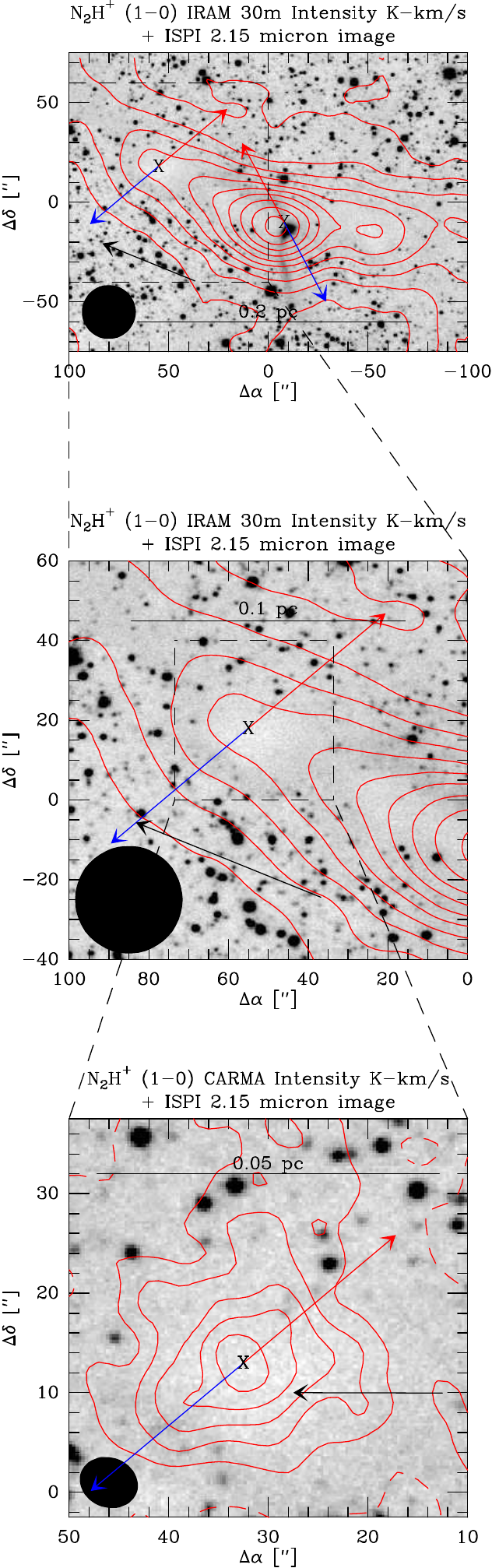}
\includegraphics[scale=0.75]{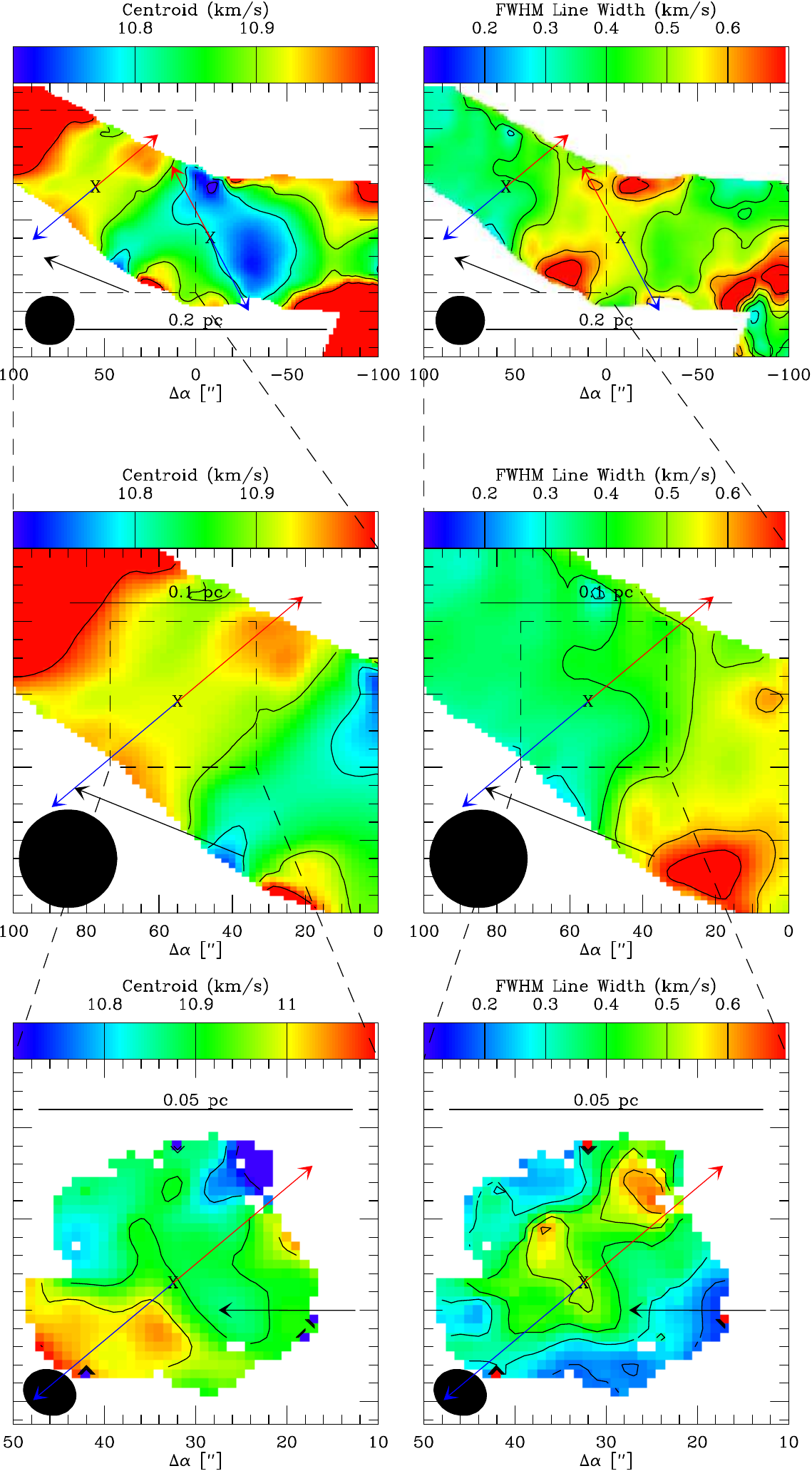}
\end{center}
\caption{HH108MMS-- Same as Figure \ref{HH108IRS}. The IRAM 30m contours
start at 3$\sigma$ with 10$\sigma$ intervals; the CARMA data start at $\pm$3$\sigma$ with 3$\sigma$ intervals.
The single-dish integrated intensity map shows HH108MMS as an extension
along a large-scale filamentary structure that connects with HH108IRS.
However, the interferometer map finds the \nthp\ peak
directly coincident with the protostar. The single-dish velocity field is fairly constant throughout the region of the 
envelope, but the large scale gradient is normal to its outflow. The CARMA velocity map then shows that there is
a gradient along the outflow and the linewidth map shows increased linewidth along the outflow,
this detail was absent in the single-dish map.}
\label{HH108MMS}
\end{figure}
\clearpage

\begin{figure}
\begin{center}
\includegraphics[scale=0.75]{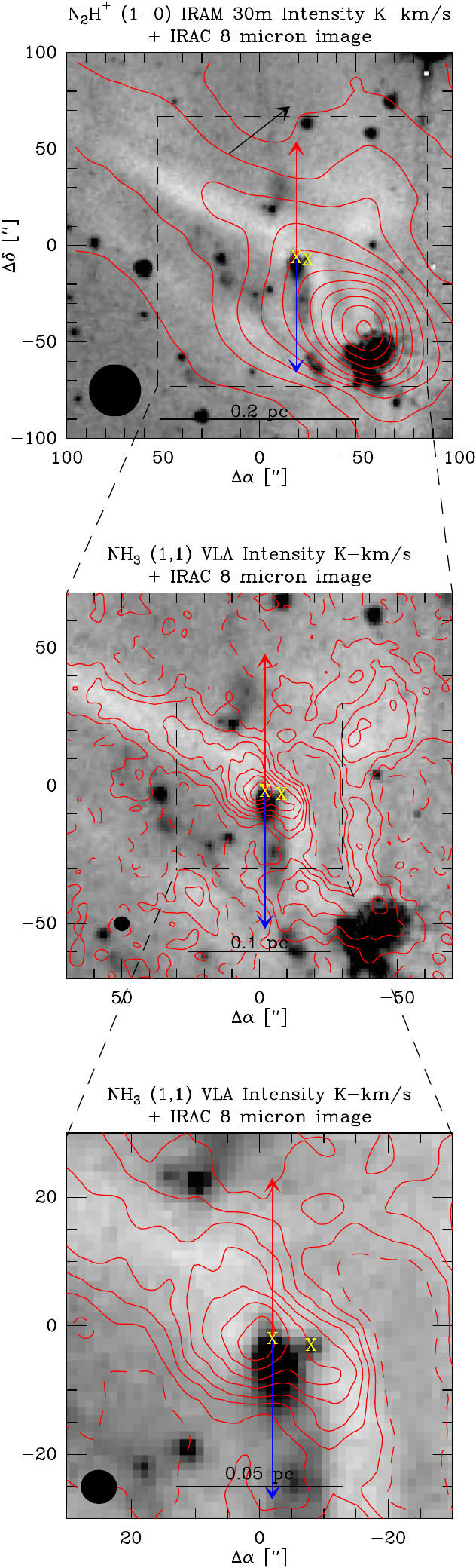}
\includegraphics[scale=0.75]{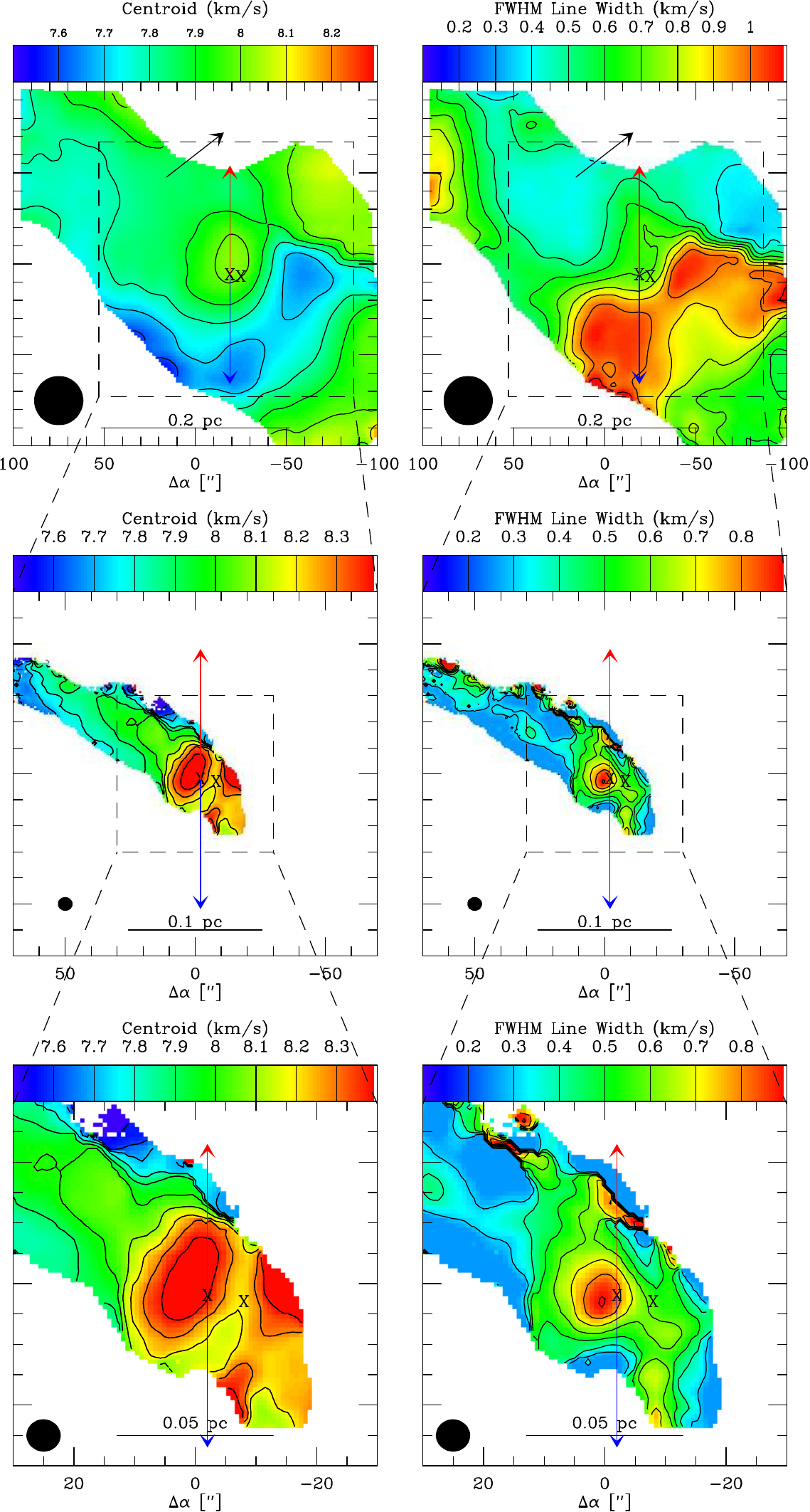}
\end{center}
\caption{Serpens MMS3-- Same as Figure \ref{L1157} but with VLA \nht\ (1,1) data and an additional zoom-in of the
\nht\ data in the bottom row.
The IRAM 30m contours start at 5$\sigma$ with 10$\sigma$ intervals; the VLA data start at $\pm$3$\sigma$ with 5$\sigma$ intervals.
 The \nthp\ peak in the single-dish
data is not on Serpens MMS3 but rather near a small cluster of young stars to the southwest, but there is an 
extension toward MMS3. The single-dish velocity field near MMS3 shows a red-shifted pocket associated with its position with 
a gradient extending along the direction of the large-scale filament shown in the top panels. The VLA \nht\ map
reveals more detail as the emission closely follows the filamentary structures in the region. The \nht\ velocity
map also shows the deep red-shifted pocket of emission next to the protostar and reveals more detail
in the gradient along the filament. There is a corresponding increase in linewidth to the east of the protostar
at the location of the red-shifted emission.}
\label{serpmms3}
\end{figure}
\clearpage

\begin{figure}
\begin{center}
\includegraphics[scale=0.75]{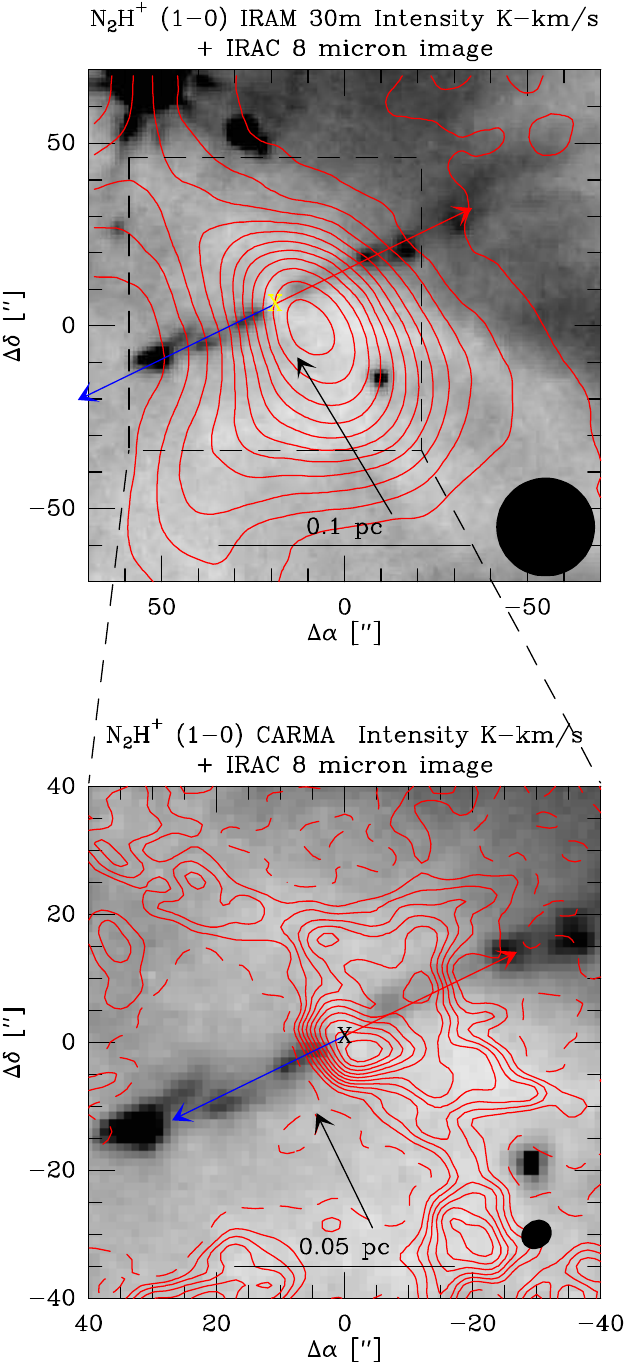}
\includegraphics[scale=0.75]{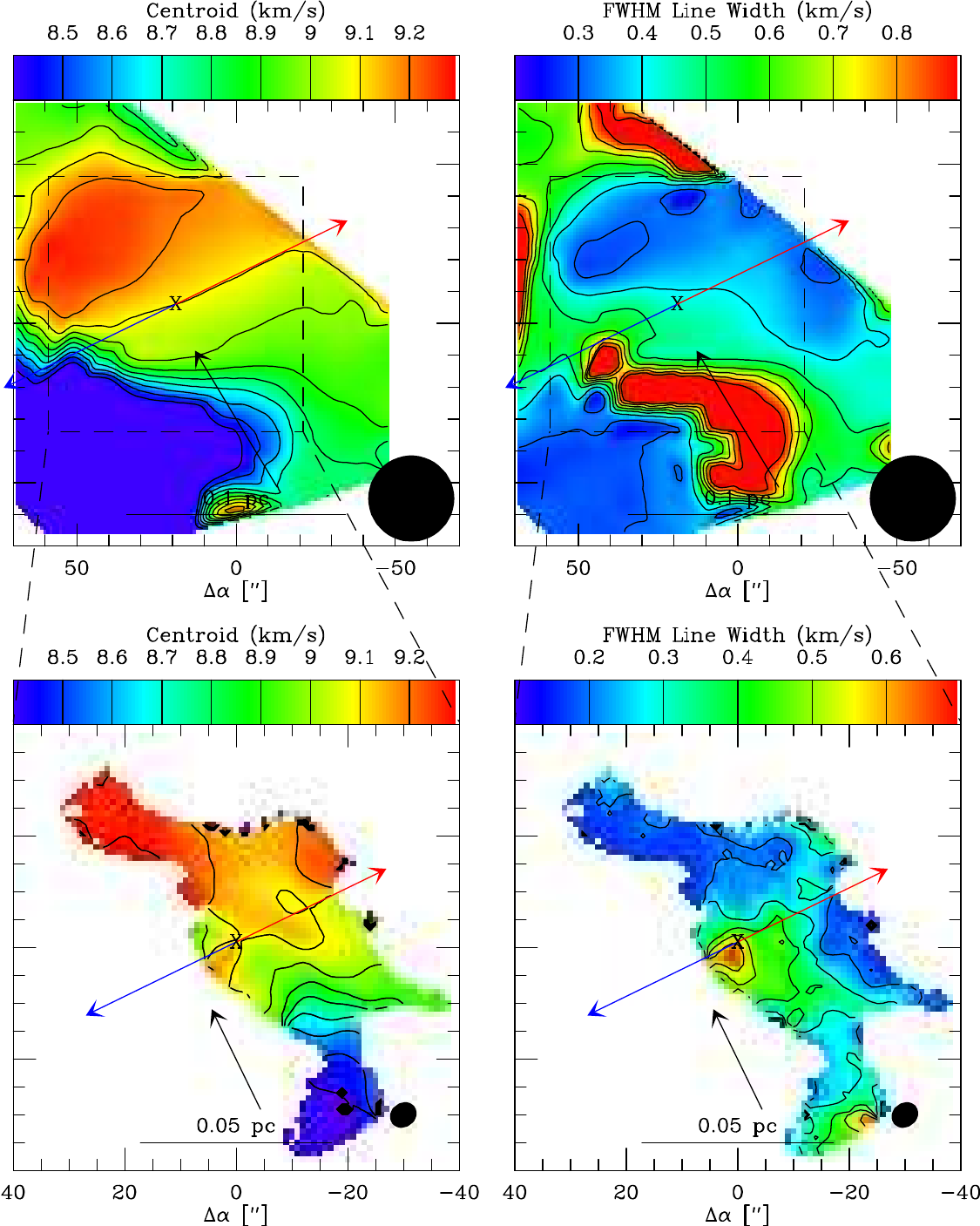}
\end{center}
\caption{HH211-- Same as Figure \ref{L1165}. 
The IRAM 30m contours start at 3$\sigma$ with 10$\sigma$ intervals; the CARMA data start at $\pm$3$\sigma$ with 3$\sigma$ intervals.
As shown in the left panels, the \nthp\ emission mapped by the 30m and CARMA 
correlates very well with the 8\mum\ extinction in the region. We also notice that the \nthp\ peak
is offset from the protostar to the southwest in the single-dish and interferometer data. The velocity field shows
a linear gradient normal to the outflow of HH211; however, south of the protostar there is another
\nthp\ velocity component blue-shifted from the rest of the gas in the region. \citet{tanner2011} referred to this
as the southwest extension. This transition region appears as artificially large
line-width in the top right panel. The velocity field in the CARMA data also shows the linear gradient with a slight increase
in linewidth near the protostar.}
\label{HH211}
\end{figure}
\clearpage

\begin{figure}
\begin{center}
\includegraphics[scale=0.75]{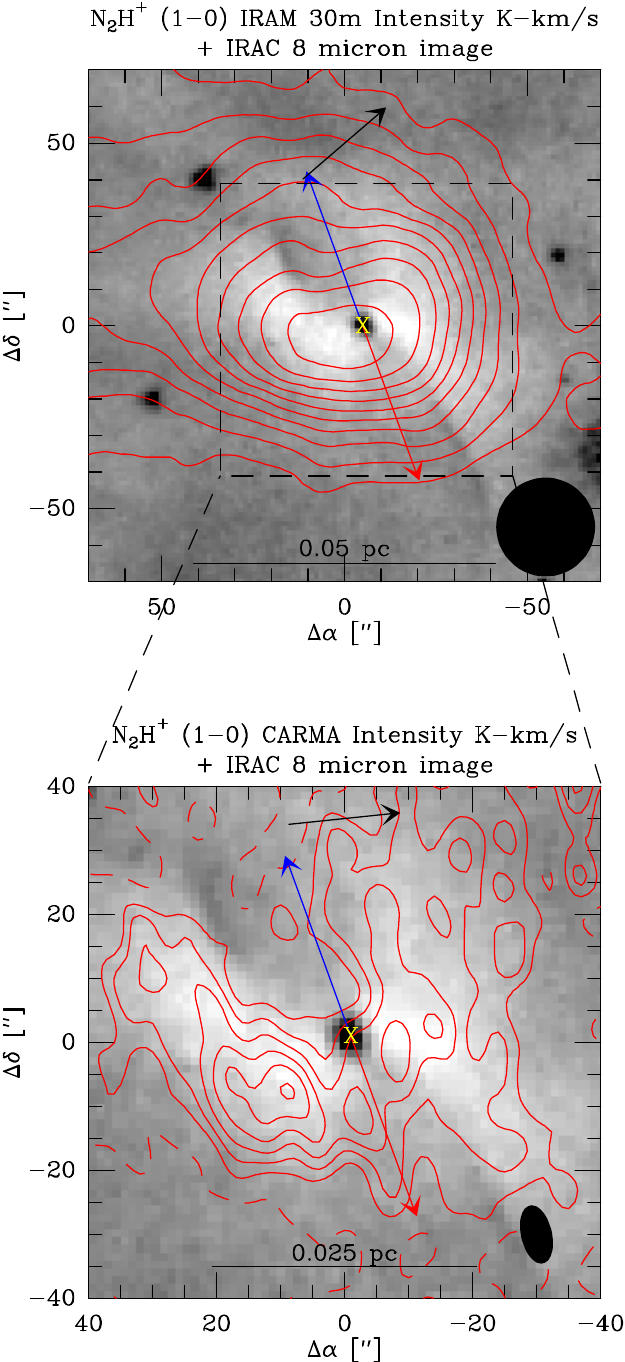}
\includegraphics[scale=0.75]{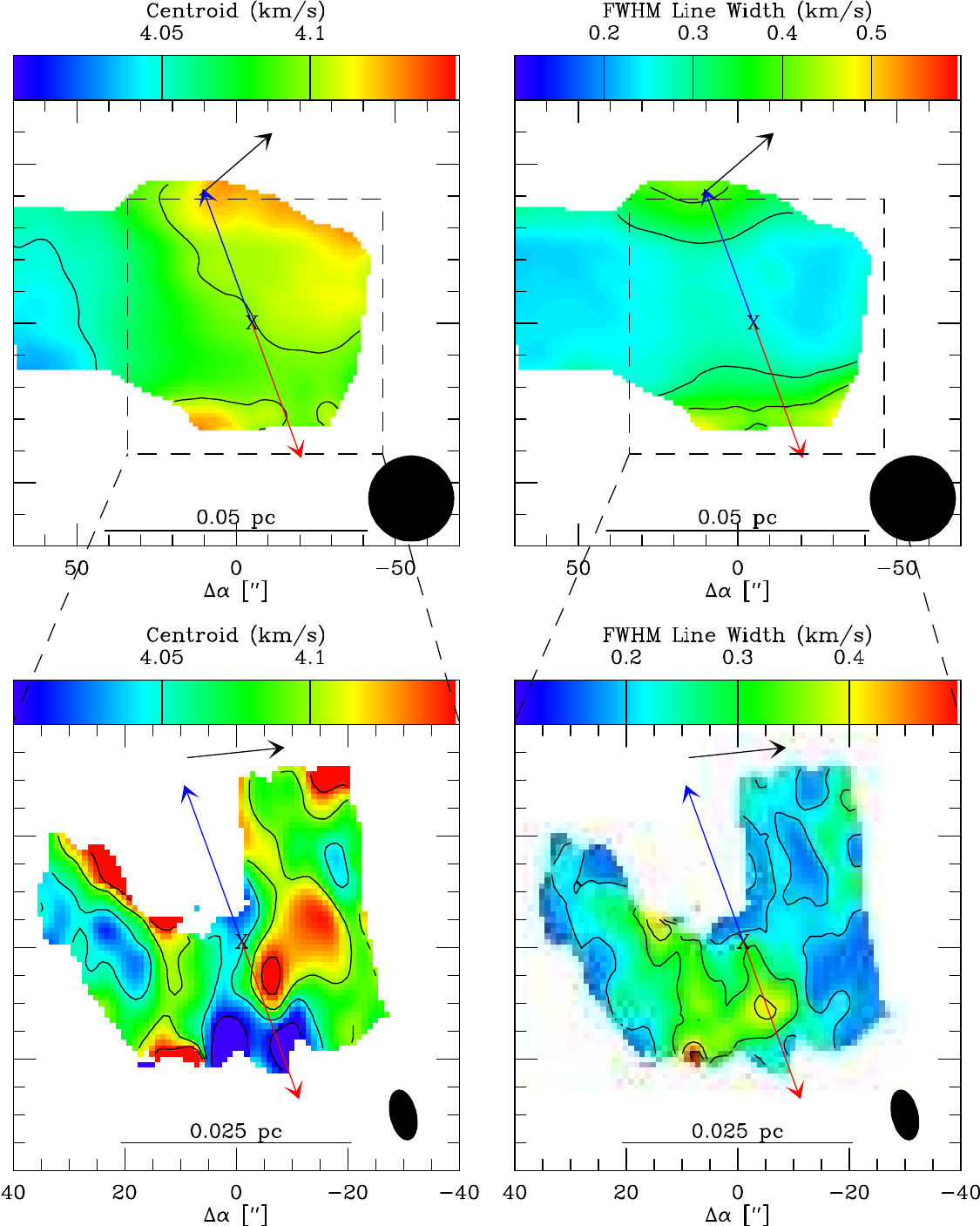}
\end{center}
\caption{IRAS 16253-2429-- Same as Figure \ref{L1165}.
Both the IRAM 30m  and CARMA contours start at 3$\sigma$ with 3$\sigma$ intervals. The single-dish \nthp\ emission traces
the roughly symmetric envelope of IRAS 16253-2429 quite well with the peak emission just east of the protostar. The CARMA \nthp\
map also traces the envelope well and the northern outflow cavity is prominent as an
evacuated region, probably due to resolved-out emission. The single-dish velocity
field shows a very small velocity gradient roughly normal
to the outflow. The CARMA velocity map shows complex structure,
but there is an overall gradient in the same direction as the single-dish map.
In addition, there is a red-shifted feature at the same location of as the outflow cavity wall.
The single-dish and interferometer linewidth maps show slight enhancements
at the same locations, where the peak emission is present, but there also
is a linewidth increase toward the edge of the envelope along
the outflow in the single-dish map. Note that the linewidth in the envelope
is extremely narrow, only $\sim$0.2 \kms\ over the regions where the outflow
could not be interacting.}
\label{IRAS16253}
\end{figure}
\clearpage

\begin{figure}
\begin{center}
\includegraphics[scale=0.75]{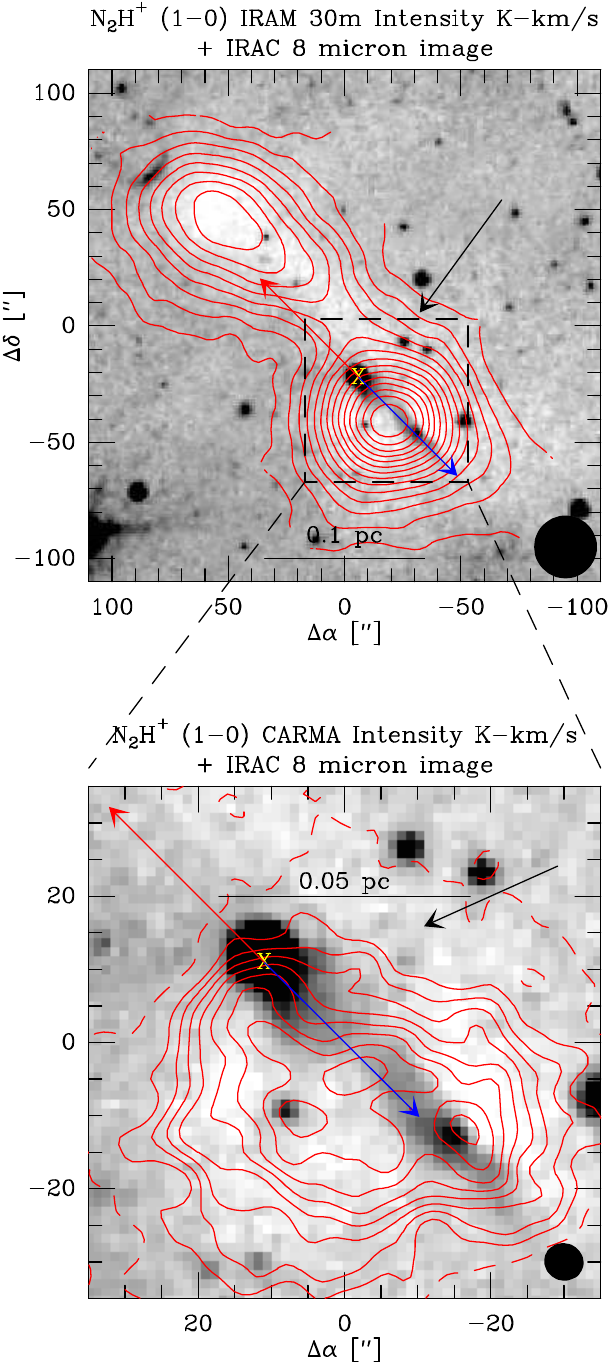}
\includegraphics[scale=0.75]{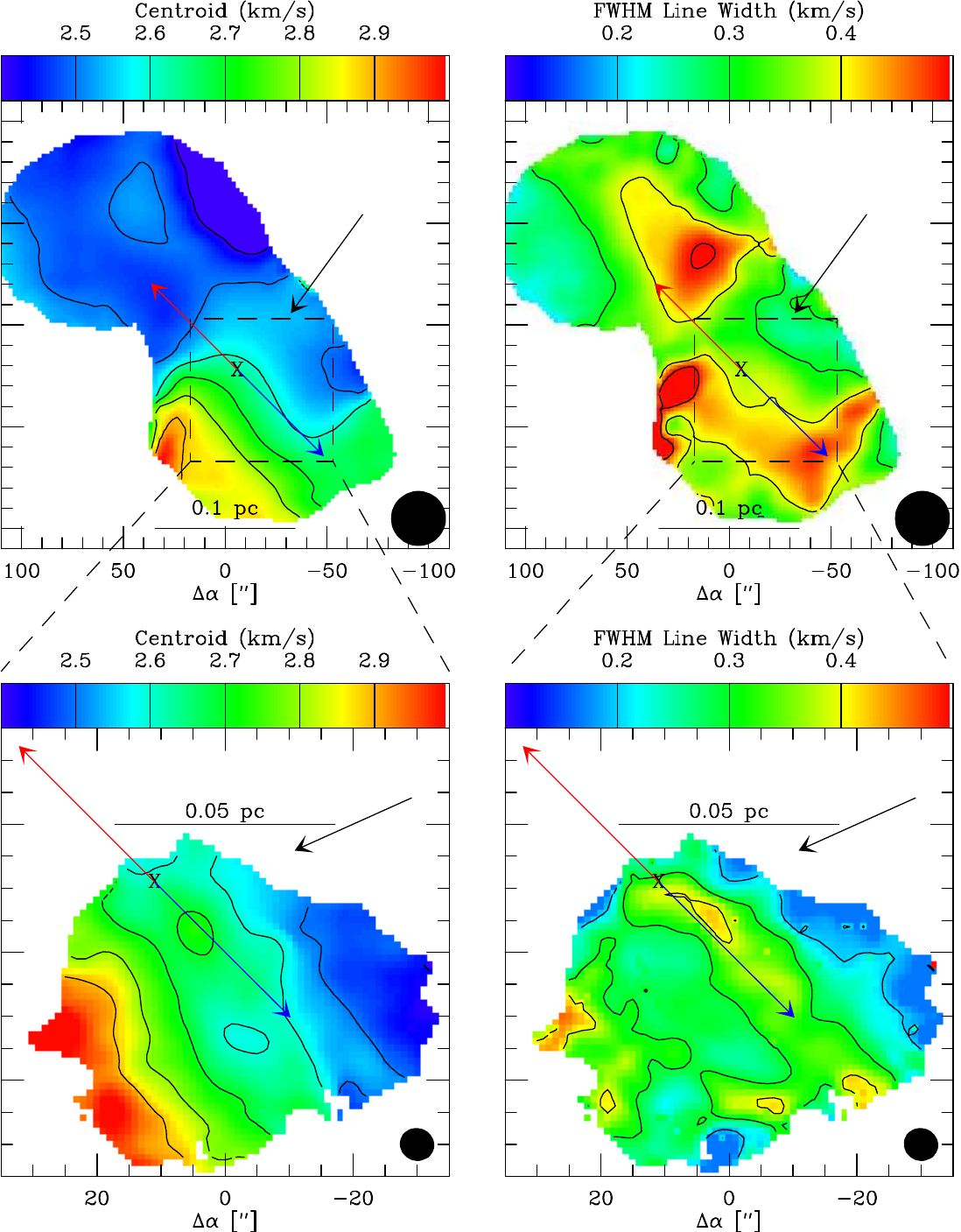}
\end{center}
\caption{L1152-- Same as Figure \ref{L1165}.
The IRAM 30m contours start at 10$\sigma$ with 10$\sigma$ intervals; the CARMA data start
at $\pm$3$\sigma$ with 3$\sigma$ intervals. The single-dish \nthp\ emission traces two connected peaks over 
$\sim$0.1 pc; one is star-less the other is adjacent to the protostar. The CARMA \nthp\ map focuses on the 
protostellar clump and clearly shows that the protostar is offset from most of the \nthp\ emission. Both velocity maps
show a strong gradient normal to the outflow, while the star-less clump does not have much velocity structure. The linewidth
maps show increases along the outflow; there is a feature in the single-dish map which is not exactly along the
outflow but quite near it. The CARMA linewidth map shows increased linewidth that correlates very strongly
with the jet-like emission shown in the 8\mum\ image.}
\label{L1152}
\end{figure}
\clearpage

\begin{figure}
\begin{center}
\includegraphics[scale=0.75]{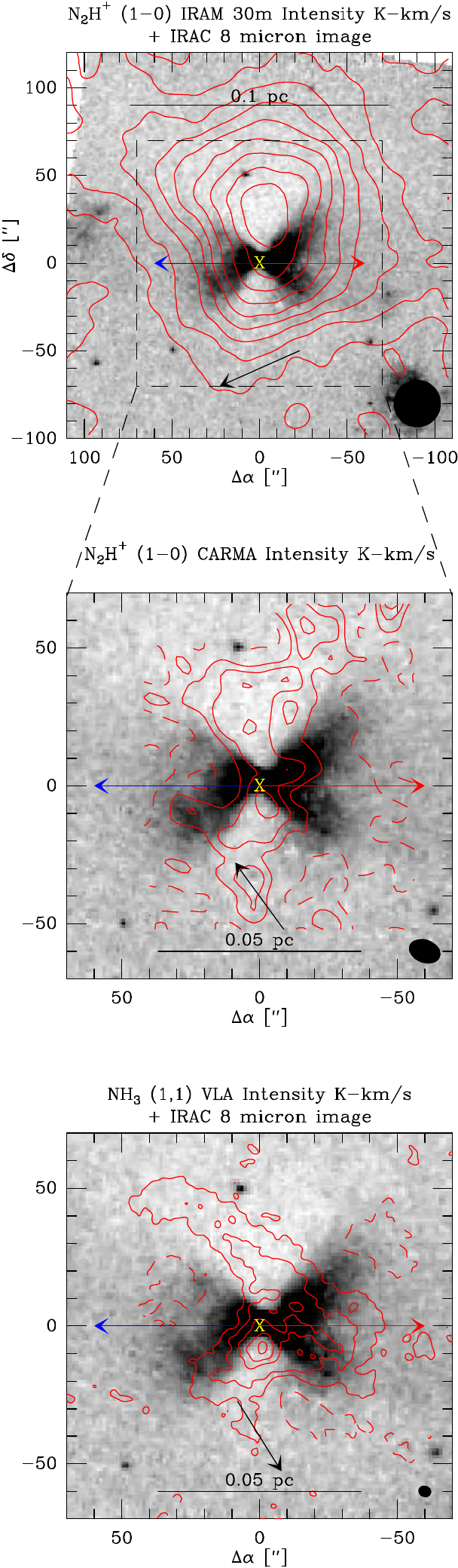}
\includegraphics[scale=0.75]{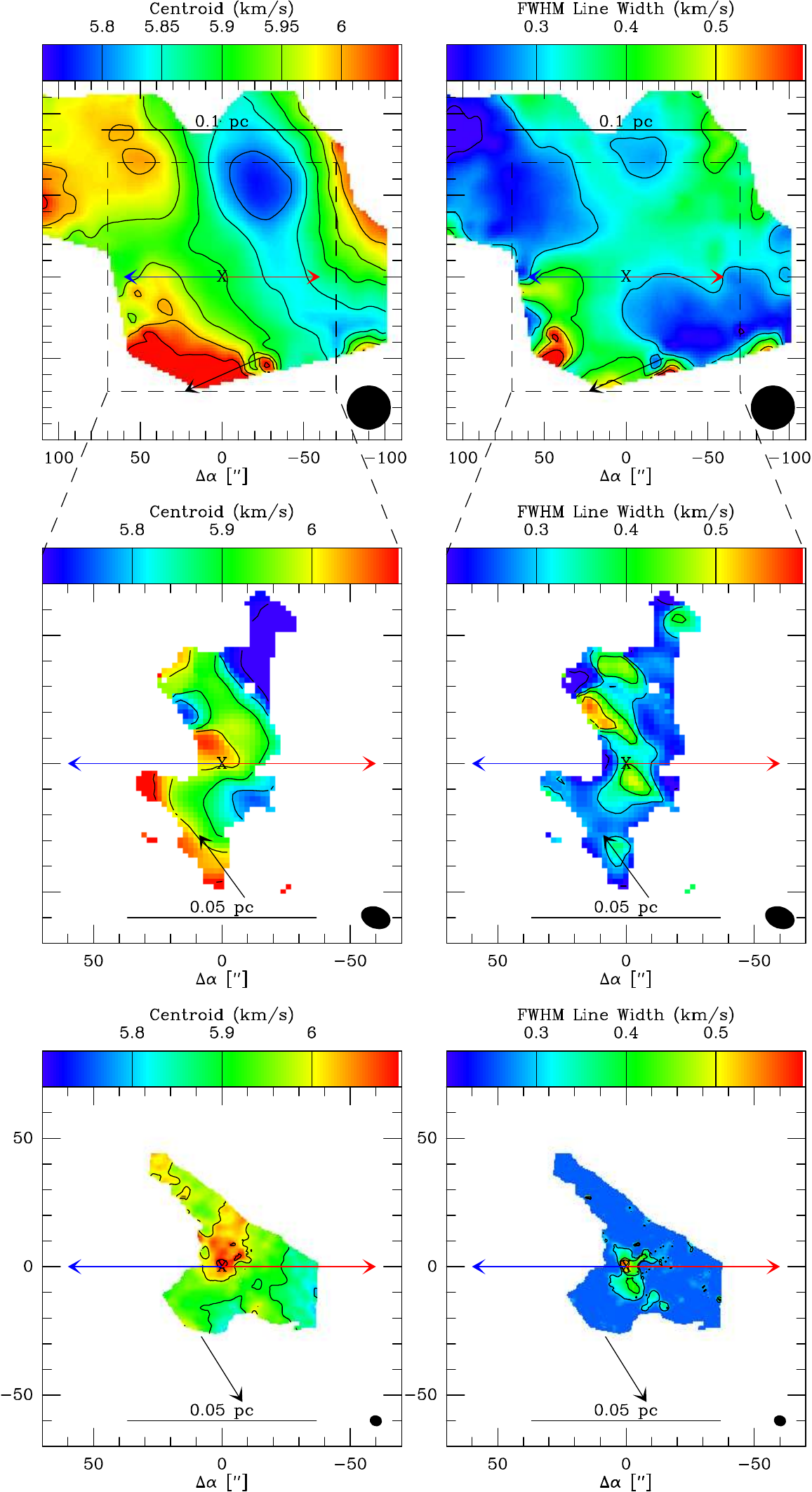}
\end{center}
\caption{L1527-- Same as Figure \ref{L1165}, but VLA \nht\ data are also shown in the \textit{bottom row}.
The IRAM 30m contours start at 5$\sigma$ with 10$\sigma$ intervals; both the
CARMA and VLA contours start at $\pm$3$\sigma$ with 3$\sigma$ intervals. We show the \nht\ as well since 
they have slightly better resolution and because the \nthp\ data only have moderate signal-to-noise. The single-dish \nthp\
data are peaked north of the protostar, which is also the case for the CARMA data, but the VLA data are peaked south
of the protostar. Both the VLA and CARMA data seem to trace a structure curving to the northeast which
may be associated with the outflow cavity. The single-dish velocity field appears to have a component
along the outflow and normal to it. The interferometer velocity maps reflect the single-dish velocities on the
largest scales, but near the protostar there is a small-scale velocity gradient in the opposite direction 
of the large-scale gradient. There is little increase in linewidth in the single-dish data,
but there appears to be increased linewidth near the protostar in both sets of interferometer data.}
\label{L1527}
\end{figure}
\clearpage

\begin{figure}
\begin{center}
\includegraphics[scale=0.75]{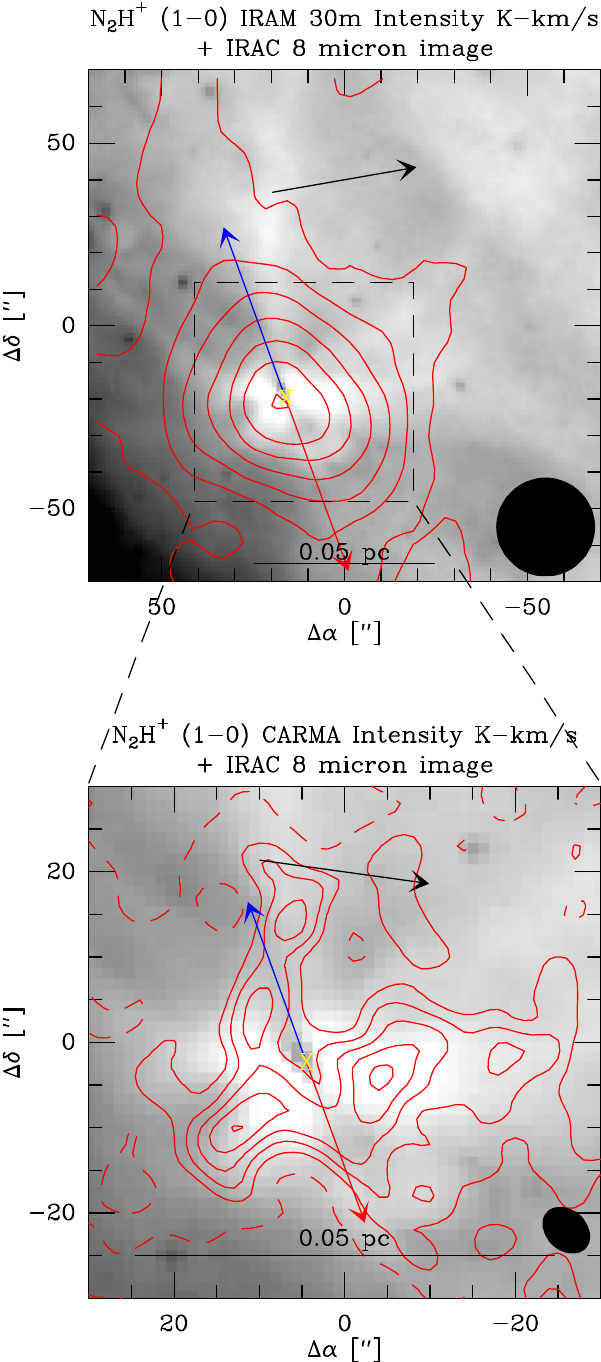}
\includegraphics[scale=0.75]{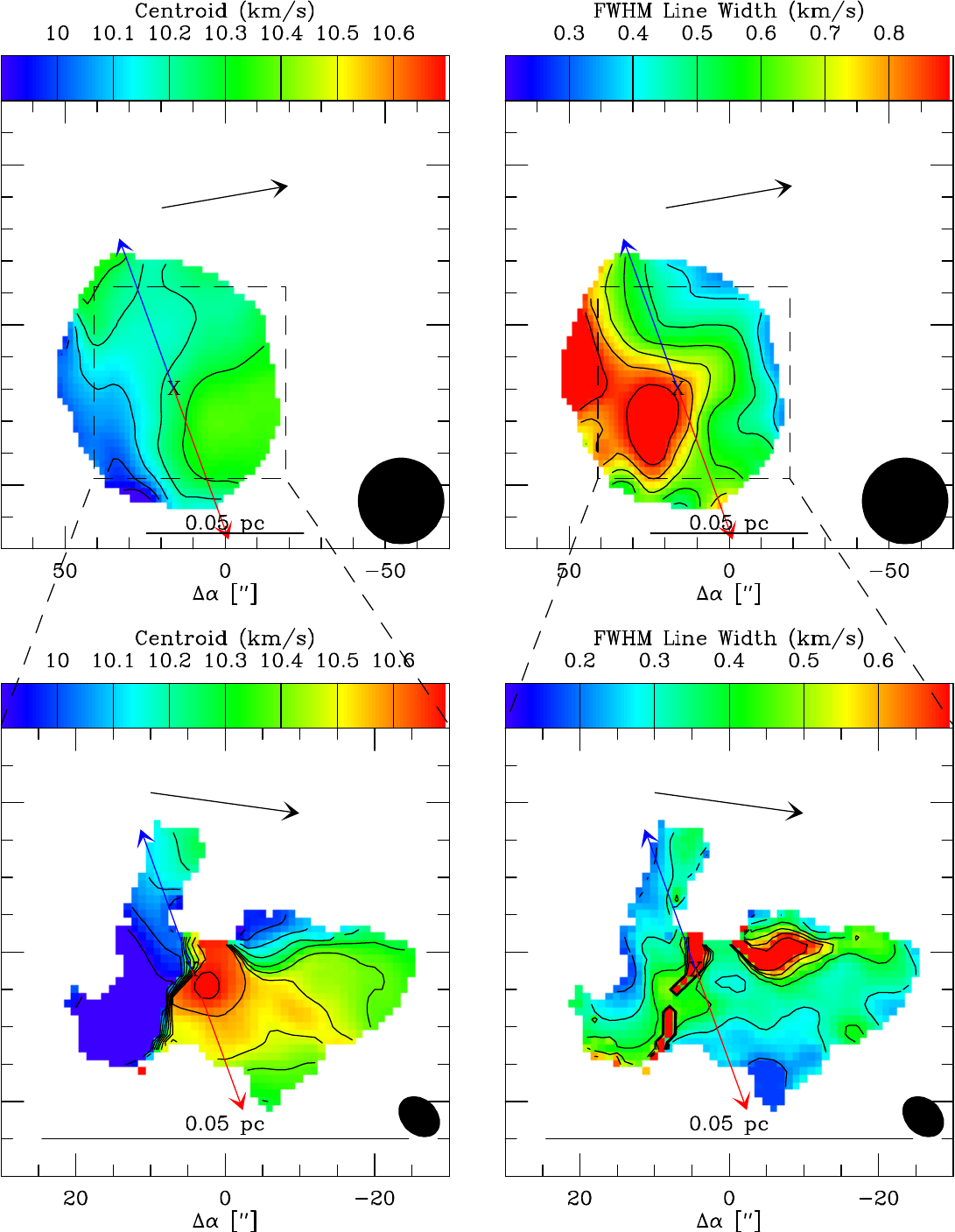}

\end{center}
\caption{RNO43-- Same as Figure \ref{L1165}. 
The IRAM 30m contours start at 3$\sigma$ with 10$\sigma$ intervals; the CARMA contours start
at $\pm$3$\sigma$ with 3$\sigma$ intervals. The single-dish intensity map appears mostly unresolved, while the CARMA
data trace the 8$\mu$m extinction structures closely with a depression of emission coincident with the protostar. The single-dish velocity
field shows a gradient mostly normal to the outflow and the interferometer data reveal a sharp velocity
shift between the east and west sides of the envelope. This appears as a line of broad linewidth (an artifact from
fitting) where the two
components are viewed on top of each other; this is associated with a region of large linewidth in the single-dish map.}
\label{RNO43}
\end{figure}
\clearpage

\begin{figure}
\begin{center}
\includegraphics[scale=0.65,angle=-90]{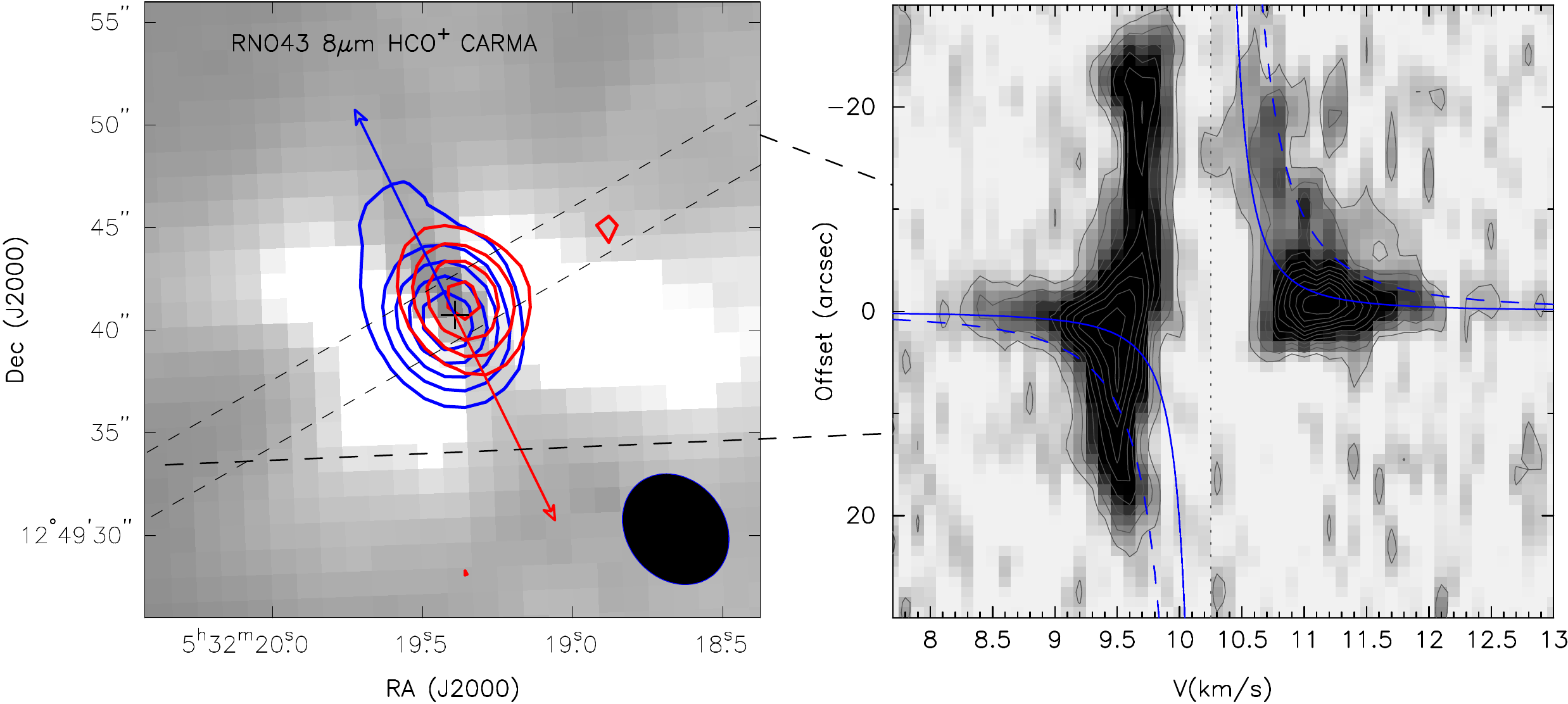}
\end{center}
\caption{RNO43-- Same as Figure \ref{L1165-pv-hco}. The contours levels are $\pm$5, 8, 11, 14, and 17$\sigma$ ($\sigma$ = 0.12 K)
 for the red and blue-shifted emission. The emission morphology of the HCO$^+$ in both the
integrated intensity map on the left and in the PV plot are quite similar to L1165. The total velocity 
extent of $\sim \pm$2 \kms\ is similar in magnitude to L1165. However, the red and blue peaks are
much closer together due to the increased distance of RNO43. The PV plot contours start at 3$\sigma$
and increase in 3$\sigma$ intervals ($\sigma$=0.2671). The \textit{solid-blue} curve represents Keplerian
rotation (or infall) for a 0.67 $M_{\sun}$ central object and the \textit{dashed-blue} curve is for
a 2.67 $M_{\sun}$ central object.}
\label{RNO43-pv-hco}
\end{figure}
\clearpage

\begin{figure}
\begin{center}
\includegraphics[scale=0.75]{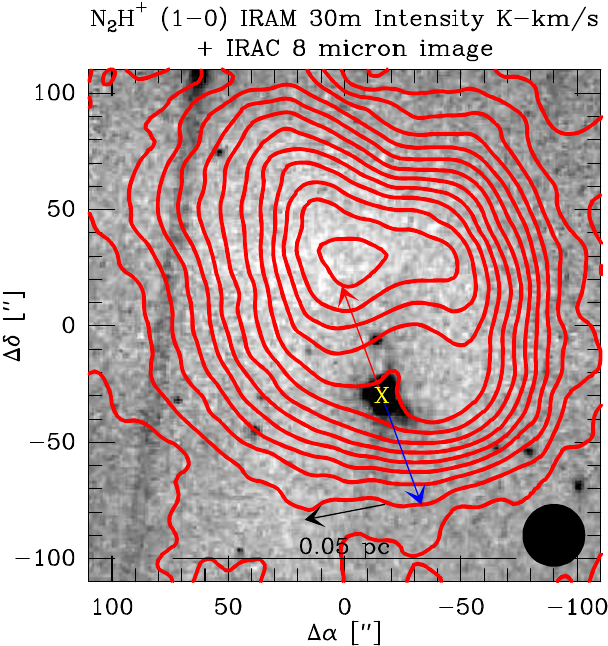}
\includegraphics[scale=0.75]{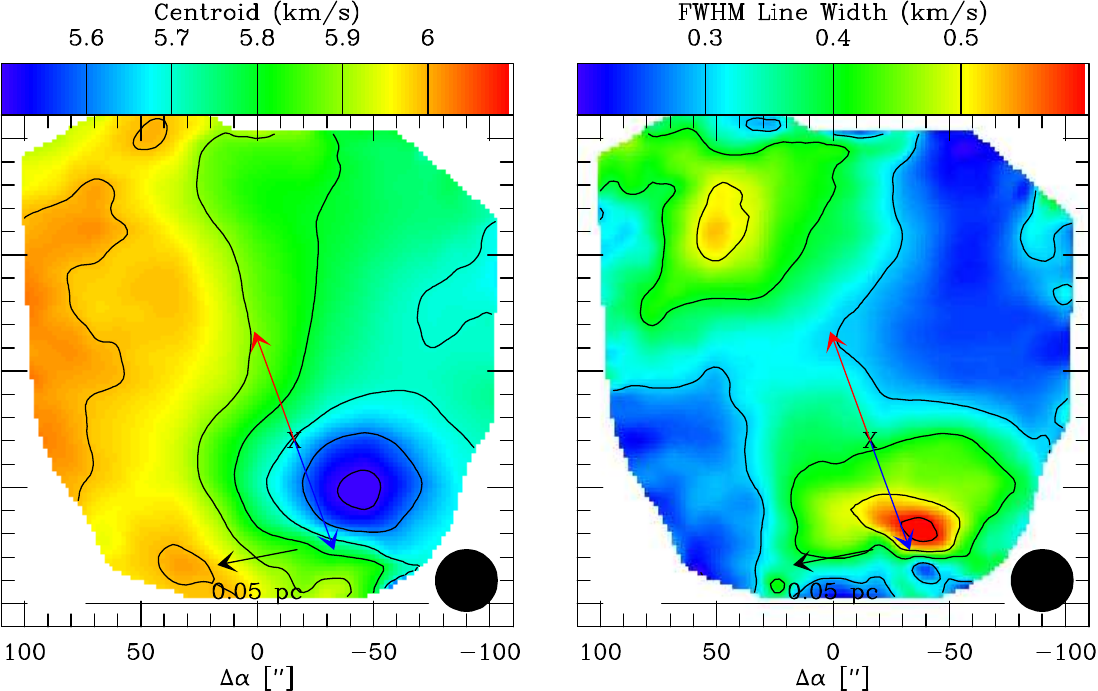}
\end{center}
\caption{IRAS 04325+2402-- Same as Figure \ref{L1157}, but we only have single-dish data for this object.
The IRAM 30m contours start at 10$\sigma$ with 10$\sigma$ intervals. The 
\nthp\ contours correlate very well with the 8$\mu$m extinction and the peak is actually on an apparent
star-less core north of the protostar. There is a tail of emission wrapping towards the protostar, but at lower
intensity levels. The velocity field shows a gradient roughly normal to the outflow and there appears to be
smaller-scale structure in the velocities near the protostar. The linewidth map does not show a large increase
near the protostar but shows two peaks that are located along the outflow indicating a possible interaction.}
\label{IRAS04325}
\end{figure}
\clearpage

\begin{figure}
\begin{center}
\includegraphics[scale=0.75]{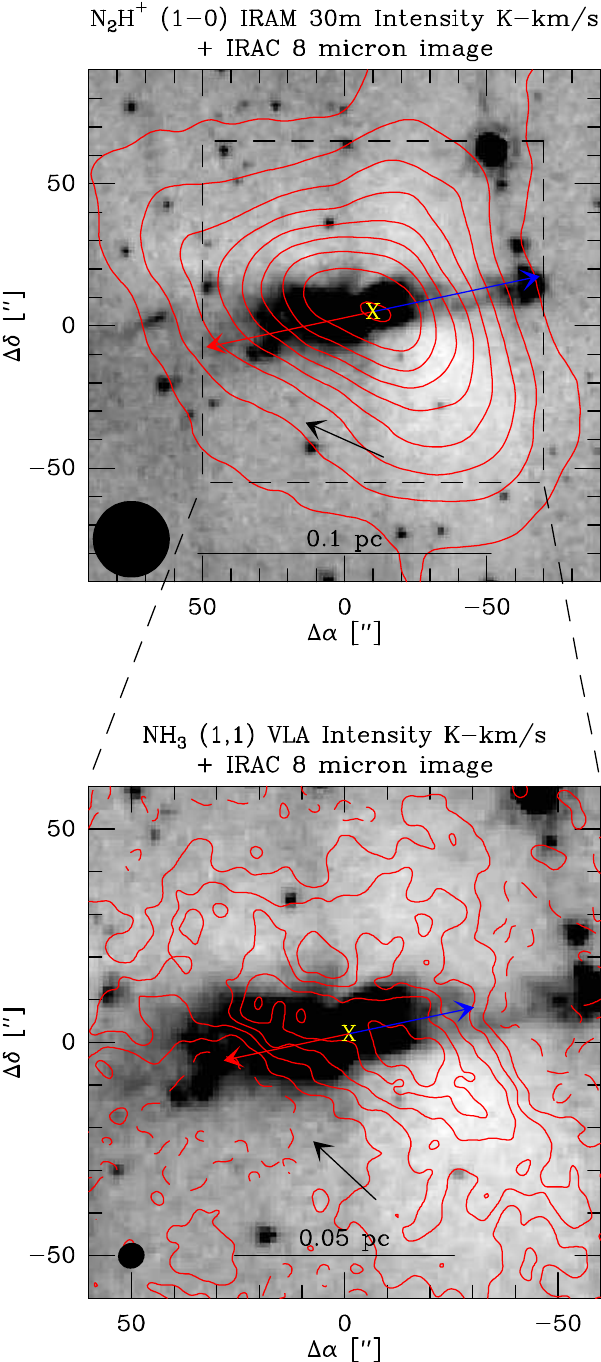}
\includegraphics[scale=0.75]{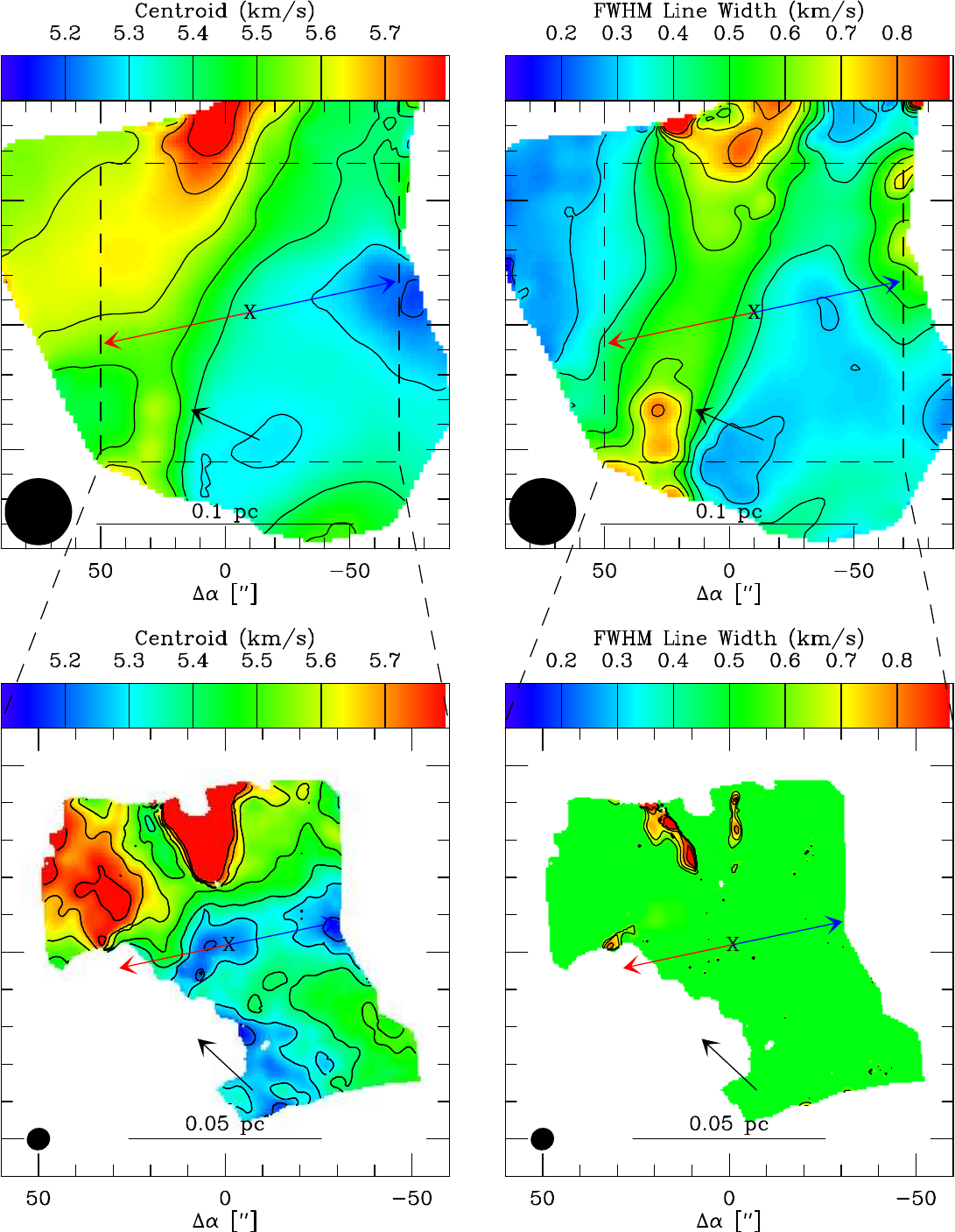}
\end{center}
\caption{L483-- Same as Figure \ref{CB230}, using archival \nht\ data from \citet{fuller2000}.
The IRAM 30m contours start at 20$\sigma$ with 20$\sigma$ intervals; the VLA contours start at $\pm$3$\sigma$
with 6$\sigma$ intervals. The \nthp\ and \nht\ emission trace the envelope seen in 8\mum\
extinction quite closely. The velocity field shows a gradient that is at an angle
45$^{\circ}$ from the outflow in the single-dish map. The VLA velocity map shows
a similar large-scale feature but there is a pocket of blue-shifted gas coincident with the
protostar and there is a red-shifted pocket directly north of the protostar. The single-dish linewidth
map shows increased linewidth at the location of the most rapid velocity changes;
the VLA linewidth map only shows a couple of features where there is a sharp
transition in the velocity components. This is due to only having 0.3 \kms\ channel width in the \nht\ data.  }
\label{L483}
\end{figure}
\clearpage

\begin{figure}
\begin{center}
\includegraphics[scale=0.75]{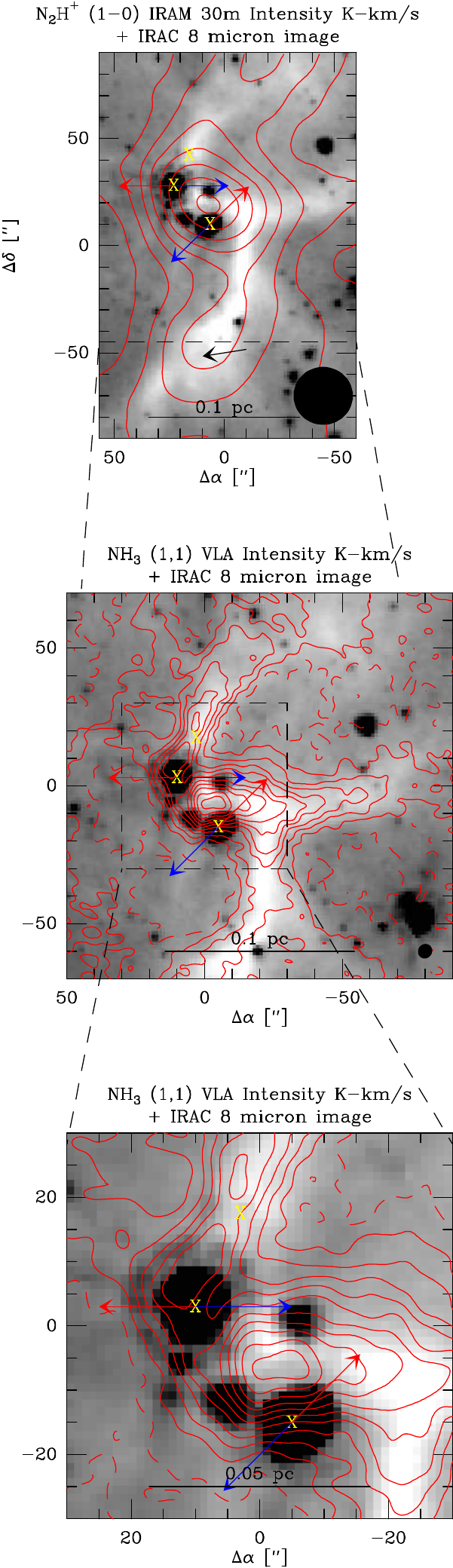}
\includegraphics[scale=0.75]{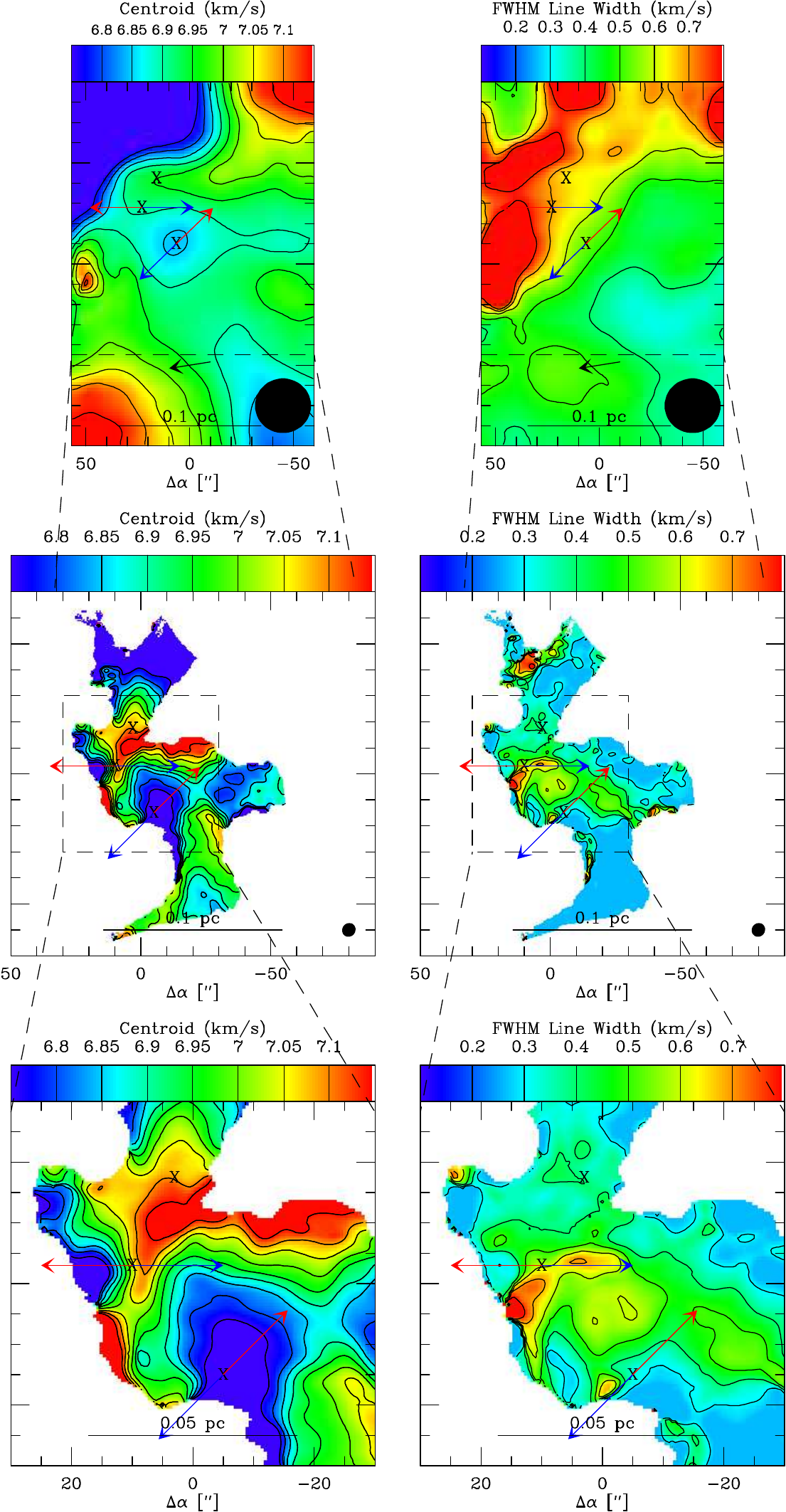}
\end{center}
\caption{L673-- Same as Figure \ref{CB230} with an additional zoom-in on the \nht\ data in the bottom row.
The IRAM 30m contours start at 20$\sigma$ with 20$\sigma$ intervals;
the VLA contours start at $\pm$3$\sigma$ with 6$\sigma$ intervals. Though only three
protostars are marked; there are likely seven in the region \citep{tsitali2010}; the southern-most
protostar marked is actually composed of three sources in Ks-band imaging (Tobin et al. in prep.).
The single-dish \nthp\ emission fills the
map, but there is increased emission where the 8\mum\ extinction is most prominent. The \nht\ map
only picks up the densest regions around the protostars. There is a gradient in the velocity map
along the filament, though there appears to be features associated with the two marked protostars in
both the single-dish and VLA maps. Both velocity maps also show a second velocity component to the north.
The linewidth maps show increased linewidth at the transition region between the two components and the 
VLA map shows increased linewidth near the protostars.
}
\label{L673}
\end{figure}
\clearpage

\begin{figure}
\begin{center}
\includegraphics[scale=0.75]{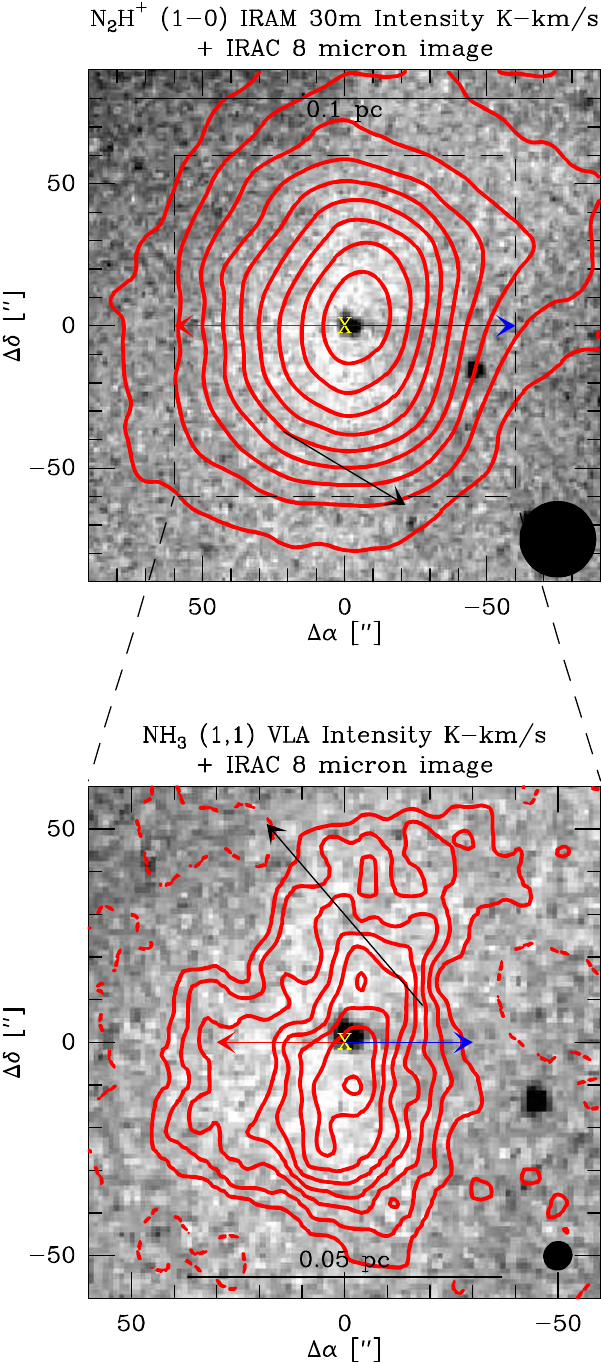}
\includegraphics[scale=0.75]{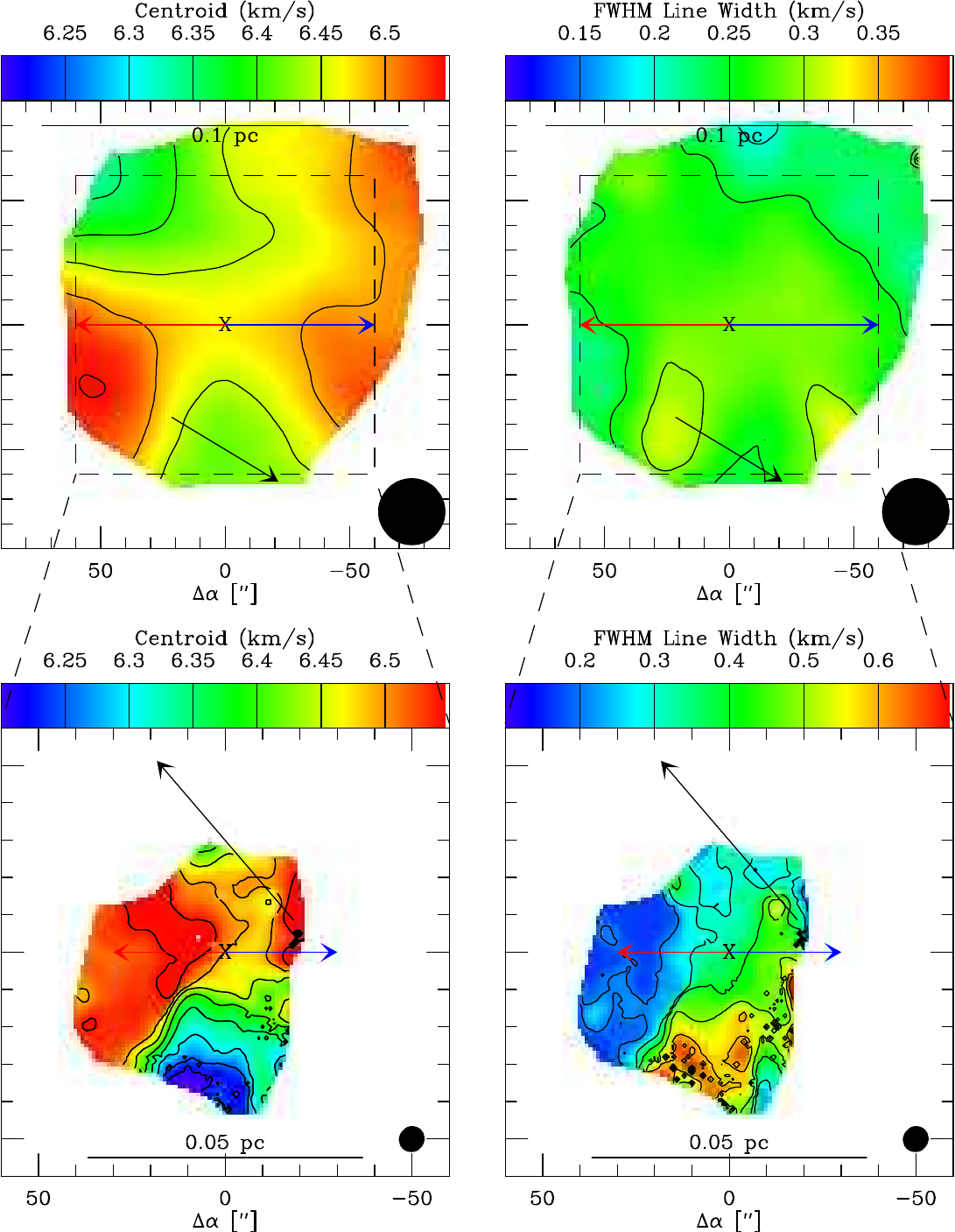}
\end{center}
\caption{L1521F-- Same as Figure \ref{CB230}.
The IRAM 30m contours start at 3$\sigma$ with 3$\sigma$ intervals; the VLA contours start
at $\pm$3$\sigma$ with 3$\sigma$ intervals. The single-dish data trace the region of 8\mum\ extinction 
well and the \nht\ data trace the inner, high density regions.
The velocity structure from the single-dish map shows weak evidence for a gradient normal to the
outflow. The \nht\ data do show a gradient, but the emission near the protostar is all red-shifted
and the blue-shifted emission is far (10\arcsec\, 1400 AU) from the protostar. However, as noted in 
the text, the \nht\ emission is optically thick and
may not fully trace the inner envelope kinematics. The single-dish linewidth map has very little structure while
the \nht\ linewidths are peaked coincident with the blue-shifted area of the velocity field.}
\label{L1521F}
\end{figure}
\clearpage

\begin{figure}
\begin{center}
\includegraphics[scale=0.75]{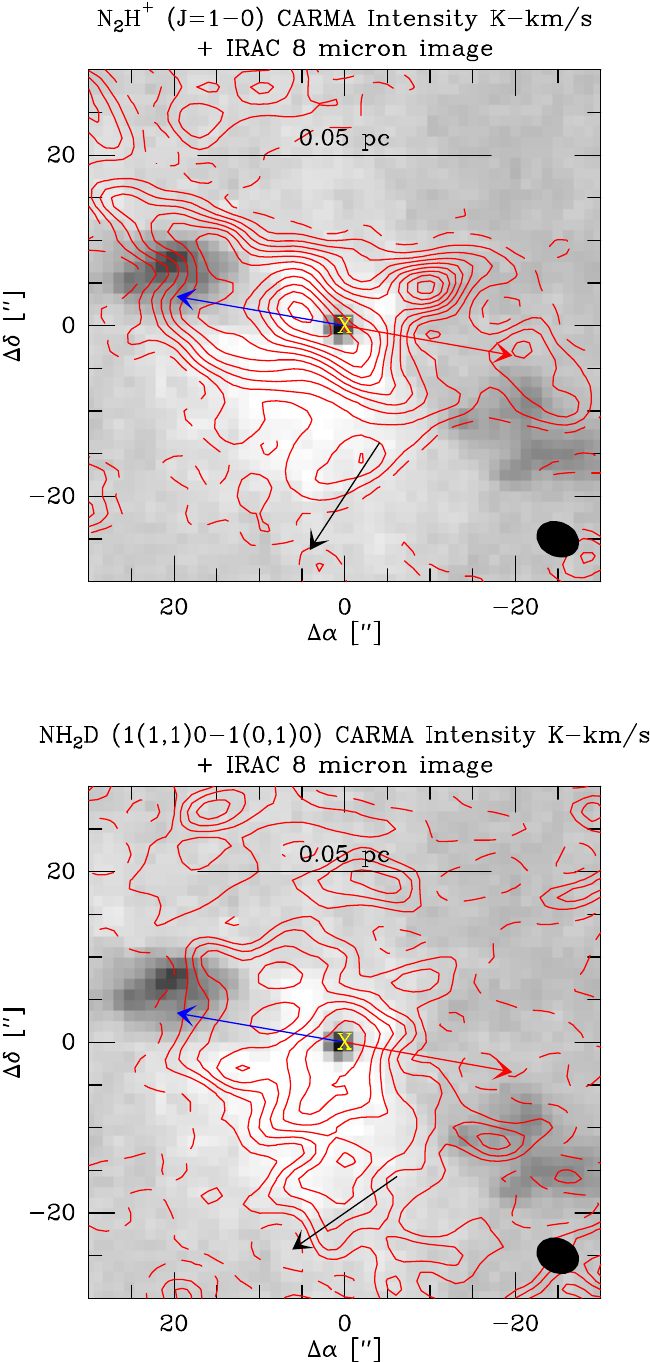}
\includegraphics[scale=0.75]{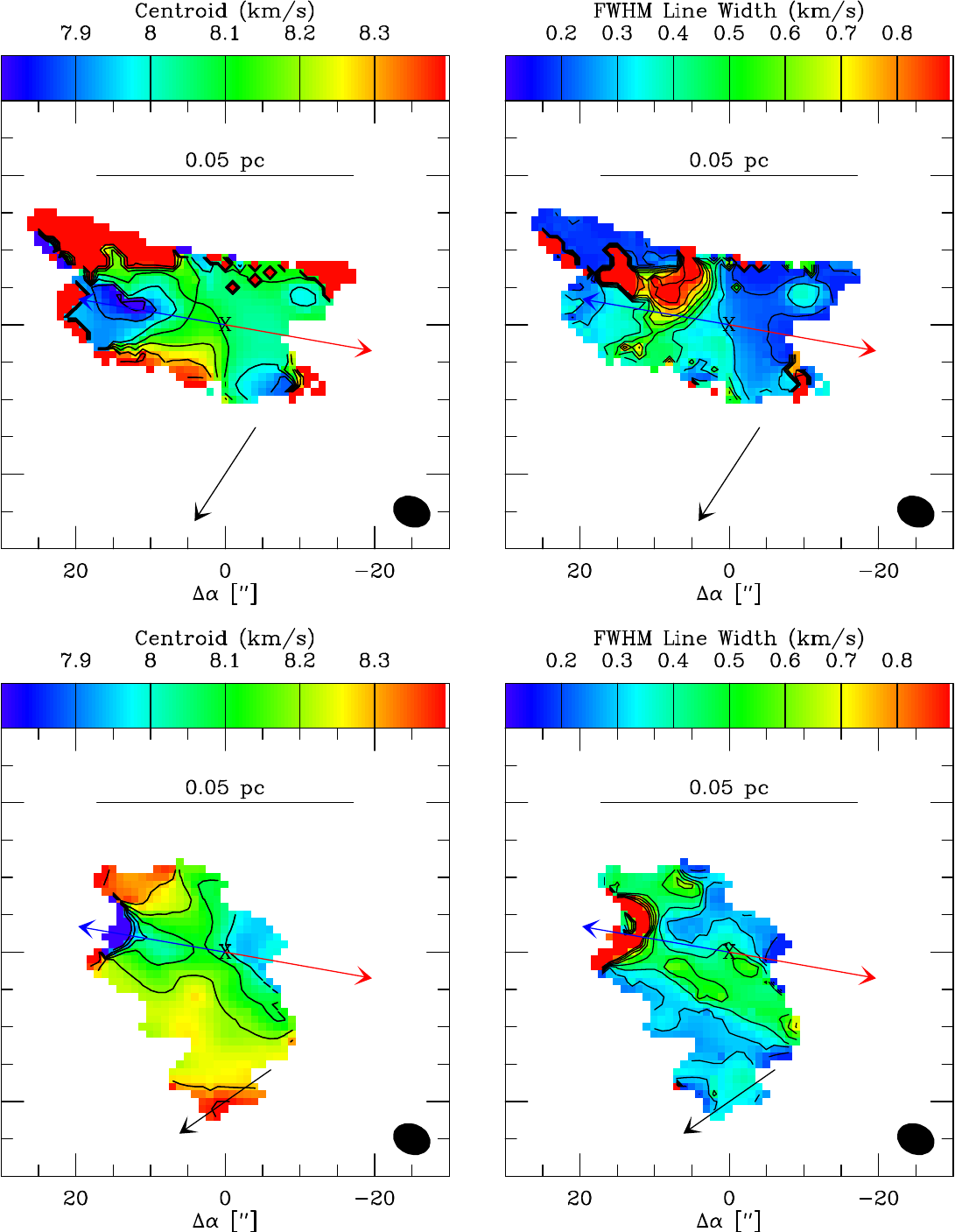}
\end{center}
\caption{Perseus 5-- The \textit{top row} shows the CARMA \nthp\ ($J=1\rightarrow0$) data and the
\textit{bottom row} shows the NH$_2$D ($1_{1,1}\rightarrow1_{0,1}$) data also from 
CARMA; the contours start at $\pm$3$\sigma$ with 6$\sigma$ intervals
for \nthp\ and $\pm$3$\sigma$ with 3$\sigma$ intervals for NH$_2$D. Both molecules have approximately the same noise level.
The integrated intensity data for both molecules
are overlaid on the IRAC 8\mum\ images. NH$_2$D appears to trace the 8\mum\ extinction
best while \nthp\ is centrally peaked. The line-center velocity of each molecular line
is shown in the center column, \nthp\ and NH$_2$D indicate that there may be a gradient normal to the
outflow, but that there is also a gradient in the direction of the outflow.
The linewidth plots in the right panels are complex, all show an enhancement away from the protostar
along the outflow; NH$_2$D in particular shows an increase through the envelope near the outflow axis.}
\label{Per5}
\end{figure}
\clearpage

\begin{figure}
\begin{center}
\includegraphics[scale=0.75]{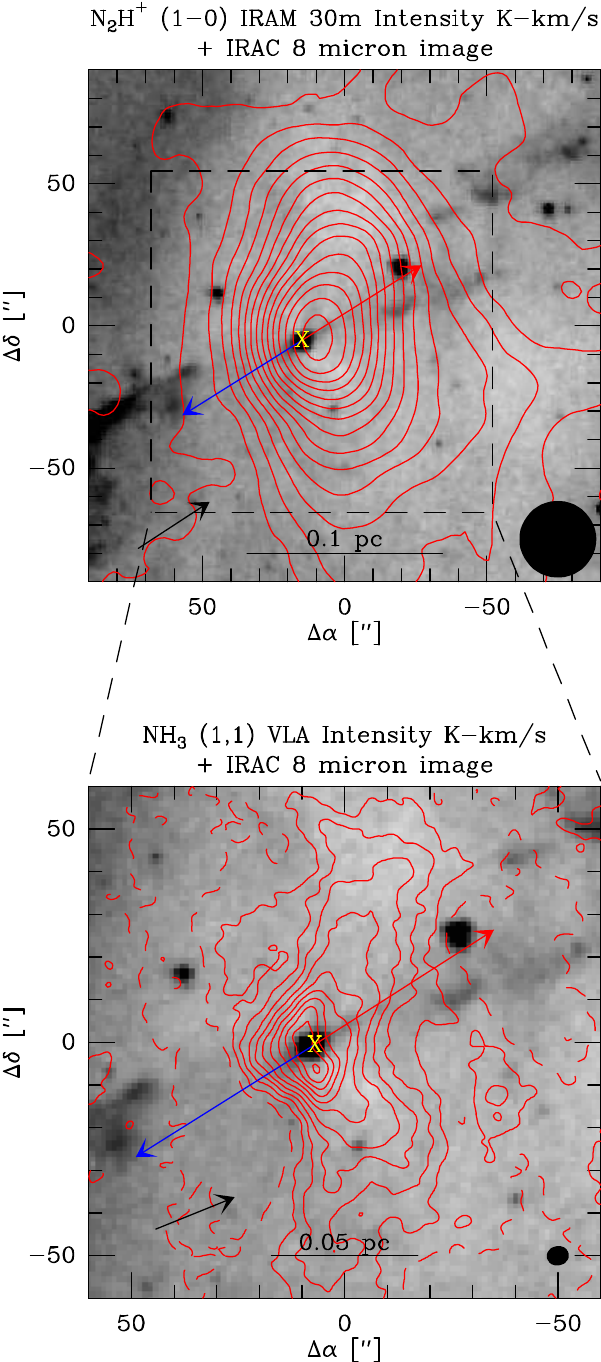}
\includegraphics[scale=0.75]{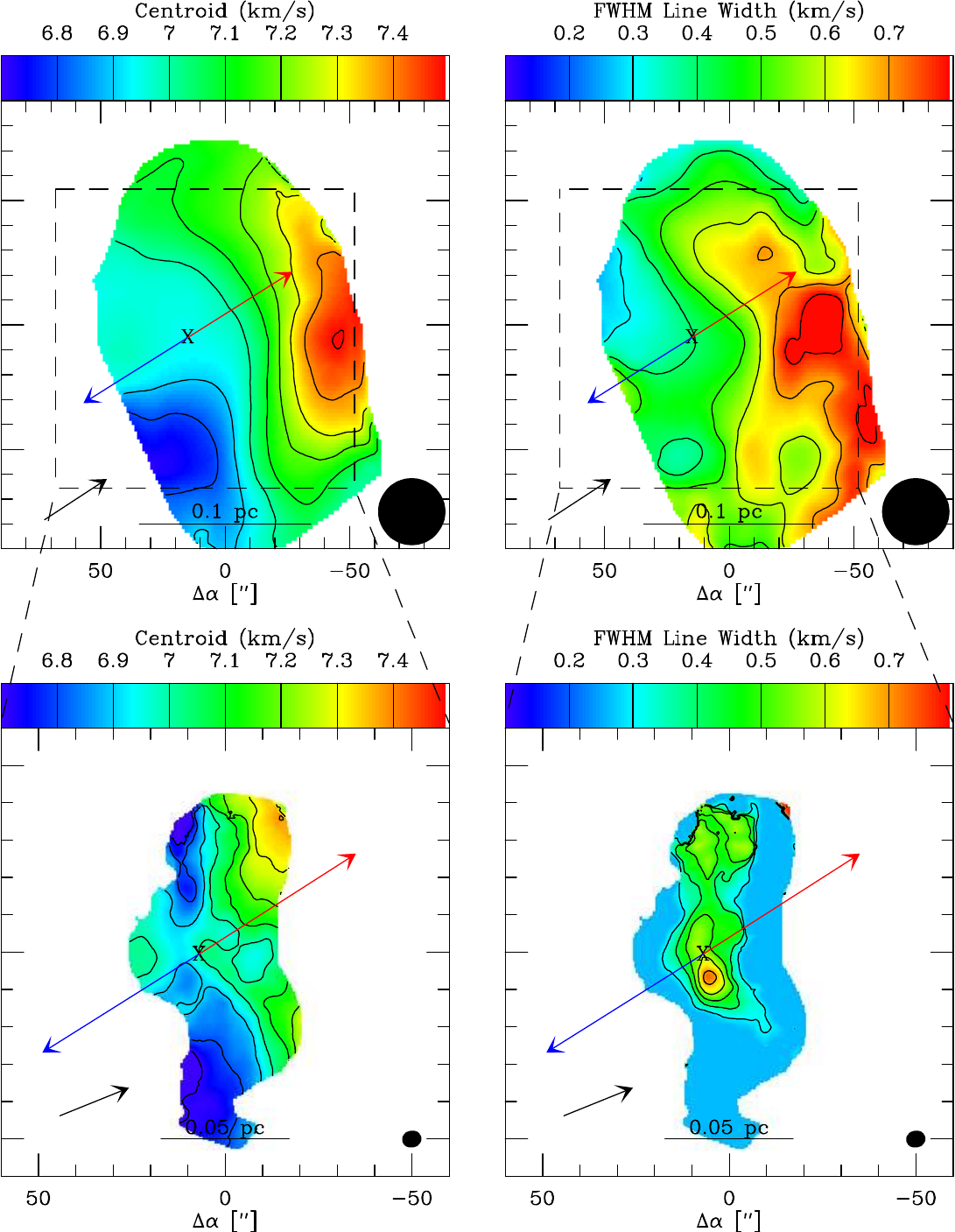}
\end{center}
\caption{IRAS 03282+3035-- Same as Figure \ref{CB230}.
The IRAM 30m contours start at 3$\sigma$ with 10$\sigma$ intervals; the VLA contours start
at $\pm$3$\sigma$ with 5$\sigma$ intervals. The single-dish \nthp\ emission closely traces
the entire area viewed in 8\mum\ extinction, peaking near the protostar. The VLA \nht\ data are picking up the densest
region near the protostar. The velocity gradient is mostly along the outflow in the single-dish and VLA \nht\ velocity map.
The linewidth in the single-dish data peaks to the west of the protostar, somewhat 
along the outflow. The \nht\ linewidth map has peaks about the protostar, normal to the outflow, and 
directly north, while the rest of the map has a small linewidth.}
\label{IRAS03282}
\end{figure}
\clearpage

\begin{figure}
\begin{center}
\includegraphics[scale=0.75,angle=-90]{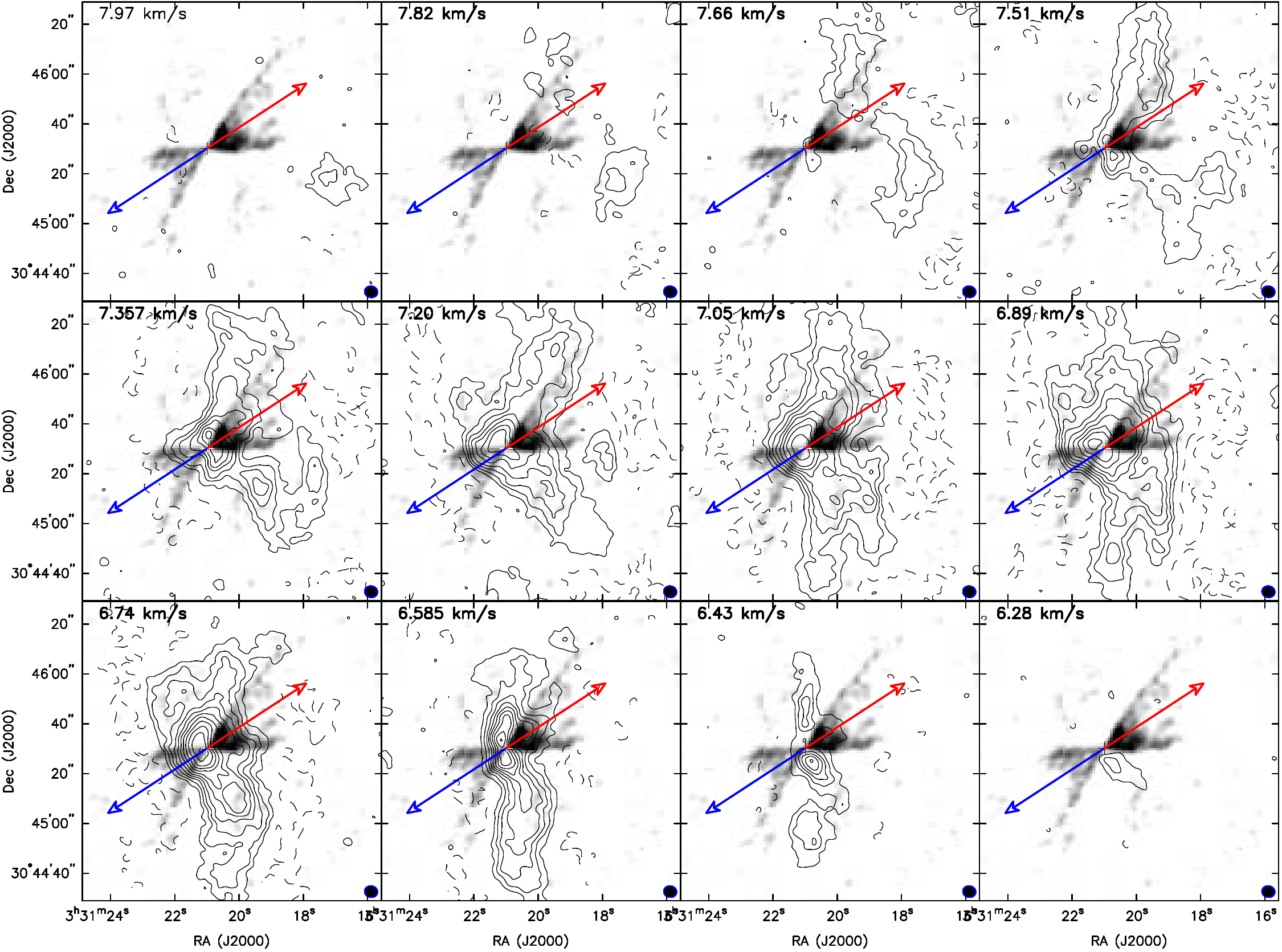}
\end{center}
\caption{Channel maps of \nht\ (1,1) emission (\textit{contours}) overlaid on CO ($J=1\rightarrow0$) 
emission from \citet{arce2006},
showing the relationship between \nht\ emission and the outflow cavity. The top and bottom rows show the red- and blue-shifted
component that are unblended; the \nht\ emission in the middle panels is blended. Contours start at 
$\pm$3$\sigma$, and then increase in 3$\sigma$ intervals. The negative contours are plotted as dashed lines, reflecting
the loss of large-scale structure. The unblended velocity channels
show very narrow structures and in the bottom row, the blue-shifted components are almost north-south in orientation. Much of the
\nht\ emission south of the protostar appears unlikely to be influenced by the outflow. 
The position of the protostar is marked with the cross and the outflow axis is marked bye the blue and red lines
denoting the outflow orientation within the plane of the sky.
}
\label{IRAS03282-chanmaps}
\end{figure}
\clearpage

\begin{figure}
\begin{center}
\includegraphics[scale=0.75]{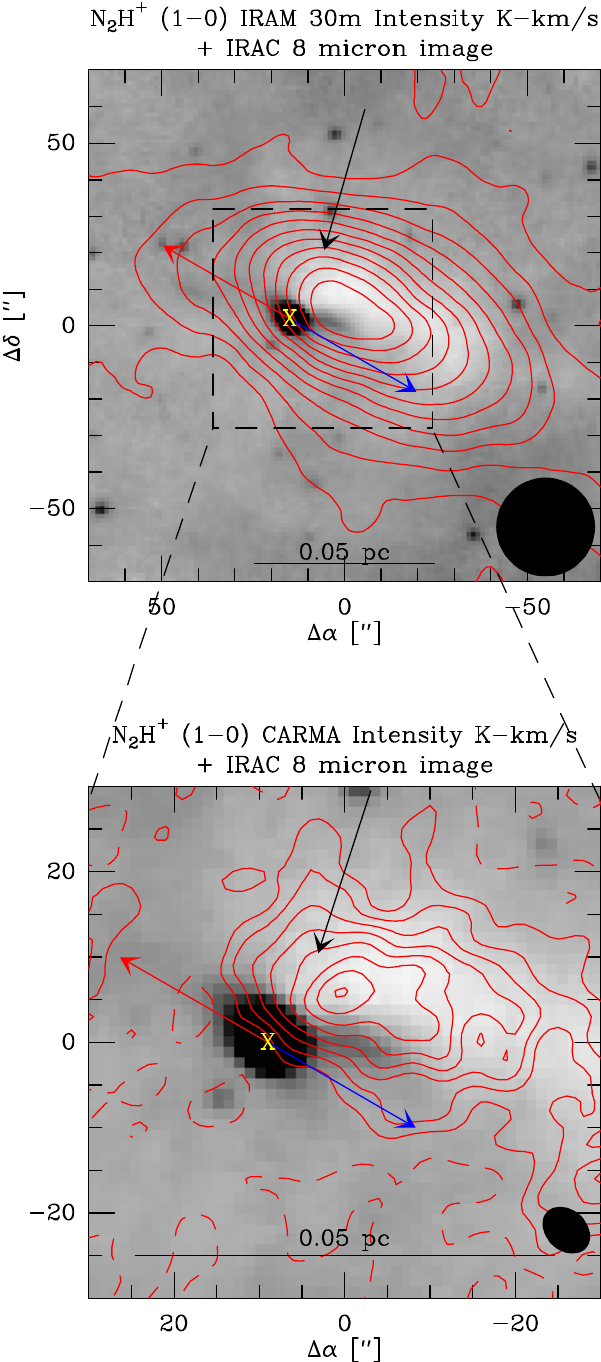}
\includegraphics[scale=0.75]{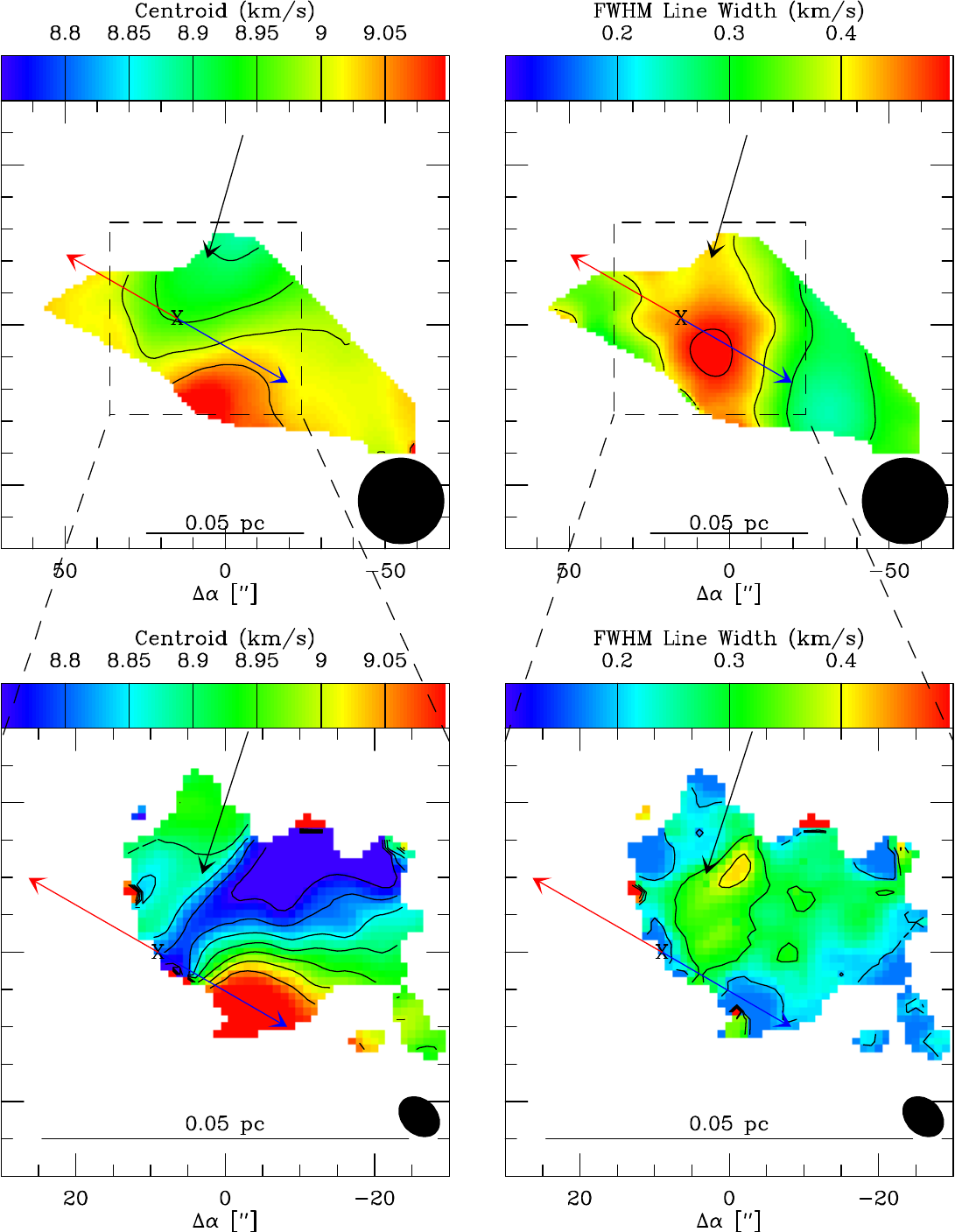}
\end{center}
\caption{HH270 VLA1-- Same as Figure \ref{L1165}. 
The IRAM 30m contours start at 3$\sigma$ with 6$\sigma$ intervals; the CARMA contours start
at $\pm$3$\sigma$ with 3$\sigma$ intervals. The single-dish map shows strong emission
coincident with the 8\mum\ extinction extended along the outflow. The CARMA \nthp\ map shows a similar morphology,
with no emission coincident with the protostar. The single-dish velocity field shows a
gradient normal to and along the outflow;
however, the CARMA velocity map shows that the gas velocity is relatively constant normal to the outflow,
but along the outflow axis there is a clear gradient with more red-shifted emissions toward the outflow axis.
The velocity contours trace
a similar shape as the outflow cavity; the faint scattered light of the outflow cavity can be seen in the 8\mum\ image.
The single-dish linewidth map shows a peak along the outflow southwest of the protostar, this appears to be the
unresolved velocity gradient along the outflow as 
the CARMA map does not have increased linewidth along at the same location.}
\label{HH270}
\end{figure}
\clearpage

\begin{figure}
\begin{center}
\includegraphics[scale=0.5]{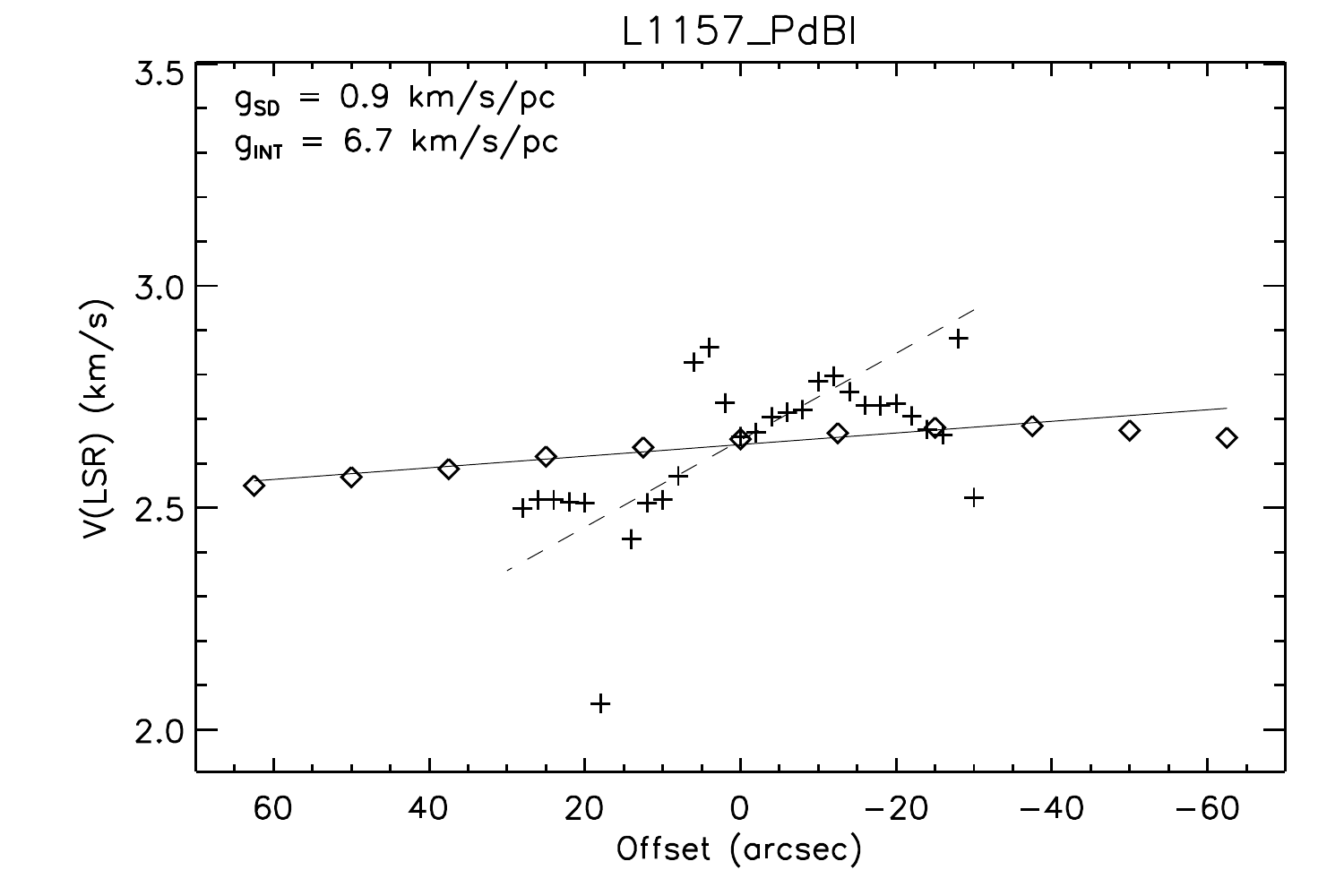}
\includegraphics[scale=0.5]{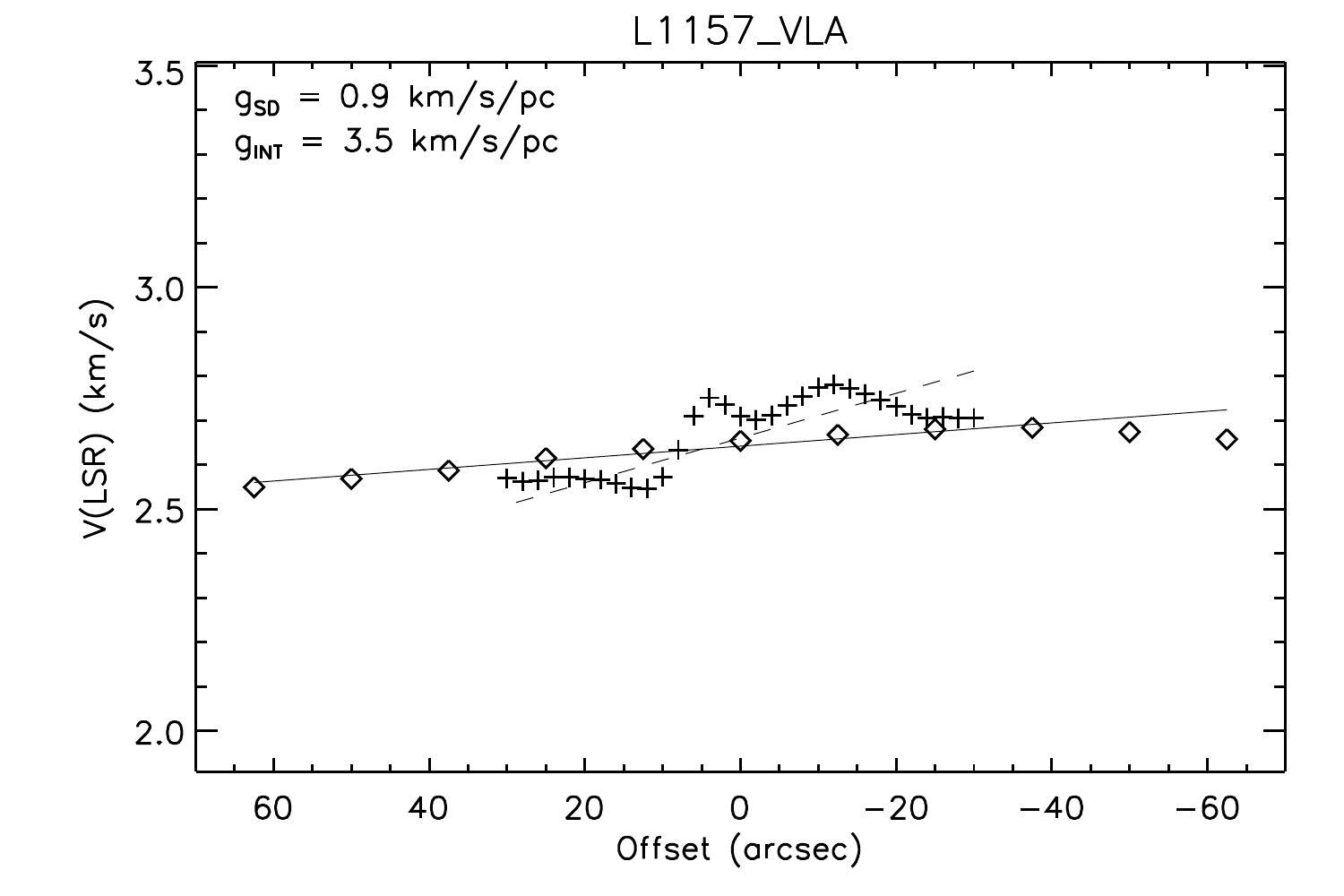}
\includegraphics[scale=0.5]{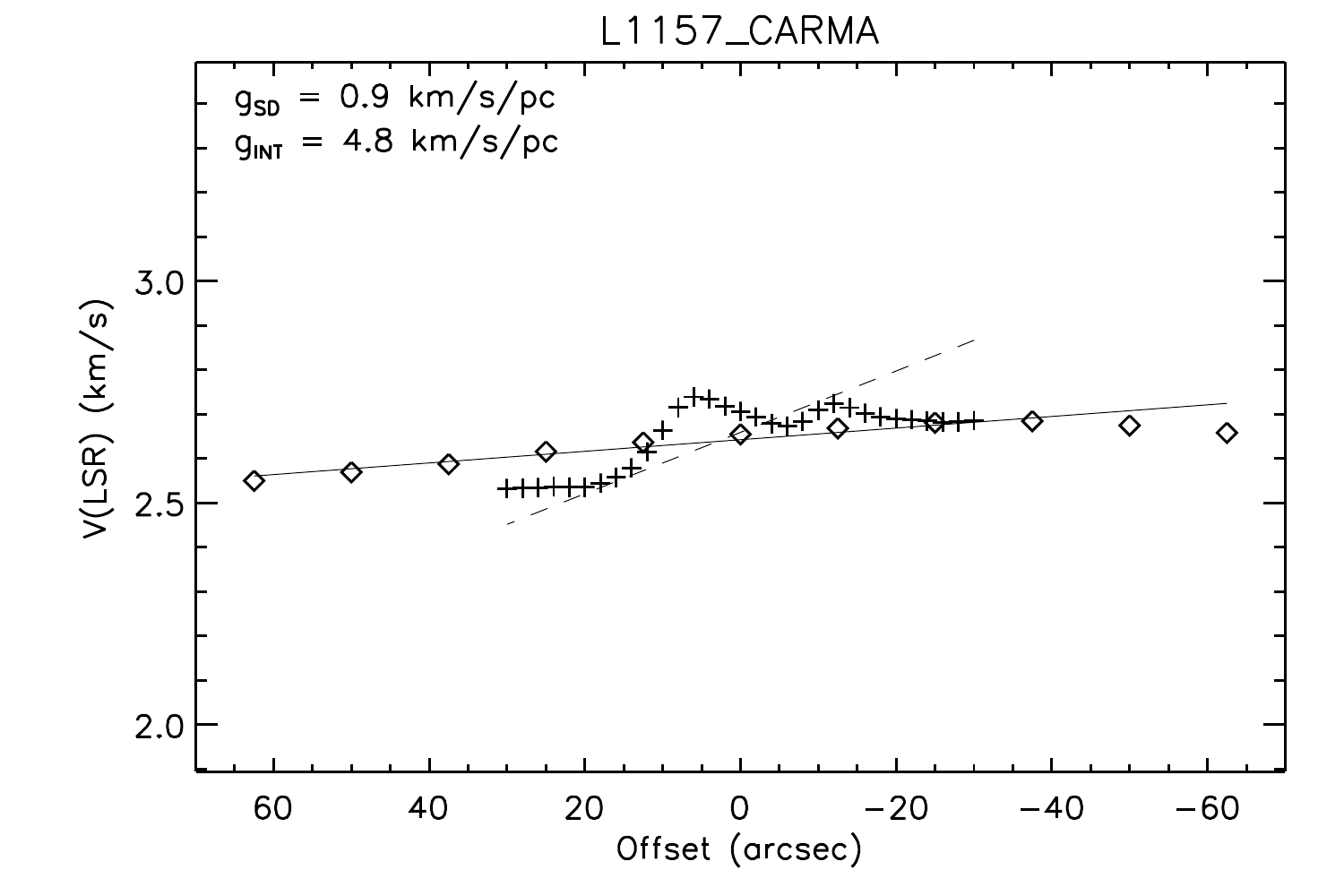}
\includegraphics[scale=0.5]{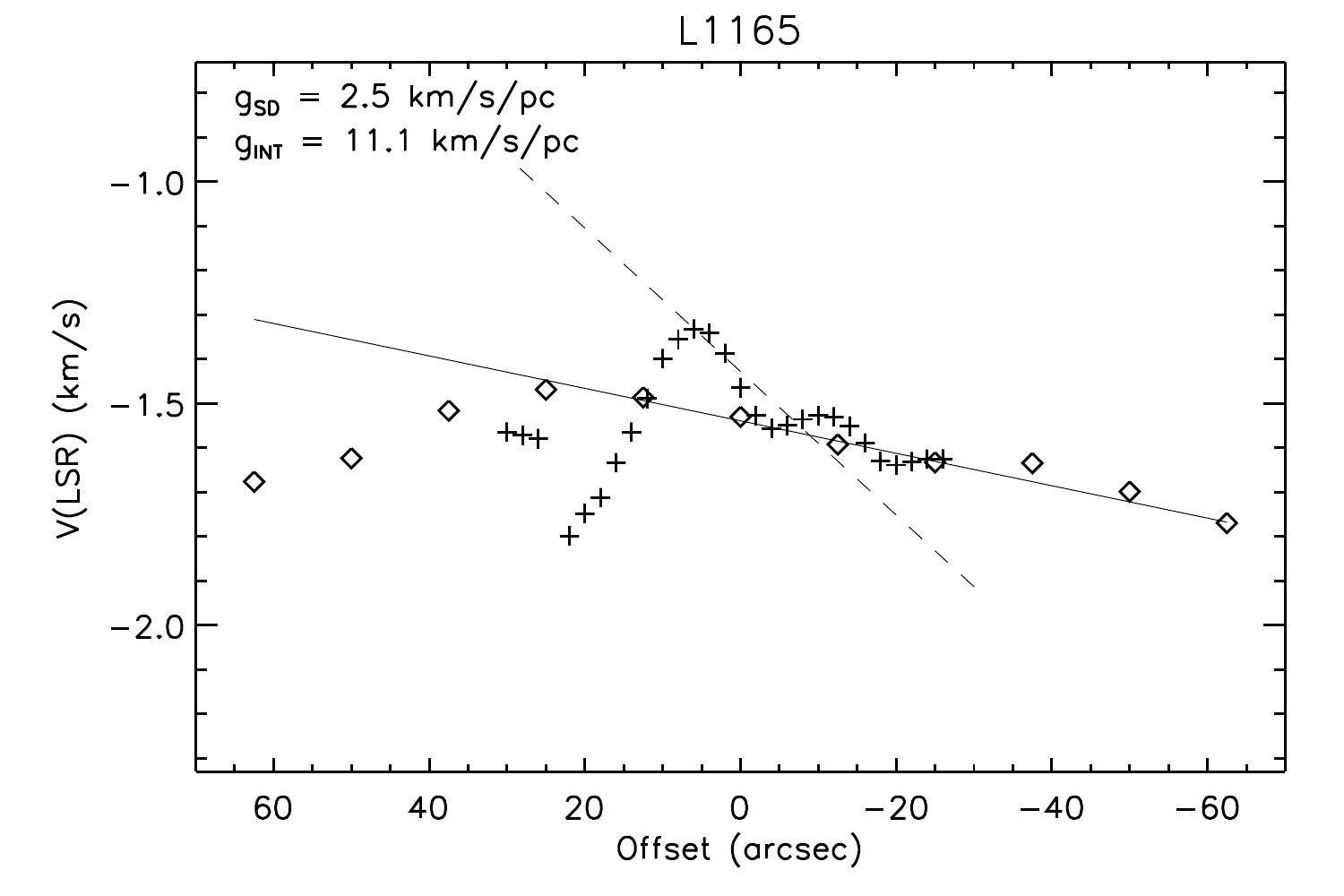}
\includegraphics[scale=0.5]{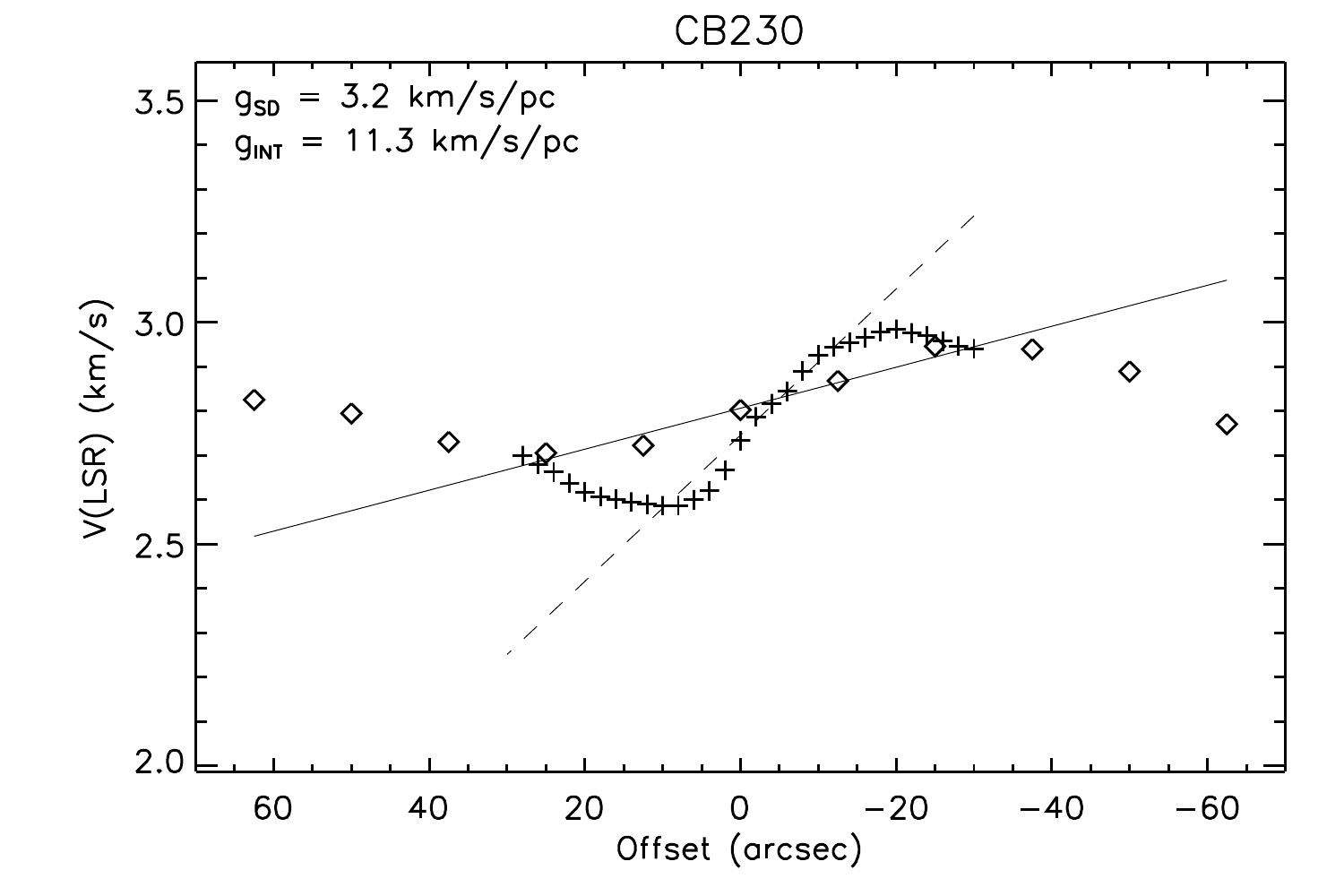}
\includegraphics[scale=0.5]{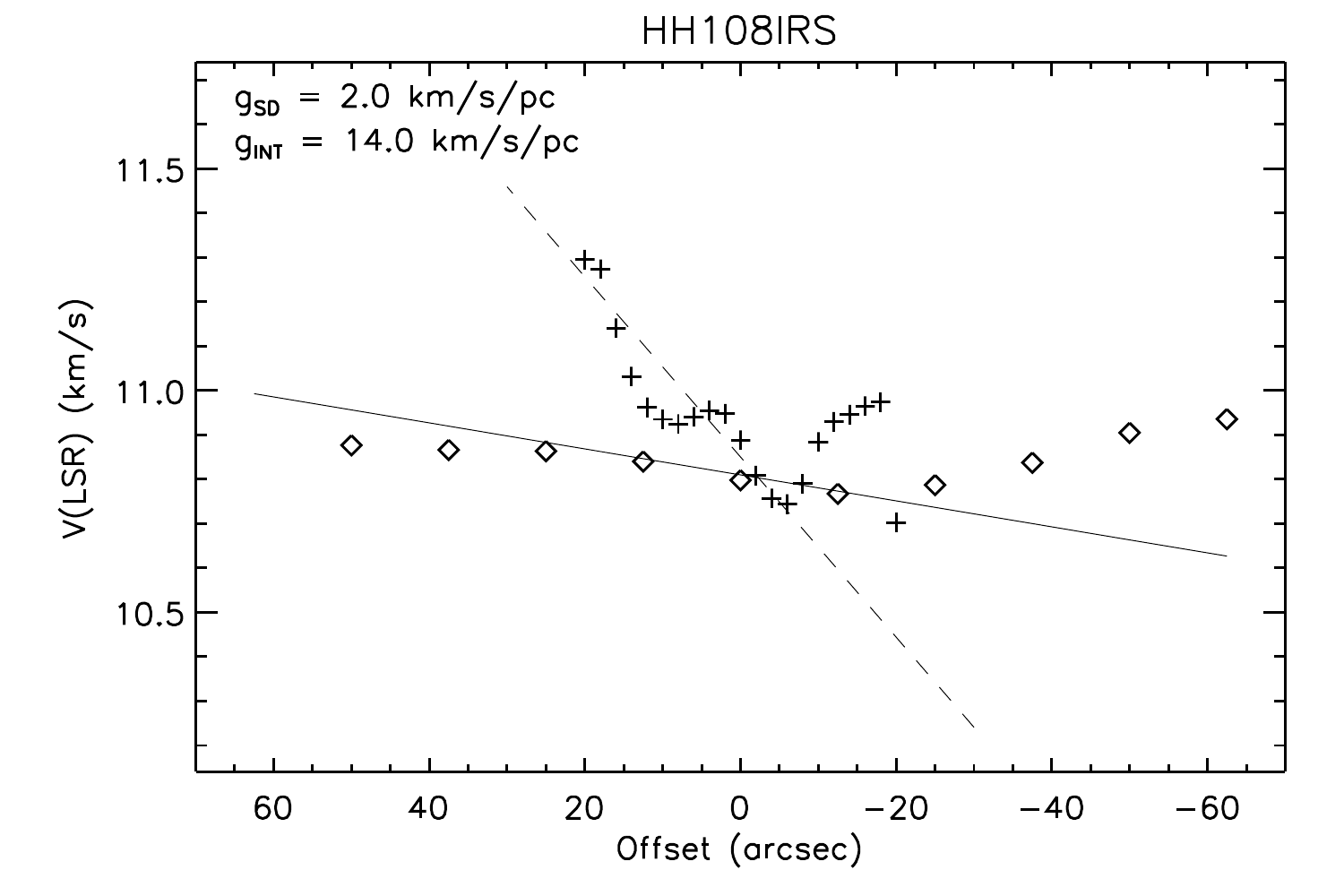}

\end{center}
\caption{Line center velocity cuts taken across each protostellar envelope, normal to the outflow. The diamonds are
the single-dish data and the plus-signs are the interferometer data. Each point for the single-dish data
is the average of the 27$^{\prime\prime}$ beam surrounding each point. Each interferometer point is the average
of points within the semi-major axis of the synthesized beam. The velocity errors at each point are generally $<$0.05 \kms,
approximately the size of the symbols. The solid-lines are the linear fits to
the single-dish velocity data between $\pm$30\arcsec\ for most sources. Serpens MMS3 was fit
between 60\arcsec\ and 0\arcsec\ and L1152 was fit between 40\arcsec\ and -20\arcsec.}
\label{profiles}
\end{figure}
\clearpage

\begin{center}
\includegraphics[scale=0.5]{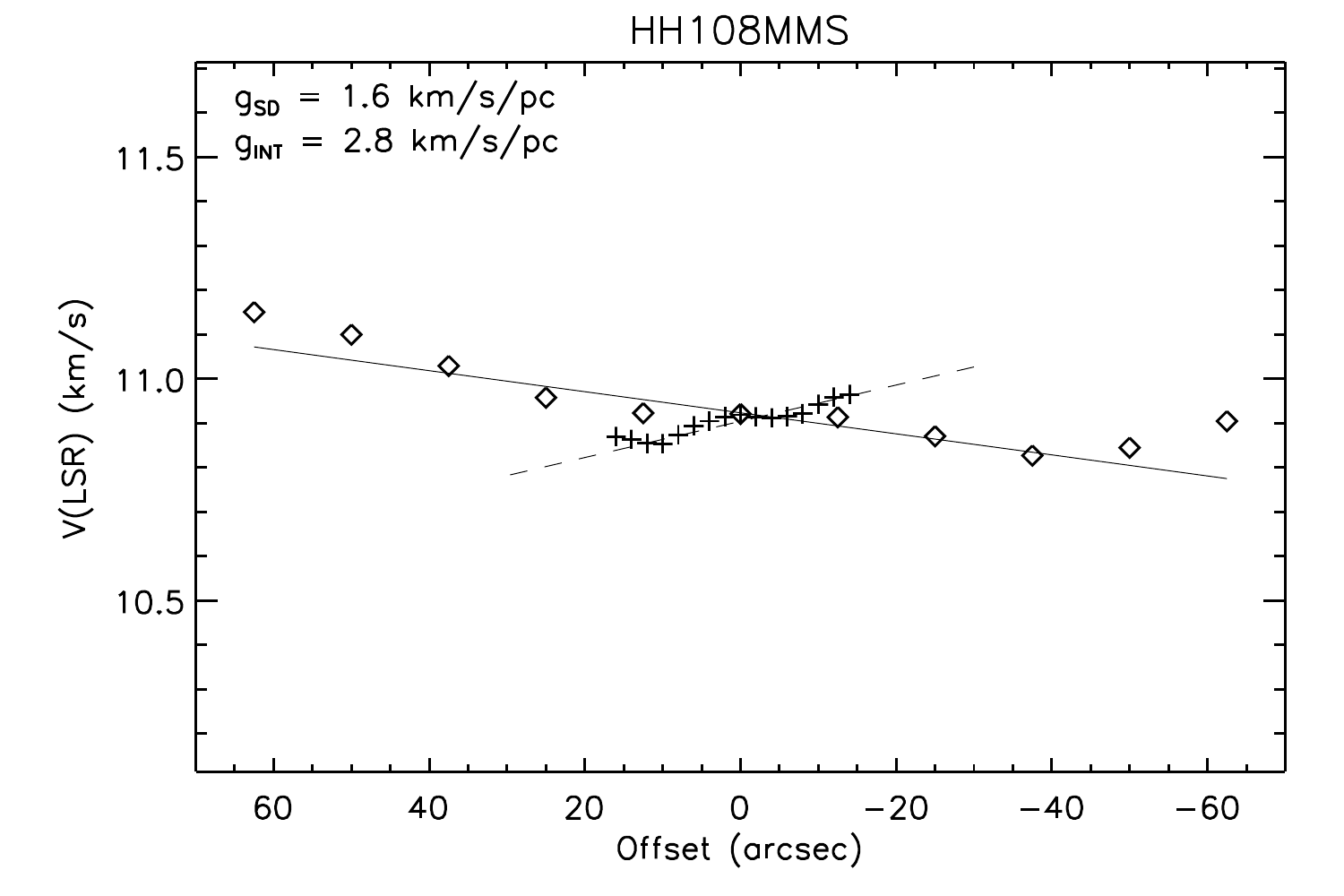}
\includegraphics[scale=0.5]{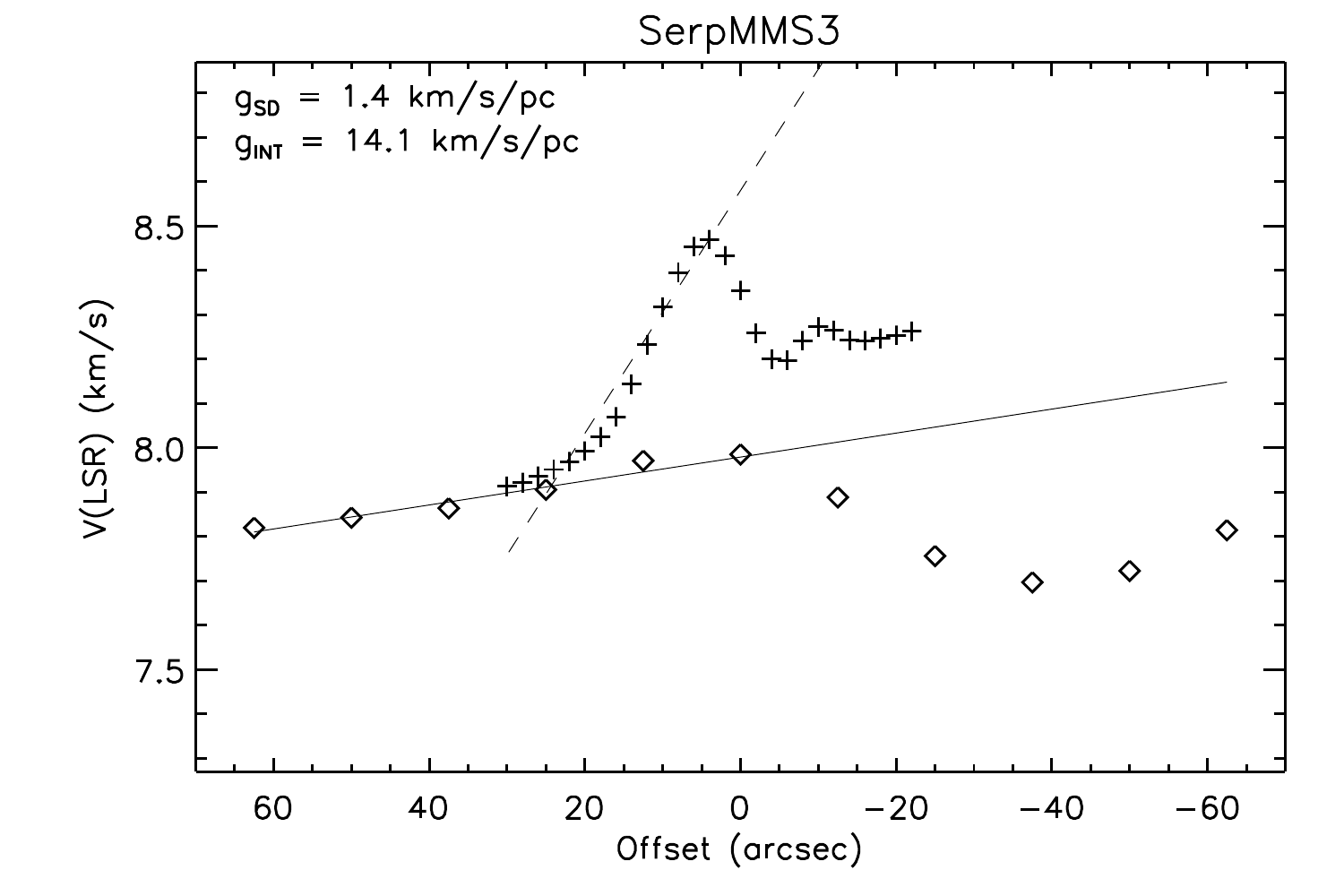}
\includegraphics[scale=0.5]{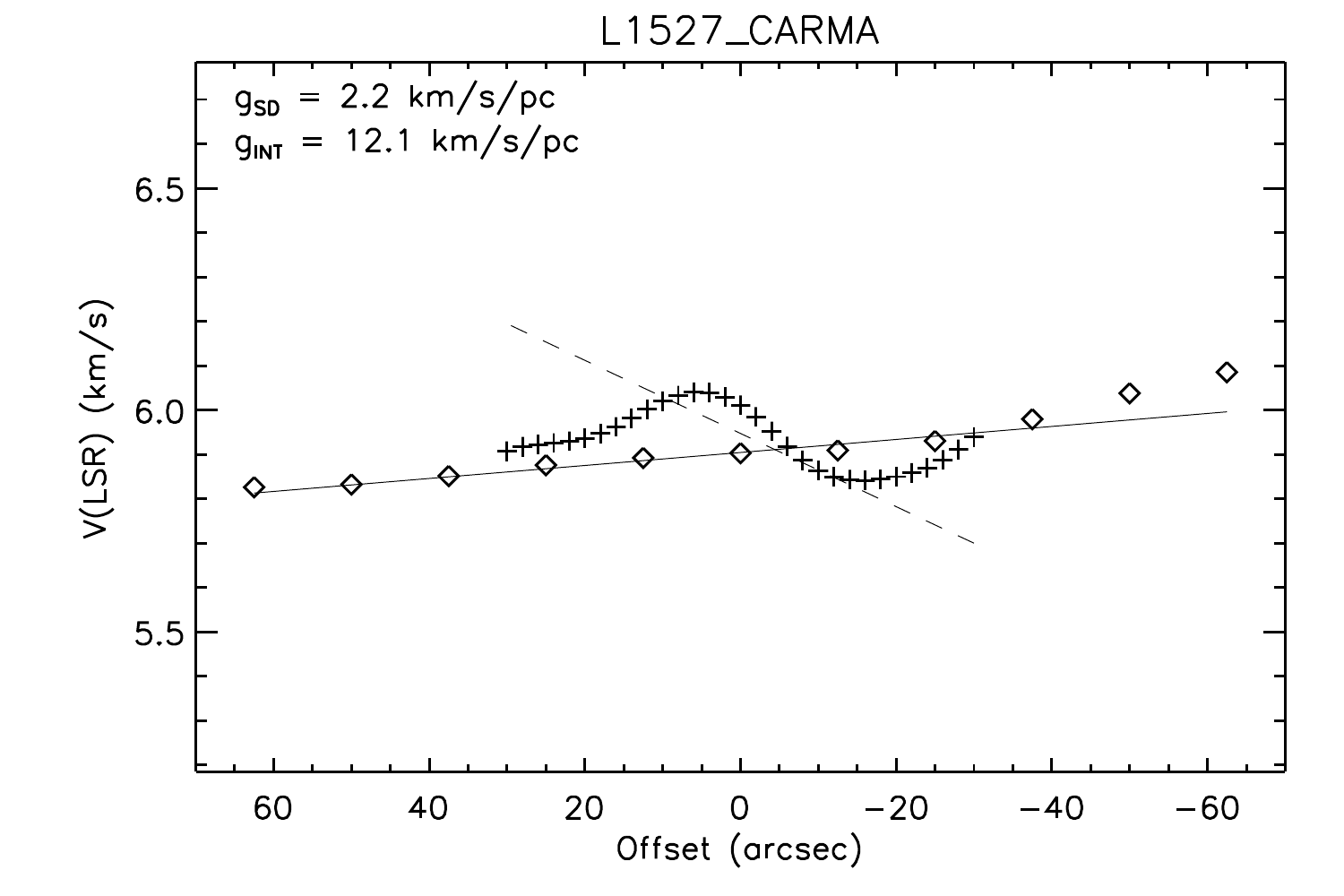}
\includegraphics[scale=0.5]{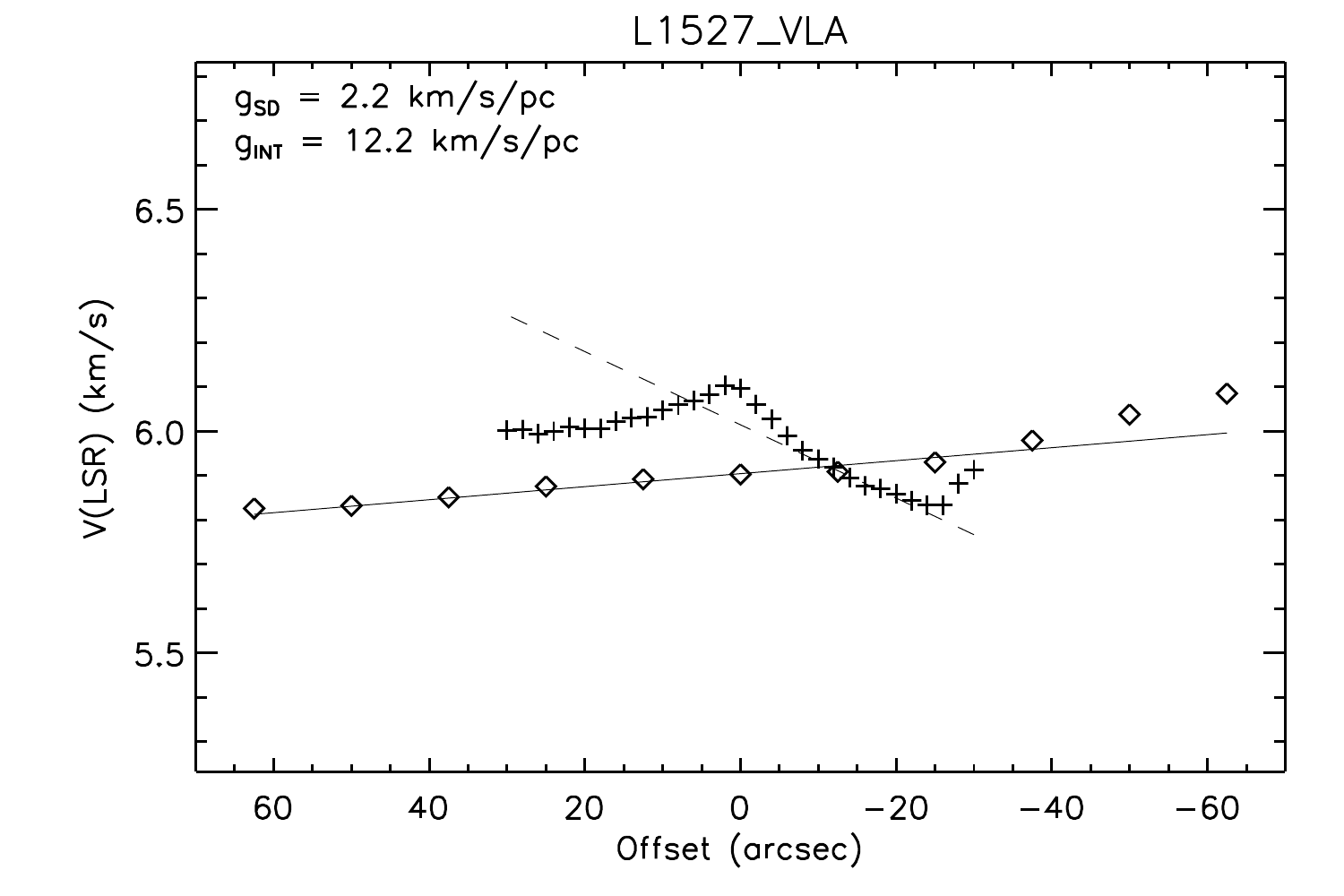}
\includegraphics[scale=0.5]{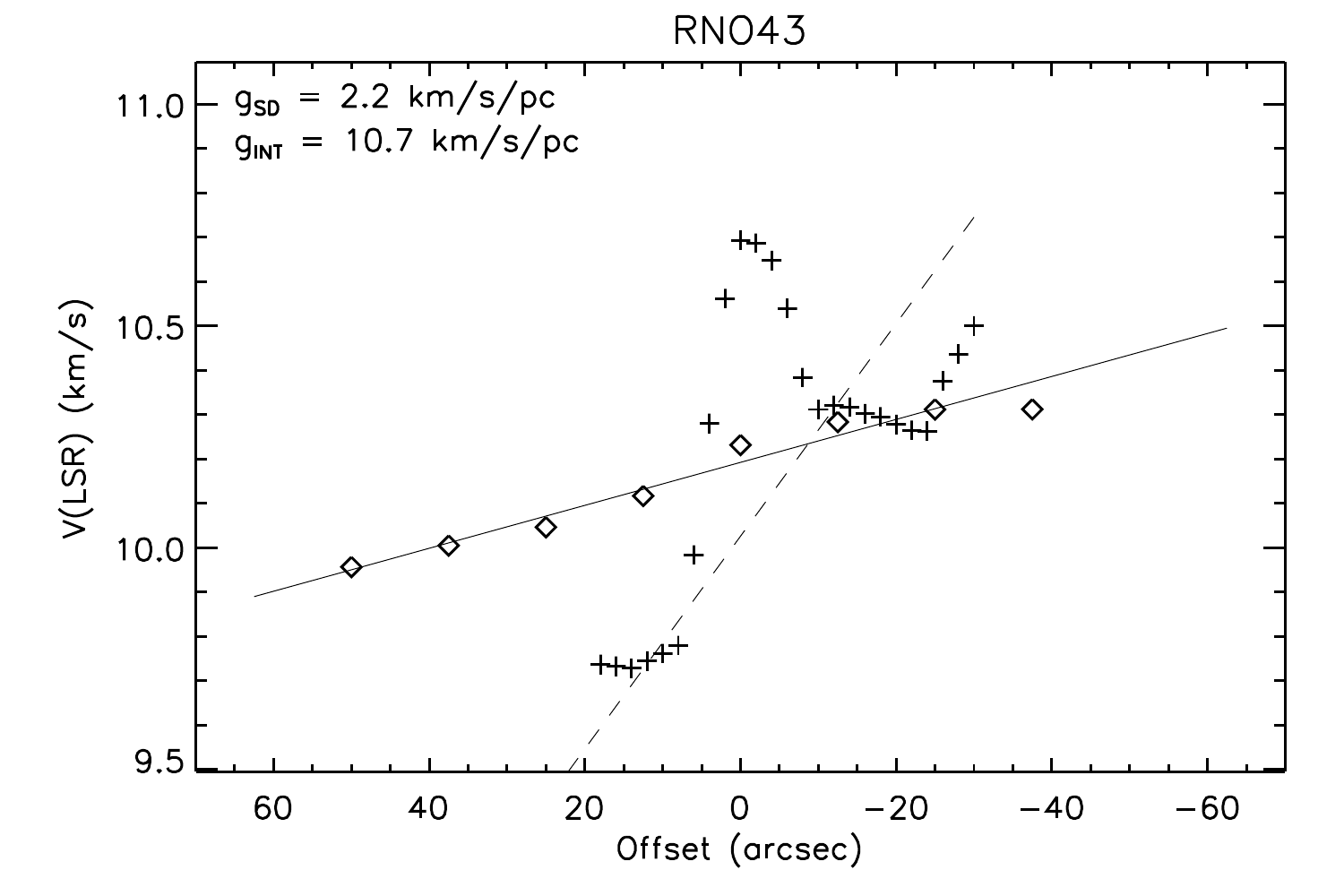}
\includegraphics[scale=0.5]{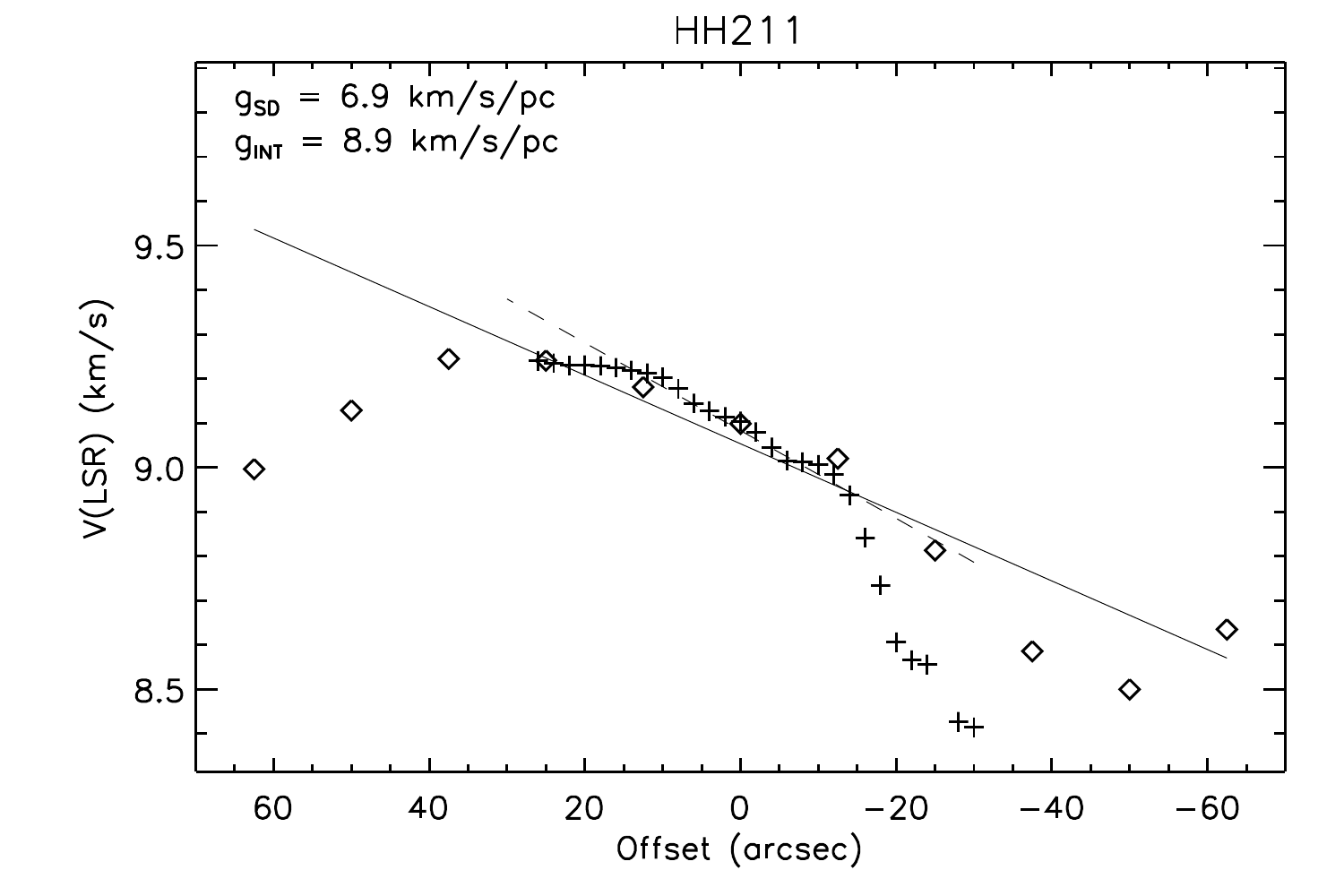}
\includegraphics[scale=0.5]{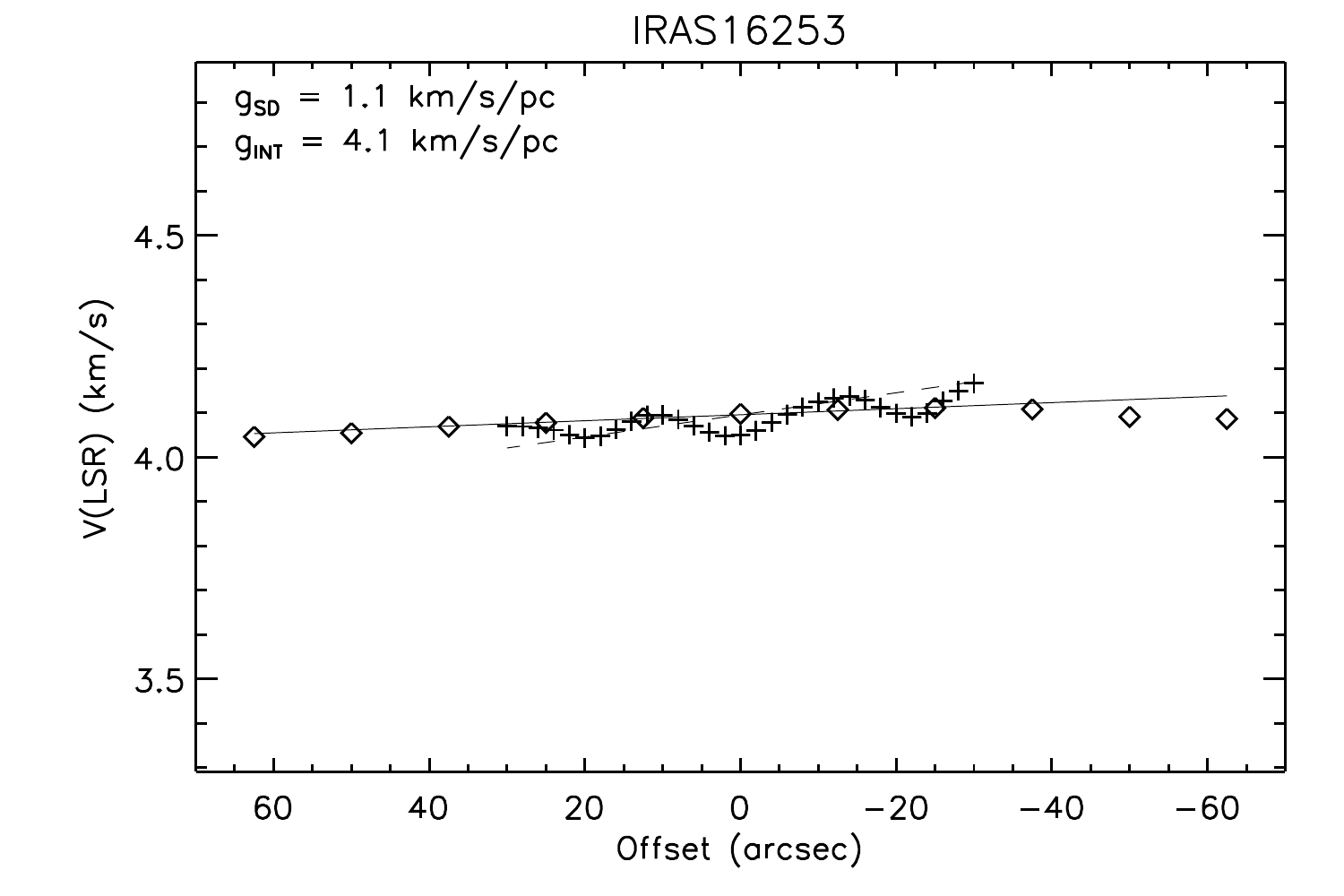}
\includegraphics[scale=0.5]{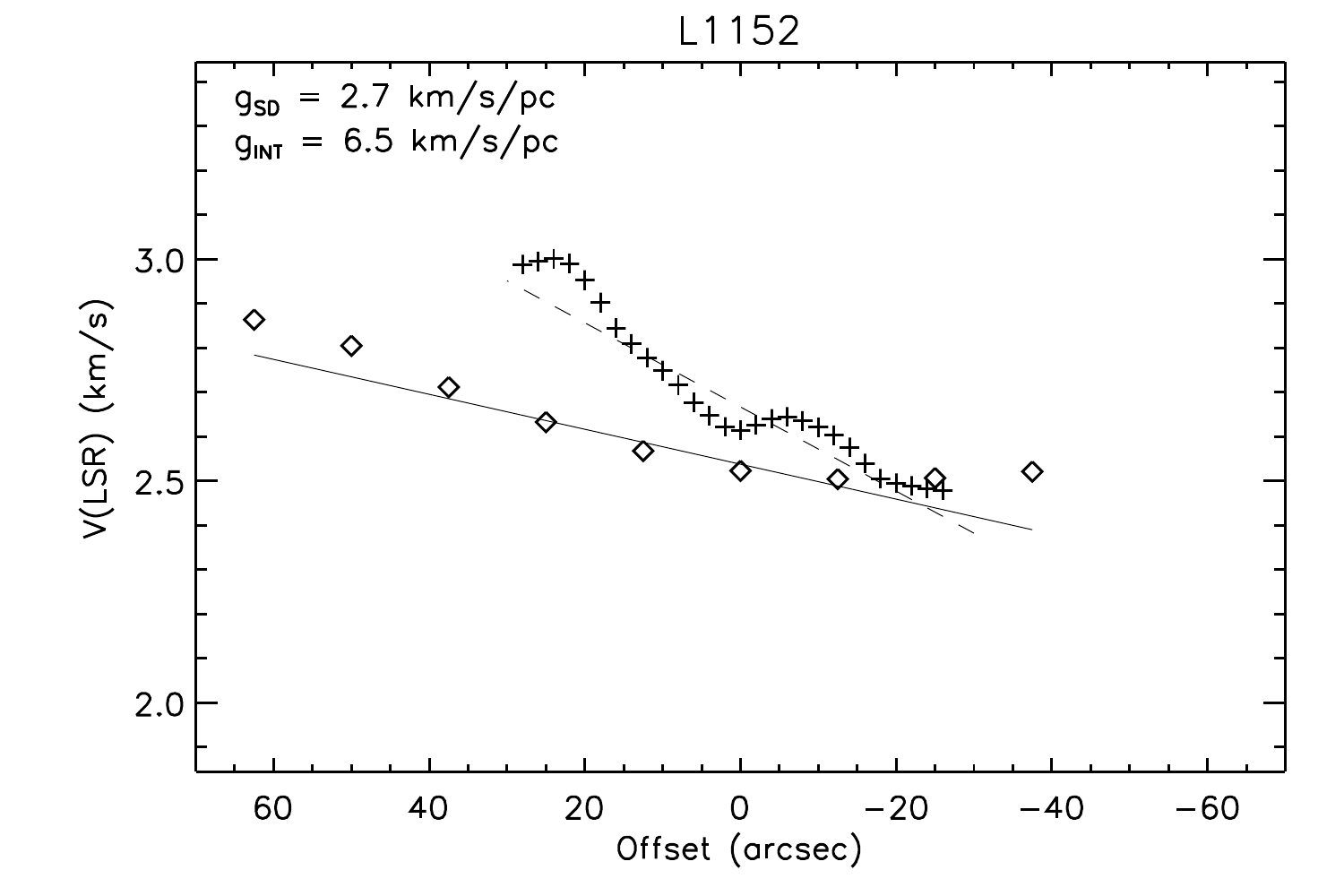}
\end{center}
\centerline{Fig. 24 --- cont'd.}

\begin{center}
\includegraphics[scale=0.5]{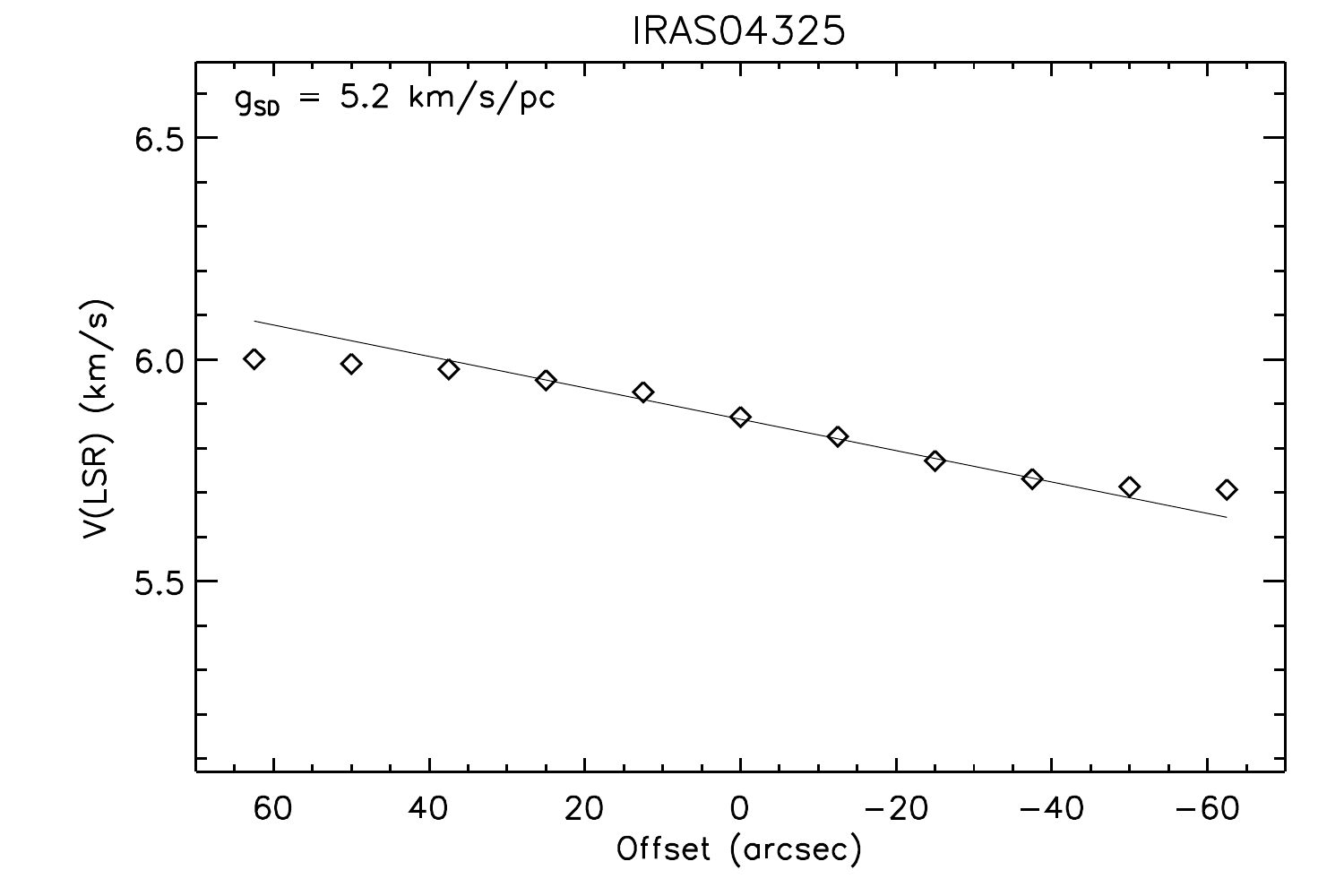}
\includegraphics[scale=0.5]{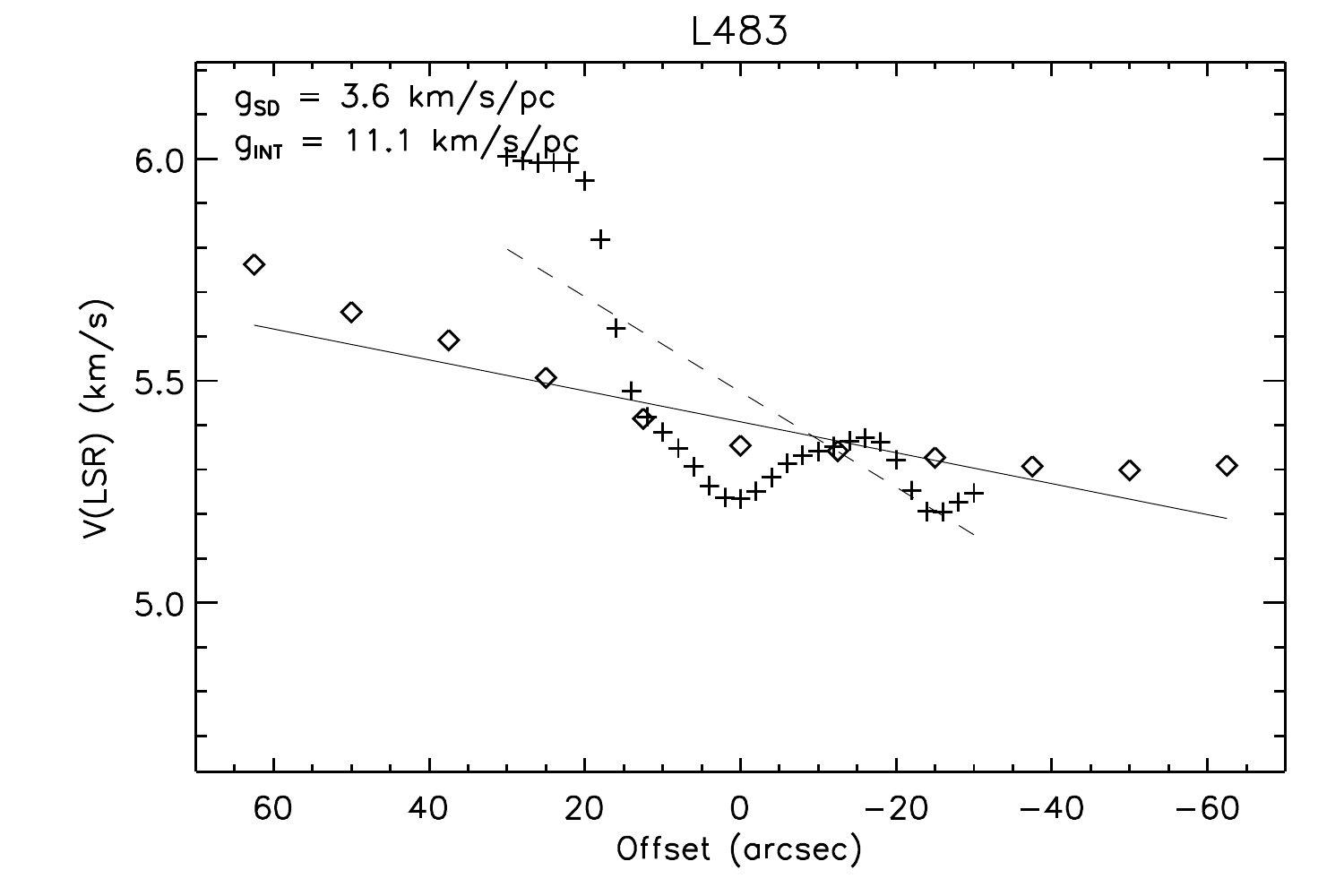}
\includegraphics[scale=0.5]{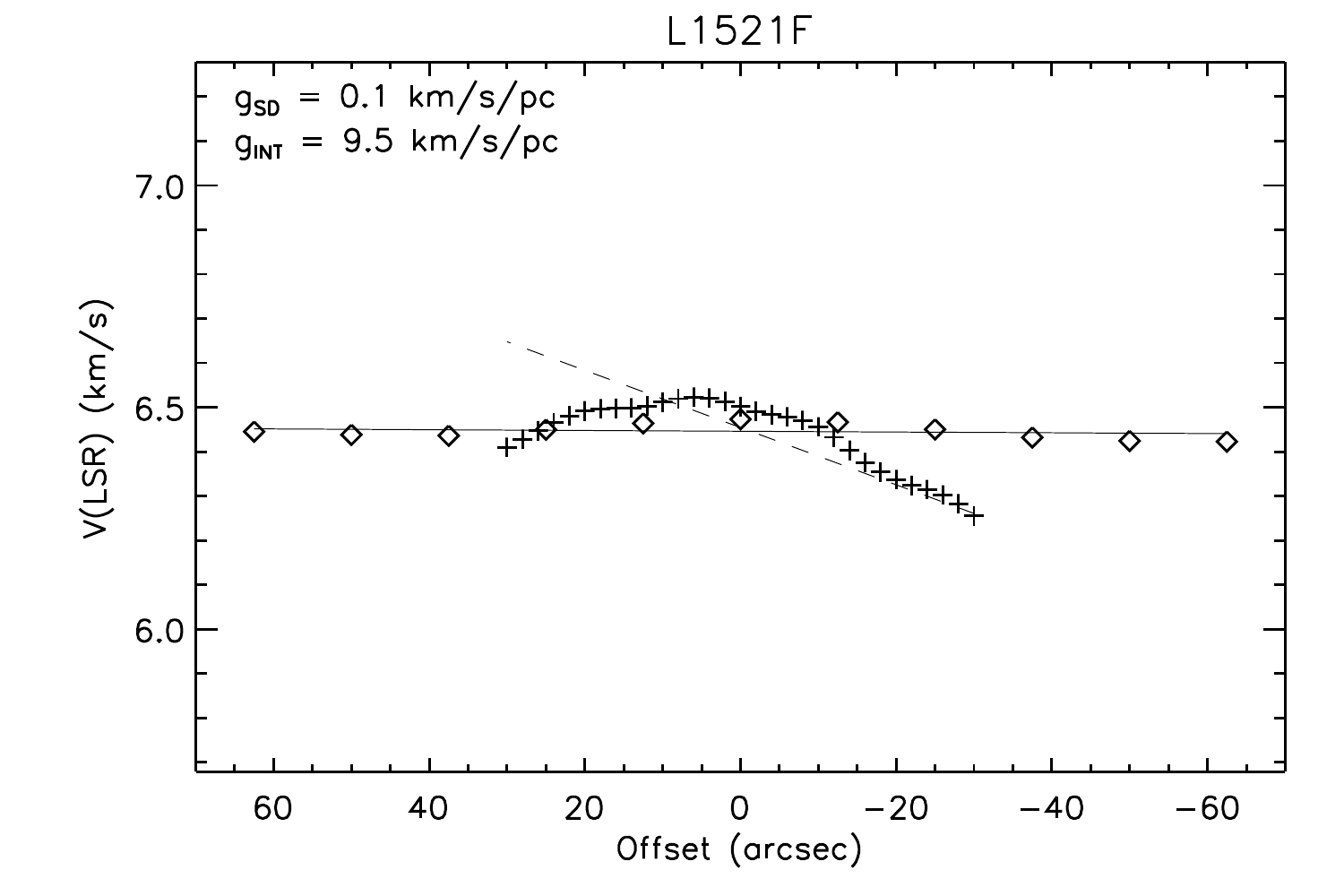}
\includegraphics[scale=0.5]{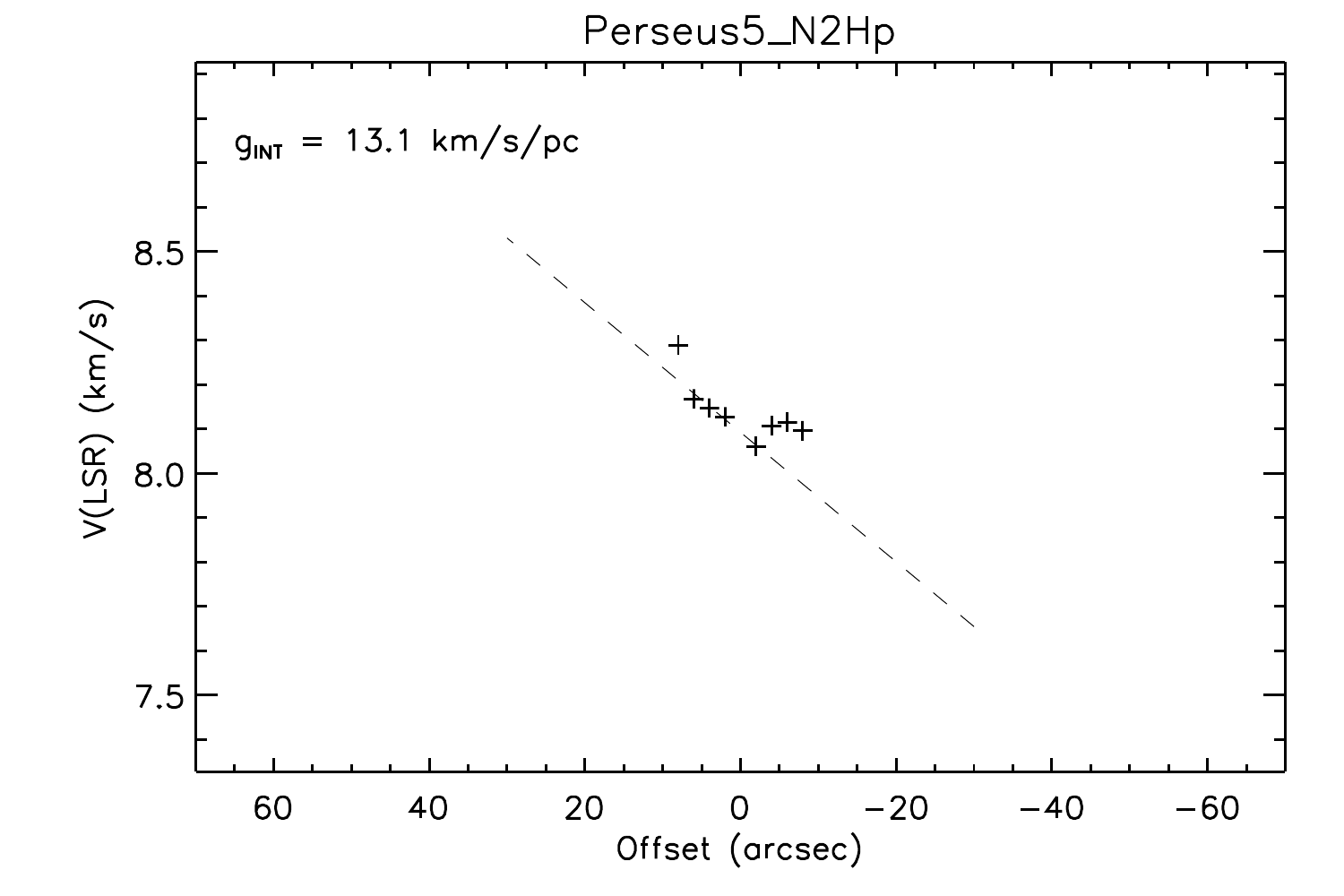}
\includegraphics[scale=0.5]{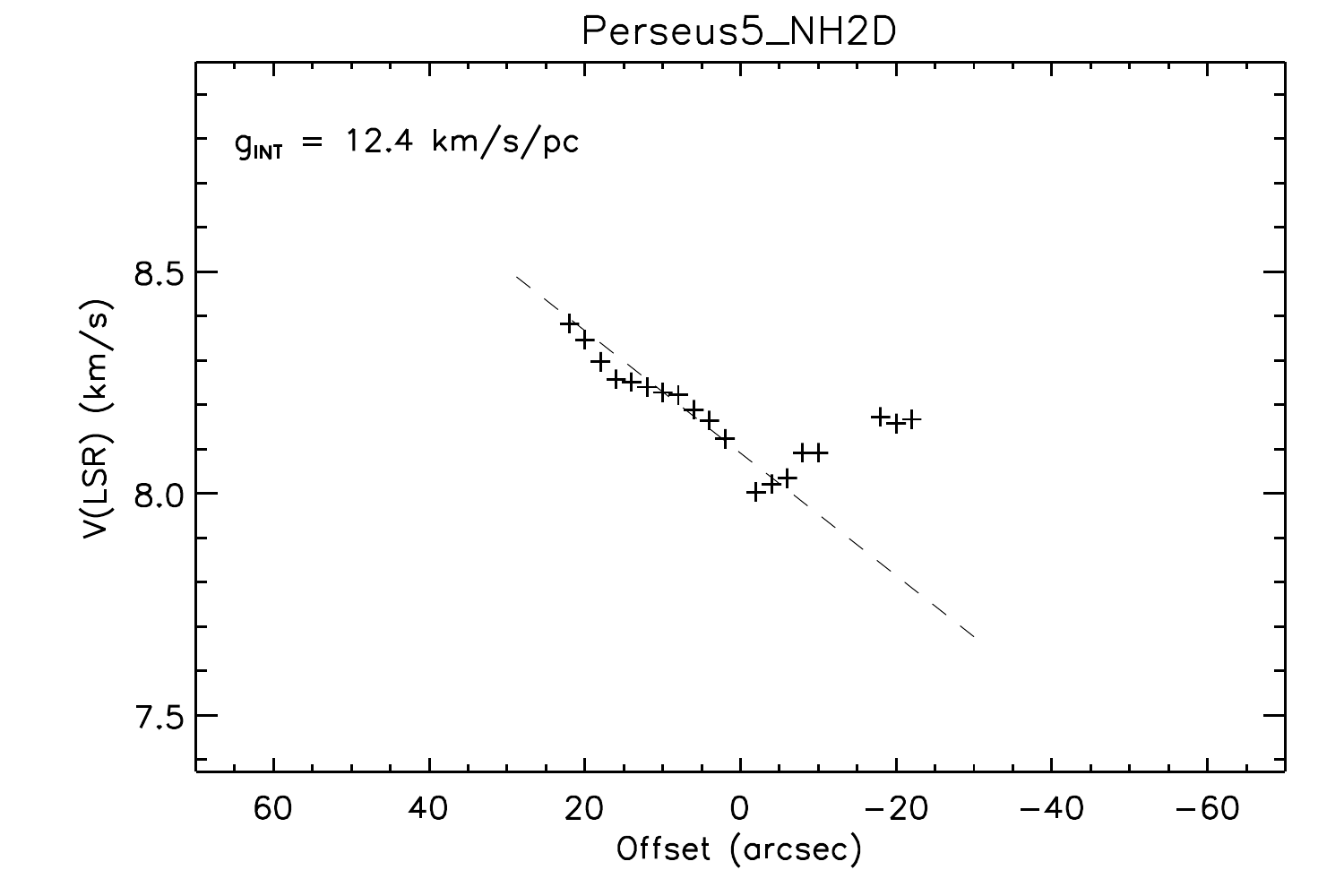}
\includegraphics[scale=0.5]{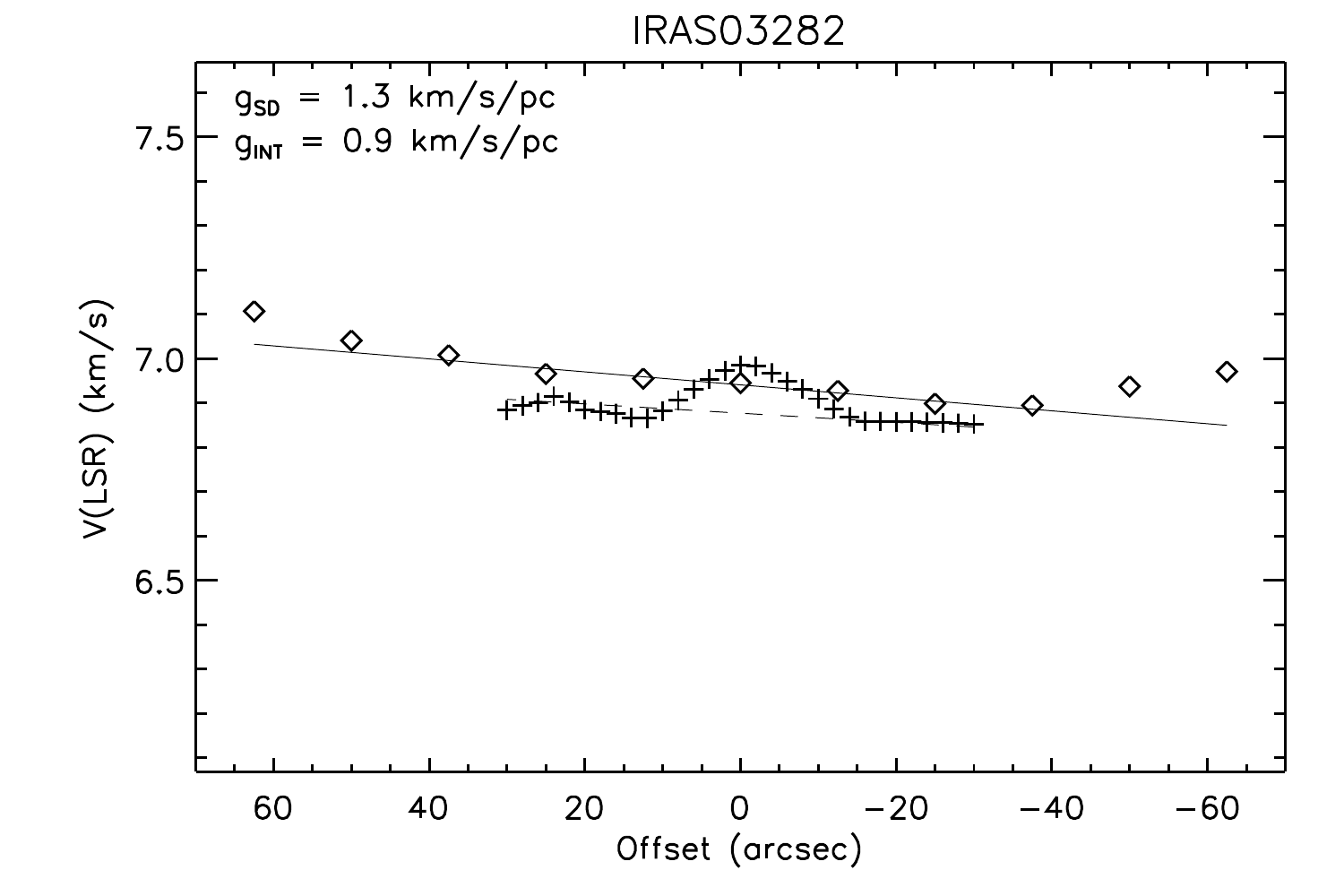}
\includegraphics[scale=0.5]{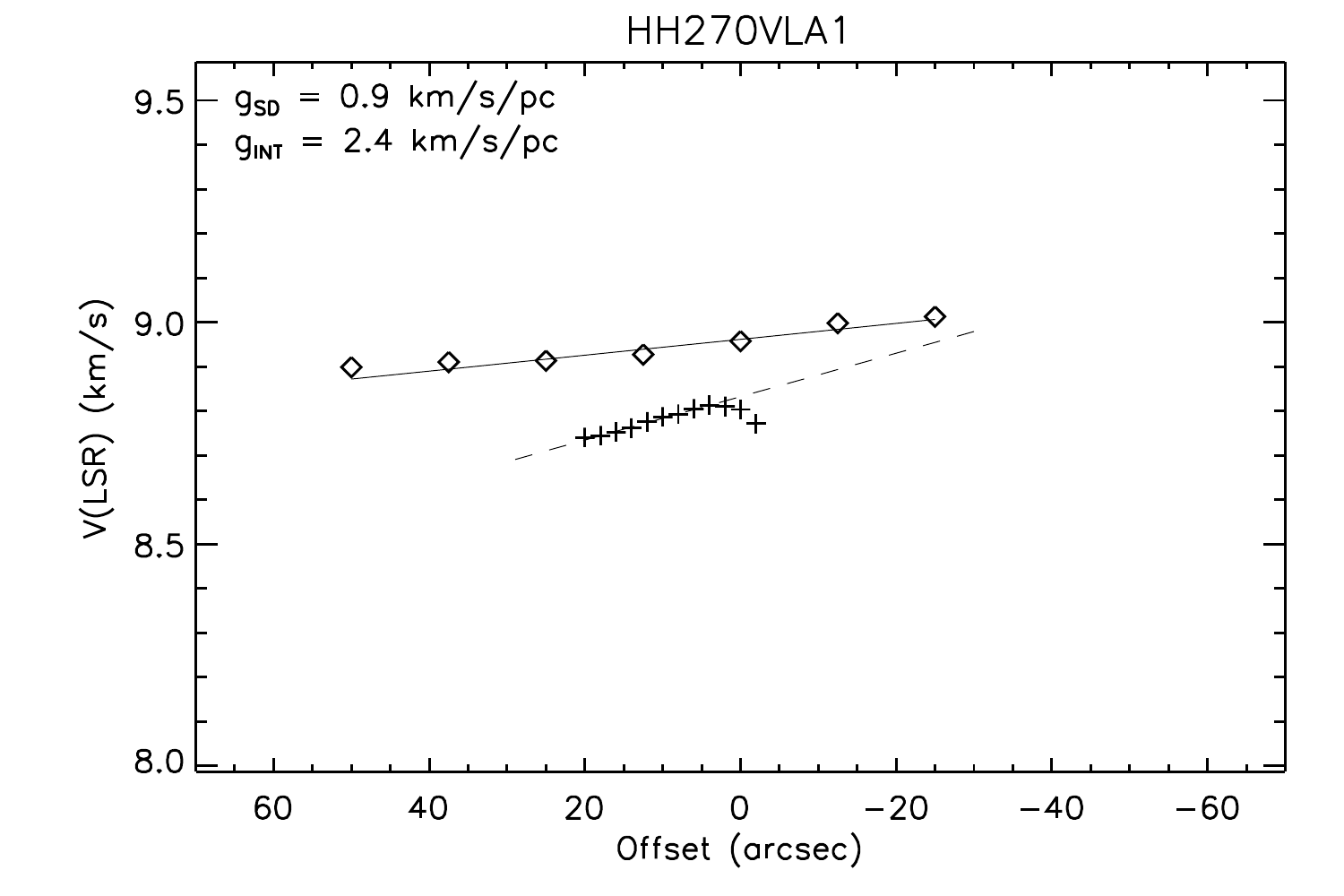}
\end{center}
\centerline{Fig. 24 --- cont'd.}

\begin{figure}
\begin{center}
\includegraphics[scale=0.35]{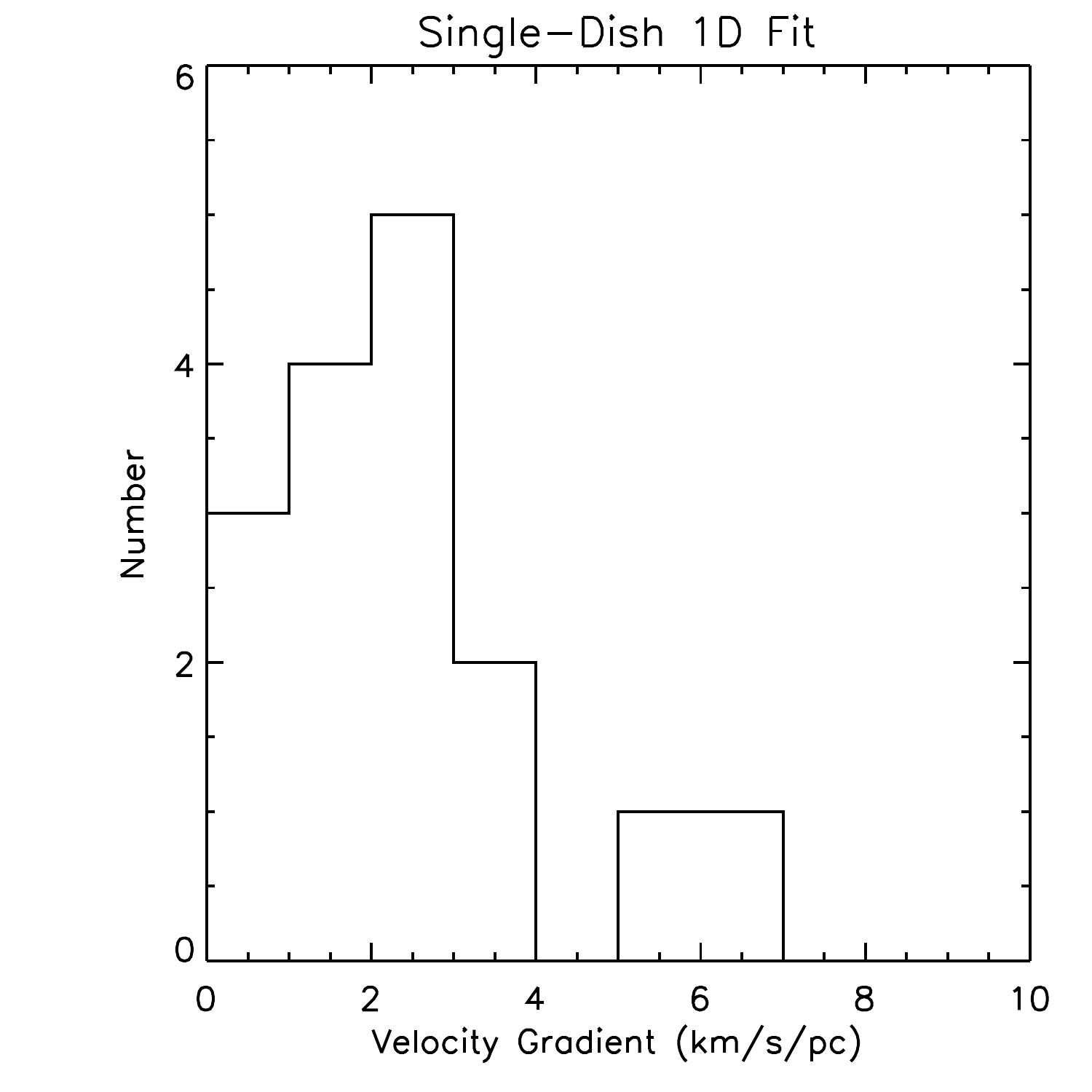}
\includegraphics[scale=0.35]{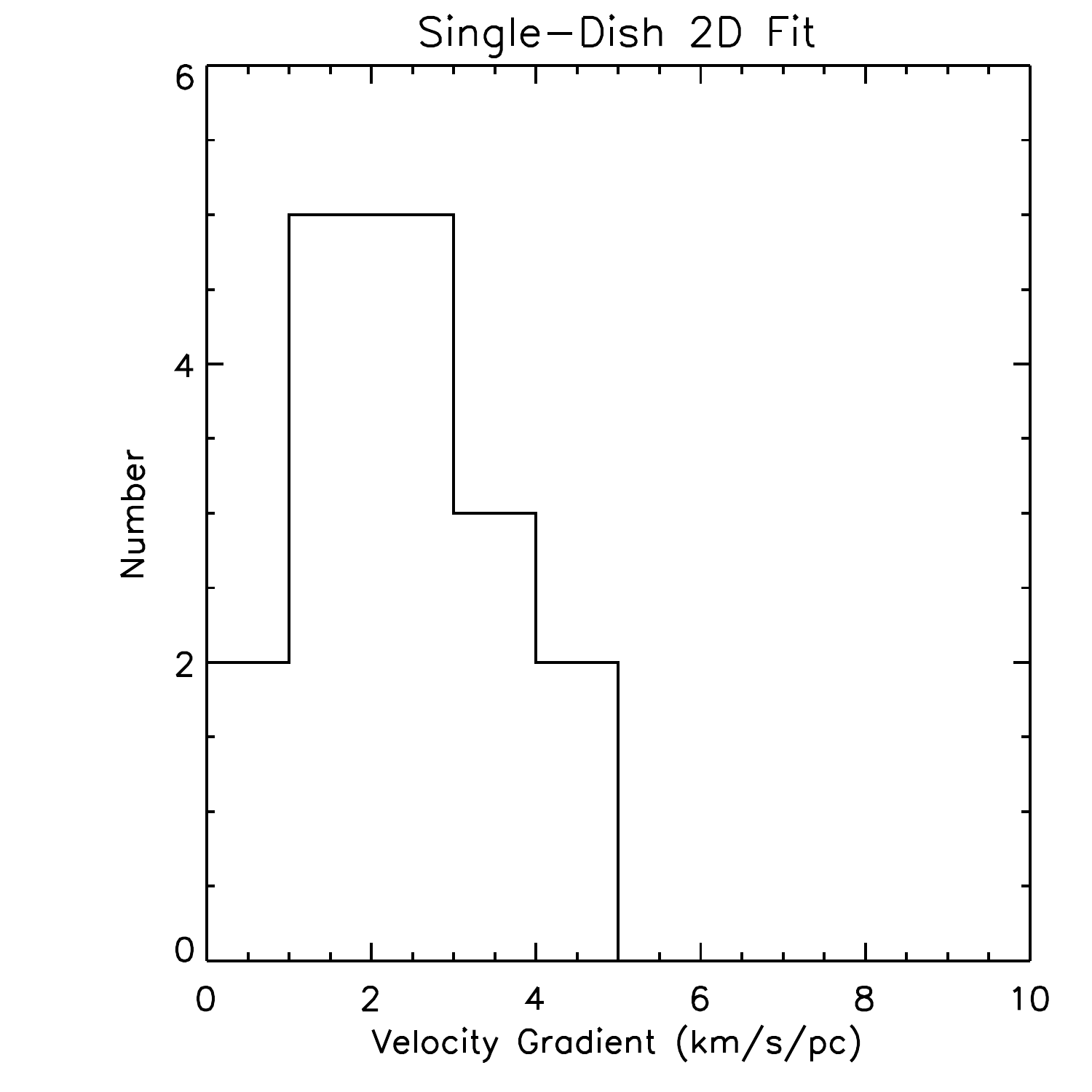}
\includegraphics[scale=0.35]{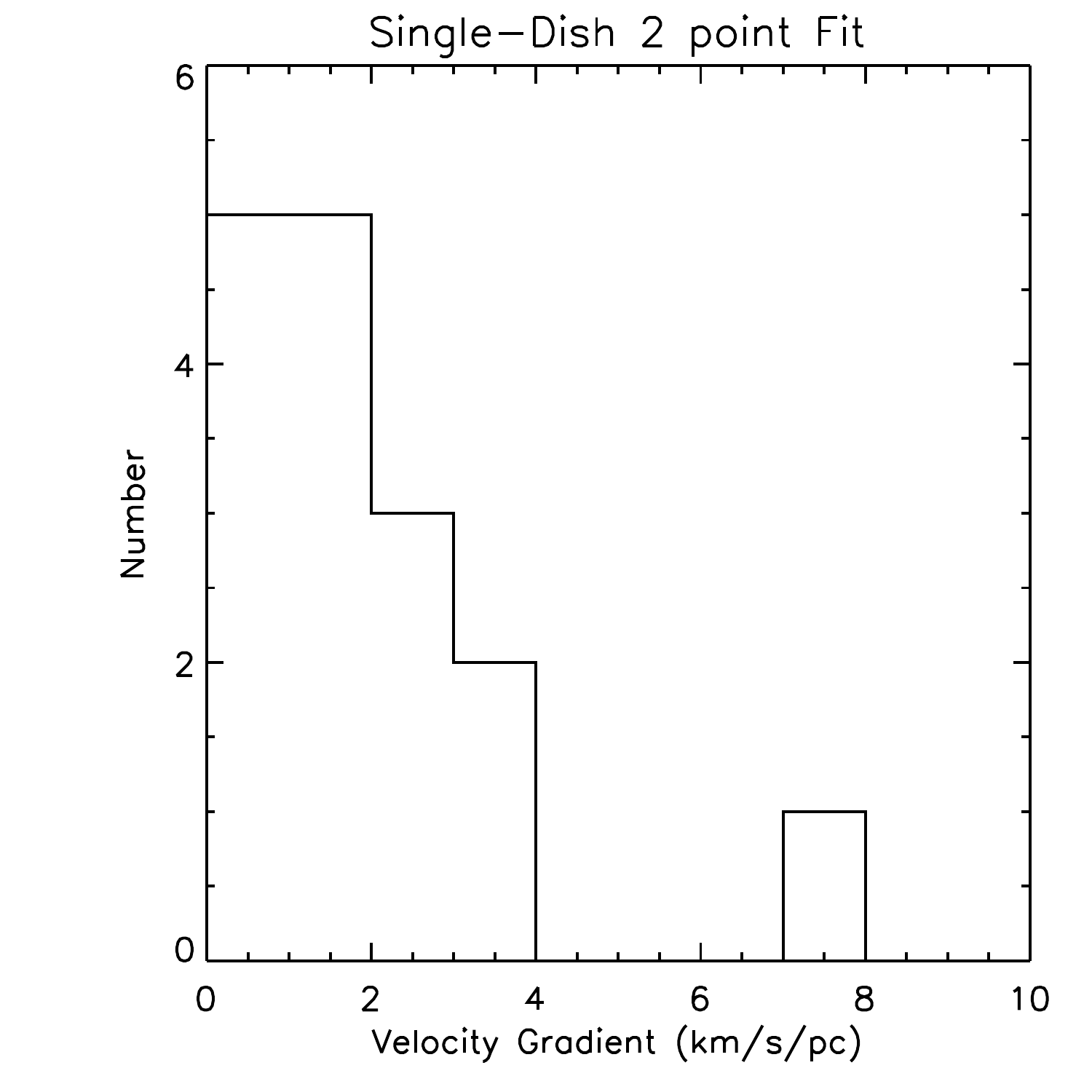}
\end{center}
\caption{Histogram plots of measured velocity gradients from single-dish data. The left panel shows the velocity
gradients derived from one-dimensional cuts across the velocity field, taken normal to the outflow. The middle
panel shows the velocity gradients derived from a two-dimensional fit to the velocity field. The right panel shows
the velocity gradients derived from the velocity difference at $\pm$10000 AU from the protostar, also normal to the outflow.
The distributions from the one- and two-dimensional fits are comparable while the two point method is skewed toward
smaller gradients; this difference is likely due to the gradients fits picking up on higher velocity emission 
that sometimes turns over toward lower velocities by $\sim$ 10000 AU. The middle panel include the two-dimensional fit
for L673 which is absent from the left and right panels.}
\label{sdgradienthistos}
\end{figure}
\clearpage

\begin{figure}
\begin{center}
\includegraphics[scale=0.5]{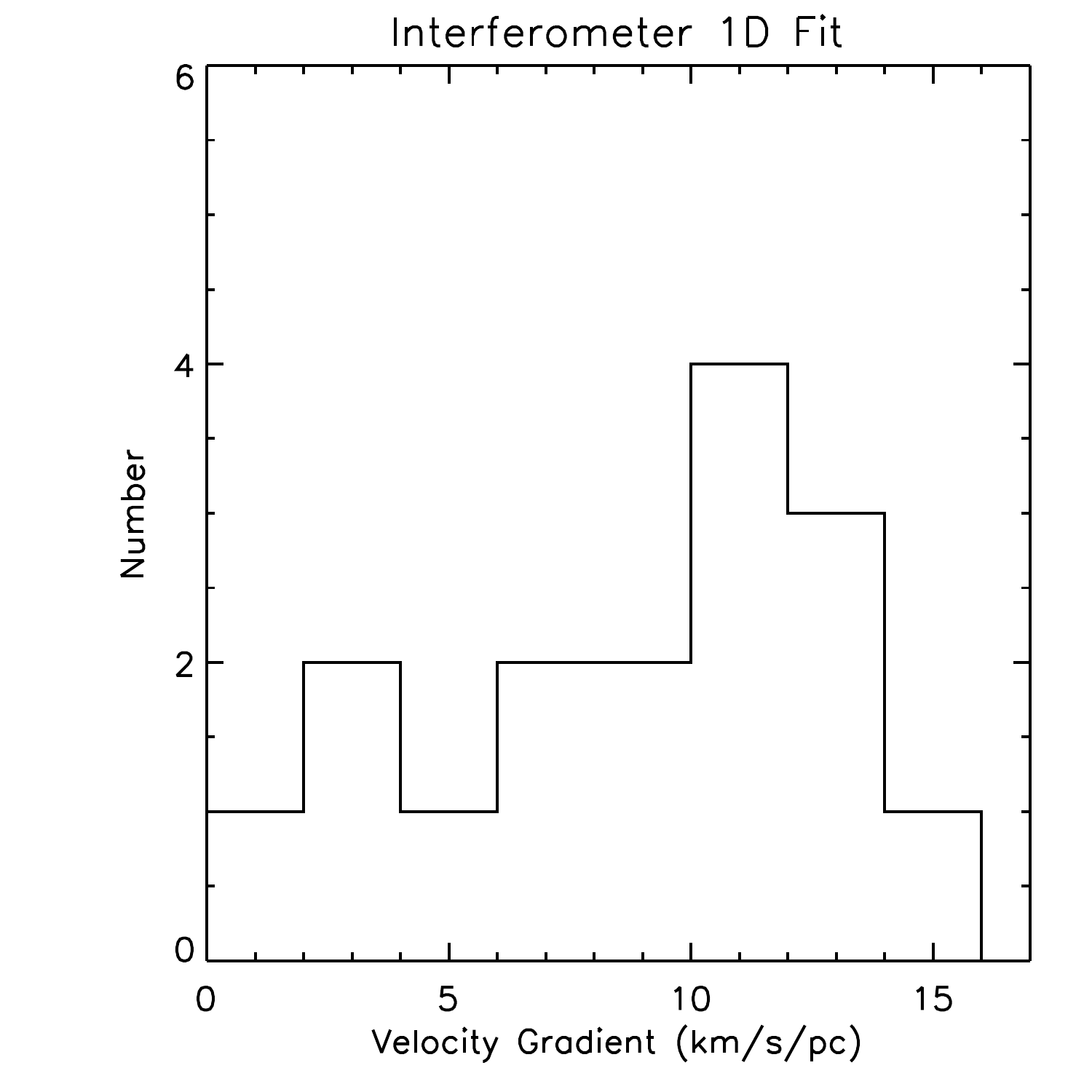}
\includegraphics[scale=0.5]{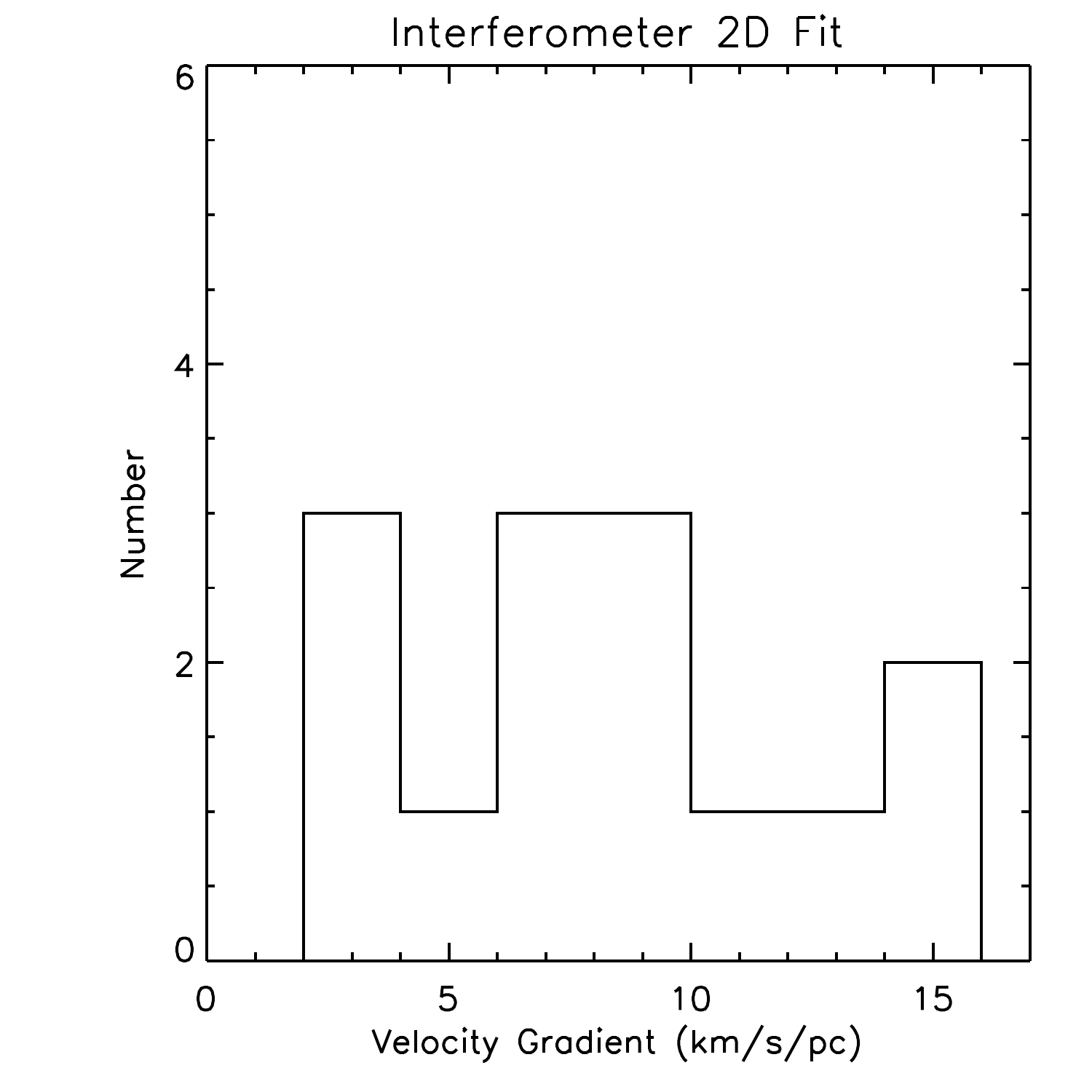}
\end{center}
\caption{Histogram plots of measured velocity gradients from interferometer data. The left panel shows the velocity
gradients derived from one-dimensional cuts across the velocity field, taken normal to the outflow. The right
panel shows the velocity gradients derived from a two-dimensional fit to the velocity field. The differences in these
distributions likely result from the two-dimensional method having to fit all the data where complexities in the velocity
field may reduce the gradient fit. The left panels include L1157 and Serpens MMS3 while they are absent from the right panels.}
\label{intgradienthistos}
\end{figure}
\clearpage

\begin{figure}
\begin{center}
\includegraphics[scale=0.5]{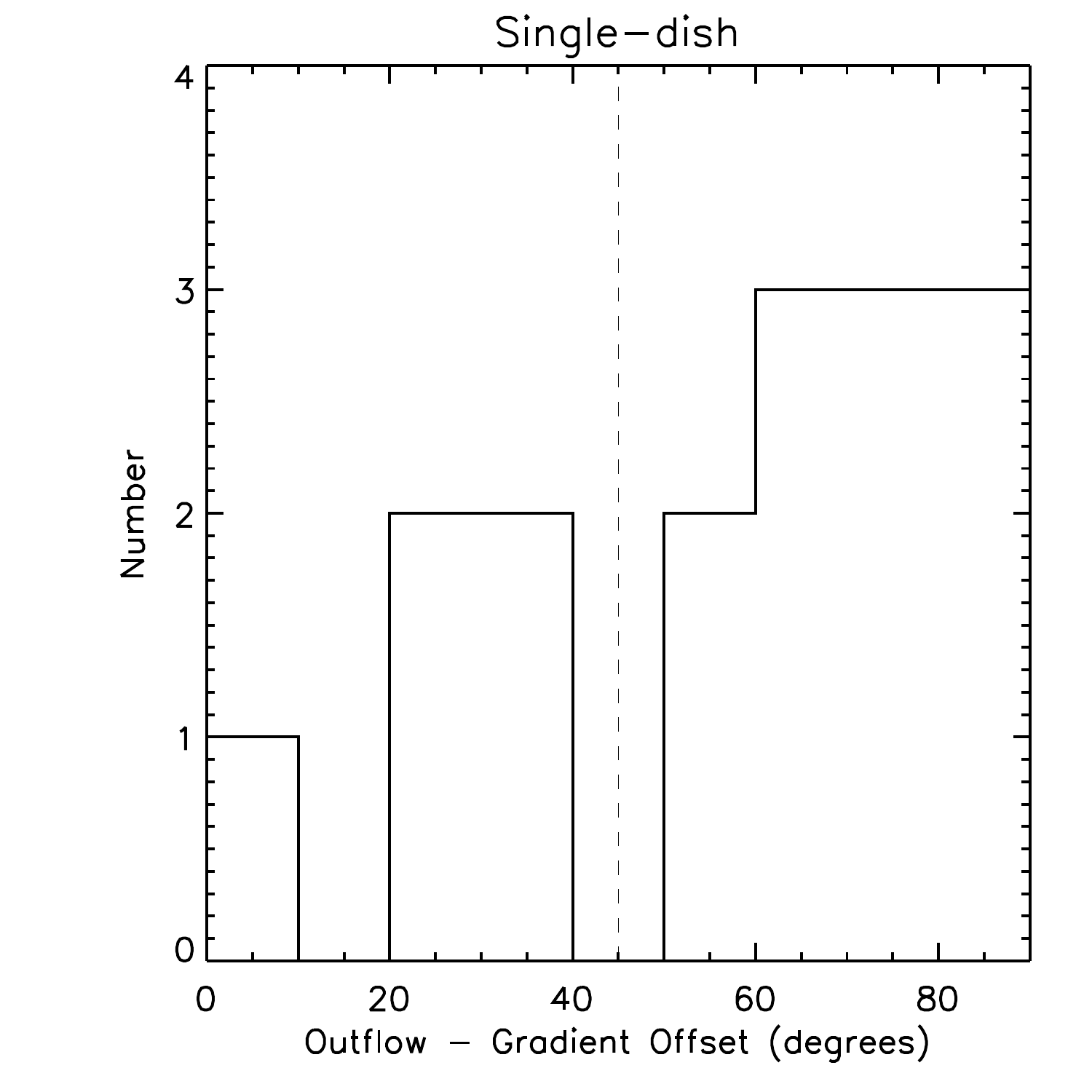}
\includegraphics[scale=0.5]{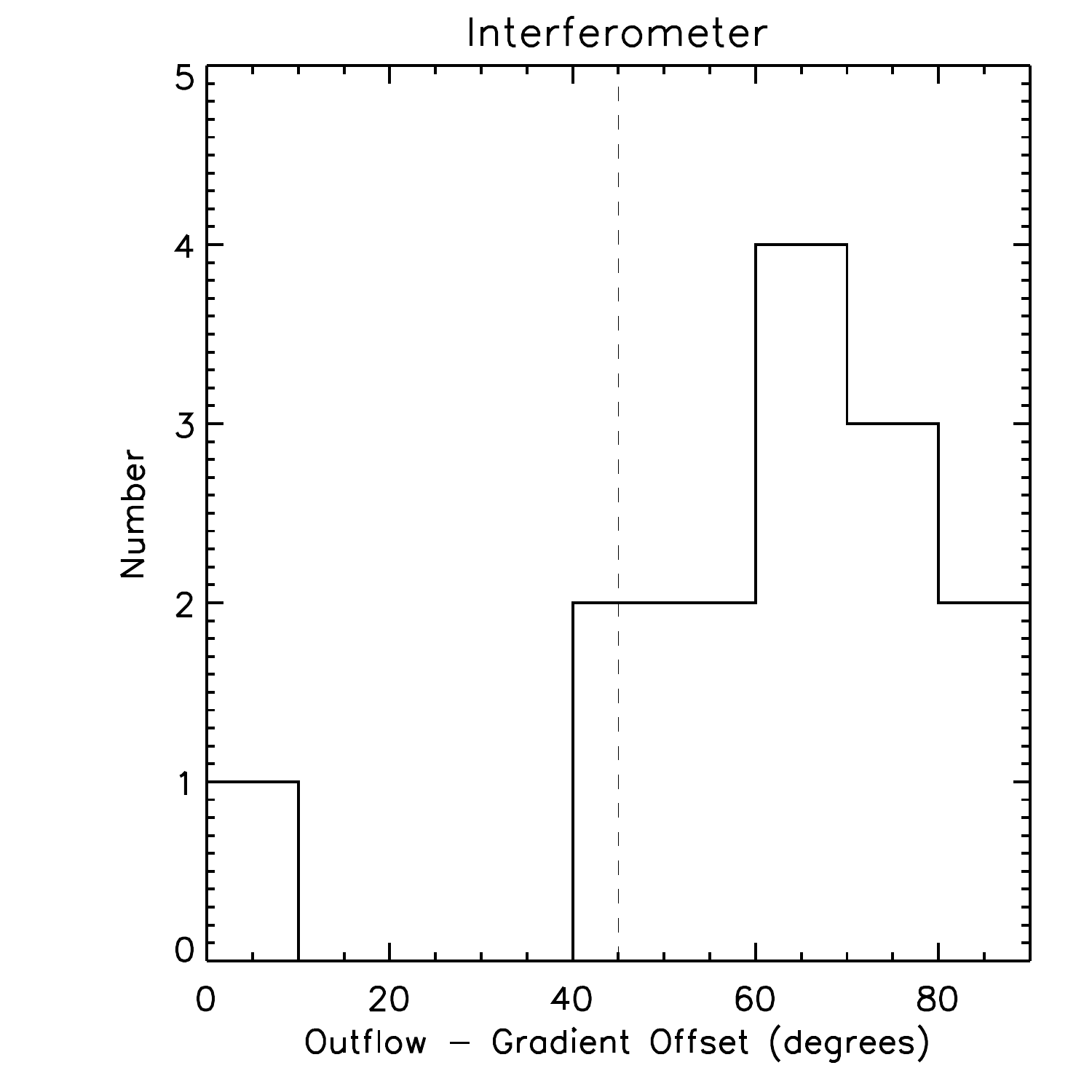}
\end{center}
\caption{Histogram plots of velocity gradient position angle offset relative to outflow position angle
for single-dish (\textit{left panel}) and interferometric (\textit{right panel})
velocity gradient measurements. An offset of 90$^{\circ}$ indicates that the gradient is normal
to the outflow and most of the velocity gradients are within 45$^{\circ}$ of normal to
the outflow (\textit{dashed line}); however, there is substantial dispersion in this relationship. The 
shift in number toward 90\degr\ in the interferometer observations may reflect that smaller-scale
motion is becoming more ordered. The left panels include L1157 and Serpens MMS3 while they are absent from the right panels.}
\label{paoffset}
\end{figure}
\clearpage

\begin{figure}
\begin{center}
\includegraphics[scale=0.5]{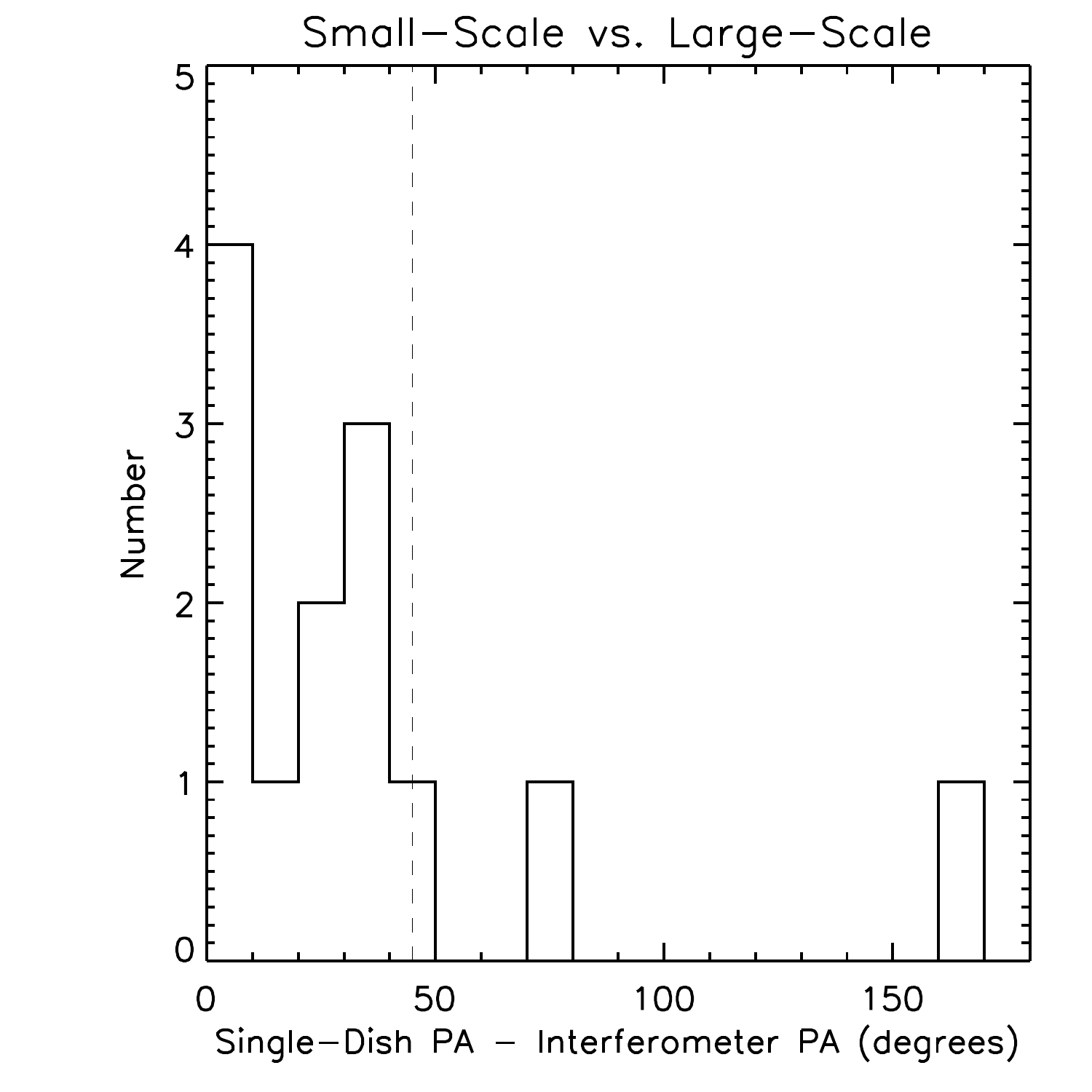}

\end{center}
\caption{Histogram plot of single-dish velocity gradient position angle minus the interferometric
velocity gradient position angle. Most velocity gradients at large and small-scales are within
45\degr\ of each other, indicating that the line of sight velocities at large-scales
reflect similar velocity structure at small-scales. This plot does not include L1157, Serpens MMS3, or L673.}
\label{sdintoffset}
\end{figure}
\clearpage

\begin{figure}
\begin{center}
\includegraphics[scale=0.75]{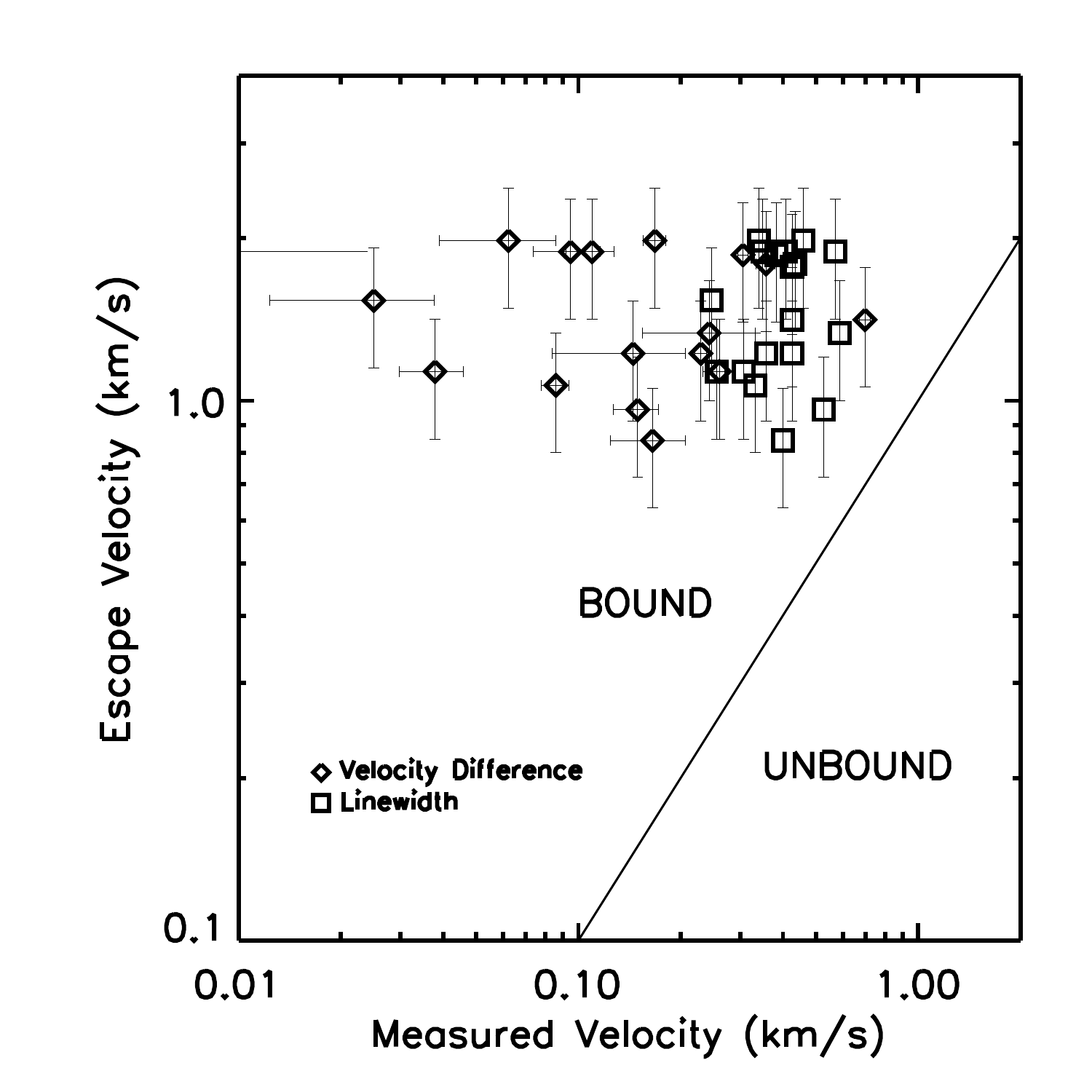}
\end{center}
\caption{Plot of escape velocity at 10000 AU versus velocity shifts measured at 10000 AU radii (\textit{diamonds}) and
the average \nthp\ FWHM from the single-dish data (\textit{squares}). The escape velocity
is calculated using the total mass of the envelope from the 8$\mu$m 
extinction data in Paper I plus a 1 $M_{\sun}$ solar mass central object; the error bars in the calculated escape
velocity reflect a 50\% uncertainty in total mass. This shows  
that the envelopes are consistent with being gravitationally bound on large-scales
and that the envelopes are not supported by rotation, turbulence, and/or thermal pressure. The protostar closest to
rotational support is HH211; its substantial rotation was also highlighted by \citet{tanner2011}.}
\label{bound}
\end{figure}

\clearpage

\begin{figure}
\begin{center}
\includegraphics[scale=0.75]{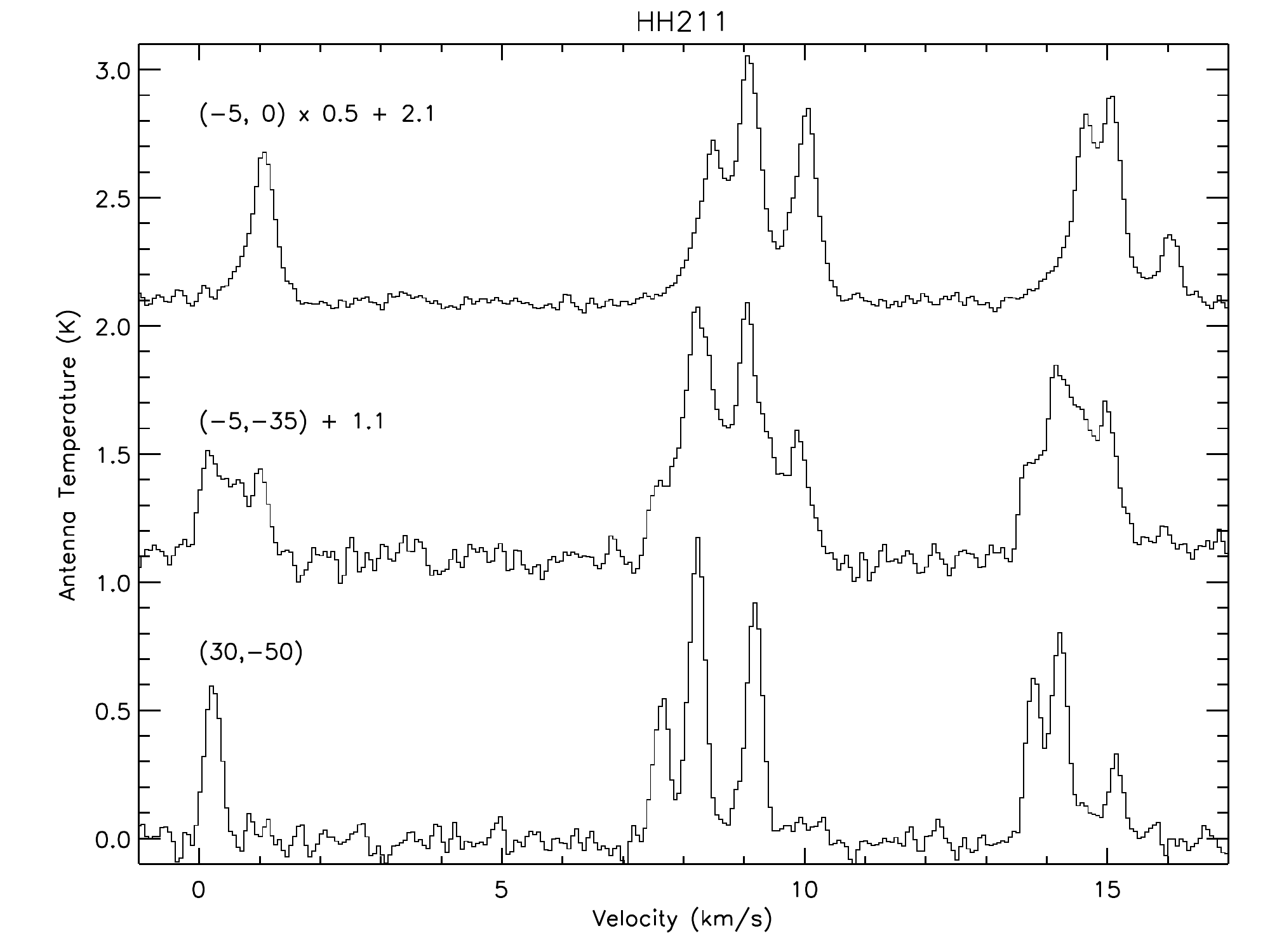}
\end{center}
\caption{Plot of single-dish \nthp\ line profiles at three positions in HH211 showing the transition between different
velocity components. The middle spectrum is in the region where the two components are blended,
demonstrating that the lines are not extremely wide.}
\label{lineprofiles}
\end{figure}

\clearpage

\begin{figure}
\begin{center}
\includegraphics[scale=0.75]{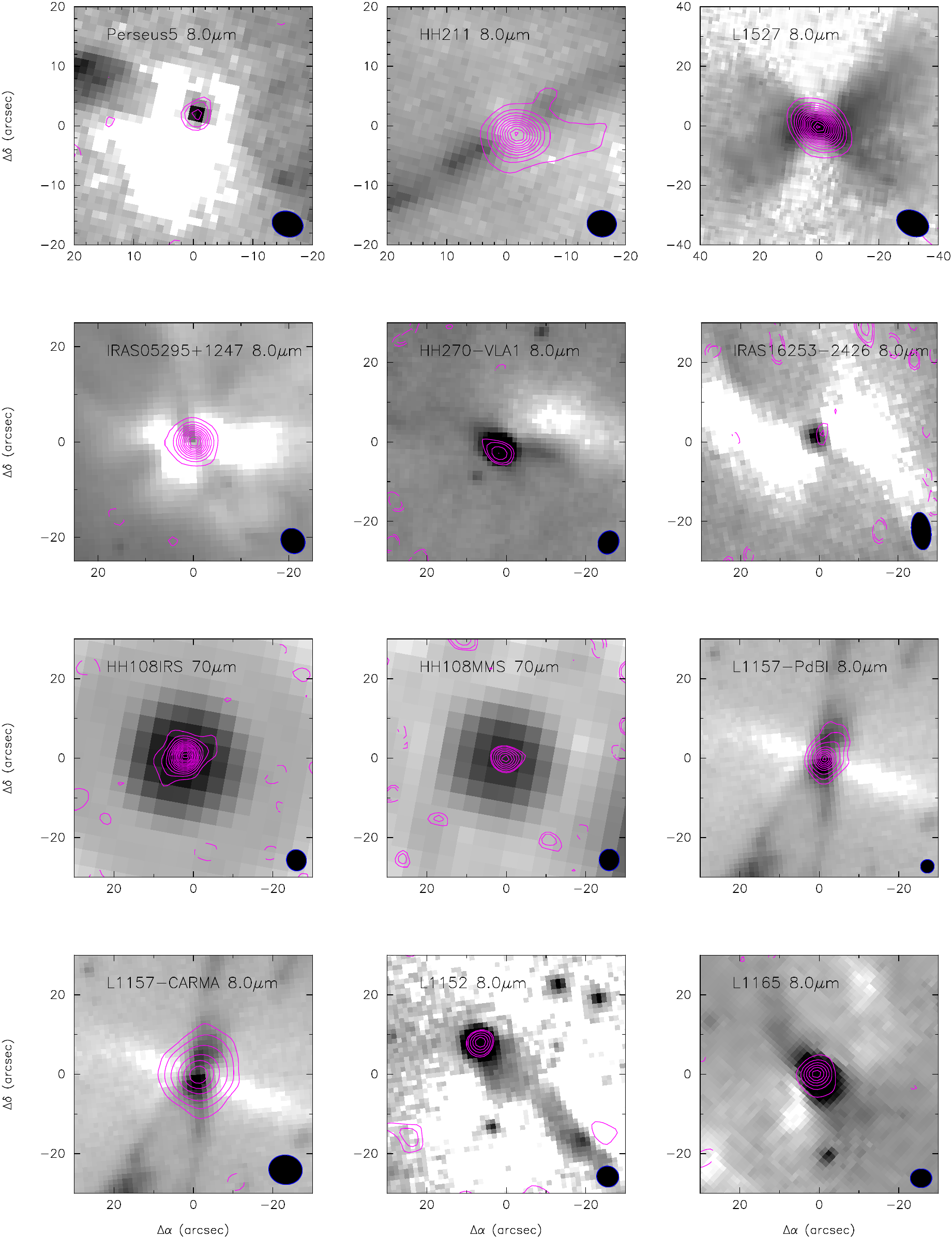}
\end{center}
\caption{IRAC 8\mum\ images of sources observed, with CARMA $\lambda$=3mm continuum contours overlaid. For most protostars, the continuum
emission is directly coincident with the 8\mum\ point sources; 70\mum\ in the case of HH108IRS/MMS.
The contours for Perseus 5, HH270, IRAS 16253-2429, HH108MMS, and L1152 are $\pm$2.5$\sigma$, 3,4,5,...,9$\sigma$ and then increase in 3$\sigma$ intervals.
The contours for HH211, L1527, RNO43, HH108IRS, and L1165 start at $\pm$3$\sigma$ and increase by 3$\sigma$.
The contours for L1157 are $\pm$3, 6, 12, 24, 36, 48, 60, 72, 84, and 96$\sigma$.
}
\label{cont8um}
\end{figure}

\clearpage

\begin{deluxetable}{llllllllllll}
\rotate
\tablewidth{0pt}
\tabletypesize{\scriptsize}
\tablecaption{Source Properties}
\tablehead{
  \colhead{Source}  & \colhead{RA} & \colhead{Dec}      & \colhead{Distance}    &  \colhead{Mass$_{8\mu m}$} & \colhead{Mass$_{submm}$\tablenotemark{*}}   & \colhead{L$_{bol}$}     & \colhead{T$_{bol}$}  &\colhead{Morphological} & \colhead{Outflow PA} & \colhead{References}\\
                    & \colhead{(J2000)} &  \colhead{(J2000)}     & \colhead{(pc)}          &  \colhead{($M_{\sun}$)}     & \colhead{($M_{\sun}$)}                      & \colhead{($L_{\sun}$)}  & \colhead{(K)}        &\colhead{Classification} & \colhead{(\degr)} & \colhead{(Distance, M$_{ref}$, L$_{bol}$,)}  \\
                    &                   &     &                                           &  \colhead{(r$<$0.05pc)}     &                                    &                                   &                      &                         &                   & \colhead{(T$_{bol}$, Outflow PA)}
}
\startdata
Perseus 5       & 03:29:51.88 & +31:39:05.7 & 230   & 2.0 & 1.24         & 0.46  & 41     & One-sided & 80  & 23, 17, 3, 3, 1\\
IRAS 03282+3035 & 03:31:21.10 & +30:45:30.2 & 230   & 2.4 & 2.2          & 1.2   & 33     & Irregular & 122 & 23, 4, 3, 3, 26\\
HH211           & 03:43:56.78 & +32:00:49.8 & 230  & 1.1 &   1.5        & 3.02  & 24     & Irregular & 116 & 23, 3, 3, 3, 6  \\
L1521F          & 04:28:39.03 & +26:51:35.0 & 140 & 2.3 &  1.0         & 0.03  & $\sim$20 & Spheroidal & 270 & 20, 10, 8, 1, 7 \\
IRAS 04325+2402 & 04:35:35.39 & +24:08:19.0 & 140   &  -   & -                          &  0.97 & 73     & Binary Core  & 200 & 20, -, 15, 15, 16 \\
L1527           & 04:39:53.86 & +26:03:09.5 & 140  & 0.8  & 2.4         & 0.9   & 56     & One-sided & 90 & 20, 4, 9, 15, 31\\
RNO43           & 05:32:19.39 & +12:49:40.8 & 460 & 2.8  & 2.6          & 12.5 & 56     & Irregular & 20 & 21, 10, 2, 15, 21 \& 1\\
HH270 VLA1      & 05:51:34.64 & +02:56:45.8 & 420  & 1.9  & -                           & 7.0  & 32     & One-sided & 240 & 22, -, 2, 1, 1\\
IRAS 16253-2429 & 16:28:21.42 & -24:36:22.1 & 125 & 0.8  & 0.98         & 0.25 & 35     & Spheroidal & 20 & 19, 10, 3, 3, 26\\
L483            & 18:17:29.93 & -04:39:39.6 & 200  & 3.5  & 1.8          & 11.5 & $<$54  & Irregular & 282 & 25, 4, 3, 15, 30 \\
Serpens MMS3    & 18:29:09.13 & +00:31:31.6 & 400  & 0.95 & 2.2         &  1.6  & 39     & Flattened & 180 & 18, 3, 14, 3, 1 \\
HH108IRS        & 18:35:42.14 & -00:33:18.5 & 300 &  -   & 4.5    & $\sim$8.0 & 28 & Flattened & 208 & 17, 14, 14, 1, 1\\
HH108MMS        & 18:35:46.46 & -00:32:51.2 & 300 &  -   & 3.6    & 0.7   & 18     & Flattened & 130 & 17, 17, 1, 1, 1\\
L673-SMM2       & 19:20:25.96 & +11:19:52.9 & 300 & 1.0  & 0.35        &  2.8  & -      & Flattened & 270, 135 & 24, -, -, 12, 1 \& 12 \\
L1152           & 20:35:46.22 & +67:53:01.9 & 300 & 3.4  & 12.0        & 1.0   & 33     & Binary Core & 225 & 5, 10, 3, 1, 13 \\
L1157           & 20:39:06.25 & +68:02:15.9 & 300 & 0.86  & 2.2        & 3.0  & 29     & Flattened  & 150 & 5, 4, 3, 15, 27\\
CB230           & 21:17:38.56 & +68:17:33.3 & 300  & 1.1  & 1.1        & 7.2   & 69     & One-sided & 0 & 5, 11, 3, 15, 28\\
L1165           & 22:06:50.46 & +59:02:45.9  & 300 & 1.1  & 0.32        & 13.9 & 46     & Irregular & 225 & 5,  1, 12, 1, 12\\

\enddata
\tablecomments{ Properties of sources observed in the single-dish and/or interferometric sample. The 8\mum\ extinction
masses are taken within 0.05 pc of the protostar and note that some of the masses have been rescaled to account 
for a different distance estimate as compared to Paper I. Positions are reflect the coordinates of the 24\mum\ 
point source from \textit{Spitzer} data or the 3mm continuum continuum position for protostars observed with CARMA.
The Outflow position axes (PA) are not well constrained since the outflows are known to precess they can have fairly large
angular width; a conservative estimate of uncertainty would be $\pm$10\degr. 
References: (1) This work, (2) \citet{tobin2010a}, (3) \citet{enoch2009}, 
 (4) \citet{shirley2000}, (5) \citet{kirk2009}, (6) \citet{lee2009}, (7) \citet{bourke2006}
 (8) \citet{terebey2009}, (9) \citet{tobin2008}, (10) \citet{young2006}, (11) \citet{kauffmann2008}
 (12) \citet{visser2002}, (13) \citet{chapman2009}, (14) \citet{enoch2007},
 (15) \citet{froebrich2005}, (16) \citet{hartmann1999}, (17) \citet{chini2001}, (18) \citet{dzib2010},
 (19) \citet{loinard2008}, (20) \citet{loinard2007}, (21) \citet{bence1996}, (22) \citet{menten2007},
(23) \citet{hirota2011}, (24) \citet{herbig1983}, (25) \citet{jorgensen2004}, (26) \citet{arce2006},
(27) \citet{gueth1996}, (28) \citet{launhardt2001}, (29) \citet{stanke2006}, (30) \citet{fuller1995},
(31) \citet{hoger1998}.} 
\tablenotetext{*}{Mass was computed with sub/millimeter bolometer data assuming an isothermal temperature.}

\end{deluxetable}

\begin{deluxetable}{lllllll}
\tablewidth{0pt}
\rotate
\tabletypesize{\scriptsize}
\tablecaption{IRAM 30m Observations}
\tablehead{
  \colhead{Source} & \colhead{RA}      & \colhead{Dec}      & \colhead{Receiver} & \colhead{Date}   & \colhead{$\sigma_T$} & \colhead{$\Delta v$} \\
                   & \colhead{(J2000)} &  \colhead{(J2000)} &                    & \colhead{(UT)}   & \colhead{(mK)}       & \colhead{(\kms)}\\
}
\startdata
IRAS 03282$+$3035 & 03:31:19.8  & +30:45:37   & EMIR E090      & 25 Oct 2009  &  60 & 0.067 \\
HH211-mm          & 03:43:55.3  & +32:00:46   & EMIR E090      & 24 Oct 2009  & 70   & 0.067 \\
L1521F            & 04:28:39.0  & +26:51:37   & EMIR E090      & 24 Oct 2009  & 200 & 0.067 \\
IRAS04325+2402    & 04:35:36.6  & +24:08:54   & EMIR E090      & 25 Oct 2009  & 60 & 0.067 \\
L1527\tablenotemark{a}& 04:39:54.0    & +26:03:22   & EMIR E090      & 26 Oct 2009  & 50  & 0.067 \\
RNO43             & 05:32:18.2  & +12:50:01   & EMIR E090      & 24 Oct 2009  & 50   & 0.067 \\
HH270 VLA1        & 05:51:33.6  & +02:56:46   & AB100          & 27 Dec 2008  & 50   & 0.067 \\
IRAS 16253-2429   & 16:28:21.9  & -24:36:22   & AB100          & 29 Dec 2008   & 170   & 0.067 \\
L483              & 18:17:30.3  & -04:39:40.6 & EMIR E090      & 25 Oct 2009  & 75 & 0.067 \\
Serpens MMS3          & 18:29:10.2  & +00:31:18   & EMIR E090      & 24 Oct 2009  & 200 & 0.067 \\
HH108             & 18:35:42.7  & -00:33:07   & EMIR E090      & 23 Oct 2009 & 80 & 0.067 \\
L673-SMM2         & 19:20:25.4  & +11:19:44   & EMIR E090    &23/24 Oct 2009   & 65 & 0.067 \\
L1152             & 20:35:47.17 & +67:53:22   & EMIR E090      & 24 Oct 2009  & 55 & 0.067 \\
L1157             & 20:39:05.6  & +68:02:13.2 & EMIR E090      & 25 Oct 2009  & 30  & 0.067  \\
CB230             & 21:17:28.7  & +68:17:42.8 & AB100          & 26 Dec 2008  & 50  & 0.067  \\
L1165             & 22:06:50.8  & +59:03:06.5 & EMIR E090      & 25 Oct 2009   & 60  & 0.067  \\

\enddata
\tablecomments{Observations of protostellar envelopes taken with the IRAM 30m telescope. The positions reflect the map
center and not necessarily the protostar position.} 
\tablenotetext{a}{Source was observed in both observing runs; however, the data taken in the 2009 EMIR run were significantly
better quality and we only use those data for analysis.}
\end{deluxetable}

\begin{deluxetable}{lllllclll}
\tablewidth{0pt}
\rotate
\tabletypesize{\scriptsize}
\tablecaption{CARMA Observations}
\tablehead{
  \colhead{Source} & \colhead{RA} & \colhead{Dec}  &\colhead{Config.} & \colhead{Date} & \colhead{Calibrators} & \colhead{Beam}  & \colhead{$\sigma_T$} & \colhead{$\Delta v$} \\
             & \colhead{(J2000)} &  \colhead{(J2000)}    & &    \colhead{(UT)}          &   \colhead{(Gain, Flux)}    & \colhead{(\arcsec)}    & \colhead{(mK)}  & \colhead{(\kms)} \\
}
\startdata
Perseus 5         & 03:29:51.84 & +31:39:06.2  & D-array  & 10 Apr 2010       & 0336+323, Uranus  & 5.2$\times$4.1  & 450 & 0.067 \\
HH211-mm          & 03:43:56.8  & +32:00:50.3  & D-array  & 16 Apr 2010       & 0336+323, Uranus  & 5.0$\times$4.3   & 400 & 0.067 \\
L1527             & 04:39:53.9  & +26:03:09.6  & E-array  & 02/05/06 Oct 2008 & 3C111, 3C84       & 11.1$\times$8.2   & 100  & 0.1 \\
RNO43             & 05:32:19.1  & +12:49:43.1  & D-array  & 18/22 Apr 2010    & 0532+075, Mars    & 5.5$\times$4.6   & 300  & 0.067  \\
..                & ..          & ..           & ..       & 2 Sept. 2010      & 0532+075, Uranus  & ..               & ..  & 0.067 \\
HH270 VLA1        & 05:51:33.6  & +02:56:47.2  & D-array  & 24/25/26 Jul 2009 & 0532+075, Mars    & 5.9$\times$5.0    & 500 & 0.1 \\
IRAS 16253-2429   & 16:28:21.6  & -24:36:23.1  & D-array  & 12 Aug 2009       & 1625-254, MWC349  & 9.3$\times$4.9    & 500  & 0.1  \\
HH108 IRS         & 18:35:42.17 & -00:33:18.3  & D-array  & 10 Apr 2010       & 1830+063, Neptune & 5.9$\times$5.0    & 500  & 0.067 \\
HH108 MMS         & 18:35:46.53 & -00:32:51.4  & D-array  & 14/15 Apr 2010    & 1830+063, MWC349  & 5.9$\times$5.0     & 450 & 0.067  \\
L1152             & 20:35:44.14 & +67:52:51.3  & D-array  & 11 Apr 2010       & 1927+739, MWC349  & 5.4$\times$5.1    & 600 & 0.067 \\
L1157             & 20:39:06.3  & +68:02:15.8  & D-array  & 19/29 Mar 2009    & 1927+739, MWC349  & 7.3$\times$6.6   & 150 & 0.1 \\
..                & ..          & ..            & E-array  &  02 Oct 2008     & 1927+739, MWC349  & ..            & .. & 0.1 \\
L1165             & 22:06:50.32 & +59:02:45.5  & D-array  & 25 Jul 2009       & 2038+513, MWC349  & 5.3$\times$4.7   & 400  & 0.1 \\
 
\enddata
\tablecomments{Observation parameters of protostars observed with CARMA, the coordinates listed for each protostar denote the
phase center of the observation, not necessarily the position of the protostar. The frequency resolution for the 2010 observations 
was 0.067 \kms; however, this was rebinned to 0.1 \kms\ to increase signal-to-noise. The observations of L1165, IRAS 16253-2429, and HH270 VLA 1
only observed the central three hyperfine \nthp\ lines, while the observation of L1157 and L1527 observed the isolated \nthp\ line. All
other observations observed all 7 \nthp\ hyperfine transitions. } 
\end{deluxetable}

\begin{deluxetable}{lllllclll}
\rotate
\tablewidth{0pt}
\tabletypesize{\scriptsize}
\tablecaption{Very Large Array Observations}
\tablehead{
  \colhead{Source} & \colhead{RA} & \colhead{Dec} &\colhead{Config.} & \colhead{Date} & \colhead{Calibrators} & \colhead{Beam}& \colhead{$\sigma_T$} & \colhead{$\Delta v$} \\
             & \colhead{(J2000)} &  \colhead{(J2000)}  & &    \colhead{(UT)}          & \colhead{(Gain, Flux)}& \colhead{(\arcsec)}    & \colhead{(mK)}  & \colhead{(\kms)}  \\
}
\startdata
IRAS 03282$+$3035 & 03:31:20.34 & +30:45:32.0 & D-array  & 12 Nov 2009       & 0336+323, 3C48    & 5.2$\times$4.6   & 240 & 0.15 \\
L1521F            & 04:28:38.9  & +26:51:35.0 & D-array  & 19 Mar 2007       & 0418+380, 3C48    & 7.1$\times$6.7   & 125 & 0.075 \\
L1527             & 04:39:53.9  & +26:03:09.6 & D-array  & 07 Mar 2003       & 0510+180, 3C48    & 4.8$\times$4.1   & 300 & 0.15  \\
L483              & 18:17:29.83 & -04:39:38.3 & D-array  & 18 Apr 1995       & 1751+096, 3C286   & 6.3$\times$5.9   & 300 & 0.3 \\
Serpens MMS3          & 18:29:09.1  & +00:31:34.0 & D-array  & 11/12 Nov 2009    & 1851+005, 3C48    & 5.6$\times$5.3   & 270  & 0.15  \\
L673-SMM2         & 19:20:26.27 & +11:20:08.2 & D-array  & 6 Nov 2009        & 1925+211, 3C286   & 5.4$\times$5.0  & 250 & 0.15 \\
L1157             & 20:39:06.3  & +68:02:15.0 & D-array  & 2/3 Jan 2010      & 1927+739, 3C48    & 4.6$\times$4.0   & 250 & 0.15  \\
CB230             & 21:17:37.59 & +68:17:38.5 & D-array  & 18/20 Oct 2009    & 2022+616, 3C48    & 4.5$\times$3.3   & 325 & 0.15  \\

\enddata
\tablecomments{Observation parameters of protostars observed with VLA, the coordinates listed for each protostar denote the
phase center of the observation, not necessarily the position of the protostar. The frequency resolution of the L1521F observation
was 0.075 \kms; however, this was rebinned to 0.15 \kms\ to increase signal-to-noise. }

\end{deluxetable}

\begin{deluxetable}{lllllclll}
\rotate
\tablewidth{0pt}
\tabletypesize{\scriptsize}
\tablecaption{Plateau de Bure Interferometer Observations}
\tablehead{
  \colhead{Source} & \colhead{RA} & \colhead{Dec} &\colhead{Config.} & \colhead{Date} & \colhead{Calibrators} & \colhead{Beam} & \colhead{$\sigma_T$} & \colhead{$\Delta v$} \\
             & \colhead{(J2000)} &  \colhead{(J2000)}    & &    \colhead{(UT)}          &  \colhead{(Gain, Flux)}    & \colhead{(\arcsec)}      & \colhead{(mK)}   & \colhead{(\kms)} \\
}
\startdata
L1157             & 20:39:05.6  & +68:02:13.2 & D-array  & 17 Jun, 07 Jul 2009 & 1927+739, MWC349  & 3.4$\times$3.3 & 80 & 0.125\\
..                & ..          & ..           & C-array  & 13 Nov 2009       & 1927+739, MWC349  & ..               & .. & 0.125 \\ 
\enddata
\tablecomments{Observation parameters of the protostar observed with the PdBI, the coordinates listed denote the
phase center of the observation, not necessarily the position of the protostar.} 

\end{deluxetable}

\begin{deluxetable}{lllllllcl}
\rotate
\tablewidth{0pt}
\tabletypesize{\scriptsize}
\tablecaption{Single-Dish \nthp\ Properties}
\tablehead{
  \colhead{Source} & \colhead{V$_{lsr}$}& \colhead{FWHM} & \colhead{I$_{max} \pm \sigma_I$\tablenotemark{\dagger}} & \colhead{$\tau_{tot}$} & \colhead{T$_{ex}$} & \colhead{N(\nthp)} & \colhead{\nthp\ Mass} & \colhead{Radius}\\
                   & \colhead{(\kms)}& \colhead{(\kms)}&  \colhead{(K \kms)} &\colhead{}             & \colhead{(K)}      & \colhead{(cm$^{-2}$)} & \colhead{(10$^{-9}$ $M_{\sun}$/X(\nthp))}& \colhead{(\arcsec)}\\

}
\startdata
IRAS 03282+3035   & 7.1 & 0.67 & 5.9$\pm$0.04  & 4.9      & 6.4   & 1.8$\times$10$^{13}$  & 1.4  & 160$\times$80\\
HH211-mm          & 9.1 & 0.5  & 5.8$\pm$0.05  & 6.9      & 5.2   & 2.3$\times$10$^{13}$  & 1.4  & 120$\times$110\\
L1521F            & 6.45 & 0.29 & 4.0$\pm$0.14  & 17.5     & 4.7   & 4.1$\times$10$^{13}$  & 1.9  & 160$\times$110\\
IRAS 04325+2402   & 5.9  & 0.45 & 5.4$\pm$0.04  & 5.4      & 6.8   & 1.9$\times$10$^{13}$  & 2.4  & 180$\times$160\\
L1527             & 5.9  & 0.36 & 3.1$\pm$0.03  & 9.3      & 4.3   & 1.7$\times$10$^{13}$  & 0.9  & 180$\times$160\\
RNO43             & 10.25  & 0.69 & 2.5$\pm$0.04  & 0.9      & 5.6   & 2.3$\times$10$^{13}$  & 0.8  & 90$\times$80\\
HH270 VLA1        & 8.9  & 0.41 & 2.5$\pm$0.04  & 4.6      & 4.2   & 6.5$\times$10$^{12}$  & 1.2  & 135$\times$65\\
IRAS 16253-2429   & 4.05  & 0.3  & 4.2$\pm$0.13   & 6.6      & 5.6   & 1.7$\times$10$^{13}$  & 0.4  & 100$\times$80\\
L483              & 5.5  & 0.51 & 10.3$\pm$0.06  & 16.1     & 5.4   & 9.1$\times$10$^{13}$  & 10.4 & 200$\times$150\\
Serpens MMS3-full & 8.5  & 0.62 & 12.5$\pm$0.12  & 6.6      & 6.9   & 6.1$\times$10$^{13}$  & 21.5 & 200$\times$150\\
Serpens MMS3-proto& 8.5  & 0.62 & 5.4$\pm$0.12   & 3.5      & 6.1   & 1.5$\times$10$^{13}$  & 1.9  & no boundary\\
HH108-full         & 11.0 & 0.54 & 4.7$\pm$0.05  & 4.0     & 5.6   & 1.1$\times$10$^{13}$  & 2.35 & 160$\times$50\\
HH108IRS          & 11.0  & 0.54 & 4.7$\pm$0.05  & 4.0     & 5.6   & 1.1$\times$10$^{13}$  & 0.9  & no boundary\\
HH108MMS          & 10.9  & 0.54 & 1.9$\pm$0.05  & 2.5     & 5.1   & 3.0$\times$10$^{12}$  & 0.4  & no boundary\\
L673-SMM2         & 7.1  & 0.48  & 7.1$\pm$0.04  & 2.6      & 8.5   & 1.6$\times$10$^{13}$  & 5.3  & $\sim$150$\times$80\\
L1152             & 2.5  & 0.47  & 5.4$\pm$0.04  & 7.3      & 5.5   & 2.4$\times$10$^{13}$  & 3.9  & 260$\times$95\\
L1157             & 2.65  & 0.39  & 4.6$\pm$0.02  & 4.5     & 4.8   & 1.4$\times$10$^{13}$  & 3.0  & 200$\times$70\\
CB230             & 2.8  & 0.5   & 4.0$\pm$0.04  & 3.6      & 5.0   & 8.9$\times$10$^{12}$  & 1.0  & 115$\times$75\\
L1165             & -1.50  & 0.42  & 2.5$\pm$0.04  & 5.6      & 4.2   & 8.3$\times$10$^{12}$  & 0.6  & 90$\times$60\\

\enddata
\tablecomments{The \nthp\ properties of each source are tabulated, column densities are scaled by 1.17 to account for beam efficiency.
 The FWHM linewidth is taken from the average single-dish
spectrum over the entire envelope. The statistical uncertainty in the linewidth is less than 0.01 \kms;
however, this value varies across the envelope and will vary depending on the region averaged. 
The optical depth, T$_{ex}$, column density are measured at the \nthp\ emission peak; however, multiple
positions are quoted for HH108 and Serpens MMS3.
The envelope mass is computed by summing the column density over the entire emitting region and assuming a 
constant \nthp\ abundance of 1.0$\times$10$^{-9}$ relative to H$_2$. The envelope radius is measured where
the \nthp\ is 10\% of the peak. The optical depths, column densities, and masses should only be 
regarded as accurate within factors of a few owing to systematic errors in calculation of optical
depth and excitation temperature.} 
\tablenotetext{\dagger}{Sum of all hyperfine line emission, multiplied by channel velocity width on the T$_A$ scale and 
$\sigma_I$ =  $\sigma_T$N$_{ch}^{1/2}\Delta v$ where N$_{ch}^{1/2}$ is the number of channels in the summed
velocity range and $\Delta v$ is the channel width; N$_{ch}$ can be calculated from the information given
in the table.}
\label{sdnthp}
\end{deluxetable}

\begin{deluxetable}{llllllllcl}
\rotate
\tablewidth{0pt}
\tabletypesize{\scriptsize}
\tablecaption{Interferometric \nthp\ Properties}
\tablehead{
  \colhead{Source} & \colhead{V$_{lsr}$} & \colhead{FWHM}   & \colhead{I$_{max} \pm \sigma_I$\tablenotemark{\dagger}}& \colhead{$\tau_{tot}$} & \colhead{T$_{ex}$} & \colhead{N(\nthp)} & \colhead{\nthp\ Mass} & \colhead{Radius\tablenotemark{c}}\\
                   & \colhead{(\kms)}    & \colhead{(\kms)} & \colhead{(K \kms)}        & \colhead{}             & \colhead{(K)}      & \colhead{(cm$^{-2}$)} & \colhead{(10$^{-9}$ $M_{\sun}$/X(\nthp))} & \colhead{(\arcsec)}\\

}
\startdata
Perseus 5                           & 8.06 & 0.29  & 6.1$\pm$0.28                  & 3.1      & 13.1   & 3.0$\times$10$^{13}$  & 0.3  & 20\\
HH211-mm                            & 9.12 & 0.39  & 8.3$\pm$0.25                  & 9.9      & 8.4    & 9.3$\times$10$^{13}$  & 1.1   & 40\\
L1527\tablenotemark{a}              & 6.01 & 0.26  & 0.9$\pm$0.05\tablenotemark{a} & 8.8      & 4.8    & 1.8$\times$10$^{13}$  & 0.2   & 57\\
RNO43                               & 9.87, 10.5 & 0.43, 0.32  & 3.0$\pm$0.19      & 5.5      & 6.7    & 1.8$\times$10$^{13}$  & 0.4  & 23\\
HH270 VLA1\tablenotemark{b}         & 8.86 & 0.41  & 6.0$\pm$0.24                  & 4.9      & 11.2   & 3.5$\times$10$^{12}$  & 0.9  & 28\\
IRAS 16253-2429\tablenotemark{b}    & 4.12 & 0.21  & 4.4$\pm$0.24                  & 7.4      & 7.6    & 3.2$\times$10$^{13}$  & 1.5   & 36\\
HH108IRS                            & 10.8 & 0.59  & 14.1$\pm$0.3                  & 8.2      & 12.5   & 8.4$\times$10$^{13}$  & 1.2   & 21\\
HH108MMS                            & 10.9 & 0.40  & 5.0$\pm$0.3                   & 8.4      & 6.4    & 4.3$\times$10$^{13}$  & 0.7  & 12\\
L1152                               & 2.63 & 0.34  & 9.6$\pm$0.36                  & 10.1     & 10.7   & 1.3$\times$10$^{14}$  & 4.3   & 29\\
L1157  (PdBI)                       & 2.7  & 0.74  & 19.8$\pm$0.06                 & 10.4     & 8.0    & 1.2$\times$10$^{14}$  & 0.7  & 25\\
L1157  (CARMA)                      & 2.7  & 1.0   & 16.7$\pm$0.06                 & 4.8      & 8.5    & 4.7$\times$10$^{13}$  & 0.8  & 41\\
L1165\tablenotemark{b}              & -1.5 & 0.38  & 6.6$\pm$0.21                  & 6.6      & 8.0    & 4.5$\times$10$^{13}$  & 0.4   & 17 \\

\enddata
\tablecomments{The \nthp\ properties of each source are tabulated. The V$_{lsr}$ and FWHM linewidth are taken from the
position of peak intensity. The statistical uncertainty in the linewidth is less than 0.01 \kms;
however, the linewidth varies across the envelope. The optical depth, T$_{ex}$, column density are
measured at the \nthp\ emission peak.
The envelope mass is computed by summing the column density over the entire emitting region and assuming a; regions of bad data toward edges
are masked out of the measurement. The optical depths, column densities, and masses should only be 
regarded as accurate within factors of a few owing to systematic errors in calculation of optical
depth and excitation temperature.} 

\tablenotetext{\dagger}{Intensity quoted is summed the central three hyperfine lines of \nthp\ and multiplied by the channel velocity width.
$\sigma_I$ =  $\sigma_T$N$_{ch}^{1/2}\Delta v$ where N$_{ch}^{1/2}$ is the number of channels in the summed
velocity range and $\Delta v$ is the channel width; N$_{ch}$ can be calculated from the information given
in the table. $\sigma_I$ is the statistical uncertainty and does not take into account the $\sim$10\% absolute
calibration uncertainty.}
\tablenotetext{a}{The values for L1527 are highly uncertain due to only observing the isolated \nthp\ line and are scaled by
a factor of 27/3 to approximate the flux from the other hyperfine components.}
\tablenotetext{b}{The column density and masses are scaled by 27/15 since the bandwidth only enabled 
the central three hyperfine components of \nthp\ to be observed.}
\tablenotetext{c}{The radii quoted are the radius of a circle which encompasses the detected regions of the envelope
above the 3$\sigma$ level in the integrated intensity map. Note that these radii do not reflect the dimensions of the 
entire core because they are highly influenced by the primary beam, uv-coverage, and sensitivity.}
\label{intnthp}
\end{deluxetable}

\begin{deluxetable}{lllllllcl}
\rotate
\tablewidth{0pt}
\tabletypesize{\scriptsize}
\tablecaption{Interferometric \nht\ Properties}
\tablehead{
  \colhead{Source} & \colhead{V$_{lsr}$} & \colhead{FWHM}  & \colhead{I$_{max} \pm \sigma_I$\tablenotemark{\dagger}}  & \colhead{Optical Depth} & \colhead{T$_{ex}$} & \colhead{N(\nht)} & \colhead{\nht\ Mass} & \colhead{Radius\tablenotemark{b}}\\
                   & \colhead{(\kms)}    & \colhead{(\kms)}   & \colhead{(K \kms)}    & \colhead{}             & \colhead{(K)}      & \colhead{(cm$^{-2}$)} & \colhead{(10$^{-8}$ $M_{\sun}$/X(\nht))} & \colhead{(\arcsec)}\\

}
\startdata
IRAS 03282+3035    & 6.98 & 0.38 & 6.4$\pm$0.13   & 10.1      & 12.7   & 1.4$\times$10$^{15}$  & 5.7 & 60\\
L1521F\tablenotemark{a} & 7.11 & 0.37 & 1.4$\pm$0.06   & 10.1     & 3.9   & 8.0$\times$10$^{15}$  & 29.3 & 49\\
L1527              & 5.95 & 0.26 & 2.6$\pm$0.16   & 1.0      & 10.45   & 3.9$\times$10$^{14}$  & 1.4 & 39\\
L483\tablenotemark{a} & 5.41 & 0.51 & 5.8$\pm$0.18   & 6.8     & 8.3   & 2.3$\times$10$^{15}$  & 27.7 & 62\\
Serpens MMS3       & 8.22 & 0.51  & 7.0$\pm$0.19  & 2.1      & 10.2   & 2.0$\times$10$^{15}$  & 7.5 & 50\\
L673-SMM2          & 6.93 & 0.50  & 8.4$\pm$0.14   & 0.53      & 33.3   & 7.2$\times$10$^{14}$  & 11.3 & 76\\
L1157              & 2.71 & 0.56 & 8.6$\pm$0.15  & 1.6      & 13.4   & 8.5$\times$10$^{14}$  & 16.0 & 87\\
CB230              & 2.78 & 0.47 & 6.5$\pm$0.22   & 1.1      & 16.9   & 6.1$\times$10$^{14}$  & 4.8 & 36\\

\enddata
\tablecomments{The \nht\ properties of each source are tabulated.  The V$_{lsr}$ and FWHM linewidth are taken from the
position of peak intensity. The statistical uncertainty in the linewidth is less than 0.01 \kms;
however, the linewidth varies across the envelope. The optical depth, T$_{ex}$, column density are
measured at the \nht\ emission peak except.
The envelope mass is computed by summing the column density over the entire emitting region and assuming a 
constant \nht\ abundance of 1.0$\times$10$^{-8}$ relative to H$_2$; regions of bad data toward edges
are masked out of the measurement. The optical depths, column densities, and masses should only be 
regarded as accurate within factors of a few owing to systematic errors in calculation of optical
depth and excitation temperature. The masses from \nht\ also appear to be overestimating the actual envelope 
masses, compared to the 8\mum\ and submillimeter dust masses; thus, the \nht\ abundance relative to H$_2$ could
be larger than assumed.  }
\tablenotetext{a}{\nht\ lines in these sources are optically thick making mass and column density measurements uncertain.}
\tablenotetext{b}{The radii quoted are the radius of a circle which encompasses the detected regions of the envelope
above the 3$\sigma$ level in the integrated intensity map. Note that these radii will not reflect the dimensions of the 
entire core in most cases because they are highly influenced by the primary beam, uv-coverage, and sensitivity.}

\tablenotetext{\dagger}{Intensity quoted is summed over the main \nht\ lines and multiplied by the channel velocity width.
$\sigma_I$ =  $\sigma_T$N$_{ch}^{1/2}\Delta v$ where N$_{ch}^{1/2}$ is the number of channels in the summed
velocity range and $\Delta v$ is the channel width; N$_{ch}$ can be calculated from the information given
in the table. $\sigma_I$ is the statistical uncertainty and does not take into account the $\sim$10\% absolute
calibration uncertainty.}

\label{intnht}
\end{deluxetable}

\begin{deluxetable}{llllll}
\rotate
\tablewidth{0pt}
\tabletypesize{\scriptsize}
\tablecaption{Single-Dish Velocity Gradients}
\tablehead{
  \colhead{Source}  & \colhead{Gradient \tablenotemark{\dagger}} & \colhead{Gradient Fit\tablenotemark{\dagger}} & \colhead{Gradient Fit} & \colhead{Gradient PA} & \colhead{Gradient PA}\\
                    & \colhead{(10000AU)}                        & \colhead{1D}                                  & \colhead{2D}           & \colhead{2D}          &   \colhead{- Outflow PA}\\
                    & \colhead{(km s$^{-1}$ pc$^{-1}$)} & \colhead{(km s$^{-1}$ pc$^{-1}$)} & \colhead{(km s$^{-1}$ pc$^{-1}$)}& \colhead{($^{\circ}$)} & \colhead{($^{\circ}$)}\\

}
\startdata
Perseus 5           & -                             & -                              & -             & -             & - \\
IRAS 03282+3035     & 1.13$\pm$0.2               & 1.31 $\pm$0.004                & 4.6$\pm$0.04  & 301$\pm$0.5   & 1\\
HH211-mm            & 7.2$\pm$0.3                  & 6.9$\pm$0.03                   & 4.2$\pm$0.07  & 31$\pm$0.9    & 85\\
L1521F              & 0.26$\pm$0.13                 & 0.13$\pm$0.002                 & 0.76$\pm$0.02 & 239$\pm$1.8   & 31\\
IRAS 04325+2402\tablenotemark{a} & 3.1$\pm$0.11          & 5.2$\pm$0.01                   & 3.5$\pm$0.01  & 79$\pm$0.3    & 59\\
L1527               & 2.7$\pm$0.29                  & 2.2$\pm$0.01                   & 2.2$\pm$0.02  & 114$\pm$0.5   & 24\\
RNO43               & 2.5$\pm$0.9                   & 2.2$\pm$0.1                    & 3.3$\pm$0.4   & 280$\pm$4.6   & 80\\
HH270 VLA1          & 0.98$\pm$0.2                 & 0.88$\pm$0.004                  & 1.4$\pm$0.04  & 166$\pm$1.3   & 74\\
IRAS 16253-2429     & 0.8$\pm$0.18                  & 1.2$\pm$0.002                  & 1.7$\pm$0.04  & 311$\pm$1.3   & 69\\
L483                & 3.7$\pm$0.24                  & 3.6$\pm$0.01                   & 2.2$\pm$0.02  &  66$\pm$0.4   & 36\\
Serpens MMS3        & 1.5$\pm$0.24\tablenotemark{b} & 1.4$\pm$0.01\tablenotemark{b}  & 1.1$\pm$0.01  & 307$\pm$0.6   & 53\\
HH108IRS            & 0.64$\pm$0.24                 & 2.0$\pm$0.004                  & 1.2$\pm$0.03 & 96$\pm$2.1    & 68\\
HH108MMS            & 1.7$\pm$0.13                  & 1.6$\pm$0.004                  & 2.0$\pm$0.05   & 67$\pm$1.6    & 62\\
L673-SMM2           & -                             & -                              & 1.0$\pm$0.02 & 100$\pm$2.1   & 10\\
L1152\tablenotemark{a} & 1.7$\pm$0.42                  & 2.7$\pm$0.004                  & 3.2$\pm$0.04  & 145$\pm$0.7   & 80\\
L1157               & 0.9$\pm$0.084                 & 0.9$\pm$0.003                  & 0.69$\pm$0.01 & 303$\pm$1.4   & 27\\
CB230               & 2.4$\pm$0.094                 & 3.2$\pm$0.01                   & 2.3$\pm$0.2   &  281$\pm$1.0  & 79\\
L1165               & 1.5$\pm$0.63                  & 2.5$\pm$0.01                   & 2.8$\pm$0.1   & 154$\pm$2.7   & 71\\
\enddata
\tablecomments{The kinematic properties of each source are tabulated. The uncertainty in gradient PA is
statistically small; however, it is subject to systematics resulting from the spatial distribution of data points. 
The outflow position angles are given in Table 1.} 
\tablenotetext{\dagger}{Gradients are measured normal to the outflow.}
\tablenotetext{a}{Gradients are measured from center of the core as the protostars are located at the edge.}
\tablenotetext{b}{Gradient was measured along the the velocity gradient, shifted slightly from being normal to the outflow.}
\end{deluxetable}

\begin{deluxetable}{lllllll}
\rotate
\tablewidth{0pt}
\tabletypesize{\scriptsize}
\tablecaption{Interferometric Velocity Gradients}
\tablehead{
  \colhead{Source} & \colhead{Gradient Fit\tablenotemark{\dagger}} & \colhead{Gradient Fit} & \colhead{Gradient PA} & \colhead{Gradient PA} &\colhead{Single-Dish PA}  & \colhead{Region Fit}\\
                   &  \colhead{1D}                                 &  \colhead{2D}          & \colhead{2D}          &   \colhead{- Outflow PA} & \colhead{ - Int. PA} & \colhead{1D, 2D}\\
                   & \colhead{(\kmspc)}                            & \colhead{(\kmspc)}     & \colhead{(\degr)}     & \colhead{(\degr)} & \colhead{(\degr)}             & \colhead{(\arcsec)}\\

}
\startdata
IRAS 03282+3035         & 0.94$\pm$0.003                & 8.7$\pm$0.01  & 294$\pm$0.04 & 9 & 7    & 30$\ge$R$\ge$10, all \\
Perseus 5 - \nthp\      & 13.1$\pm$0.05                 & 8.6$\pm$0.1   & 147$\pm$0.8  & 66 & -    & 10$\ge$R, 10$\ge$R\\
Perseus 5 - NH$_2$D     & 12.4$\pm$0.06                  & 17.4$\pm$0.13   & 125$\pm$0.4  & 43 & - & 30$>$R, 30$>$R\\
HH211-mm                & 8.9$\pm$0.02                  & 6.6$\pm$0.03   & 26$\pm$0.3   & 89 & 5   & 20$>$R, 20$>$R\\
L1521F                  & 9.5$\pm$0.02                  & 5.7$\pm$0.01   & 41$\pm$0.2   & 41  & 161 & 30$>$R, all\\
L1527-CARMA             & 12.1$\pm$0.14                 & 13.5$\pm$0.2   & 36$\pm$ 0.7  & 54 & 78  & 15$\ge$R, 20$>$R\\
L1527-VLA               & 12.2$\pm$0.13                 & 15.2$\pm$1.7   & 212$\pm$4.6  & 58 & 98  & 15$\ge$R, 20$>$R\\
RNO43                   & 10.7$\pm$0.09                 & 15.9$\pm$0.04  & 262$\pm$0.14 & 62 & 18  & 20$\ge$R$\ge$7, 15$>$R\\
HH270 VLA1              & 2.4$\pm$0.06                  & 6.0$\pm$0.03   & 162$\pm$0.2  & 77 & 4   & 20$>$R, all\\
IRAS 16253-2429         & 4.1$\pm$0.02                  & 3.5$\pm$0.06   & 279$\pm$1.0  & 79 & 33  & 30$>$R, 20$>$R\\
L483                    & 11.1$\pm$0.16                 & 9.3$\pm$0.03   & 45$\pm$0.2   & 57 & 21  & 20$\ge$R$\ge$10\\
Serpens MMS3            & 14.1$\pm$0.14\tablenotemark{b}& -              & -            & -  & -   & +30$\ge$R$\ge$+5\\
HH108IRS                & 14.0$\pm$0.07                 & 15.1$\pm$0.05  & 140$\pm$0.1  & 68 & 44  & +30$>$R$>$-10, +25$>$R$>$-5 \\
HH108MMS                & 2.7$\pm$0.02                  & 2.3$\pm$0.7    & 90$\pm$0.1    & 40 & 22 & 15$\ge$R, all\\
L1152\tablenotemark{a}  & 6.5$\pm$0.010                 & 6.3$\pm$0.01   & 114$\pm$0.1  & 69 & 30  & 25$>$R, 25$>$R\\
L1157-PdBI              & 6.2$\pm$0.01                  & -   & -  & - & -& 20$>$R, ...\\
L1157-CARMA             & 4.8$\pm$0.02                  & 10.5$\pm$0.03  & 179$\pm$0.1          & 29 & 124 & 20$>$R, 25$>$R\\
L1157-VLA               & 3.5$\pm$0.01                  & 3.5$\pm$0.06   & 260$\pm$0.9    & 70 & 43 & 20$>$R, 25$>$R\\
CB230                   & 11.3$\pm$0.03                 & 3.2$\pm$0.1   & 278$\pm$1.3  & 82 & 4    & 15$>$R, 20$>$R\\
L1165                   & 11.1$\pm$0.03                  & 11.2$\pm$0.1   & 120$\pm$0.4  & 75& 34 & 20$>$R, 10$\ge$R\\
\enddata
\tablecomments{Same as Table 9, but for the interferometer data.} 

\tablenotetext{\dagger}{Gradients are measured normal to the outflow.}

\tablenotetext{a}{Gradients are measured from center of the core as the protostar is located
at the edge.}
\tablenotetext{b}{Gradient was measured along the the velocity gradient, shifted slightly from being
normal to the outflow.}

\end{deluxetable}

\begin{deluxetable}{llll}
\rotate
\tablewidth{0pt}
\tabletypesize{\scriptsize}
\tablecaption{3mm Continuum Fluxes}
\tablehead{
  \colhead{Source} & \colhead{RA} & \colhead{Dec} & \colhead{$S_{\nu}$} \\
             & \colhead{(J2000)} &  \colhead{(J2000)}  & \colhead{(mJy)} \\
}
\startdata
Perseus 5   & 03:29:51.876 & 31:39:05.7 & 4.0$\pm$1.0\\
HH211       & 03:43:56.78  & 32:00:49.8 & 43.4$\pm$2.0\\
L1527       & 04:39:53.86 & 26:03:09.5  & 32.6$\pm$6.0\\
HH270VLA1   & 05:51:34.64 & 02:56:45.8  & 6.3$\pm$1.0\\
RNO43       & 05:32:19.39 & 12:49:40.8  & 9.4$\pm$3.5\\
IRAS 16253-2429 & 16:28:21.42 & -24:36:22.1 & 1.8$\pm$1.0\\
HH108MMS    & 18:35:46.46 & -00:32:51.2 & 10.3$\pm$2.3\\
HH108IRS    & 18:35:42.14 & -00:33:18.5 & 35.7$\pm$4.9\\
L1165       & 22:06:50.46 & 59:02:45.9  & 12.1$\pm$1.7\\
L1152       & 20:35:46.22 & 67:53:01.9  & 15.2$\pm$2.8\\
L1157-PdBI  & 20:39:06.25 & 68:02:15.9  & 37.5$\pm$6.1\\
L1157-CARMA & 20:39:06.22 & 68:02:16.05  & 55.1$\pm$8.3\\
\enddata
\tablecomments{The 3mm continuum fluxes for sources observed with CARMA and the PdBI. The flux is measured
from the the naturally weighted continuum image in a box centered on the source with dimensions four times the
size of the FWHM listed in Tables 3 and 5. The errors on the measured fluxes are statistical only and do not
consider the 10\% absolute calibration error. The difference in continuum flux between the CARMA and PdBI observations 
can be attributed to the CARMA observations including shorter uv-spacing data from CARMA E and D-arrays as compared
to D-array at the PdBI.} 

\end{deluxetable}

\end{small}
\end{document}